\documentclass{ws-ijmpd}
\usepackage{graphicx} 

\usepackage[super,compress]{cite}
\usepackage{subfig}
\usepackage{physics}
\usepackage{subfloat}
\usepackage[normalem]{ulem}
\usepackage{tikz}
\usepackage{slashed}
\usepackage{tensor}
\pdfoutput=1

\usepackage[utf8]{inputenc}
\usepackage{hyperref}
\usepackage{color}
\usepackage{overpic}
\usepackage{xcolor}

\definecolor{pink1}{rgb}{0.858, 0.188, 0.478}

\let\normalcolor\relax

\def\be{\begin{equation}}
\def\ee{\end{equation}}
\def\ba{\begin{eqnarray}}
\def\ea{\end{eqnarray}}

\def\nn{\nonumber}

\begin{document}

\markboth{Robert B. Mann}
{Black Hole Chemistry: the first 15 years}

%
\catchline{}{}{}{}{}
%

\title{Black Hole Chemistry: the first 15 years}

\author{Robert B. Mann}

\address{Department of Physics and Astronomy, University of Waterloo, \\
Waterloo, Ontario N2L 3G1
Canada \\
rbmann@uwaterloo.ca}

\maketitle

\begin{history}
\received{Day Month Year}
\revised{Day Month Year}
\end{history}

\begin{abstract} 
The introduction of thermodynamics into gravitational physics began 5 decades ago with the discovery that black holes behave like
thermodynamic systems once semiclassical quantum effects are taken into account. Notions of temperature, entropy, work, and phase changes that
were introduced into gravitational physics and originally applied to black holes, were later extended to cosmological horizons and other settings as well.
A major development occurred 15 years ago with the introduction of pressure in the form of a cosmological constant. By extending the thermodynamic
phase space to include this term, along with its conjugate volume, black holes were found to exhibit a broad variety of phase transitions that
resembled phenomena seen in chemistry labs.  Black hole thermodynamics has become Black Hole Chemistry, which has led to a wealth of insights into the nature of black holes, introducing  concepts such as Van der Waals fluids, reentrant phase transitions, and triple points into gravitational physics.  I discuss the origins of Black Hole Chemistry and its basic features covered in an earlier review \cite{Kubiznak:2016qmn}, and then go on to describe  developments in the subject that have taken place since then.  Examples include multicritical behaviour, polymeric transitions, superfluid transitions, scalar hair,  heat engines, NUT-charge, acceleration thermodynamics, the Joule-Thompson expansion, holography, complexity, central charge criticality, microstructure, thermodynamic tension, phase dynamics, and thermodynamic topology.  This wealth of new phenomena suggest that we likely still have  a lot to learn from Black Hole Chemistry.
\end{abstract}


\ccode{PACS numbers:}


\section{Prologue}

The first introduction of the notion of a black hole came  in 1783 with the consideration of a dark star \cite{Michell:1783}: an object so dense that its escape velocity is larger than the speed of light.  Such stars would be detectable by distant observers only from their gravitational influence on nearby luminous objects; they could not be observed directly via emission of any phenomena, and so would be invisible. 

However the gravitational physics of these objects is in conflict with the laws of thermodynamics.  Their inability to emit anything due to their extreme gravitational attraction indicates they must be at the absolute zero of temperature. If a hot body ventured too close, it would be absorbed by the dark star, which would have to remain at zero temperature. This would mean that a hot body in contact with a cold one would not `warm it up', in violation of  the second law of thermodynamics: the initial entropy of the system (hot body and cold body) would be larger than the entropy of the final, more massive, cold body.

It would take nearly 200 years before the paradoxical nature of this gravito-thermodynamic conflict would be appreciated. This in part was because
there were no theoretical constraints known at the time for preventing objects from  moving faster than the speed of light. The advent of relativity over 100 years ago indicated this constraint was a law of nature, and its incorporation into gravitational physics eventually returned us to the concept of a dark star, now known as a black hole.  The paradox of the second law emerged fifty years ago, and was resolved by Jakob Bekenstein  \cite{Bekenstein:1973ur}, who realized that the size the horizon of a black hole must in some way correspond to its thermodynamic entropy.  The notion of a temperature for a black hole  came shortly afterward due to the work of Stephen Hawking \cite{Hawking:1974rv}, and the subdiscipline of black hole thermodynamics was born.

Conspicuous by its absence was any notion of thermodynamic pressure.  Ignored for quite some time, this omission made black hole thermodynamics very much unlike any thermodynamics with which we have familiarity \cite{Dolan:2010ha}, such as takes place in chemistry labs worldwide.  Four decades after the birth of black hole thermodynamics, a notion of pressure was found to occur in the form of vacuum energy manifest via a cosmological constant \cite{Kastor:2009wy}. Subsequent further study indicated that black holes, depending on their properties, behave like  thermodynamic systems studied by chemists. Over the following 15 years, black hole thermodynamics became Black Hole Chemistry. 

This review describes the origins, birth, and phenomena of Black Hole Chemistry.  A review of this subject appeared over 8 years ago \cite{Kubiznak:2016qmn}, but there have been many new developments since then.  Phenomena such as  polymeric transitions, superfluidity, 
scalar hair, multicriticality, heat engines, Joule-Thompson expansions, NUT-charge, acceleration thermodynamics, holographic complexity, central charge criticality, microstructure, phase dynamics,  thermodynamic topology and more have been the subject of many investigations since.

This review is primarily focussed on these new developments.  While enough contextual and background material is provided so that newcomers to the subject can orient themselves to this discipline, the primary focus is on how black hole chemistry has developed and matured since the original review \cite{Kubiznak:2016qmn}.  It is my hope that this review will spur further developments in this rich and exciting subject.

\section{Introduction}

The four laws of thermodynamics \cite{1985Guggenheim} are perhaps the strongest of the established pillars of the physical sciences.  Succinctly stated,
they are\\

\begin{enumerate}
\item{\bf 0th Law}  A system is thermal equilibrium with itself if the temperature within the system is spatially uniform and temporally constant.\\

\item{\bf 1st Law} When energy enters or leaves a system (as work, heat, or matter), the internal energy $U$ of the system changes in accordance with the law of conservation of energy: $\Delta U = \Delta Q - \Delta W$, where $\Delta Q$ is the heat supplied to the system and $\Delta W$ is the work done by the system.\\

\item{\bf 2nd Law}  In any thermodynamic process, the sum of the entropies of interacting thermodynamic systems never decreases.\\ 

\item{\bf 3rd Law}  It is not possible for any process, no matter how idealized, to reduce the entropy of a system to its 
value at the absolute-zero of temperature in a finite number of operations.\\
\end{enumerate}

These laws have alternate versions of equivalent meaning.  The 0th law provides an independent definition of temperature, and 
is tantamount to stating that, as Maxwell put it, `all heat is of the same kind' \cite{Maxwell_2011}.  
It is also equivalent to the statement that if two thermodynamic systems are both in
thermal equilibrium with a third system, then they are in thermal equilibrium with each
other \cite{fowler1965statistical}.   The first law is a statement of the conservation of energy stemming from the realization that heat has motive power \cite{carnot1890}.  The second law expresses the assertion that  heat always flows spontaneously from hotter regions of a system to colder ones. The third law was originally expressed by Nernst as a heat theorem \cite{Nernst1926}, 
stating that the entropy change  for any chemical or physical transformation approaches zero 
as absolute zero is approached. It later was expressed as the unattainability principle, understood as 
the statement that 
cooling a system to absolute zero  requires either an infinite number of steps or an infinite amount of time.

Perpetual motion machines are prohibited by the  first and second laws.  They imply that it is not possible to build any device
that produces work with no energy input (the first kind of perpetual motion machine), nor any device that  spontaneously converts thermal energy into mechanical work (the second kind of perpetual motion machine).

Black holes originally seemed to be very far removed from thermodynamics.  
The physical interpretation of the original Schwarzschild solution \cite{Schwarzschild:1916} was poorly understood for decades.  Some clarity was obtained when Oppenheimer and Snyder \cite{Oppenheimer:1939ue} showed that the equations of general relativity predict that a collapsing ball of dust (a form of stress-energy with density but no pressure),  matched to Schwarzschild's spherically symmetric solution, will   shrink to a size smaller than   $r=2GM/c^2$ in a finite amount of time measured by an observer at the edge of the dust.   
 The notion of such a gravitationally completely collapsed object was replaced with the term `black hole' by Wheeler in 1967 at a talk
 he gave at the NASA Goddard Institute of Space Studies \cite{WheelerFord1998}, and the term has been used ever since. 
 
The connection with thermodynamics took several more years to emerge. Physically, a black hole is  an object (or more generally, a spatial region) whose escape velocity is greater than the speed of light. Insofar as experiment indicates that  nothing can travel faster than light, any object crossing the boundary of a black hole (its horizon) cannot return to the region outside.  
Essentially, the gravity of a black hole is so strong that escape is impossible.   It is here that a connection with thermodynamics presents itself.  Since any object at finite temperature radiates heat (in the form of either  photons or particles or some other quanta), a black hole is presumably an object at an absolute zero of temperature since its no-escape property means that it cannot emit anything.   
 
However, further reflection on this situation indicates that black holes violate the second law. Any hot object crossing the horizon will not be able to escape the black hole, leaving it at absolute zero temperature.  But this means the hot body has not warmed up the cold body, in violation of the second law: the initial entropy of the system (hot body plus black hole) is greater than the final entropy of the system (the remaining black hole). It seems that the very notion of a black hole is in contradiction with the laws of thermodynamics.

Consider, for example, a cup of hot tea   poured into a black 
hole\footnote{This example is taken from a conversation between John Wheeler  and Jakob Bekenstein, who was Wheeler's PhD student at the time
\cite{Oppenheim:2015}.}.  
Whereas the hot tea has entropy (since its many possible microstates yield the same macroscopic description)
the black hole does not, since  it is described only by macroscopic quantities such as mass, electric charge, and angular momentum.
The hot tea should warm up the black hole according to the second law, but since the final state is still a black hole, it does not warm and so the total entropy of the system evidently decreases.   

This contradiction was resolved by Jakob Bekenstein  \cite{Bekenstein:1973ur}, who realized that in any such process the mass of the black hole must increase due to energy conservation.  
  After absorbing the tea, the black hole has a larger mass.  Consequently it has a larger size:   its horizon must get bigger, since
  the area of a black hole horizon is an increasing function of mass; for example
the area of the horizon is proportional to the square of the mass for  a Schwarzschild black hole.  
Relying on a  theorem of Hawking \cite{Hawking:1971tu}, that  the area of a black hole can only increase,  
Bekenstein reasoned that a black hole must have  thermodynamic entropy that is an increasing function of its area $A$, and further 
proposed that this entropy  was proportional to $A$. This is because an application of the second law to 
a pair of merging black holes indicates that the entropy must scale at least  linearly with area;   changing the area by some minimal amount implies that the scaling is not more than linear.

This proportionality of entropy to area suggested that black holes in fact did obey  the second law of thermodynamics, since the area can never decrease.  The entropy of the black hole, after absorbing the tea, increases enough to offset the loss of the entropy of the tea after it has been poured into the black  hole.

But how can an object  remain at zero temperature whilst its entropy increases?  It would seem that black holes should have a temperature after all, and thus be able to radiate particles.  This indeed was shown to be the case by  Stephen Hawking \cite{Hawking:1974rv}, who originally set out to refute Bekenstein's ideas, but instead discovered that quantum effects
imply that a black hole   can indeed radiate particles.  In natural units with all physical constants set equal to unity, the temperature of the black hole is  equal to the surface gravity $\kappa_H$ at its event horizon divided by $2\pi$; restoring units $k_{\rm B} T = \frac{\hbar c \kappa_H}{2\pi }$, where $k_{\rm B}$ is Boltzmann's constant, $\hbar$ is Planck's constant  divided by $2\pi$, and $c$ is the speed of light. 
 This in turn indicated that the entropy $S$ of a black hole is equal to one-quarter of the area of its horizon; restoring units
 $S= \frac{k_{\rm B} c^3}{4\hbar G_N}A$.  The presence of $\hbar$ in these formulae  indicates that quantum physics -- semiclassically -- has now entered the realm of gravitational physics.
 
These considerations led to the formulation of the four laws of
black hole mechanics \cite{Bardeen:1973gs}\\
\begin{enumerate}
\item{\bf 0th law}  The surface gravity $\kappa$ of a stationary black is constant on the black hole horizon.   \\
\item{\bf 1st law}  A change $dM$ in the mass of a black hole with charge $Q$ and angular momentum $J$ is
\begin{equation}\label{bh1law}
dM = \frac{\kappa}{8\pi} dA + \Omega_H dJ + \Phi_H dQ +\cdots
\end{equation}
in natural units, where $\Omega_H$ is the angular velocity of the   horizon of the black hole and
 $\Phi_H$ is the electromagnetic potential at the horizon; the dots refer to possible additional work terms.\\
 
\item{\bf 2nd law} The area $A$ of the event horizon of a black hole never decreases in any physical process. \\

\item{\bf 3rd law}  The surface gravity of a black hole cannot be reduced to zero in a finite number of steps via any physical procedure \cite{Israel:1986gqz}. \\
\end{enumerate}

The parallel with the laws of thermodynamics is striking.  Stationary black holes, having constant surface gravity, are analogous to equilibrium states in thermodynamics, with the temperature $T$ of an equilibrium state being constant.  The mass $M$ of  the black hole corresponds to the thermodynamic energy of the system, and differences in mass between nearby solutions are equal to differences in horizon area times horizon surface gravity plus additional work terms, analogous to the respective heat and work terms in 
the first law of thermodynamics.  The second law, which is a statement of the area theorem $dA \geq 0$ in general relativity\cite{Hawking:1971tu}, is
fully analogous to the increase in entropy as per the second law of thermodynamics, as noted by Bekenstein\cite{Bekenstein:1973ur}.
The third law for black hole mechanics came somewhat later: it means that if some procedure for decreasing the surface gravity of a black hole can be found,  then the 3rd law states that  for $n$ repetitions of this process 
 $\kappa_n > \kappa_{n+1} > 0$ if $n$ is finite.  It is likewise analogous to the third law of thermodynamics
 and recently has played an important role\cite{Sorce:2017dst}  in terms of violating weak cosmic censorship \cite{Penrose:1969pc,Wald:1997wa},  which is the conjecture that all singularities arising from gravitational collapse must be hidden within black holes. 
 
 Another noteworthy feature that emerged from the laws of black hole mechanics was the {\it Gibbs--Duhem relation}
 \begin{equation}\label{oldSmarr}
M=2 (TS+ \Omega J) + \Phi Q\,,
\end{equation}
generally referred to as the Smarr relation, since it was Smarr who first pointed this out \cite{Smarr:1972kt}. 
 It  expresses a relationship between the intensive $(T,\Omega,\Phi)$ and extensive $(M,J,Q)$ thermodynamic variables.  
  
 For these reasons the laws of black hole mechanics became firmly established as the laws of black hole thermodynamics, and  the subject of black hole thermodynamics was born. 
 A whole new set of techniques were subsequently developed  for analyzing the behaviour of black holes that in turn yielded  deep insights concerning the relationship between gravity and quantum physics.  Black hole entropy was shown to be the Noether charge associated with diffeomorphism symmetry \cite{Wald:1993nt}. 
 The laws of gravitation were proposed to be deeply connected with the laws of thermodynamics \cite{Jacobson:1995ab, Padmanabhan:2009vy}.   Deep connections between the geometric structure of spacetime  \cite{Bianchi:2012ev,Gruber:2016mqb} and  the quantum information concept of entanglement entropy were discovered \cite{Ryu:2006bv}, and  the linearized Einstein equations were later shown to follow from the first law of entanglement entropy  \cite{Faulkner:2013ica}.  
 
 Perhaps the strangest finding to emerge is that the resolution
 of one paradox (violation of the second law) gave rise to another, more perplexing one: the loss of information.  The same quantum physics that allows a black hole to have a temperature violates its own self-consistency insofar as the evaporation of a black hole
apparently converts pure states into mixed states, violating unitary evolution, a cornerstone of  quantum physics  \cite{Giddings:1995gd,Mathur:2009hf}.  This {\it information paradox} has yet to be resolved   \cite{Mathur:2005zp,Almheiri:2012rt, Hawking:2016msc}, though recent work employing new geometric methods \cite{Almheiri:2020cfm}
suggests a path forward.

Notwithstanding these considerations,  50 years on, black hole thermodynamics remains our strongest connection linking 
 gravitational physics with quantum physics, giving us important clues as to how to unite these two conceptually different 
 paradigms  \cite{Wald:1999vt, Traschen:1999zr, Grumiller:2014qma, Carlip:2014pma}.
.

\section{The Birth of Black Hole Chemistry}

Chemistry is about the interplay of matter.  It is about the structure, properties, and behaviour of different substances, and particularly about how a substance  changes in a reaction.  Chemical thermodynamics \cite{1985Guggenheim} in particular is concerned with exploring the relationship between work, heat, energy, and the behaviour of  molecules and atoms.  It  employs concepts like enthalpy and free energy, employing
them to understand chemical reactions, equilibrium, and substances in different states or phases (solid, liquid, gas), and in changes from one phase to another. 

Black hole thermodynamics originally seemed somewhat removed from chemical thermodynamics, being concerned primarily with
temperature and entropy, and how these were related to the basic properties of a black hole.  Furthermore, there is no pressure-volume term in the first law \eqref{bh1law}, a term  conspicuous in chemistry.  Indeed, it is quite striking that  the first observation \cite{Dolan:2010ha} that the black hole mechanics/thermodynamics correspondence  is completed with this term came nearly four decades after  the original arguments \cite{Bardeen:1973gs}.
  
Not quite 10 years after the advent of the laws of black hole mechanics, a discovery was made indicating that  black holes could exhibit phase behaviour \cite{Hawking:1982dh}.  The presence of a negative cosmological constant $\Lambda$ was necessary for this to occur.  A static black hole of mass $M$ in such a setting -- known as a Schwarzschild Anti de Sitter (AdS) black hole -- had a temperature that depended on both
$M$ and $\Lambda$.  For any given temperature above a certain threshold a black hole could exist in one of two states: large or small.
Large black holes had greater surface area than small ones, and so were the thermodynamically preferred state as they had greater entropy.  This was the first notion that black holes could have two different states, or phases.  More surprisingly, a computation of
the Gibbs free energy of the black hole indicated that at a temperature below a certain value (but above the threshold), the thermodynamically favoured state was not a black hole but instead was  thermal AdS. In other words, the thermodynamically stable state  at low temperatures is an AdS spacetime filled with radiation.  But at a sufficiently high temperature, it is thermodynamically favourable for this radiation to gravitationally collapse into a black hole.  
This was the first hint that  a black hole could change phases analogous to the way chemical systems can change phases.

Despite this intriguing finding, much of the literature on black hole thermodynamics was restricted to asymptotically flat spacetimes.
This in part was motivated by a procedure for computing the mass of a black hole (and other conserved quantities) in such settings.
However apart from flatness not always being an appropriate  idealization (and in reality not ever satisfied), problems nonetheless emerged. For example, if  the temperature is fixed at infinity,  a Schwarzschild black hole has negative  heat capacity \cite{DeWitt1979} and the formal expression for the partition function is not logically consistent \cite{DeSabbata:1992cb}.  Consequently
there was motivation to  develop a theoretical framework for  black hole
thermodynamics that did not rely on the assumption of asymptotic
flatness  \cite{York:1986it,Whiting:1988qr,Brown:1992bq,Brown:1992br}.

This quasilocal approach to black hole
thermodynamics allowed one to study and compute quantities associated with gravitational and matter fields 
within a finite, bounded spatial region, so the
asymptotic behaviour of the gravitational field became
irrelevant.   Black hole spacetimes that were  asymptotically curved or black holes in spatially
closed universes could be investigated for their thermodynamic properties.  In the asymptotically flat case,
with temperature fixed at the finite spatial boundary, 
the black hole partition function was now consistent \cite{York:1986it}.  

It was in the quasilocal context that the first studies of the  thermodynamic properties of asymptotically AdS black holes
were properly carried out \cite{Brown:1994gs}. Earlier investigations were not completely correct, since the thermodynamic internal energy was identified with the conserved mass at infinity \cite{Hawking:1982dh,BanadosEtal:1992} and the temperature identified
with $\kappa_H/2\pi$. Both quantities, however, depend on the normalization of a timelike Killing vector field, and in the absence of an asymptotically flat region there is no physically preferred choice.

Shortly afterward it became clear that this formalism could be employed to incorporate the cosmological constant into the first law as a thermodynamic variable \cite{Creighton:1995au},  though the implications of this possibility were not considered. This notion was reintroduced a few years later for the specific case of Kerr-Newman-AdS black holes \cite{Caldarelli:1999xj}, where a range of possible black hole phases was delineated and a generalization of the Smarr formula introduced. However the notion 
of the cosmological constant as thermodynamic pressure and its proper association with a conjugate black hole volume was achieved 
nearly a decade later when the laws of black hole mechanics \cite{Bardeen:1973gs}
 were generalized to asymptotically AdS spacetimes \cite{Kastor:2009wy} in $D$ spacetime dimensions. 

The 
 first law of black hole thermodynamics \eqref{bh1law} generalizes to 
 \be\label{firstBH}
\delta M= T\delta S +\sum_i^{N} \Omega_i\delta J_i\, {+  V \delta P }
+\sum_j \Phi_j \delta Q_j\,,
\ee
whose derivation is reproduced in  \ref{appA}, where  
 \be\label{eq:press}
P = -\frac{\Lambda}{8\pi} = \frac{(D-1)(D-2)}{16\pi \ell^2}
\ee
is interpreted as thermodynamic pressure, and
\be\label{Vdef}
V\equiv \left(\frac{\partial M}{\partial P}\right)_{S,Q,J}
\ee
is its conjugate, interpreted as the thermodynamic  volume of the black hole. The quantities $M$  and $J_i$ are the conserved charges  respectively associated with the time-translation and rotational Killing vectors of the spacetime, properly defined in the AdS setting
\cite{Kastor:2009wy}.  Note that in dimensions $D>4$ multiple planes of rotation can exist, and the black hole can
have $N=\lfloor\frac{D-1}{2}\rfloor$ angular momenta \cite{Myers:1986un}, with conjugate horizon angular velocities $\Omega_i$.
The quantities $Q_j$, with respective conjugate potentials $\Phi_j$, are conserved charges  associated with any Abelian gauge fields (Maxwell fields) that couple to gravity. As in the asymptotically flat case, the entropy $S=A/4$, where $A$ is the area of the black hole event horizon, and the temperature   $T=\kappa/2\pi$ with $\kappa$ its surface gravity.

The Smarr-Gibbs-Duhem relation likewise is completed with a pressure volume term:
\be
\frac{D-3}{D-2}M=TS+\sum_i \Omega_iJ_i  -\frac{2}{D-2}PV
+\frac{D-3}{D-2} \sum_j \Phi_j  Q_j\,,
\label{smarrBH}
\ee
in $D$ dimensions \footnote{It is possible to sensibly interpret these formulae in both $D=3$ and $D=2$ dimensions   \cite{Frassino:2015oca}.}. 

 The $PV$ term here is crucial;   AdS black holes do not satisfy the  relation  \eqref{oldSmarr}.   
 The  relation \eqref{smarrBH} follows from the homogeneity of the mass $M=M(A,\Lambda,Q_j,J_i)$ as a function
of the other extensive variables.  For homogeneous functions 
$$
f(x,y,\dots,z) \rightarrow f(\alpha^p x, \alpha^q y,\dots,\alpha^rz)   = \alpha^{s} f(x,y,\dots,z)
$$
and the derivative with respect to $\alpha$  implies
\be\label{Euler}
sf(x,y,\dots,z) = p \left(\frac{\partial f}{\partial x}\right) x + q \left(\frac{\partial f}{\partial y}\right) y+\dots+
r \left(\frac{\partial f}{\partial z}\right) z\,,
\ee
which is Euler's formula. Replacing $f$ with $M$ yields \cite{Caldarelli:1999xj,Kastor:2009wy} 
\be\label{Euler3}
(D-3)M=(D-2)\frac{\partial M}{\partial A}A+(D-2)\sum_i \frac{\partial M}{\partial J_i}J_i+(-2)
\frac{\partial M}{\partial \Lambda}\Lambda+(D-3)\sum_j\frac{\partial M}{\partial Q_j}Q_j 
\ee
since the respective scaling dimensions of $(M,Q^j)$ are  $D-3$, $(A,J^i)$ are $D-2$,  and $\Lambda$ is $-2$. From the first law \eqref{firstBH} we have
$$
\frac{\partial M}{\partial A}=\frac{\partial M}{\partial S} = T \qquad 
\frac{\partial M}{\partial J_i}= \Omega_i  \qquad \frac{\partial M}{\partial Q_j}= \Phi_j 
\qquad \frac{\partial M}{\partial \Lambda}= 8\pi V
$$
upon identifying  $S$ with $A/4$ and $P$ with $-\Lambda/(8\pi)$; insertion of the above into \eqref{Euler3}
yields  \eqref{smarrBH}. We see that the inclusion of the $PV$ term is {\it required} for  \eqref{smarrBH}  to hold.
The thermodynamic volume $V$ can be interpreted as the change in $M$ under variations in $\Lambda$, with the black hole entropy, angular momenta, and charges held fixed.

Further investigation has indicated that the Smarr relation has very broad applicability \cite{Kubiznak:2016qmn}.
Essentially any dimensionful constant can be regarded as a thermodynamic variable, with an appropriate conjugate variable
obtained from the homogeneity relations above.  Examples include asymptotically de Sitter spacetimes \cite{Dolan:2013ft}
(with $\Lambda > 0$), Born-Infeld electrodynamics \cite{Gunasekaran:2012dq}, Lovelock gravity \cite{Kastor:2010gq}, 
asymptotically Lifshitz spacetimes \cite{Brenna:2015pqa}, and more exotic black objects \cite{Caldarelli:2008pz, Altamirano:2014tva,Hennigar:2014cfa}.  Sometimes one encounters alterative Smarr relations \cite{Bertoldi:2009dt,Bertoldi:2009vn, Dehghani:2010kd,Liu:2014dva, Berglund:2011cp,Dehghani:2011hf,Dehghani:2013mba, Way:2012gr}
 (none incorporating a notion of volume), but these have all been shown \cite{Brenna:2015pqa} to be special cases of \eqref{smarrBH}.
 
Black hole chemistry  completes the correspondence between gravitational thermodynamics and chemical thermodynamics,
as illustrated in the following table:
\begin{align}  
&\textrm{Table 1: Comparison of Thermodynamics with Black Hole Mechanics} \nonumber \\
&\begin{array}{|l|c|l|c|}
\hline
\multicolumn{2}{|c|}{\mbox{Thermodynamics}} & \multicolumn{2}{|c|}{\mbox{Black Hole Mechanics}} \\
\hline
\mbox{Enthalpy} &  H=E+PV & \mbox{Mass} & M\\
\hline
\mbox{Temperature} & T & \mbox{Surface Gravity} & \frac{  \kappa}{2\pi}\\
\hline
\mbox{Entropy}  &S &\mbox{Horizon Area} & \frac{A}{4} \\
\hline
\mbox{Pressure}  &P & \mbox{Cosmological Constant}  &  -\frac{\Lambda}{8\pi} \\
\hline
\mbox{First Law}  &\delta H= T \delta S +V \delta P + \ldots  & \textrm{First Law} & \delta M= \frac{\kappa}{8\pi} \delta A  +V \delta P + \ldots\\
\hline
\end{array}
\nonumber
\end{align}
where the dots represent the work terms 
$\sum_i \Omega^i \delta J_i + \sum_j \Phi^j \delta Q^j\,$  for multiply charged and spinning black hole solutions.

Some care must be taken in computing the conjugate potentials for charge and angular momentum in asymptotically AdS spacetimes \cite{Gibbons:2004ai, CveticEtal:2010}. 
For the electric (and magnetic) $U(1)$ charges, $\Phi_j=\Phi_{j+}-\Phi_{j \infty}$, allowing for both non-trivial (gauge independent) potentials on the horizon $\Phi_{j +}$ and at infinity $\Phi_{j \infty}$. Likewise $\Omega_i=\Omega_{i +}-\Omega_{i \infty}$, where the latter term   allows for the possibility of a rotating frame at infinity \cite{Gibbons:2004ai}. 
 
\section{Black Hole Chemistry Basics}
\label{4s}

Black hole chemistry   provides a new perspective on black hole thermodynamics.  We will see that it has remarkable consequences whose net effect indicates that black holes are objects having properties and behaviour akin to phenomena observed in a chemistry lab.

The basic quantity of interest in black hole chemistry is the free energy.  There are a variety of expressions for the free energy, depending on the thermodynamic ensemble.  The one most commonly employed is the {\it Gibbs free energy}
\be\label{GibbsFE}
G=M-TS=G(P,T,J_1,\dots,J_N,Q_1,\dots, Q_n) 
\ee
which corresponds to an ensemble where all extensive parameters are fixed.  This is easily seen by employing
the first law \eqref{firstBH}
\be
dG = dM-TdS - SdT =  -SdT +\sum_i^{N} \Omega_i dJ_i\, +  V dP 
+\sum_j \Phi_j dQ_j 
\ee
which indicates that if all extensive variables $(P, J_i, Q_j)$ are kept fixed, the Gibbs free energy $dG=-SdT$  and so can be regarded as a function of the temperature $T$.  The equilibrium state corresponds to the global minimum of $G$.

Another useful quantity is the specific heat at constant pressure
\be
C_P\equiv C_{P,J_1,\dots, J_N,Q_1,\dots,Q_n}=T\Bigl(\frac{\partial S}{\partial T}\Bigr)_{P,J_1,\dots,J_N,Q_1,\dots, Q_n}
\ee
whose sign is indicative of the stability of a state.  Negative specific heat means that as temperature increases, the entropy of
the state decreases, signifying instability.  Conversely, positive specific heat indicates stability.

These conceptual tools provide a means of  constructing  {\it phase diagrams}, obtaining {\it critical points} and {\it critical exponents}, and for determining whatever other interesting transitional behaviour might arise.

 \subsection{The chemistry of Hawking-Page transitions}
 \label{4p1}
 
The perspective offered by black hole chemistry is readily illustrated by considering
Schwarzschild-AdS black holes immersed in a bath of radiation, which led to the first discovery of a phase transition for
black holes \cite{Hawking:1982dh}.

The metric for a Schwarzschild-AdS (SAdS) black hole is
\be\label{Schw-Ads}
ds^2  = - f dt^2+\frac{dr^2}{f}+r^2  d\Omega_{D-2,k}^2\,,
\ee
where
\be\label{dOmegak}
d\Omega_{D-2,k}^2 = d\theta^2 +\frac{\sin^2(\sqrt{k}\theta)}{k}d\Omega^2_{D-3,1}
\ee 
is the metric on a compact $(D-2)$-dimensional space $\Sigma_k$ of constant curvature of
sign  $k$, with $k=0$  corresponding to a torus,  $k=-1$  to 
a compact hyperbolic space, and   $k=1$ to a $(D-2)$-sphere \cite{Aminneborg:1996iz,Mann:1996gj}.
For this latter case  
\be \label{volsphere}
\Omega_{D,k=1} \equiv \int d\Omega_{D,1} = \frac{2\sqrt{\pi^{D+1}}}{\Gamma\left(\frac{D+1}{2}\right)}
\ee
is the volume of this transverse space, yielding the well known $\Omega_{2,k=1} = 4\pi$.

The metric function 
\be\label{FSchw}
f=k-\left(\frac{r_0}{r}\right)^{D-3}+\frac{r^2}{\ell^2}\, 
\ee
in $D$ dimensions, where $r_0$ is a constant corresponding to the mass.

Setting $D=4$ and $r_0=2M$,  we can express the relevant thermodynamic quantities as
functions of $(r_+,\ell)$, where $f(r_+)=0$ determines the location of the horizon.  This gives
\be\label{SAdSTherm}
M = \frac{r_+ A_k}{8}\Bigl(k + \frac{r^2_+}{\ell^2}\Bigr)\,, \quad S=\frac{\pi A_k}{4} r^2_+\,,\quad
T = \frac{f'(r_+)}{4\pi} = \frac{k \ell^2 + 3r^2_+}{4\pi \ell^2 r_+}\,,\quad V =   \frac{\pi A_k}{3} r^3_+\,,
\ee
making use of the formulas above.  Here $\pi A_k$ is the area of the constant-curvature space: $A_{k=1} = 4$ (sphere) 
 $A_{k=0} = A B$ (where $A$ and $B$ and the sides of the torus). There is no simple formula for $A_{k=-1}$.

\begin{figure}
\centering
\includegraphics[width=0.49\textwidth,height=0.3\textheight]{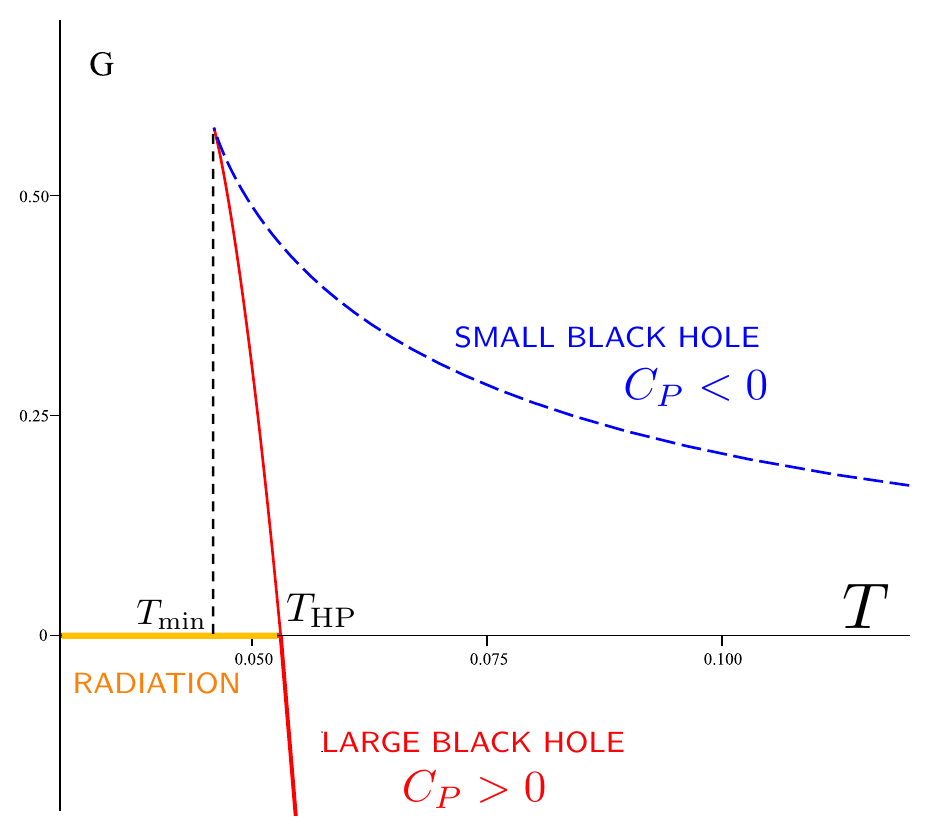} 
\includegraphics[width=0.49\textwidth,height=0.3\textheight]{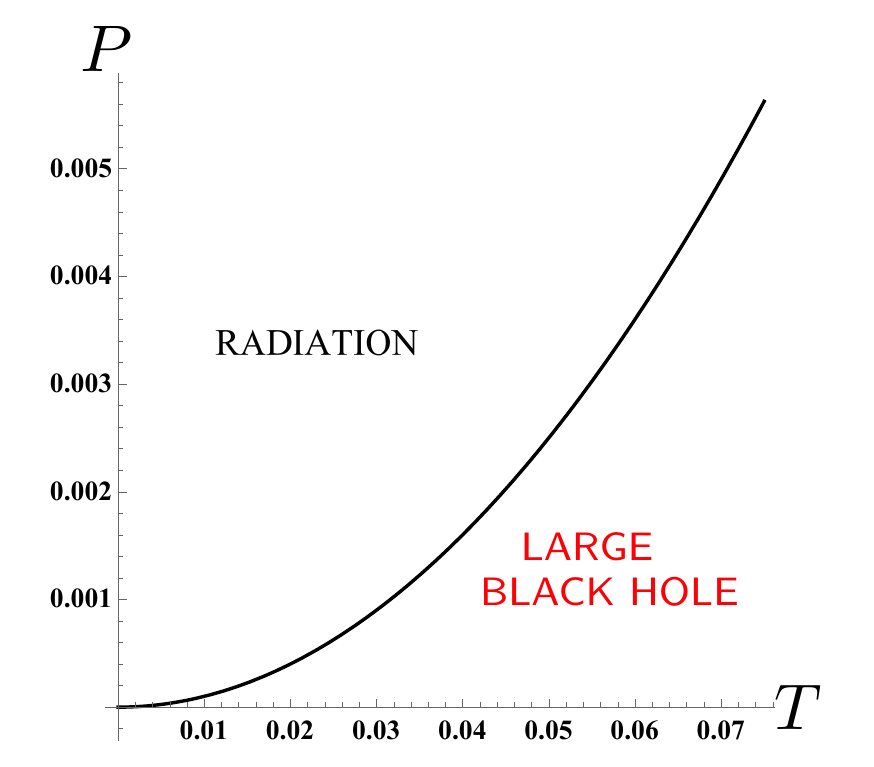}
\caption{{\bf Hawking--Page transition as a liquid/solid transtion.} {\it Left}. The Gibbs free energy 
as a function of temperature for fixed pressure $P=1/(96\pi)$ is shown for a Schwarzschild-AdS black hole.  
The upper branch (blue, dashed) corresponds to  small black holes; these have  negative specific heat and so are  thermodynamically unstable. The lower branch (red, solid) of large black holes has positive specific heat and 
for $T>T_{\mbox{\tiny  HP}}$ is the global minimum of $G$, thus 
 corresponding  to the  thermodynamically preferred state. The free energy of the radiation (orange, solid)  is normalized to zero; we see for 
 $T<T_{\mbox{\tiny  HP}}$ it is the  thermodynamically preferred state.  There is a  discontinuity in the first derivative of the radiation/black hole Gibbs free energy at $T_{\mbox{\tiny  HP}}$, characteristic of a first order phase transition. {\it Right.} The phase diagram
 shows that at any given pressure $P$ there is a high-entropy phase  at large $T$ (corresponding to a large black hole) and a low-entropy
 phase at small $T$ (corresponding to radiation). The coexistence line is of infinite length and is reminiscent of a solid/liquid phase diagram.
}
\label{Fig:HP}
\end{figure}

The Gibbs free energy $G=M-TS$ for the SAdS black hole with $k=1$ is shown in the left panel of Fig.~\ref{Fig:HP}. 
At any given temperature above $T=T_{\textrm{min}}$, small black holes ($r_+<\ell/\sqrt{3}$) are 
thermodynamically unstable black holes with negative specific heat, shown on the upper branch.  The 
 lower branch corresponds to  large black holes ($r_+ > \ell/\sqrt{3}$) with positive specific heat.  The two branches of black holes   meet at a cusp at $T=T_{\textrm{min}}$; no black hole solutions exist at smaller temperatures.
 For  $T > T_{\mbox{\tiny  HP}}$ (or  $r_+ > r_{\mbox{\tiny  HP}}=\ell$) the large black holes 
 have negative Gibbs free energy, which is lower than that of an AdS space filled with hot radiation, and are the 
 thermodynamically  preferred state.  There is a first order phase transition between thermal radiation and large black holes
 at  $T=T_{\mbox{\tiny  HP}}=1/(\pi \ell)$, known as a {\it Hawking--Page} \cite{Hawking:1982dh} transition. 
In the context of the AdS/CFT correspondence conjecture \cite{Maldacena:1997re}, this phase transition can be interpreted as a confinement/deconfinement phase transition in the dual quark 
gluon plasma \cite{Witten:1998zw}.

From the perspective of black hole chemistry, the thermal radiation/large black hole behaves as a solid/liquid phase transition \cite{Kubiznak:2014zwa}.
It is perhaps a bit counterintuitive to think of radiation as corresponding to the solid phase, but it is of lower entropy (and higher free energy) than a large black hole for $T > T_{\mbox{\tiny  HP}}$.  We can understand this by considering the 
{\it coexistence line}, which can be obtained by setting $G=0$:
\be\label{HPcoexistence}
P|_{\mbox{\tiny  coexistence}}=\frac{3\pi}{8} T^2 
\ee
and whose  slope satisfies the {\it Clausius--Clapeyron} equation \cite{Kubiznak:2016qmn}
\be
\frac{dP}{dT}\Bigr|_{\mbox{\tiny  coexistence}}=\frac{\Delta S}{\Delta V}=\frac{S_{bh}-S_r}{V_{bh}-V_{r}}=\frac{S_{bh}}{V_{bh}}
=\frac{3}{4\ell}
\ee
(taking $S_r\approx 0$ and $V_r\approx 0$).  The coexistence curve on the  $P-T$ phase diagram (right panel of
Fig.~\ref{Fig:HP})  has no terminal point,
indicating that this phase transition is present for all pressures. This is reminiscent of a solid/liquid phase transition \cite{Kubiznak:2014zwa}.

Replacing $\ell$ in \eqref{SAdSTherm} with pressure (using \eqref{eq:press}), we obtain
\be\label{HPstate}
P=\frac{k_{\rm B} T}{v}- \hbar c\frac{k l_P^2}{2\pi v^2}
\ee
which can be regarded as the  ``{\it equation of state}'' for the system, where (temporarily restoring units, where the Planck length  $l_P = \sqrt{\hbar G/c^3}$) the quantity $v$ 
\begin{equation} 
\label{specvol}
v =2r_+l_P^2=2\Bigl(\frac{3V}{4\pi}\Bigr)^{1/3}=6\frac{V}{N}
\end{equation}
is the `{\it specific volume}' \cite{Kubiznak:2012wp, Altamirano:2014tva}, 
given by the thermodynamic volume $V$ divided by the `number of states' associated with the horizon, $N=A/l_P^2$.   We see for planar black holes ($k=0$) that  the {\it ideal gas law}, $T=Pv$, is recovered.

\subsection{Black Holes as Van der Waals fluids}
\label{4p2}

A more interesting situation ensues for charged AdS black holes, whose metric and gauge field strength are
\ba\label{HDRN}
ds^2 &=& -f(r) dt^2 + \frac{dr^2}{f(r)} + r^{2} d\Omega^2_{D-2,k=1}\,,\nonumber\\
\textsf{F}&=&d\textsf{A}\,,\quad \textsf{A}=-\frac{Q}{r}dt\,,
\ea
which is an exact solution to the Einstein-Maxwell equations that follow from variation of the action
\be\label{EinMaxact}
 \frac{1}{16 \pi G_N} \int d^Dx \sqrt{-g} \left( R -2 \Lambda - \frac{1}{4} \textsf{F}^2 \right)
\ee
with $\Lambda = -3/\ell^2$ and $G_N$ Newton's constant.  The gauge field strength  $\textsf{F} = d\textsf{A}$ and  $f(r)$  is given by
\be\label{metfunction}
f = 1 - \frac{16 \pi G_N M}{(D-2)\Omega_{k=1} r^{D-3}} +  \frac{8 \pi G_N}{(D-2)(D-3)}\frac{Q^2}{ r^{2D-6}} + \frac{r^2}{\ell^2} 
\ee
and $d\Omega_{D-2,k=1}^2$ is the metric for the standard element on $S^{D-2}$. 
The parameters $M$ and $Q$ are respectively the ADM mass 
and   total charge of the black hole. The thermodynamic quantities can be written in terms of $r_+$, $\ell$, and $Q$, yielding
\cite{Chamblin:1999tk, Chamblin:1999hg}
\ba\label{TDchgBH}
M&=\frac{1}{2}\Bigl(r_+  + \frac{Q^2}{r_+  } + \frac{r_+^{3} }{\ell^2}\Bigr)\; , \qquad
S=\frac{A}{4}=\pi r_+^2 \\
 \Phi&=\frac{Q}{r_+}\,,\quad T = \frac{f'(r_+)}{4\pi}
= \frac{1}{4\pi r_{+}} \left(1 - \frac{Q^2}{r_+^{2}} + 3\frac{r_+^2}{\ell^2} \right) 
\label{TDchgBH2}
\ea
in $D=4$, where $f(r_+)=0$, which locates the outer (event) horizon at $r=r_+$. 
Taking the variation of $M$, we obtain
\ba
\delta M &=& \frac{1}{2}  \left(1 - \frac{Q^2}{r_+^{ 2} }  + 3\frac{r_+^{ 2} }{l^2}  \right) \delta r_+
 -  \frac{r_+^{3} }{\ell^3} \delta \ell  +   \frac{Q}{r_+  }\delta Q \nonumber\\
 &=& \frac{4\pi r_+ T}{2}\frac{\delta S}{2\pi r_+}  + \frac{4\pi r_+^3}{3} \delta P + \Phi \delta Q \nonumber \\
 &=& T {\delta S} + V\delta P + \Phi \delta Q\, ,
 \label{FlawRN}
\ea
where  \eqref{eq:press} has been used and
\be\label{volrp}
V  = \frac{4\pi r_+^{3}}{3}\,,
\ee
can be inferred  from \eqref{Vdef}.  Equation \eqref{FlawRN} is a particular case of the first law \eqref{firstBH}.

It is also straightforward to show that
\be\label{RNSmarr}
 M-2TS- \Phi Q=- \frac{r_+^3}{\ell^2}=-2PV\,,
\ee
upon using \eqref{volrp}.  This is a particular case of   the Smarr relation \eqref{smarrBH}; clearly it would not
be satisfied  without the $PV$ term.

We can take the expression for the temperature $T$ in \eqref{TDchgBH2} and use
\eqref{eq:press} to replace $\ell$ with $P$.  
 Solving for $P$ and using \eqref{specvol} yields the equation of state:
\be\label{RNstate}
P=\frac{T}{v}-\frac{1}{2\pi v^2}+\frac{2Q^2}{\pi v^4}
\ee
which qualitatively reproduces the behaviour of the Van der Waals equation
\be\label{VdWstate}
\Bigl(P+\frac{a}{v^2}\Bigr)(v-b)=T \Rightarrow P = \frac{T}{v-b} - \frac{a}{v^2}
\ee
where the parameter $b$ corresponds to the ``volume of fluid particles'' and  $a>0$ measures the attraction between particles.

\begin{figure*}
\centering
\begin{tabular}{cc}
\includegraphics[width=0.49\textwidth,height=0.3\textheight]{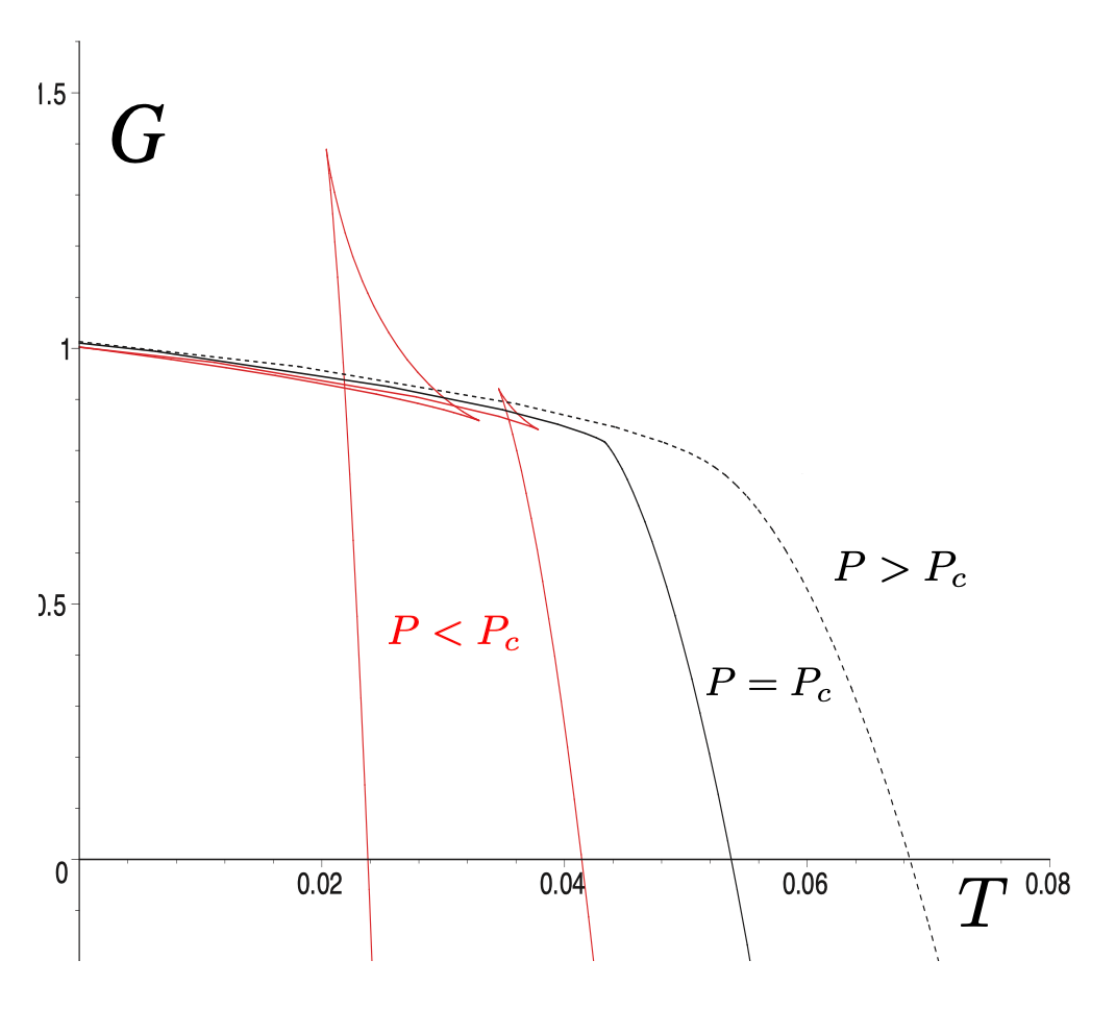} &
\includegraphics[width=0.49\textwidth,height=0.3\textheight]{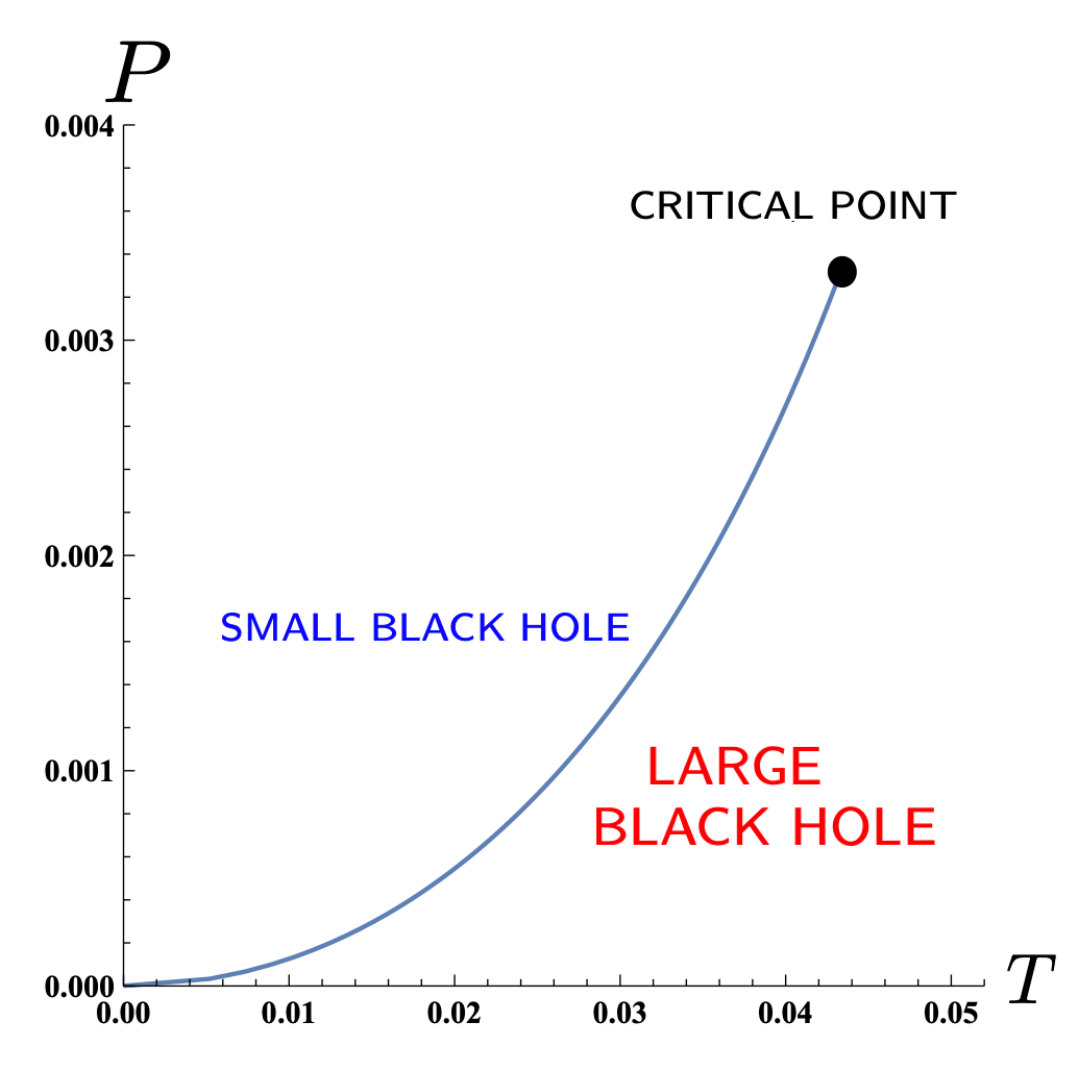}
\end{tabular}
\caption{{\bf Black Hole Van der Waals behaviour.} {\it Left.} Plots of the Gibbs free energy of a charged-AdS black hole 
for several different values of the pressure $P$,  displayed for fixed $Q=1$. For sufficiently low pressures $P<P_c$, a characteristic swallowtail emerges, shown by the red lines.
{\it Right.} The $P-T$  phase diagram illustrates  first order phase transition behaviour from small to large black holes 
as the temperature increases for fixed pressure. The coexistence line is analogous to
a liquid/gas phase transition, and terminates at a critical point where the phase transition is of   second order.
All quantities are in Planckian units.
}
\label{Fig:Swallow}
\end{figure*}

Plotting in Fig.~\ref{Fig:Swallow} 
the Gibbs free energy
\be\label{GibbsRN}
G=M-TS=\frac{l^2r_+^2-r_+^4+3Q^2l^2}{4l^2r_+}\,
\ee
we observe swallowtail behaviour at sufficiently low pressures, shown in the left panel  of Fig.~\ref{Fig:Swallow}. Such swallowtails are
 characteristic of first-order phase transitions.  As $P$ approaches the critical value $P_c$ from below, the swallowtail shrinks, terminating at a {\it critical point}.  This can be found from \eqref{RNstate} by setting $\partial P/\partial v = \partial^2 P/\partial v^2 = 0$, yielding
 \be\label{critRNAdS}
P_c=\frac{1}{96\pi Q^2}\,, \quad v_c=2\sqrt{6}Q\,, \quad T_c=\frac{\sqrt{6}}{18\pi Q}\,,
\ee
for the respective critical pressure, specific volume, and temperature, 
at which point the phase transition becomes  {\it second-order}.  The right panel of Fig.~\ref{Fig:Swallow} depicts 
the {\it coexistence line}  in the  $P-T$ phase diagram for the system, along with the critical point. 

Charged AdS black holes were expected to admit first order {\it small-black-hole/large-black-hole} (SBH/LBH) phase transitions
in a canonical (fixed charge) ensemble \cite{Chamblin:1999tk,Chamblin:1999hg,Cvetic:1999ne,Cvetic:1999rb}.The perspective
of black hole chemistry gave a proper identification between intensive and extensive variables \cite{Dolan:2011xt, Kubiznak:2012wp},
completing the analogy between this kind of black hole and a  Van der Waals fluid \footnote{It is possible to construct black holes whose
equation of state is exactly that of a Van der Waals fluid, but such solutions in Einstein gravity require exotic matter \cite{Rajagopal:2014ewa,Delsate:2014zma}.}.

The analogy can be further extended by investigating the critical exponents of the system.  These quantities 
parametrize the dependence of various thermodynamic quantities on  $t=T/T_c-1$, where $T_c$ is the critical temperature, and
  characterize the behaviour of various physical quantities in the vicinity of a critical point, obtained by expanding
  the equation of state about that point.   They are defined as follows \cite{Kubiznak:2012wp}:
\begin{itemize}
\item
The behaviour of the specific heat at constant volume is governed by the exponent $\alpha$. 
\be
C_V=T \frac{\partial S}{\partial T}\Big|_{V}\propto |t|^{-\alpha}\,.
\ee
\item The {\it order parameter} $\mathfrak{M}=V_l-V_s$ 
\be
\mathfrak{M} =V_l-V_s\propto |t|^\beta
\ee
quantifies the  difference between the volume of a large black hole $V_l$ and   a small black hole $V_s$  on a given isotherm, and is governed by the exponent $\beta$.  This order parameter could alternatively be defined  as the difference $\mathfrak{m}=v_l-v_s$
between the specific volumes.
\item
The exponent $\gamma$ 
\be
\kappa_T=-\frac{1}{V}\frac{\partial V}{\partial P}\Big|_T\propto |t|^{-\gamma}
\ee
governs the behaviour of the {\it isothermal compressibility} $\kappa_T$.
\item Finally, the exponent $\delta$ governs the  behaviour of the pressure in terms of the volume
\be
|P-P_c|\propto |V-V_c|^\delta
\ee
on the critical isotherm $T=T_c$.
\end{itemize}
For the charged black hole these exponents can be obtained by expanding 
about the critical point, yielding
\begin{equation}\label{eqn:meanfield}
\alpha = 0\,, \quad \beta = \frac{1}{2}\,, \quad \gamma=1\,, \quad \delta=3\,.
\end{equation}
which are the standard exponents from mean field theory, and are the same as for a Van der Waals fluid. One curious
fact is the  critical ratio $P_c v_c/T_c = 3/8$ is exactly that of a Van der Waals fluid. This equality appears to be a 
coincidence;  for dimensions $D>4$, it does not hold, and $P_c v_c/T_c = (2D-5)/(4D-8)$ \cite{Gunasekaran:2012dq}.

The coexistence curve can be analytically determined for the 4D charged-AdS black hole \cite{Mo:2016sel}. Along this curve the 
size of the black hole discontinuously changes from a 
small radius, $r_s$, to a large one, $r_l$,  where
\be \label{rsrl}
r_s=\frac{1}{2}\Bigl(\sqrt{\ell^2-2Q\ell}-\sqrt{\ell^2-6Q\ell}\Bigr)\, \quad r_l=\frac{1}{2}\Bigl(\sqrt{\ell^2+2Q\ell}+\sqrt{\ell^2-6Q\ell}\Bigr)\,,
\ee
and both the temperature and  Gibbs free energy are the same: $T(r_s)=T(r_l)$ and $G(r_s)=G(r_l)$. From these equations
we find
\be
T|_{\mbox{\tiny  coexistence}}=\frac{r_s^2-Q^2}{4\pi r_s^3}+\frac{3r_s}{4\pi \ell^2}\in(0,T_c)\,,
\ee
for the coexistence curve, where $\ell$ is given in terms of the pressure from \eqref{eq:press} and
  $r_s$ is  therefore a function of both $(Q,P)$.  The slope of this curve
  \be\label{CCeqn}
\frac{dP}{dT}\Bigr|_{\mbox{\tiny  coexistence}}=\frac{\Delta S}{\Delta V}=\frac{S_l-S_s}{V_l-V_s}\,,
\ee
yields the {\it Clausius--Clapeyron} equation, previously  verified using an approximate coexistence formula \cite{Wei:2014qwa}.

\begin{figure*}
\centering
\includegraphics[width=0.49\textwidth,height=0.3\textheight]{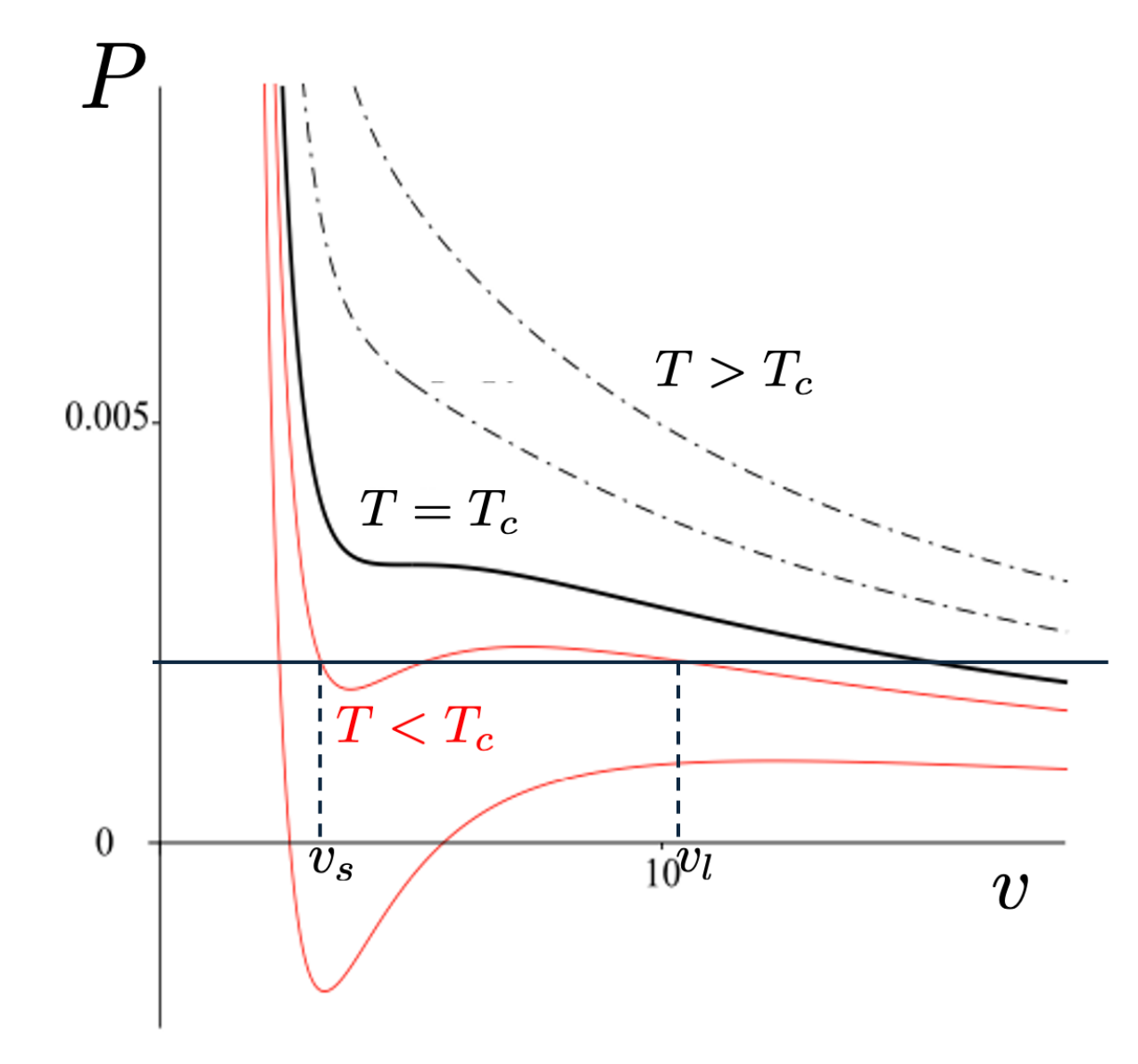} 
\caption{{\bf $P$-$v$ Diagram} The graph is a plot of pressure as a function of specific volume at various temperatures,
obtained from \eqref{RNstate} for a fixed $Q=1$.   Maxwell's equal area law is qualitatively displayed: 
the upper red isotherm $T<T_c$ intersects the black isobar  such that the areas above and below the isobar are equal. 
The quantitatively correct area law holds in the   $P-v$ diagram for both black holes and 
Van der Waals fluids. For volumes $v_s < v < v_l$ both the SBH (liquid) and LBH (gas) states coexist.
}
\label{Fig:pv}
\end{figure*}

The coexistence curve can alternatively be obtained  by imposing   {\it Maxwell's equal area law} \cite{Spallucci:2013osa}, which states for a line of constant pressure drawn through a $P-V$ curve,   the two phases coexist when the areas above and below this line
are equal (see Fig.~\ref{Fig:pv}). Note that  the equal area law is qualitatively but not quantitatively correct  in the $P-v$ plane \cite{Lan:2015bia, Xu:2015hba}, since  $v\propto V/N$ where $N$ is no longer a constant but $N=N(r_+)$.

In general the SBH/LBH coexistence line must be computed numerically for a  typical black hole  \cite{Wei:2014qwa, Wei:2015ana, Cheng:2016bpx, Wei:2015iwa}.   There is latent heat $\Delta Q=T\Delta S$ at the transition, which vanishes at the critical point where the phase transition is  second order. The validity of {\it Ehrenfest's equations} at the critical point
 \cite{Mo:2013ela, Mo:2014wca, Mo:2014mba} can be verified, similar to the Clausius--Clapeyron equation \cite{Zhao:2014fea}.  

Recently it has been shown that it is possible to study the dyanmics of black hole phase transitions \cite{Li:2020nsy}.   The thermodynamic stability of
each of the phases of a charged AdS black hole is  determined by the topography of the underlying free energy landscape.  Using a particular form
of the Fokker-Planck equation known as the Smoluchowski equation, the  probability that a large (small) black hole can transit to a small (large) black hole due to  thermal fluctuations can be calculated. It is also possible to compute how fast the black hole system undergoes such a stochastic process for the first time, known as the first passage time. Both the mean first passage time and its fluctuations are determined from the temperature-dependent barrier heights of the free energy landscape.  Dynamics of black hole phase transitions is now an active area of research \cite{Wei:2020rcd,Li:2020spm,Li:2021vdp,Lan:2021crt,Li:2021zep,Yang:2021ljn,Mo:2021jff,Kumara:2021hlt,Li:2021tpu,Xu:2021usl,Du:2021cxs,Li:2022ylz,Li:2022oup,Li:2023ppc,Safir:2023thg,Liu:2023sbf,Wang:2024zbp,Wu:2024zig,Li:2024tyk}.

\subsection{Reentrant phase transitions} 
\label{4p3}

A third example of more novel chemical behaviour is that of a  {\it reentrant phase transition (RPT)}. This refers to a situation
in which a monotonic variation of any thermodynamic quantity results in two (or more) phase transitions such that the initial state and the final state are macroscopically similar. They have been observed in multicomponent fluid systems, gels, ferroelectrics, liquid crystals, and binary gases, where the reentrant behaviour often emerges as a consequence of two (or more) `competing driving mechanisms' 
\cite{narayanan1994reentrant}.  In 1904  Hudson \cite{Hudson:1904} first observed this phenomenon  in a nicotine/water mixture.  
For a sufficient fixed percentage of nicotine, at high temperatures the water and nicotine mix.  As the  temperature of the mixture 
cools,    the homogeneous mixed state separates into distinct nicotine/water phases.  Further cooling returns the system to 
the homogeneous state due to polar bonding of the water with the nicotine.

 \begin{figure*}
\centering
\begin{tabular}{cc}
\includegraphics[width=0.49\textwidth,height=0.3\textheight]{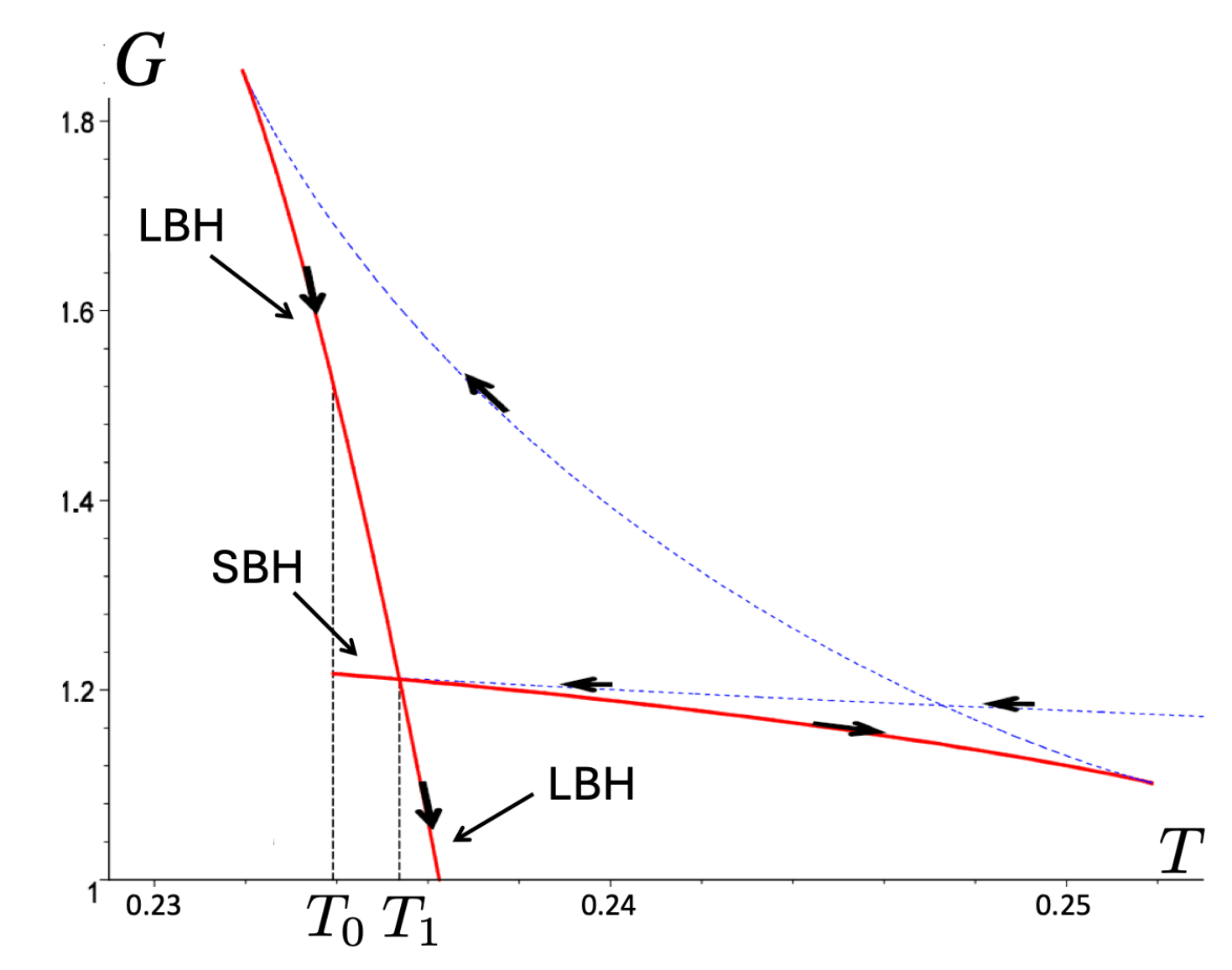} &
\includegraphics[width=0.49\textwidth,height=0.3\textheight]{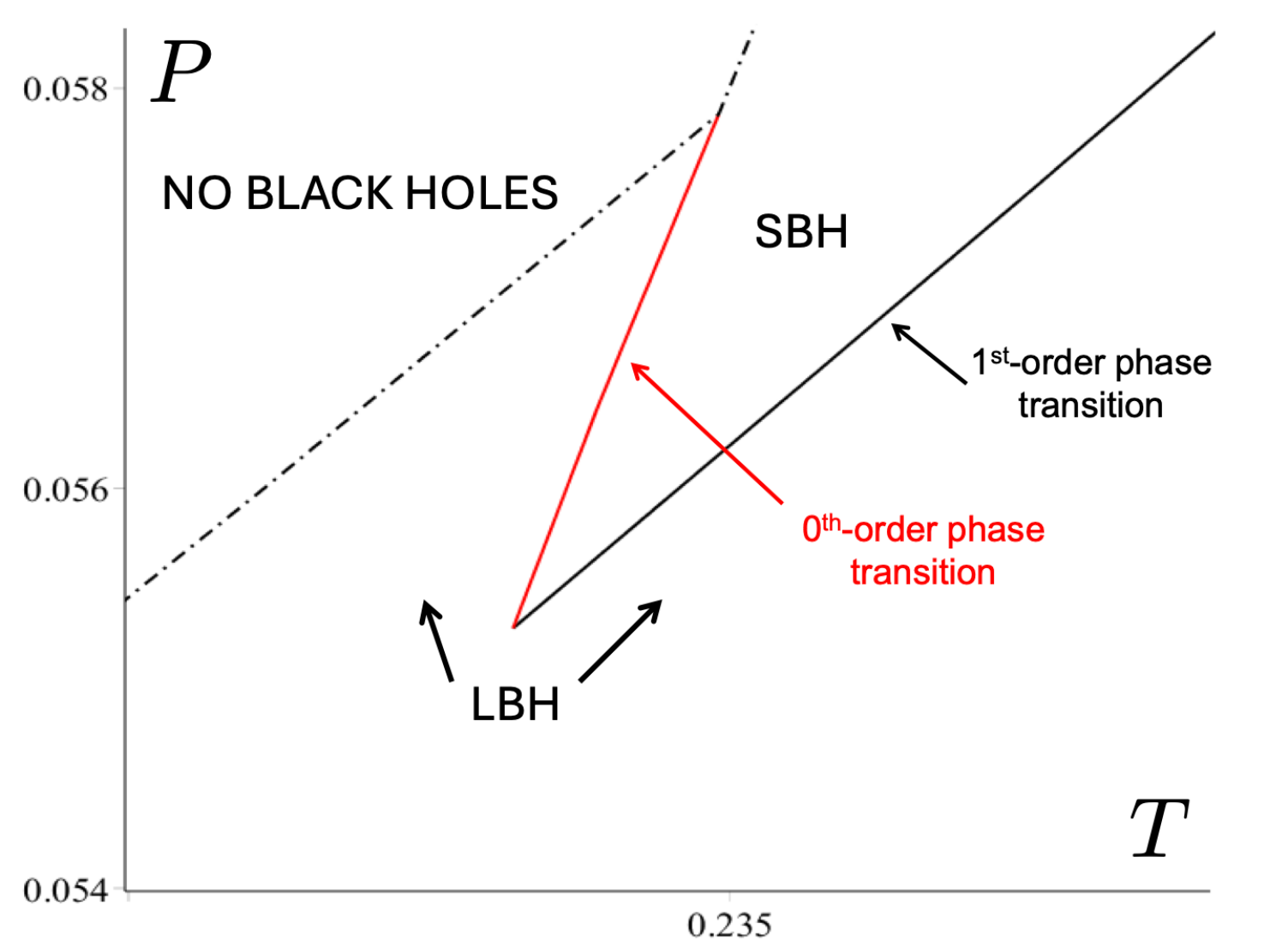}
\end{tabular}
\caption{{\bf Reentrant phase transition in a $D=6$ singly spinning Kerr-AdS black hole } {\it Left.}  The diagram illustrates the 
behaviour of $G$ when an RPT takes place, which is for the range $P\in(0.0553,0.0579)$. The arrows on the curves indicate the direction of   increasing $r_+$.  At any point $T>T_1$, the stable thermodynamic state is an LBH. Decreasing the temperature from this point, at $T=T_1$ there is a first order LBH/SBH phase transition after which the SBH is the stable state.  As $T$ is further decreased,  the system follows the red curve until $T=T_0$, where there is a cusp at which the lower blue dashed curve and the red curve join.
Here $G$ has a discontinuity at its global minimum; for smaller $T$ the system jumps to the
uppermost vertical red curve and becomes an LBH again. This corresponds to a zeroth order SBH/LBH phase transition. 
{\it Right.} The  $P-T$ diagram corresponding to the left panel  illustrates 3 possible phases in the range the range $P\in(0.0553,0.0579)$: an LBH region, an SBH region, and a region where no black hole solutions exist. The LBH and SBH 
are separated by  coexistence lines corresponding to 1st-order (black curve) and 0th-order (red curve) phase transitions. The 1st-order coexistence line eventually terminates at a critical point (not displayed). The angular momentum $J=1$.}
\label{Fig:RPT}
\end{figure*}

The first   RPT in gravitational thermodynamics was observed for $D=4$ black holes in Einstein gravity coupled to Born--Infeld
electrodynamics \cite{Gunasekaran:2012dq}. RPTs were soon discovered in many other black hole systems in $D>4$ dimensions \cite{Altamirano:2013ane, Altamirano:2013uqa, Altamirano:2014tva, Wei:2015ana}, as well as higher curvature gravity
 \cite{Frassino:2014pha,Wei:2014hba,Hennigar:2015esa, Sherkatghanad:2014hda} and for higher-curvature black holes 
 with scalar hair \cite{Hennigar:2015wxa,Astefanesei:2021vcp}.
 
 For a singly spinning black hole in $D$ spacetime dimensions the metric is  \cite{GibbonsEtal:2005}
\ba
ds^2&=&-\frac{\Delta}{\rho^2}(dt-\frac{a}{\Xi}\sin^2\!\theta d\varphi)^2+
\frac{\rho^2}{\Delta}dr^2+\frac{\rho^2}{\Sigma}d\theta^2\nonumber\\
&+&
\frac{\Sigma \sin^2\!\theta}{\rho^2}[adt-\frac{(r^2+a^2)}{\Xi}d\varphi]^2+r^2\cos^2\!\theta d\Omega_{D-2}^2\,,\qquad
\ea
where 
\ba
\Delta&=&(r^2+a^2)(1+\frac{r^2}{\ell^2})-2mr^{5-D}\, \quad \Sigma=1-\frac{a^2}{\ell^2}\cos^2\!\theta\,,\nonumber\\
\Xi&=&1-\frac{a^2}{\ell^2}\ \quad \rho^2=r^2+a^2\cos^2\!\theta\,,
\ea
and $d\Omega_{D-2}^2$ is given by \eqref{dOmegak} with $k=1$.
The associated thermodynamic quantities   (in Planck units) are \cite{Das:2000cu,Gibbons:2004ai}
\begin{eqnarray}
M&=&\frac{\Omega_{D-2}}{4\pi}\frac{m}{\Xi^2}\left(1+\frac{(D-4)\Xi}{2}\right)\,, \label{BHOM} \\
J&=&\frac{\Omega_{D-2}}{4 \pi}\frac{ma}{\Xi^2}\,,\quad  {\Omega}_H=\frac{a}{l^2}\frac{r_+^2+\ell^2}{r_+^2+a^2}\,,\quad
\label{BHJ}\\
T&=&\frac{1}{2\pi}\Bigr[r_+\Bigl(\frac{r_+^2}{\ell^2}+1\Bigr) \left(\frac{1}{a^2+r_+^2}+
\frac{D-3}{2 r_+^2}\right)-\frac{1}{r_+}\Bigr]\,,\qquad \label{BHT} \\
S&=&\frac{\Omega_{D-2}}{4}\frac{(a^2+r_+^2) r_+^{D-4}}{\Xi}=\frac{A}{4}\,, \label{BHS}
\end{eqnarray} 
where $r_+$ is the black hole horizon radius (the largest positive real root of $\Delta=0$)
and $\Omega_{D-2}$ is given by \eqref{volsphere}. 
The   Gibbs free energy $G=G(T,P,J)$ is   
\ba
G\!&=&\!M-TS =\!\frac{\Omega_{D-2}r_+^{D-5}}{16\pi \Xi^2}\Bigl(3a^2+r_+^2-\frac{(r_+^2\!-\!a^2)^2}{\ell^2}+\frac{3a^2r_+^4+a^4r_+^2}{\ell^4}\Bigr)
\nonumber
\ea
and  depends on the angular momentum $J$. 

The free energy $G$ is plotted in the left panel of  Fig.~\ref{Fig:RPT}. We see that  as the temperature is monotonically decreased
beginning with some $T>T_1$, there is an LBH/SBH/LBH reentrant phase transition \cite{Altamirano:2013ane}. At $T_0$ there is a
discontinuity in the global minimum of the Gibbs free energy, referred
to as a zeroth-order phase transition \cite{Gunasekaran:2012dq}. This  phenomenon has been seen  in superfluidity and superconductivity \cite{maslov2004zeroth}.

We see again a situation in chemistry that has a parallel in black hole thermodynamics:
\be
\begin{array}{|c|c|c|}
\hline
{\mbox{Low $T$}} & \mbox{Medium $T$} & \mbox{High $T$} \\
\hline
{\mbox{mixed}} & \mbox{water/nicotine} & \mbox{mixed} \\
\hline
{\mbox{LBH}} & \mbox{SBH} & \mbox{LBH} \\
\hline
\end{array}
\ee
Multiple RPTs have been observed in a number of settings \cite{Frassino:2014pha, Hennigar:2015esa, Sherkatghanad:2014hda}. They
need not be associated with a zeroth-order phase transition; it is common for an RPT to occur via a succession of two first order phase transitions.

\section{Phenomenology of Black Hole Chemistry}
\label{5s}

The perspective of  black hole chemistry has led to a panoply of results concerning the thermodynamic properties of black holes.
Rather than review the  thermodynamic phenomenology -- which is quite vast -- the main results will be highlighted.

\subsection{Triple Points} 
\label{5p1}

The triple point of a pure substance in chemistry is the combination of pressure and temperature at which three phases exist in thermodynamic equilibrium.  These can be three of any possible phases the substance can be in, though the term most
commonly  refers to the point where a substance’s solid, liquid, and gas phase coexist in equilibrium.  The triple point of water is at 
a pressure of 611.7 pascals (or 0.0060373057 atm) and a temperature of 273.16 ${}^o$K  (0.01 °C).  Many substances such as
acetylene, ammonia, butane, carbon, chloroform, ethylene, hydrogen chloride, isobutane, methane, nitric oxide, palladium, sulfur dioxide, uranium hexaflouride, and more all have well-measured triple points.  The triple point is usually the minimum temperature at which the liquid form of a substance can exist. 

\begin{figure*}
\centering
\begin{tabular}{cc}
\rotatebox{0}{
\includegraphics[width=0.49\textwidth,height=0.3\textheight]{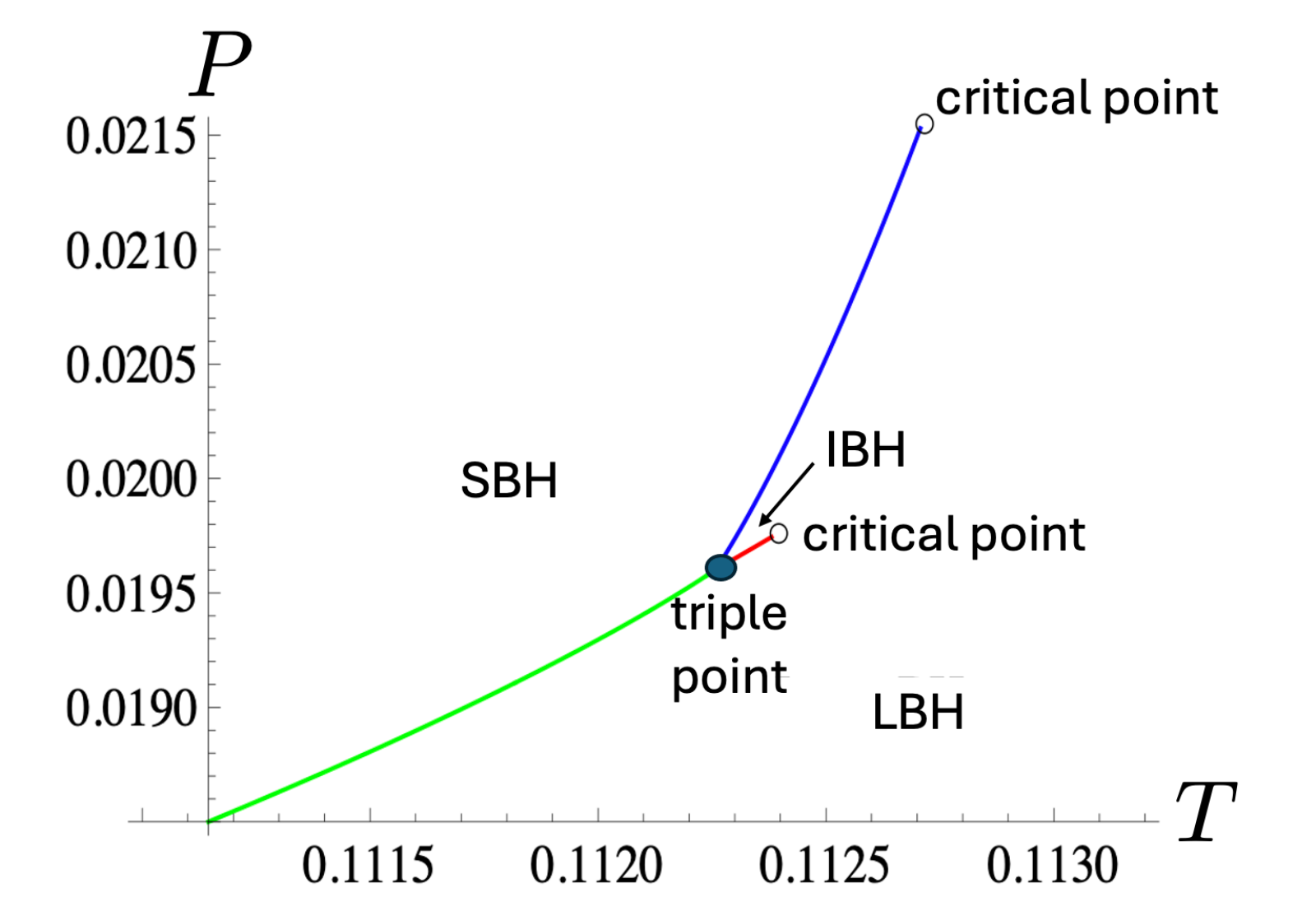}} &
\rotatebox{0}{
\includegraphics[width=0.49\textwidth,height=0.3\textheight]{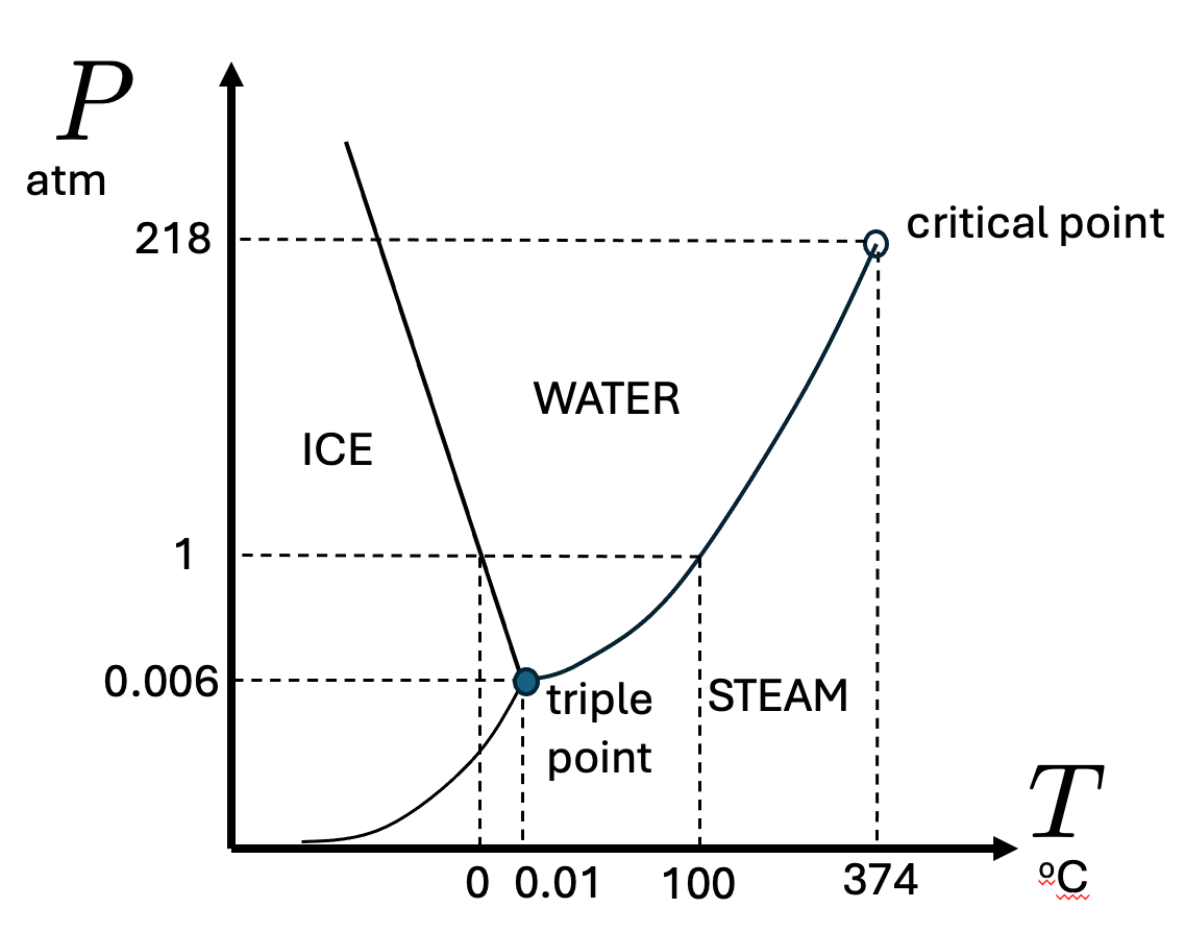}}\\
\end{tabular}
\caption{{\bf Triple Point}  {\em Left}:   Phase diagram for a charged Einstein-Gauss-Bonnet black hole in
$D=6$ dimensions with $Q=1$ \cite{Wei:2014hba}.  The small (SBH), large (LBH), and intermediate (IBH) black holes are in
thermodynamic equilibrium at the triple point.  There are two critical points: one where LBH and IBH 
become indistinguishable; the other, at a higher pressure, where SBH becomes indistinguishable from
the LBH/IBH phase.  {\em Right}: Phase diagram for water. There is a single critical point at
$P=218$ atmospheres and $T=374$${}^o$ C.
}\label{Fig:Triple}
\end{figure*}

One of the earlier discoveries of black hole chemistry was that black holes also can have triple points.  They were
first observed in doubly spinning Kerr-AdS black holes in $D=6$ dimensions\cite{Altamirano:2013uqa} and shortly afterward in Einstein-Maxwell-Gauss-Bonnet gravity \cite{Wei:2014hba,Frassino:2014pha, Hennigar:2015esa}.  A comparison of the latter case with water is shown in Fig.~\ref{Fig:Triple}.  For the black hole (left panel)
three phases are evident: small, intermediate, and large, all meeting at a triple point, fully analogous to the triple point of water
(right panel). One noteworthy distinction between the two cases is the absence of a semi-infinite coexistence line in the black hole
case, which has two critical points, in contrast to the single critical point of water.  For water, to go from solid
to liquid requires melting the ice, whereas in going from the SBH to the IBH phases  `melting' can be avoided:
it is possible to `go around' the upper critical point by raising the pressure, then the temperature, and then lowering the pressure.
 
 Triple points have since been studied in a number of contexts, including higher curvature gravity \cite{Frassino:2014pha,Wei:2014hba,Wei:2021krr}, massive gravity \cite{Dehghani:2020blz}, quasitopological  electromagnetism \cite{Li:2022vcd,Mou:2023nrx}, exotic black holes \cite{Hull:2022xew,Hull:2021bry},
and solitons \cite{Quijada:2023fkc}. 
Dynamics of phase transitions at a black hole triple point have been a  subject of recent study \cite{Wei:2021bwy,Cai:2021sag}.
 
\subsection{Multicriticality}
\label{5p2}

Most substances have more than a single triple point because their solid phase can assume different forms (or allotropes). 
If there are $p$ phases of matter, the number of triple points is $p!/(p-3)!3!$.  This raises the possibility of multiple phases
of matter being in equilibrium at some pressure and temperature: quadruple points, quintuple points, etc. Such multi-critical points 
have been observed in colloidal polymers and other heterogeneous systems \cite{Akahane2016,Garcia2017,Sun:2021gpr}. 

For quite some time multicritical points did not seem to be present in black holes.  However they were discovered 
two years ago  for charged AdS black holes in non-linear electrodynamics \cite{Tavakoli:2022kmo}, and have since been
observed in Lovelock gravity  \cite{Wu:2022plw,Wu:2022xmp}, and even for multiply rotating black holes in Einstein gravity \cite{Wu:2022bdk}. 
This latter case demonstrates that multicriticality requires neither higher curvature nor matter of any kind.

The family of multiply-rotating $D$-dimensional Kerr-AdS black holes \cite{Gibbons:2004js,Gibbons:2004uw} 
\ba \label{KAdSmetric}
ds^2&=&-W\Bigl(1+\frac{r^2}{\ell^2}\Bigr)d\tau ^2+\frac{2m}{U} \Bigl(W d\tau -\sum_{i=1}^{N} \frac{a_i \mu_i ^2 d\varphi _i}{\Xi _i}\Bigr)^2\nonumber\\
&+&\sum_{i=1}^{N} \frac{r^2+a_i^2}{\Xi _i} \mu_i ^2 d\varphi _i^2+\frac{U dr^2}{F-2m}+\sum_{i=1}^{N+\epsilon}\frac{r^2+a_i ^2}{\Xi _i} d\mu _i ^2 \nonumber\\
&-&\frac{l^{-2}}{W (1+r^2/\ell^{2})}\Bigl(\sum_{i=1}^{N+\epsilon}\frac{r^2+a_i ^2}{\Xi _i} \mu_i d\mu_i\Bigr)^2 
\ea
with metric functions 
\ba\label{KAdSmetricfunctions}
W&=&\sum_{i=1}^{N+\epsilon}\frac{\mu _i^2}{\Xi _i} \qquad U=r^\epsilon \sum_{i=1}^{N+\epsilon} \frac{\mu _i^2}{r^2+a_i^2} \prod _j ^N (r^2+a_j^2) \nonumber\\
F&=&r^ {\epsilon -2} \Bigl(1+\frac{r^2}{\ell^2}\Bigr) \prod_{i=1}^N (r^2+a_i^2) \qquad \Xi_i=1-\frac{a_i^2}{\ell^2} 
\ea
where the coordinates $\mu_i$ obey  $\sum_{i=1}^n \mu^2_i = 1$, 
have increasing numbers of
distinct rotation parameters $a_i$, and hence distinct angular momenta, as the dimension of spacetime gets larger.  This introduces an increasing number of  thermodynamic conjugate pairs to the system, allowing for more phases 
 than the small/intermediate/large  ones  seen for doubly rotating black holes  \cite{Altamirano:2013uqa} and the other cases noted
 in the previous subsection.  Multiple phases separated by first order phase transitions  for sufficiently high pressure and appropriate angular momenta are possible.    These phases   merge at a single pressure and temperature
as the pressure is lowered. For pressures below the multi-critical value  only the smallest and  largest  black hole phases remain, separated by a first order phase transition.  

The maximum number of independent rotations is $N=\frac{1}{2}(D-1-\epsilon)$ with $\epsilon=0/1$ for odd/even spacetime dimensions.
The thermodynamic parameters are \cite{Wu:2022bdk}
\ba 
M&=&\frac{m \omega _{D-2}}{4\pi (\prod_j \Xi_j)}(\sum_{i=1}^{N}{\frac{1}{\Xi_i}-\frac{1-\epsilon }{2}}) \qquad
J_i = \frac{a_i m \omega _{D-2}}{4\pi \Xi_i (\prod_j \Xi_j)} \quad \Omega_i=\frac{a_i (1+\frac{r_+^2}{\ell^2})}{r_+^2+a_i^2}
\nonumber\\
V &=& \frac{r_+ A}{D-1} + \frac{8\pi}{(D-1)(D-2)}\sum_{i=1}^n a_i J_i \qquad
S = \frac{\omega _{D-2}}{4 r_+^{1-\epsilon}}\prod_{i=1}^N \frac{a_i^2+r_+^2}{\Xi_i} 
\label{TD} \\
T&=&\frac{1}{2\pi }\Bigr[r_+\Bigl(\frac{r_+^2}{\ell^2}+1\Bigr)
\sum_{i=1}^{N} \frac{1}{a_i^2+r_+^2}-\frac{1}{r_+}
\Bigl(\frac{1}{2}-\frac{r_+^2}{2\ell^2}\Bigr)^{\!\epsilon}\,\Bigr] \nonumber
\ea
where  $A$ is the area of the outermost horizon with radius
$r_+$  (obtained from $F(r_+)-2m=0$) and the pressure is given by \eqref{eq:press}.
 
Taking the variation of the Gibbs free energy $G=M-TS$  and using the first law yields
\begin{align}
      dG&=\sum_{i=1}^{N} \Omega_i dJ_i - SdT + V dP
\end{align}
and so  $dG=-SdT$ for constant $P$ and $J_i$, in which case the extrema of $G(r_+)$ and $T(r_+)$ occur at the same $r_+$ values. The existence and distribution of swallowtails in the $G$-$T$ plot is therefore determined solely by $T$ for any fixed $(P,J_i)$,
with the swallowtail cusps corresponding to the zeros of $T^\prime = \frac{\partial T}{\partial r_+}$.   
It is possible to fix each of the $J_i$ so that $T^\prime$ has a local maximum and local minima for each rotation. The 
locations of these extrema can be changed by adjusting the pressure.  Maximal multicriticality (the maximum number of coexistent phases) occurs   when $T^\prime(r_+)$ has a root between every local extremum.  This can also be obtained from a consideration of Maxwell's equal area rule \cite{Lu:2023hgu}.

\begin{figure}
 \centering
	\includegraphics[width=0.49\textwidth,height=0.3\textheight]{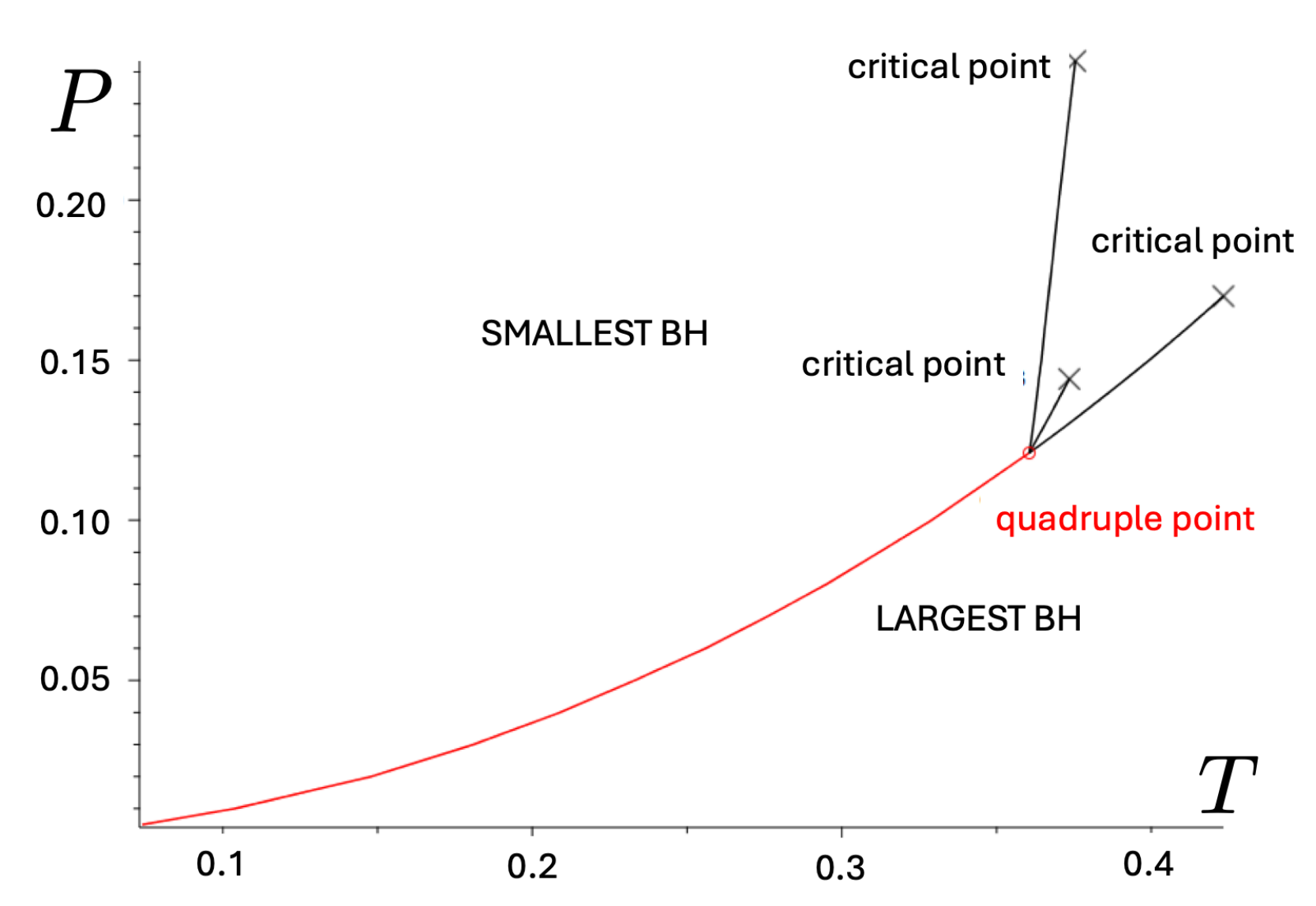}

	\caption{\textbf{Quadruple point} The $P--T$ phase diagram for a triply-rotating black hole in $D=8$ illustrates
	the existence of a quadruple point in Einstein gravity.  The angular momenta are  $J_1 = 7.967$, $J_2 = 1.24$, $J_3 = 0.12798$. For  pressures below the value $P=P_q$ at the quadruple point there is only a single first-order transition between the smallest and largest black holes  (red curve). At $P=P_q$, four phases (smallest, small, large, largest) coexist at $T_q\approx0.3606$. For 
pressures slightly larger than	$P_q$, a succession of three stable first order phase transitions  between four phases is exhibited.
 All three coexistence curves for $P> P_q$ terminate at their respective critical points. 	
	}
	\label{Fig:3rotation_PT} 
\end{figure}

The phase diagram for a quadruple point with 3 rotations in $D=8$ is shown in Fig.~\ref{Fig:3rotation_PT}.
At large values of the pressure $P$, no phase transitions are present, and the $G$--$T$ diagram (not shown) has no swallowtails.  Decreasing $P$, a critical pressure $P_{c_1} \approx 0.24335$ is attained, below which there are now two distinct phases;  the $G$--$T$ plot now has a single swallowtail. Two more critical pressures emerge as the pressure is lowered further: one at $P_{c_2}\approx 0.169948$ (where two swallowtails appear),  and  then at  $P_{c_3} \approx 0.144097$ (where a third swallowtail emerges). 
For $P < P_{c_3}$,   four distinct phases exist, characterized by the size of the horizon radius. Each is  
separated by a first order phase transition (for a total of three) for  $P_{c_3} > P > P_q = 0.121$.  The three swallowtails merge at a single  quadruple critical point  $P=P_q$, where all four phases coexist. Only one first order phase transition between the largest and the smallest black hole is present for $P<P_q$; the remaining phases are  in a thermodynamically unstable region and eventually disappear for smaller $P$.   

Multi-critical points 
in multiply rotating black holes are unlike those found in the context of non-linear electrodynamics \cite{Tavakoli:2022kmo} with  regards to the Gibbs phase rule
\cite{Sun:2021gpr}:
\be
\textsf{F}=\textsf{W}-\textsf{P}+1\, , 
\ee
which relates the degrees of freedom $\textsf{F}$ in a simple thermodynamic system to the number of coexistence phases  $\textsf{P}$ and the number 
of thermodynamic conjugate pairs $\textsf{W}$.  It governs the number of multicritical points in a system.  
For black holes in in non-linear electrodynamics, the  $n$-tuple points   have at minimum $n$ degrees of freedom and require 2 additional conjugate pairs for each new phase \cite{Tavakoli:2022kmo}. However in  the Kerr-AdS case the $n$-tuple points  always have a lower bound of $\textsf{F}=2$, with only one added rotation needed for a new phase. This disparity 
is because $T$ depends linearly on the coupling constants in non-linear electrodynamics, whereas it   depends nonlinearly on the angular momenta $J_i$.

As noted above, it is quite remarkable to observe multicriticality in vacuum Einstein gravity, as it demonstrates that no additional matter sources are required for the phenomenon to be present. Multicriticality can even occur for asymptotically flat black holes in Lovelock gravity  \cite{Wu:2022xmp}, where the higher curvature coupling is varied \cite{Mo:2018rks} and plays a role somewhat analogous to pressure. Dynamics of the phase transition near a quadrupole point were recently studied, and found to have features distinct from the triple point \cite{Yang:2023xzv}. 
The necessary and sufficient conditions for multi-criticality are not known at present.

\subsection{Polymeric Phase Transitions}\label{5p3}

Most black holes have critical points characteristic of mean field theory, as given in \eqref{eqn:meanfield}. However this is not always the case.
There exist black holes that undergo polymer-like transitions at certain {\it isolated critical points}  \cite{Frassino:2014pha,Dolan:2014vba}.
The critical exponents for these black holes are \cite{Dolan:2014vba,Hennigar:2015esa}
 \begin{equation}\label{Widrush}
\alpha = 0\,, \quad \beta = 1\, ,  \quad \gamma=K-1\,, \quad \delta=K , 
\end{equation}
and clearly differ from those in \eqref{eqn:meanfield}. However, as with the mean-field case, they satisfy the Widom scaling relation and the Rushbrooke inequality
\be\label{Widrush2}
\gamma=\beta(\delta-1) \qquad \alpha+2\beta+\gamma\geq 2 
\ee
which can both be derived   from general thermodynamic considerations. The quantity $K>2$ quantifies the degree of higher curvature
in Lovelock gravity, or in more general higher-curvature quasi-topological theories gravity \cite{Oliva:2010eb,Myers:2010ru} (see~\ref{appB})
Note that the   latter inequality in \eqref{Widrush2} is not saturated. 

\begin{figure*}
\centering
\begin{tabular}{cc}
\includegraphics[width=0.49\textwidth,height=0.3\textheight]{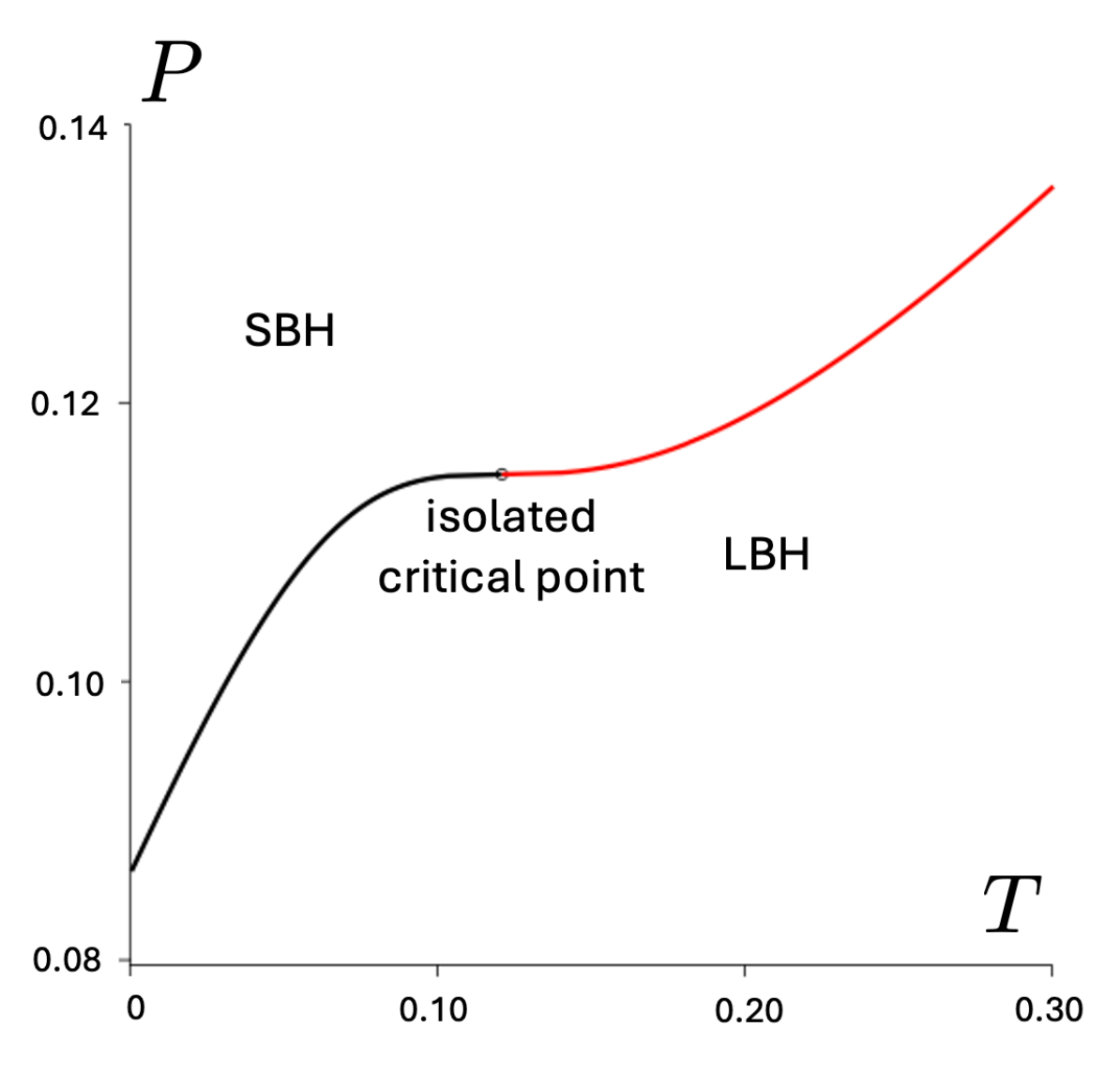} &
\end{tabular}
\caption{{\bf Isolated critical point} The figure illustrate a  $P-T$ diagram that displays two phases of black holes: large and small, separated by two
first order phase transitions (with coexistence lines are denoted by red and black curves) that meet at a single
isolated critical point where the phase transition becomes second order and is characterized by non-standard critical exponents.
This example is from 3rd order Lovelock gravity in $D=7$ \cite{Frassino:2014pha}.
}
\label{Fig:Isolated}
\end{figure*}

Isolated critical points occur when the endpoints of the coexistence lines of two different first order phase transitions meet in a single point. 
The result is a line of first order transitions punctuated by a single point at which the   phase transition is of second-order.  The situation is
illustrated in Fig.~\ref{Fig:Isolated}. Under these conditions, the Gibbs free energy develops two swallowtails whose tips coincide. 
 
The reason for referring to such transitions as polymeric is that their Prigogine--Defay ratio \cite{prigogine1974chem}, which
characterizes discontinuities in the isobaric heat capacity $\Delta C_P$, isothermal compressibility $\Delta \kappa_T$, and  isobaric thermal expansion coefficient $\Delta \alpha_P$, is less than unity.  This indicates that the phase transition has more than one order parameter and is
reminiscent of polymer or glassy transitions  \cite{gupta1976prigogine,gundermann2011predicting}.  This ratio is given by
\be\label{PiK}
\Pi=\frac{1}{VT}\Bigl(\frac{\Delta C_P\Delta \kappa_T}{(\Delta \alpha_P)^2}\Bigr)_{T}=\frac{1}{K} < 1 \, ,
\ee
where the latter equality holds for the special class of Lovelock black holes noted above \cite{Dolan:2014vba}.

Isolated critical points  for spherical black holes have also been observed in black holes with scalar hair in quasi-topological gravity
\cite{Dykaar:2017mba}, the first examples of this type.

\subsection{Superfluidity}
\label{5p4}

Superfluid transitions refer to   second order phase transitions that form a curve in phase space rather than a single point.  These kinds of transitions signal the onset of superfluidity in liquid helium~\cite{RevModPhys.71.S318}, and for this reason the same terminology is applied to black holes. They are manifest as a line of second order (continuous)  transitions in a phase diagram.

This class of phase transitions was first observed \cite{Hennigar:2016xwd} in Lovelock gravity conformally coupled to scalar fields \cite{Oliva:2011np}.  Analytic  black hole solutions with scalar hair were subsequently obtained \cite{Giribet:2014bva}, evading no-go results that had been reported previously~\cite{nogo_hairy}.    These solutions  are of inherent interest in holography 
because of the role played by scalar hair in descriptions of holographic superconductors and superfluids~\cite{Hartnoll:2008vx, Nie:2015zia}.  In addition to superfluidity, they  have reentrant phase transitions  and other interesting thermodynamic properties~\cite{Giribet:2014fla, Galante:2015voa, Hennigar:2015wxa, Chernicoff:2016jsu}.

Consider a line element of the form \eqref{Schw-Ads}, where the metric function $f$ obeys
\begin{align}\label{mastereqn} 
\sum_{j=0}^{j_{\rm max}} \alpha_j \left(\frac{\sigma-f}{r^2} \right)^j =& \frac{16 \pi G M}{(D-2)\Omega_{k} r^{D-1}} + \frac{H}{r^D} - \frac{8 \pi G}{(D-2)(D-3)}\frac{Q^2}{ r^{2D-4}} \, 
\end{align}
in Lovelock gravity conformally coupled to a scalar field and electromagnetism,
where
\be\label{defn_of_H} 
H = \sum_{j=0}^{j_{\rm max}} \frac{(D-3)!}{(D-2(j+1))!}b_j k^j N^{d-2j} \, 
\ee
is the ``hair parameter" 
and the $\alpha_j$ are the Lovelock coupling constants. The scalar field is 
\be 
\phi = \frac{N}{r}
\ee
and its equations of motion imply
\ba 
\sum_{j=1}^{j_{\rm max}} j b_j \frac{(D-1)!}{(D-2j-1)!} k^{j-1}N^{2-2j} &=& 0  
\nn\\
\sum_{j=0}^{j_{\rm max}} b_j \frac{(D-1)! \left(D(D-1)+4j^2 \right)}{(D-2j-1)!} k^j N^{-2j} &=& 0 \, ,
\label{N-constr}
\ea
providing constraints on $N$ and the $b_j$ parameters in \eqref{defn_of_H}.

The thermodynamic parameters for this solution are \cite{Hennigar:2016xwd}
\begin{align}
\label{temp_etc} 
M &= \frac{(D-2) \Omega_k}{16 \pi G} \sum_{k=0}^{k_{\rm max}} \alpha_k \sigma^k r_+^{D - 2k -1} - \frac{(D-2)\Omega_k H}{16 \pi G r_+} + \frac{\Omega_k Q^2}{2(D-3) r_+^{D-3}}
\nn\\
T & = \frac{1}{4 \pi r_+ \textsf{D}(r_+)} \left[\sum_k \sigma \alpha_k(D-2k-1) \left(\frac{\sigma}{r_+^2}\right)^{k-1} + \frac{H}{r_+^{D-2}} - \frac{8 \pi G Q^2}{(D-2) r_+^{2(D-3)}} \right]
\nn\\
S &= \frac{\Omega_k}{4 G}   \left[ \sum_{k=1}^{k_{\rm max}} \frac{(D-2) k \sigma^{k-1}  \alpha_k  }{D-2k} r_+^{D-2k}  - \frac{D}{2\sigma (D-4)} H\right] \, \quad \textrm{if } b_k = 0 \,\,\,  \forall k > 2\; ,
\end{align}
where $\textsf{D}(r_+) = \sum_{j=1}^{j_{\rm max}} j \alpha_j (k r_+^{-2})^{j-1} $.  The Smarr formula and thermodynamic
first law can be shown to hold  provided variations of the Lovelock couplings are considered~\cite{Kastor:2010gq}. 

Superfluidity can be demonstrated to hold for actions cubic in the curvature, where $\alpha_j = 0$ for $j \geq 4$. Introducing the dimensionless parameters
\begin{align}
r_+ &= v\alpha_3^{1/4} \, , \quad  T = \frac{t\alpha_3^{-1/4}}{D-2} \, , \quad  H =\frac{4\pi h}{D-2}\alpha_3^{\frac{D-2}{4}}\nn\\
Q &= \frac{q}{\sqrt{2}}\alpha_3^{\frac{D-3}{4}} \, , \quad  m = \frac{16\pi M}{(D-2)\Omega_k\alpha_3^{\frac{D-3}{4}}} \nn\\
p&=\frac{\alpha_0(D-1)(D-2)\sqrt{\alpha_3}}{4\pi} \, , \quad \alpha=\frac{\alpha_2}{\sqrt{\alpha_3}}\, 
\end{align}
the equation of state is  
\begin{align}\label{eos}
p = &\frac{t}{v}-\frac{k(D-3)(D-2)}{4\pi v^2}+\frac{2\alpha k  t}{v^3}-\frac{\alpha(D-2)(D-5)}{4\pi v^4}+ \frac{3t}{v^5}\nn\\&-\frac{k (D-7)(D-2)}{4\pi v^6}+\frac{q^2}{v^{2(D-2)}}-\frac{h}{v^D}\, ,\end{align}
obtained by solving \eqref{temp_etc} for the pressure $p$.  

\begin{figure*}[!htp]
\centering
\includegraphics[width=0.49\textwidth,height=0.3\textheight]{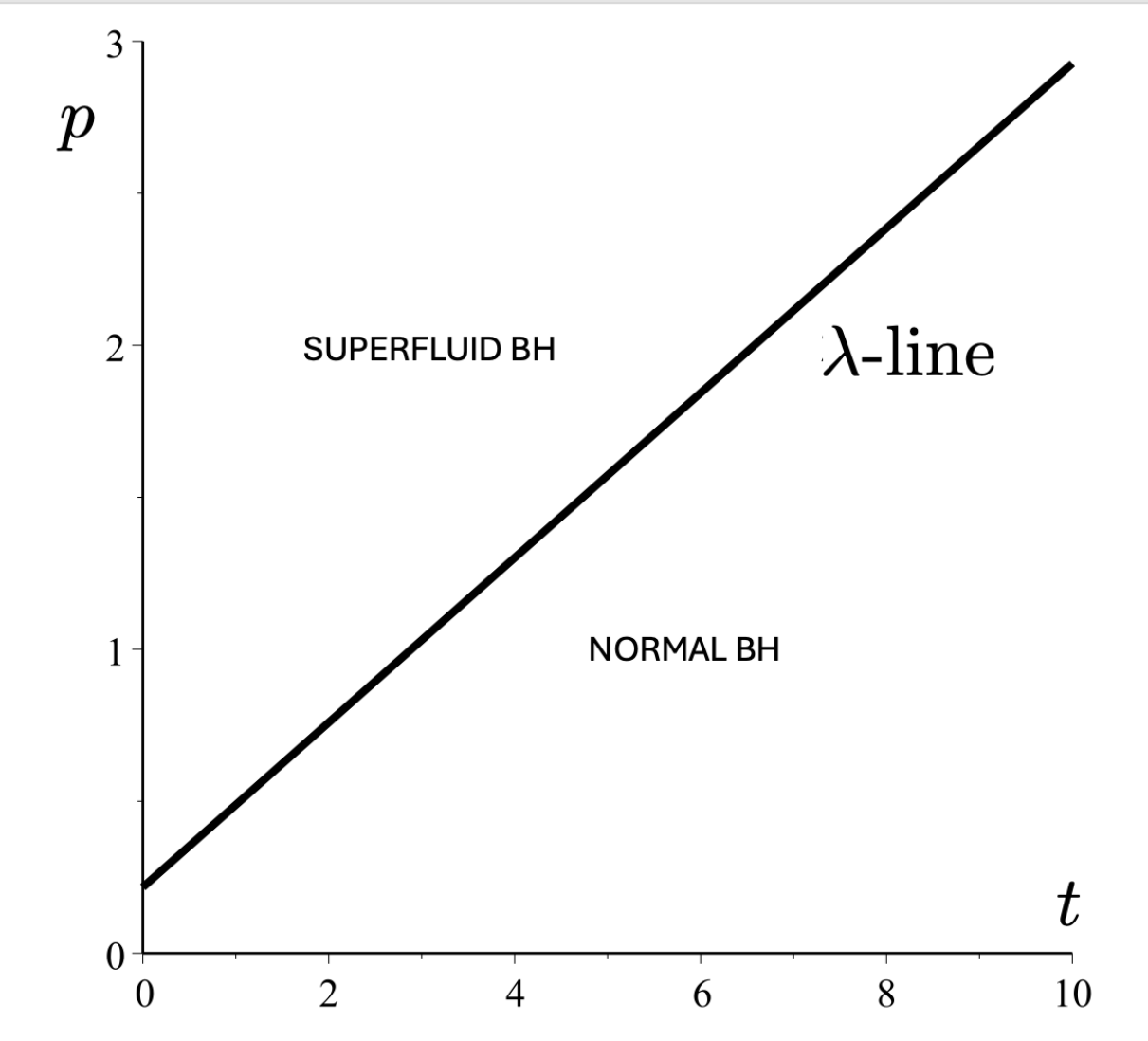} 
\caption{{\bf Phase Diagram for a $D=7$ superfluid black hole}: The black line shows a line of second-order phase transitions known as the 
$\lambda$-line in the context of superfluidity. }
\label{lamline}
\end{figure*}

The conditions for a critical point are
\be\label{eqn:cp_condition} 
\frac{\partial p}{\partial v}=\frac{\partial^2 p}{\partial v^2}=0  
\ee 
since $v$ is proportional to the specific volume in \eqref{specvol}.  These equations have the solution
\be \label{supercrit}
p_c  = \left[ \frac{8 \sqrt[^4]{3375}}{225} \right]t_c + \frac{\sqrt{15} (11D-40)(D-1)(D-2)}{900 \pi D}
\ee
provided $k=-1$, $\alpha = \sqrt{5/3}$, and 
\be 
h=\frac{4(2D-5)(D-2)^2v_c^{D-6}}{\pi D (D-4)} \qquad q^2= \frac{2(D-1)(D-2) v_c^{2D-10}}{ \pi (D-4)} 
\ee
where  $v_c = \sqrt[^4]{15}$.  In other words, the critical pressure $p_c$ is, from \eqref{supercrit}, a linear function
of the critical temperature $t_c$!   There is no first order phase transition but rather a line of second order phase transitions, 
in the $(t,p)$ plane,  illustrated in Fig.~\ref{lamline} for $D=7$. This black hole system with scalar hair exhibits \textit{infinitely many critical points}: in the $p-v$ plane, every isotherm is a critical isotherm, with an inflection point at $v= \sqrt[^4]{15}$.

\begin{figure*}[!htp]
\centering
\includegraphics[width=0.48\textwidth,height=0.3\textheight]{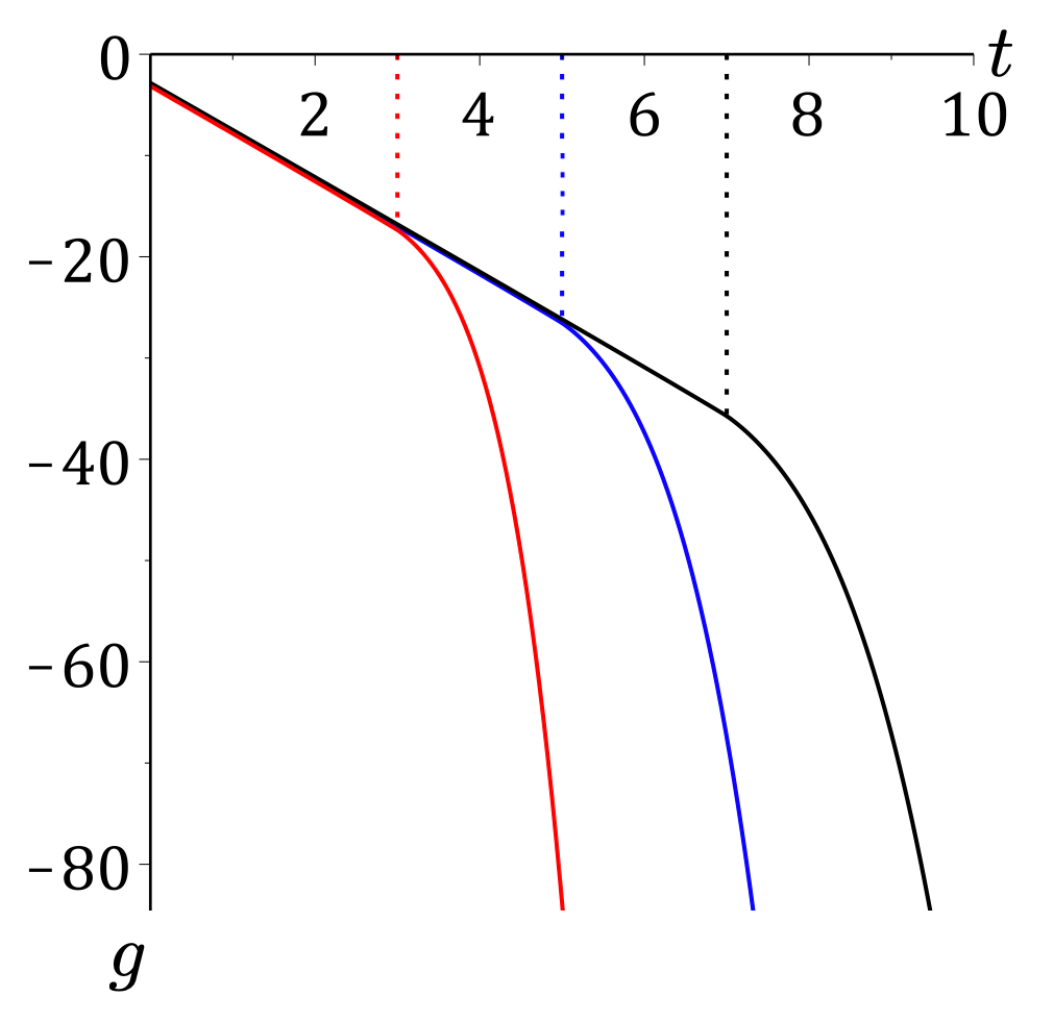}  
\includegraphics[width=0.48\textwidth,height=0.3\textheight]{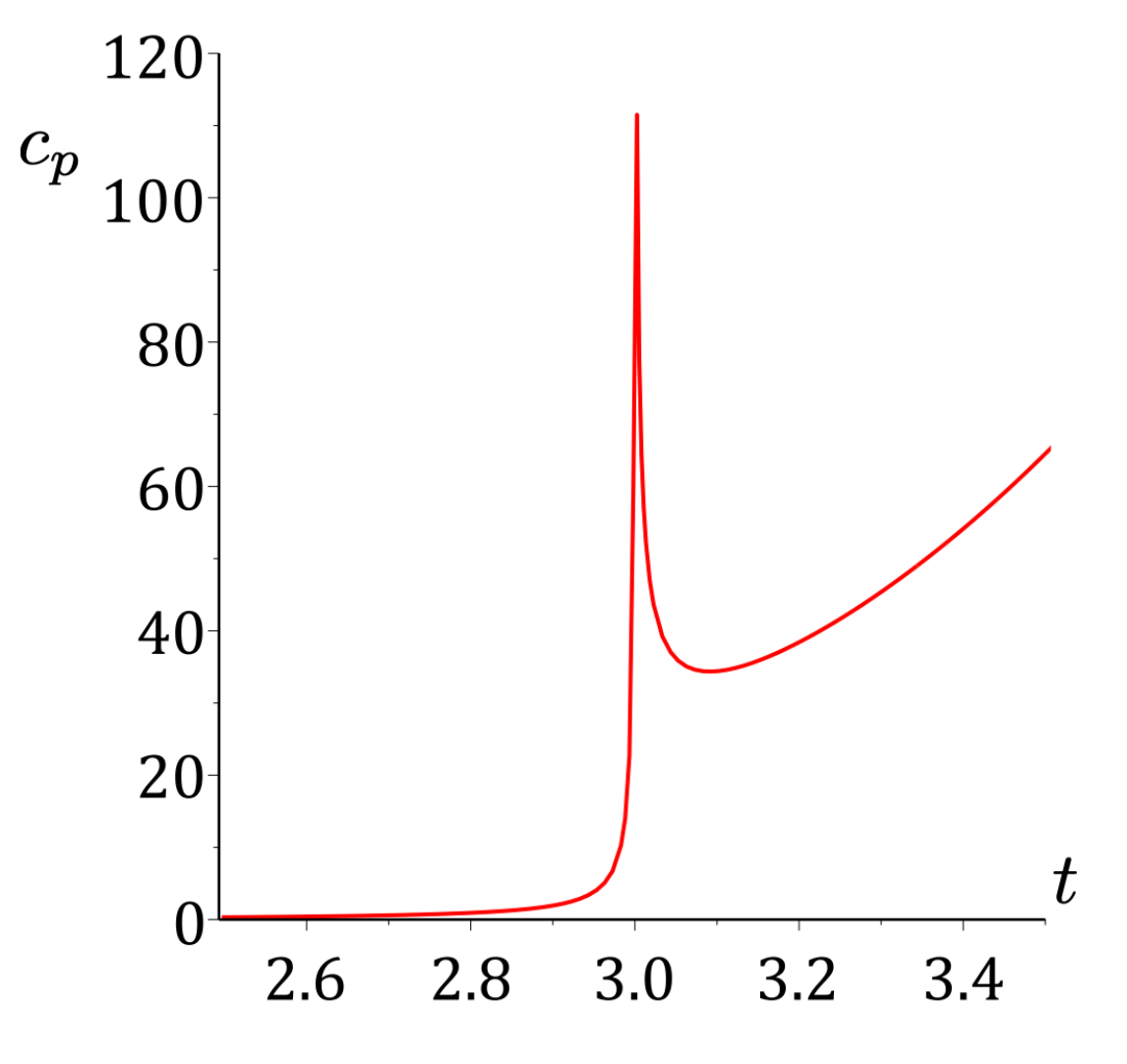}
\caption{{\bf Thermodynamic behaviour near $\lambda$ transition in $D=7$}: \textit{Left}: The dimensionless Gibbs free energy vs. temperature for three distinct pressures, chosen so that critical temperatures are  $t_c=3,5,7$ respectively corresponding to the red, blue and black curves.  Dotted lines highlight the points where the second derivative of the Gibbs free energy diverges.  \textit{Right}: A plot of the specific heat $c_p=-t\frac{\partial^2 g}{\partial t^2}$ for the case $t_c = 3$. }
\label{cl_diverginggibbs}
\end{figure*}

A plot of the dimensionless Gibbs free energy
\be
g =   \frac{M- TS}{\alpha_3^{\frac{(D-3)}{4}} \Omega_k} 
\ee
along with the specific heat $c_p = -t \, \partial ^2 g / \partial t^2$ is given in Fig.~\ref{cl_diverginggibbs}.  The latter diagram is quite striking: it fully resembles the empirical
$\lambda$ shape seen in superfluid $^4$He~\cite{RevModPhys.67.279}. 

The essential feature needed for obtaining black hole superfluidity is that the conditions \eqref{eqn:cp_condition}  for a critical point can be satisfied without fixing the temperature.  For an equation of state of the form,
\be 
P = a_1(V, \varphi_i) \,  T + a_2(V, \varphi_i)
\ee
where $a_1$ and $a_2$ are functions (and where
the $\varphi_i$ represent additional constants in the equation of state), the superfluidity criterion is satisfied provided a nontrivial solution for the following equations exists:
\begin{align}\label{supercriteqs}
\frac{\partial a_i}{\partial V} &= 0 \, , \quad  \frac{\partial^2 a_i}{\partial V^2} = 0 \,  \quad i=1,2   
\end{align}
which requires at least four  free parameters. For the black hole solution \eqref{mastereqn} these are the 
parameters $(v, \alpha, q, h)$). All cubic-and-higher Lovelock theories with conformal scalar hair can satisfy these requirements \cite{Hennigar:2016xwd}.  

 Neither  higher order Lovelock gravity without scalar hair nor rotating black holes in $5$-dimensional minimal gauged super-gravity~\cite{Chong:2005hr} admit superfluidity \cite{Hennigar:2016xwd}.  However  quasi-topological black holes with scalar hair do exhibit superfluid transitions for both the $k=-1$ hyperbolic and $k=1$ spherical horizon geometries \cite{Dykaar:2017mba}.  Recently superfluid phase transitions have been observed for charged AdS black holes in the context of holographic black hole chemistry \cite{Bai:2023woh}, discussed below in section~\ref{6p3}.
 Clearly finding the necessary and sufficient conditions for satisfying \eqref{supercriteqs} is not trivial, and as of this writing are not known.

\subsection{NUT charge}
\label{5p5}

NUT charged black holes are amongst the more intriguing vacuum solutions of general relativity.  These solutions 
have a gravitational charge analogous to the magnetic charge of a magnetic monopole.  The associated metric, the Taub-NUT metric \cite{Taub:1950ez, Newman:1963yy}  has two Killing horizons, a Kruskal extension \cite{Miller:1971em},  and no curvature singularity, though it does have  a string singularity on the polar axis (known as a  Misner string singularity) with spacetime regions having closed timelike curves in its vicinity. For these reasons Misner suggested imposing periodicity of the time coordinate so as to render the string unobservable \cite{Misner:1963fr},
analogous to what is done for the Dirac string of a magnetic monopole.  This has the pathological effect of introducing closed timelike curves throughout the spacetime, and furthermore renders the maximal extension of the spacetime problematic \cite{Misner:1963fr,HawkingEllis:book,Hajicek:1971}.   

Not long ago it was pointed out that for freely falling observers the original Taub-NUT spacetime is both geodesically complete and  free from causal pathologies \cite{Clement:2015cxa,Clement:2015aka}. Indeed the Misner string is completely transparent for  geodesics,
and no closed timelike or null geodesics exist in the spacetime (provide some restrictions are imposed on the parameters). Though 
non-geodesic observers can in principle violate causality, it is conjectured that their energy will induce a   backreaction  that will modify the spacetime so as to preserve causality \cite{Clement:2015cxa,Clement:2015aka}. 
For these reasons  imposing periodicity on  the time coordinate was   abandoned and   exploration of their
thermodynamics was initiated \cite{Hennigar:2019ive,Bordo:2019slw,BallonBordo:2019vrn}.  

The Lorentzian Taub-NUT-AdS solution is \cite{Hawking:1998ct,Clement:2015cxa}
\ba\label{ltna1}
ds^2&=&-f\Bigl[dt+2n(\cos\theta +\sigma) d\phi\Bigr]^2+\frac{dr^2}{f}  +(r^2+n^2)(d\theta^2+\sin^2\!\theta d\phi^2) \\
f&=&\frac{r^2-2mr-n^2}{r^2+n^2}- \frac{3n^4-6n^2r^2-r^4}{\ell^2(r^2+n^2)}\,,
\ea
where $m$ is the mass parameter and $n$ is the NUT charge, with $\ell$ defined in \eqref{eq:press}. The constant $\sigma$ determines the position of the Misner string: for $\sigma = - 1$ the north pole axis is regular, for $\sigma = +1$ the south pole axis is regular,
and for $\sigma = 0$ there is a string on both\footnote{Identifying the time $t\sim t+8\pi n$ renders all such strings  unobservable \cite{Misner:1963fr},  but with the price of introducing the pathologies noted above.}.  The spacetime is geodesically complete for any value of $\sigma$, but  absence of closed timelike and null geodesics requires \cite{Clement:2015cxa} $|\sigma|\leq 1$ .

In computing the thermodynamic parameters for the metric \eqref{ltna1}, the temperature
\be
T =\frac{f'(r_+)}{4\pi}=\frac{1}{4\pi r_+}\Bigl(1+\frac{3(n^2+r_+^2)}{\ell^2}\Bigr)
\ee
is straightforwardly computed, where $r_+$ is the horizon, obtained from \eqref{ltna1} by setting $f(r_+)=0$.  The mass and angular momentum can be determined using the method of conformal completion  \cite{Ashtekar:1999jx,Das:2000cu}, which yields \cite{Hennigar:2019ive}
\be
M = m\,,\qquad J = 3\sigma mn
\ee
and we see that the angular momentum of the spacetime is determined by the positioning of the Misner strings. It has been pointed out that
 NUT charge is a kind of thermodynamic multi-hair insofar as it simultaneously has both rotation-like and electromagnetic charge-like characteristics
 \cite{Wu:2019pzr,Wu:2022xpp}.

Setting $\sigma=0$ (for which closed timelike/null geodesics are absent \cite{Clement:2015cxa}), yields
\be\label{F}
F=\frac{m}{2}-\frac{1}{2\ell^2}(3n^2r_++r_+^3) 
\ee
for the free energy as determined from the a computation of the Euclidean action\footnote{This involves Wick rotating both the time coordinate $t\to i\tau$ and NUT parameter $n\to i \nu$, assuming the periodicity $\tau\sim \tau+\beta$, and then Wick-rotating both back ($\nu\to -in$).}.    

Assuming that the entropy is one-quarter of the horizon area,
the thermodynamic parameters of the solution are  
\begin{align}\label{Nut-thermo}
M &=m\qquad J=0\qquad  T=\frac{1}{4\pi r_+}\Bigl(1+\frac{3(n^2+r_+^2)}{\ell^2}\Bigr) \nn\\
V&=\frac{4}{3}\pi r_+^3\Bigl(1+\frac{3n^2}{r_+^2}\Bigr) \qquad 
S =\pi (r_+^2+n^2)
\end{align}
along with the thermodynamic conjugates
\be\label{quantities}
\psi=\frac{1}{8\pi n}\,,\quad N=-\frac{4\pi n^3}{r_+}+\frac{12\pi r_+n^3}{\ell^2}\Bigl(1-\frac{n^2}{r_+^2}\Bigr)
\ee
that  together satisfy\footnote{An alternate approach toward obtaining the first law regards $M-\psi N$ as the internal energy \cite{Awad:2022jgn,Awad:2023lyt}.} 
\be\label{flawnut}
d M=T dS+\psi dN+V dP 
\ee
generalizing the   first law \eqref{firstBH} to include NUT charge \cite{Durka:2019ajz}.
A generalization to include non-zero angular momentum can also be carried out \cite{Rodriguez:2021hks}.
 These quantities also satisfy the   modified Smarr formula:
\be\label{Smarr2}
M=2(TS-VP+\psi N) 
\ee
and are consistent with the dimensional scaling argument in \eqref{Euler}.  
The free energy \eqref{F} is
\be\label{F2}
F=M-TS-\psi N 
\ee
with $\psi N$ 
analogous  to the $\Omega J$ term in the grand-canonical ensemble.  

One of the peculiar features of this set of thermodynamic variables is that $\psi$ diverges as $n\to 0$, though the product $\psi N$ remains finite in this limit. 
This can be avoided by requiring a different definition $F=M-TS$ for the free energy.  Compatibility of this expression with \eqref{F} yields
\be\label{SNC}
S=S_{\tiny{\mbox{NC}}}=\frac{\pi(3r_+^4+12n^2r_+^2+r_+^2\ell^2-n^2\ell^2-3n^4)}{3n^2+\ell^2+3r_+^2} = S + S_{\tiny{\mbox{MS}}}
\ee
where $S$ is the entropy of the black hole in \eqref{Nut-thermo}, and 
\be\label{SMS}
S_{\tiny{\mbox{MS}}}= \frac{2\pi n^2(3r_+^2-3n^2 - \ell^2)}{3n^2+\ell^2+3r_+^2} 
\ee
is the entropy of the Misner string \cite{Mann1999,Frodden:2021ces}, obtained by Noether charge (NC) methods \cite{Garfinkle:2000ms}.  This in turn implies
\be\label{psi-N1}
\psi_{\tiny{\mbox{NC}}}=-\frac{n(\ell^2+3n^2-3r_+^2)}{2(3n^2+\ell^2+3r_+^2)}\,, \quad N_{\tiny{\mbox{NC}}}=\frac{n}{r_+}+\frac{3n(n^2+r_+^2)}{r_+\ell^2}\,,
\ee
for the remaining conjugate variables.  These variables are both well-defined in the large-$\ell$ (flat space) and $n\to 0$ limits. They are also both dimensionless, in which case the Smarr relation \eqref{Smarr2} becomes $M=2(TS-VP)$, as is easily checked.

Geometric arguments have been put forward to contend that 
the surface gravity of the black hole and its conjugate areal quantity should respectively correspond to the temperature and entropy
of the Taub-NUT black hole   \cite{Bordo:2019tyh}.  For general values of $\sigma$, the  variables $(\psi^\prime_{\textsf{N/S}},N^\prime_{\textsf{N/S}})$ respectively correspond to  the surface gravity and Misner charge  of the Misner strings, with N/S referring to the north/south polar axes.  As in
the $\sigma=0$ case,   one of $\psi^\prime_{\textsf{N/S}}$ diverges at some finite value of $n$.   It is also unclear if  the $\psi^\prime_{\textsf{N/S}}$ should be interpreted as temperatures associated with the strings, and consequently  $N^\prime_{\textsf{N/S}}$ with the corresponding string entropies  \cite{Bordo:2019tyh}. 

The geometric  \cite{Bordo:2019tyh} and Noether-charge \cite{Garfinkle:2000ms} approaches are connected by the relation
\be\label{S-rel}
 S_{\tiny{\mbox{NC}}} =  S  + \frac{\psi^\prime_{\textsf{N}} N^\prime_{\textsf{N}}  +  \psi^\prime_{\textsf{S}} N^\prime_{\textsf{S}}}{T}
\ee
which is easily checked in the $\sigma=0$ case from \eqref{Nut-thermo}, \eqref{SMS}, and \eqref{psi-N1}. 
Under analytic continuation of periodic identification of the temperature \cite{Mann:2004mi},   $\psi^\prime_{\textsf{N}} = \psi^\prime_{\textsf{S}} \propto T$
(this would give $T=1/8\pi n$ for $\sigma=0$), and the Noether charge entropy would equal the entropy from the horizon plus 
$ N^\prime_{\textsf{N}}  +  N^\prime_{\textsf{S}}$ of the strings \cite{Bordo:2019tyh},  suggesting these quantities can be regarded as string entropies.

When both electric and magnetic charge are present,   there is further ambiguity in the choice of thermodynamic potentials   \cite{Bordo:2019slw}.
The metric has the same form as in \eqref{j1}, but with
\be
f = \frac{r^2-2mr-n^2+4n^2 g^2+e^2}{r^2+n^2}- \frac{3n^4-6n^2r^2-r^4}{\ell^2(r^2+n^2)} 
\ee
and with a gauge potential
\be
\textsf{A} = -\left(\frac{er + g(r^2-n^2)}{r^2+n^2}\right)\left(dt+2n\cos\theta d\phi\right)
\ee
with electric charge $e$ and magnetic charge $g$.
 These electromagnetic charges both depend (via Gauss' law) on the radius of the sphere over which the field strength and its dual are integrated.  One can either take the magnetic charge to be the value at the horizon and the electric charge to be that at infinity, or vice-versa.  Each yields a distinct possible version of the thermodynamic first law, with differing  thermodynamic NUT charges, related to each other by
electromagnetic duality \cite{Bordo:2019slw,Chen:2019uhp}.  

From the geometric perspective  \cite{Bordo:2019tyh} the free energy is given by \eqref{F2} in
the fixed charge ensemble, where $h(r_+) = 0$ is imposed.  This condition is called the `purely electric' condition,
since it determines the magnetic charge $g$ in terms of $e$, with the total charge $Q=e$ determined to be that at infinity.  
The Gibbs free energy becomes
\be\label{FG}
G=\frac{m}{2}-\frac{(3n^2r_++r_+^3)}{2\ell^2} + \frac{e^2 (r_+^2+ n^2) }{(r_+^2 - n^2)} 
\ee
The remaining thermodynamic variables remain the same as in \eqref{Nut-thermo} and \eqref{quantities}, except for
\begin{align}
T &=\frac{1}{4\pi r_+}\Bigl(1+\frac{3(n^2+r_+^2)}{\ell^2} -  \frac{e^2 (r_+^2+ n^2) }{(r_+^2 - n^2)^2} \Bigr)  \nn\\
N &= -\frac{4\pi n^3}{r_+}\left( 1+ \frac{3(n^2-r_+^2)}{\ell^2}  -  \frac{e^2 (3r_+^2+ n^2) }{(r_+^2 - n^2)^2} \right)
\label{TNcharged}
\end{align}
 
\begin{figure*}[!htp]
\centering
\includegraphics[width=0.48\textwidth,height=0.3\textheight]{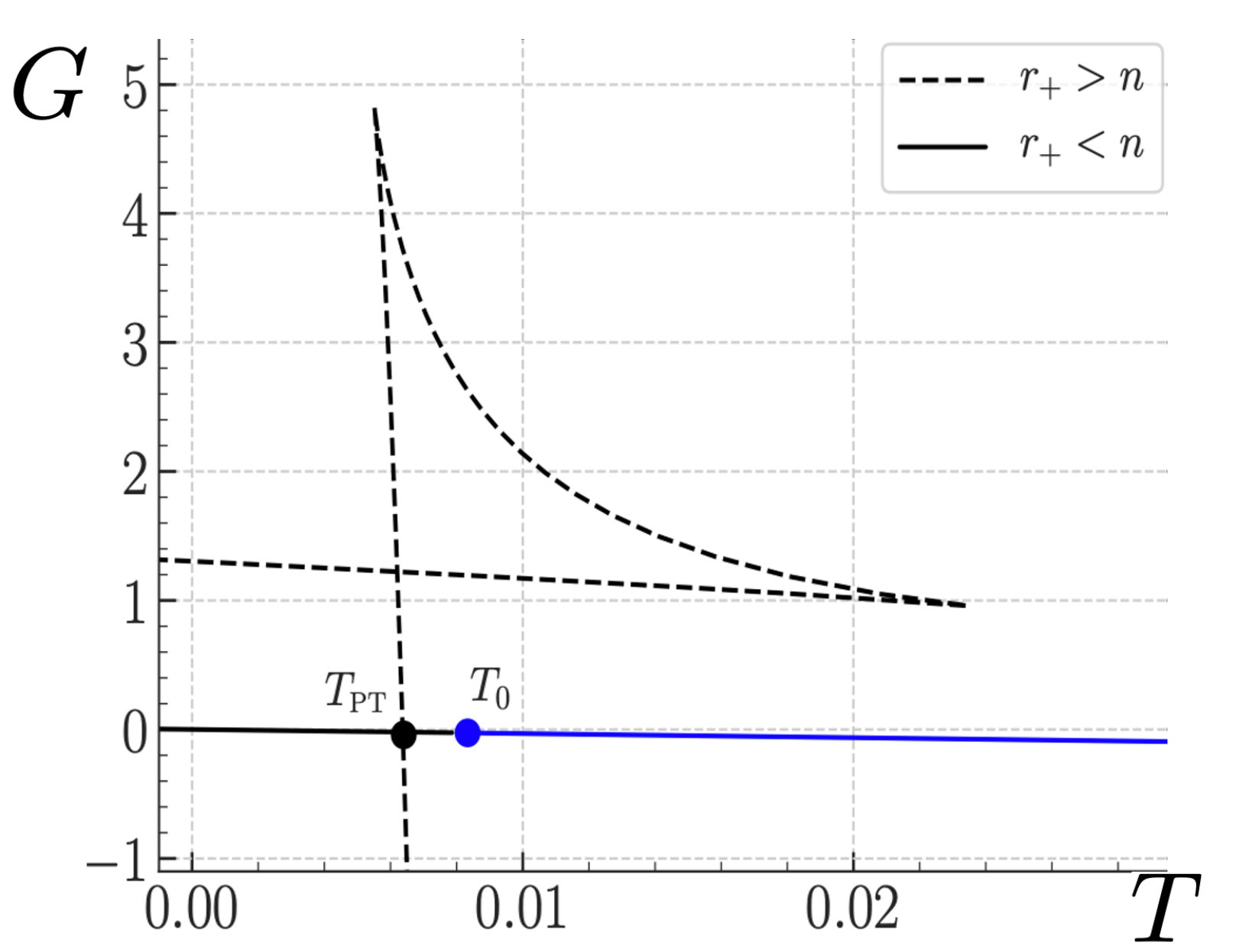}  
\includegraphics[width=0.48\textwidth,height=0.3\textheight]{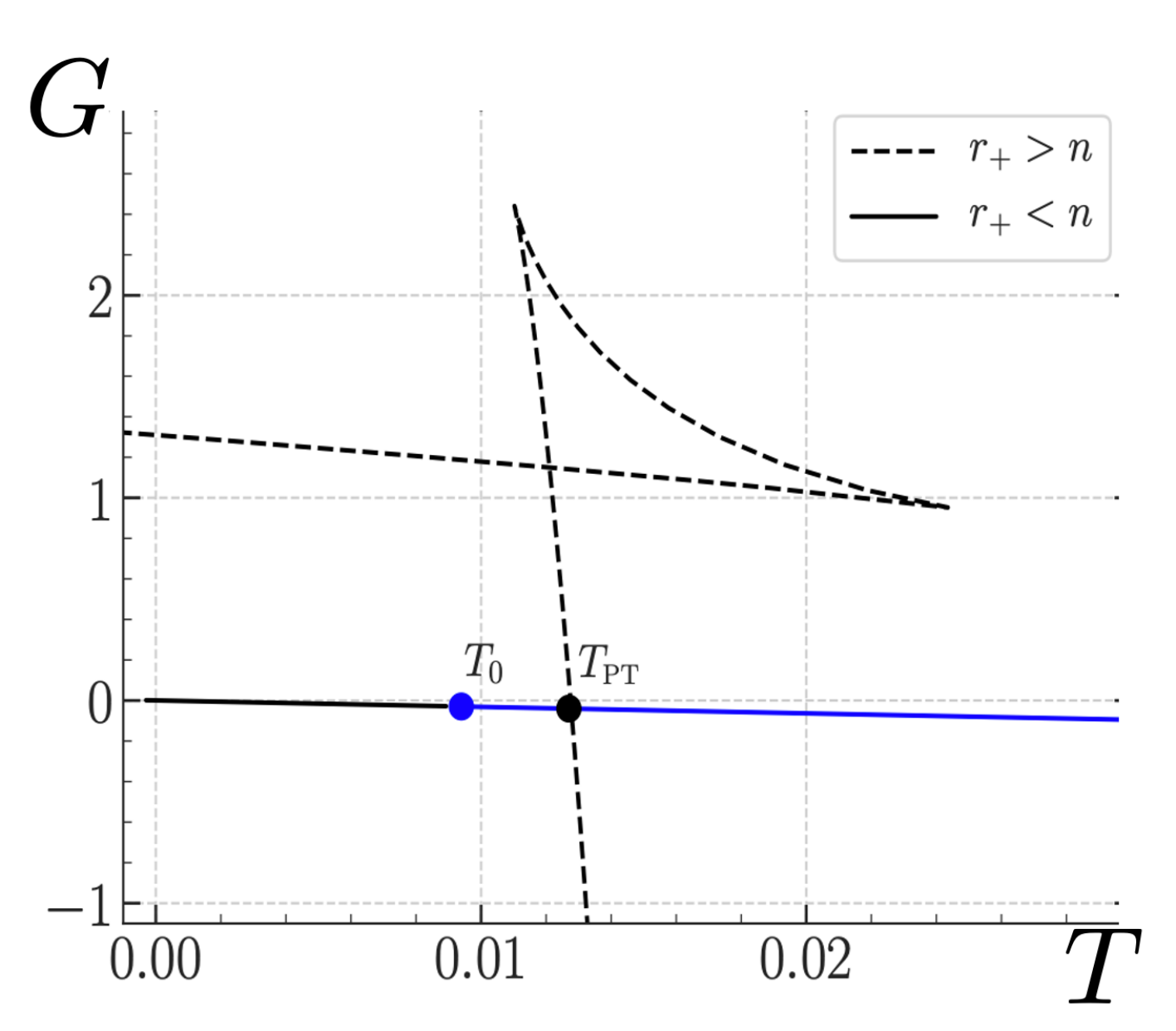}
\caption{{\bf  Free Energy of the Purely Electric Taub-NUT Black Hole}\\ 
For both panels $n = 1.006$ and $Q = 1$, with $P_t <0$. The left panel A  has $\ell=50$ (lower pressure) and the right panel B has
$\ell=25$ (higher pressure).  In both cases there are two Taub– NUT branches: a 
  standard swallowtail branch (dashed curve) of $r_+ > n$ positive mass black holes, and a lower branch (solid curve) of  $r_+ < n$
  black holes, whose radius increases from right to left. For $T > T_0$ these have negative mass (blue solid curve); the mass $M = 0$ at $T=T_0$. 
 At low pressures (left panel A) $T_0 > T=T_{\textrm{PT}}$, and there is a single large-small first-order phase transition.
 At higher pressures (right panel B) $T_0 < T=T_{\textrm{PT}}$, this transition  no longer exists if negative mass solutions are not considered;
 Instead, a zeroth order phase transition occurs at $T_0$ as the system transits from small Taub-NUT black holes on the lower branch to intermediate ones on the swallowtail branch. Further increasing the temperature,  a standard intermediate to large first-order phase transition occurs at the  swallowtail intersection.}
\label{geoGT}
\end{figure*}
 
The free energy  \eqref{FG} is plotted in Fig.~\ref{geoGT} for two different values of the pressure. For pressures below a critical value 
$P_c$ swallowtail behaviour is present (dashed black line), similar to that of charged AdS black holes in Fig.~\ref{Fig:Swallow}, indicative of a first-order
phase transition from a large Taub–NUT black hole to a small one as the temperature decreases.  However for pressures greater than
$P_t = \frac{Q^2-n^2}{8\pi n^4}$ there is another (lower) branch of small Taub–NUT solutions, whose radius decreases as the temperature increases.
If $P_t<0$ this lower branch always exists.  Most of these black holes on this branch have negative mass (shown by the blue line in the figure), but for sufficiently low temperatures the mass  becomes positive (solid black line).  The negative mass branch terminates at $T=T_0$ (at which point $M=0$), which may be greater or smaller than the intersection of the swallowtail branch with the lower branch at $T=T_{\textrm{PT}}$   \cite{Bordo:2019tyh}.

\begin{figure*}[!htp]
\centering
\includegraphics[width=0.48\textwidth,height=0.3\textheight]{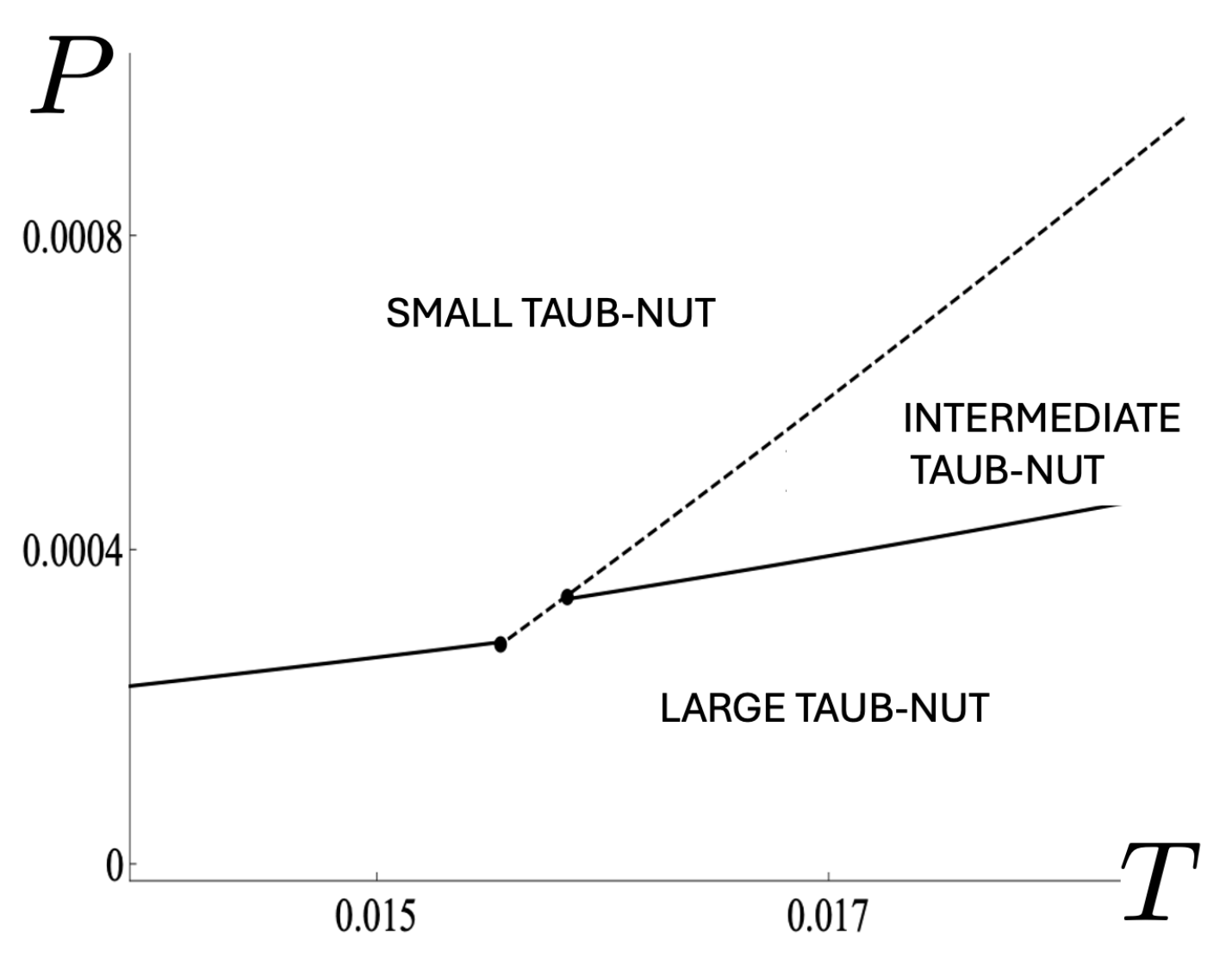}  
\caption{{\bf  Phase Diagram of the Purely Electric Taub-NUT} 
For  $P_t < P <P_c$,  $Q=0.98$ and  $n=1$ the large-intermediate-small behaviour is visible if negative mass solutions are not admitted.  
First/zeroth order phase transitions are indicated by the solid/dashed   lines.  The lower solid line corresponds to the first order transition
at  the intersection of the two branches at $T = T_{\textrm{PT}} < T_0$, shown in the left panel of Fig.~\ref{geoGT}.  For  $P_t < 0$ 
this   transition extends all the way to $T=0$. The zeroth order transition occurs when $T_{\textrm{PT} } > T_0$, shown in the right panel of Fig.~\ref{geoGT}.  
Increasing temperature from $T=0$, there is  a jump from the lower small Taub–NUT branch to the upper intermediate  branch of the swallowtail.
Further increasing the temperature, a  first order  standard swallowtail  phase transition occurs (upper solid line);
this branch  eventually terminates at a critical point.}
\label{NUTPT}
\end{figure*}

For low pressures $P_c > P > 0 > P_t$, $T_0 > T_{\textrm{PT}}$, and 
 there is a first-order phase transition from the large Taub-NUT branch to the small one of
positive mass, which takes place at the temperature $T_{\textrm{PT}}$ shown in the left panel of Fig.~\ref{geoGT}. All black holes on the swallowtail part 
for $T< T_{\textrm{PT}}$ are thermodynamically unstable.   As the pressure increases (right panel of Fig.~\ref{geoGT}), $T_0 < T_{\textrm{PT}}$,
and new phase behaviour can occur, assuming negative mass solutions (blue line) are unphysical. Beginning at high temperatures,
the Van der Waals  first order transition takes place at the intersection point on the (dashed) swallowtail, as the temperature decreases;
there is no transition at $T = T_{\textrm{PT}}$.  Further decreasing the temperature, a second transition of 0th order  occurs from the upper
(dashed) branch to the lower (solid) one at $T=T_0$.  In other words there is a large-intermediate-small Taub-NUT sequence of transitions
at any fixed pressure within  the range above the upper solid line  in the phase diagram  shown in Fig.~\ref{NUTPT}.  If negative mass solutions are admitted, this
large-intermediate-small behaviour does not occur; the only transition is a large-small one at $T=T_{\textrm{PT}}$   \cite{Bordo:2019tyh}.

A thorough study comparing this case, with entropy $S$ given in \eqref{Nut-thermo}, to that with entropy $S_{\tiny{\mbox{NC}}}$ given 
in \eqref{S-rel} for charged black holes, was carried out for a variety of thermodynamic ensembles \cite{Abbasvandi:2021nyv}:
\begin{enumerate}
\item Fixed electric and magnetic charges, fixed $N_{\tiny{\mbox{NC}}}$
\item Fixed electric and magnetic charges, fixed $\psi_{\tiny{\mbox{NC}}}$
\item Fixed electrostatic potential, fixed magnetic charge, fixed $N_{\tiny{\mbox{NC}}}$
\item Fixed electrostatic potential, fixed magnetic charge, fixed $\psi_{\tiny{\mbox{NC}}}$
\item Fixed magnetostatic potential, fixed electric charge, fixed $N_{\tiny{\mbox{NC}}}$
\item Fixed magnetostatic potential, fixed electric charge, fixed $\psi_{\tiny{\mbox{NC}}}$
\end{enumerate}
A broad variety of novel phenomena were observed including interrupted swallowtails (in which the usual large/small transition becomes unstable, and instead a large/tiny first order phase transition takes place as the temperature decreases), breaking swallowtails (in which a swallowtail is replaced by a cusp
with  second order critical behaviour occurring as the pressure increases), charge-changing transitions (in which at fixed electric potential,
large positively charged black holes undergo a first order swallowtail transition to small negatively charged ones, and vice-versa, depending on the
parameter choice), and more.   None of this behaviour provided further criteria that were  significant in distinguishing which choice
of the entropy, $S$ or $S_{\tiny{\mbox{NC}}}$ is preferable.

A further examination of the 
 thermodynamics of Euclidean dyonic Taub-NUT-AdS  black holes \cite{Mann:2020wad} indicated a relationship between  gauge field regularity conditions and thermodynamic relations. In order that  both regularity and the first law of thermodynamics are satisfied, the norm of the gauge field is required to vanish at the horizon, provided it is of non-zero size.  This regularity condition in turn yields a constraint on the magnetic and electric  charges, reducing
 the  cohomogeneity of the system. Removing the Misner string singularity \cite{Misner:1963fr}   further reduces cohomogeneity.  Solutions   with increasing electric charge have positive heat capacity, but dyonic solutions (with both electric and magnetic charge)  have both positive and negative heat capacity.  The extremal solution has finite-temperature-like behaviour, with the electric potential playing a role similar to temperature \cite{Mann:2020wad}.
 
  Thermodynamics of NUT-charged spacetimes have also been studied in the context of the more general Plebanski solution \cite{Liu:2022wku},
 higher dimensions \cite{Wu:2022mlz}, non-linear curvature corrections \cite{Chen:2024knw}, 
  braneworlds \cite{Siahaan:2022ecb}, conformal electrodynamics \cite{BallonBordo:2020jtw,Zhang:2021qga}, cosmic censorship  \cite{Yang:2023hll,Yang:2020iat}, inclusion of additional topological invariants in the action \cite{Ciambelli:2020qny}, planar horizon geometries \cite{Cano:2021qzp}, background magnetization \cite{Ghezelbash:2021lcf,Siahaan:2021uqo}, and scalarization \cite{Liu:2024bzh}. 
 The thermodynamics of Lorentzian NUT charged black holes is 
 still under active research with their physical relevance still very much an open question.  Presumably some deeper quantum gravitational description will indicate whether or not the degrees of freedom of their fundamental microstructure are associated only with the horizon.

\subsection{Heat Engines}
\label{5p6}

A heat engine in thermodynamics is a system that extracts work from two  reservoirs at different temperatures.  
The hot reservoir at temperature $T_H$ is a heat source of effectively infinite heat capacity, providing  thermal energy (or heat) to the working substance (the part of the system doing the work). Not all of the supplied heat gets converted into work;  part of it is dumped into the cold reservoir, which is at temperature $T_C$.  The rest of the heat is used to 
carry out the desired work. During its operation, the heat engine passes through a series of thermodynamic processes, and so completes a cycle.
The fact that not all heat can be converted into work implies that a given heat engine has an efficiency
\be\label{effic}
\eta = \frac{W}{Q_H} = 1 - \frac{Q_C}{Q_H}
\ee 
defined as the ratio of the
work done by the engine to the heat energy extracted from the hot reservoir. The maximally efficient heat engine is a Carnot engine, whose cycle
consists of adiabatic paths connecting the two systems at their respective temperatures (and so is fully reversible), and for which 
\be\label{maxeff}
\eta = 1 -  \frac{T_C}{T_H}
\ee

The introduction of pressure into black hole thermodynamics naturally suggests the notion of a heat engine, and a few years after the advent of  black hole chemistry, these were introduced~\cite{Johnson:2014yja}.  They are described by cycles in the pressure-volume space that extract work from AdS black holes used as the working material. 

\begin{figure}
\centering
\begin{tabular}{cc}
\includegraphics[width=0.49\textwidth,height=0.3\textheight]{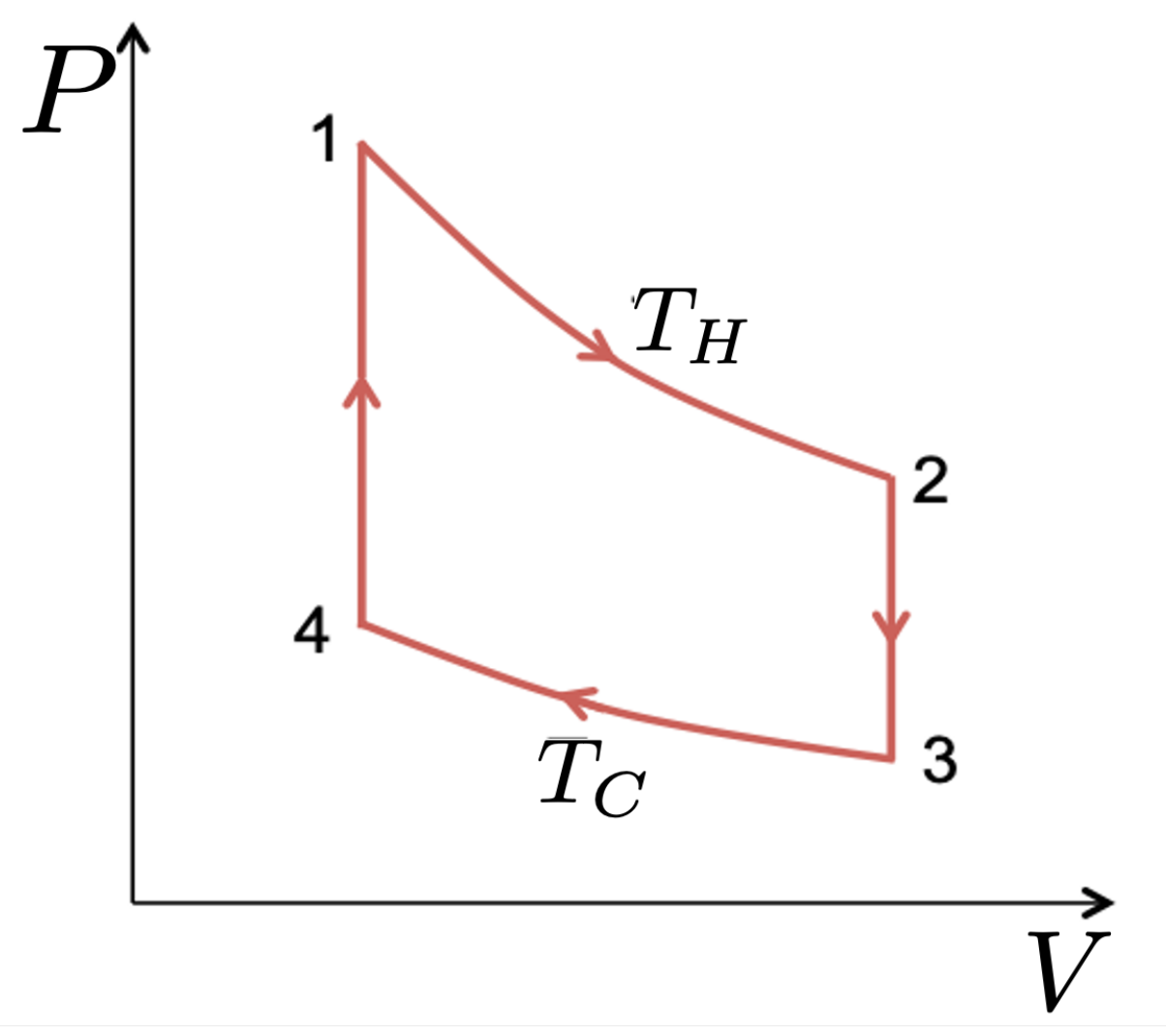} 
\includegraphics[width=0.49\textwidth,height=0.3\textheight]{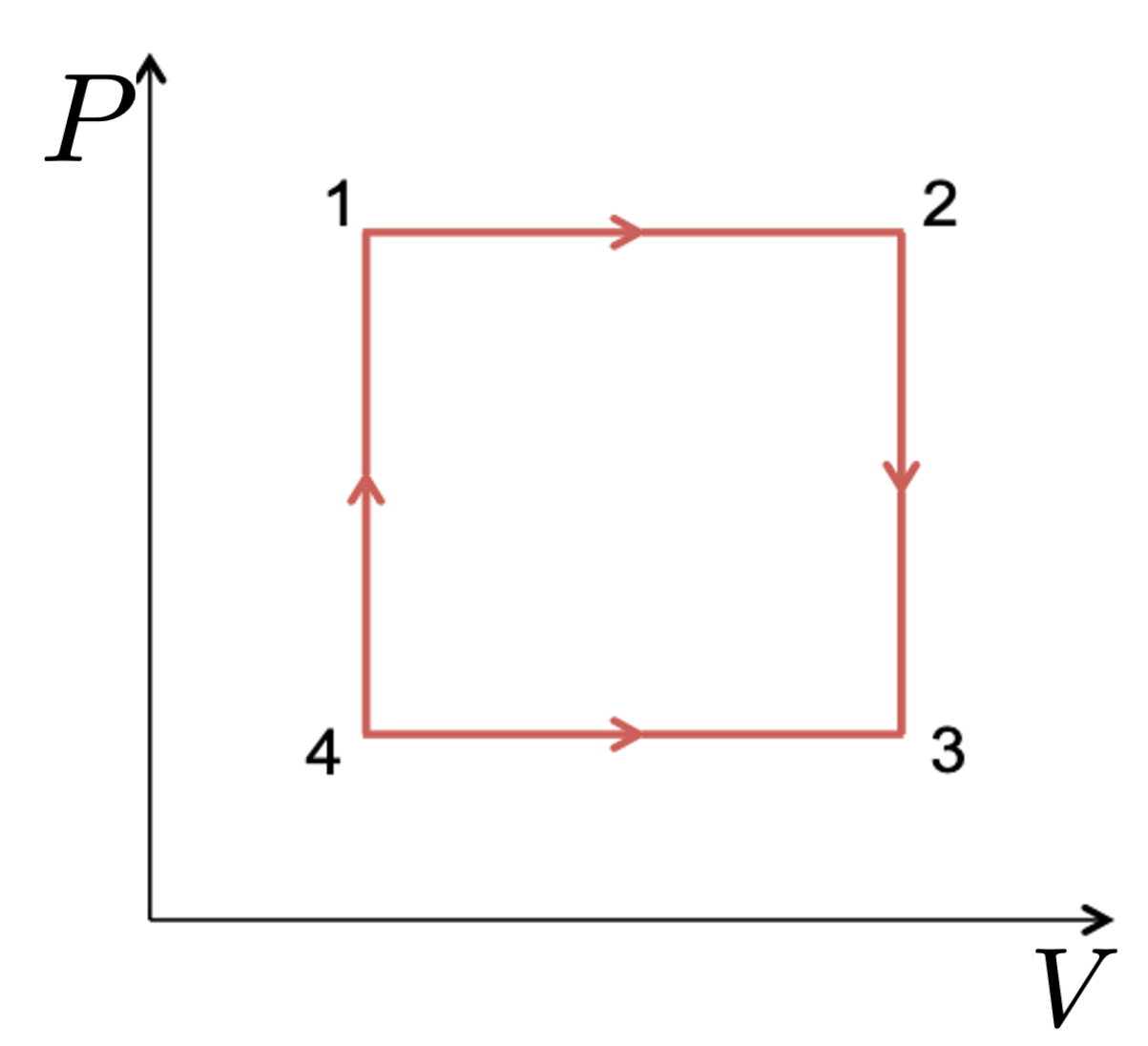} 
\end{tabular}
\caption{{\bf $P-V$ diagram of thermodynamic cycles.} {\it Left.} A diagram of the Carnot cycle. The isothermal paths are 12 and 34; the adiabatic
paths are 23 and 41.  For a black hole this is also a Stirling cycle since the adiabatic paths are also constant volume (isochoric) paths.
{\it Right.} A rectangular cycle.
}
\label{Fig:cycles}
\end{figure}

While the actual engineering of a black hole heat engine would seem to be an engineering task for an advanced civilization, it is possible to obtain
a number of interesting results.  For static black holes, whose entropy and volume are not independent, 
 the traditional maximally efficient  Carnot engine is also a Stirling engine~\cite{Johnson:2014yja} (whose cycle consists of isochoric paths), shown at left 
 in Fig.~\ref{Fig:cycles}.
 A number of subsequent results for black hole heat engines considered  the effects of higher curvature corrections~\cite{Johnson:2015ekr},
 non-linear electrodynamics~\cite{Johnson:2015fva,Balart:2021glm}, dilatons~\cite{Bhamidipati:2016gel}, rotation~\cite{Sadeghi:2016xal,EslamPanah:2019szt,Debnath:2020zdv,Roy:2023qqy,Zhong:2023lhc}, acceleration
 \cite{Zhang:2018hms} and 
 more~\cite{Zhang:2016wek,Wei:2016hkm, Setare:2015yra, Sadeghi:2015ksa,Mo:2017nhw,Liu:2017baz,Johnson:2017ood,Mo:2017nes,Hendi:2017bys,Wei:2017vqs,Zhang:2018vqs,Johnson:2018amj}.  
 Other cycles, including Otto cycles (whose cycles have  heat exchange only on isochoric curves in the left panel of Fig.~~\ref{Fig:cycles}), Brayton cycles (consisting of two adiabatic curves and two isobaric curves), and Diesel cycles (consisting of two adiabatic curves, an isobaric curve, and an isochoric curve),  have more recently been studied~\cite{Johnson:2019olt}.

Different kinds of black holes will have different efficiencies, depending on their parameters, and so a notion of benchmarking was developed
to compare various types of  black hole engine cycles~\cite{Chakraborty:2016ssb}.   Originally such considerations were restricted to 
vanishing specific heat at constant volume, $C_V = 0$, which necessarily excludes effects of rotation.  This was subsequently remedied
and a `benchmark catalogue' produced~\cite {Hennigar:2017apu}.

Consider a rectangular cycle, shown at right in Fig.~\ref{Fig:cycles}. The combinations $(V_i, P_j)$ denote coordinates of the corners, where
the subscripts $L$, $R$, $T$, and $B$ respectively refer to  ``left" , ``right", ``top" and ``bottom".   Along an isobar
\begin{align} 
\delta Q &= TdS = dM - VdP
\Rightarrow Q_{isobar} = \int_a ^b d M = M_b - M_a 
\end{align}
whereas along an isochore
\begin{align} 
\delta Q &= TdS = dU + PdV 
\Rightarrow Q_{isochore} = \int_a ^b d U = U_b - U_a 
\end{align}
where $U = M - PV$ is the relation between the internal energy $U$ and the enthalpy (mass) $M$. All  additional work terms in the first law
\eqref{firstBH} are assumed to be fixed along the isobars and isochores. Taking $(a,b)$ to be the appropriate corners of
the rectangle in Fig.~\ref{Fig:cycles} yields
\begin{align}
Q_C &=  \Delta M_T  - \Delta P V_R  
\qquad
Q_H =  \Delta M_T  - \Delta P V_L  
\end{align}
where $\Delta M_T = M(V_R, P_T)  - M(V_L, P_T)$ and  $\Delta P = P_T - P_B$.  Inserting these expressions into \eqref{effic} yields
\be \label{rect-exact}
\eta_{\sc{RECT}} = \frac{\Delta V \Delta P}{\Delta M_T  + \Delta U_L} 
\ee
where $U_L = U(V_L, P_T) - U(V_L, P_B)$. The only assumptions made in obtaining \eqref{rect-exact} are that the first law \eqref{firstBH} holds and
that the cycle is carried out quasi-statically;   consequently  the efficiency \eqref{rect-exact} of a rectangular cycle applies to any AdS black hole.

An exact result can also be obtained for elliptical cycles.  The parametric equations
\begin{align}
P(\theta) &= P_0(1+  p \sin \theta )\qquad V(\theta) = V_0(1+  v \cos \theta )\, .
\end{align}
describe an ellipse in the $(V,P)$ plane, where the  dimensionless quantities
 $p$ and $v$  correspond to the size of the axes of the ellipse.  The contributions to $Q_H$ and $Q_C$ come
 from the respective top and bottom parts of the ellipse,
 \begin{align} \label{QCell}
Q_{C \atop H}  &= \Delta M -  \pm \int_{\theta = 0}^{\theta = \pi } V_0(1 + v \cos \theta)P_0p \cos \theta d\theta 
 = \Delta M \mp P_0 V_0 \frac{\pi p v}{2} 
\end{align}
yielding 
\be\label{eqn:exact_circle}
\eta = \frac{2 \pi }{\pi + \frac{2}{pv} \frac{\Delta M }{P_0V_0}} 
\ee
from \eqref{effic}, 
where $\Delta M = M\left(V_0 (1 + v), P_0\right) - M(V_0 (1 - v), P_0)$. 

The exact  expression \eqref{eqn:exact_circle} for the efficiency assumes that $C_V = 0$, which allows a tiling of the ellipse with infinitesimally small rectangles, implying in turn \eqref{QCell}.   If this does not hold then the limits of integration  in determining the heat become unknown, and one must  integrate $TdS$ numerically. 

A lower bound on the efficiency \eqref{eqn:exact_circle} is easily obtained by recognizing that $\Delta M  <  M_R = M(V_0 (1 + v), P_0)$, and an upper bound can be obtained by noting that  the largest permissible value of $p$ is $p=1$ for any elliptical cycle.   It can then be shown that
\cite{Hennigar:2017apu}
\be 
   \eta_{\rm min } = \frac{2 \pi }{\pi + 2 M_R/P_0 p V_0 v}  \leq \eta \leq  \frac{2 \pi }{\pi + 4} 
\ee 
where equality on the left side is obtained in the limit $v = 1$ ( the cycle is as large as possible).  If $M_R<0$ (as can be the case for   some hyperbolic black holes~\cite{Mann:1997jb}) then the lower bound is zero.   The upper bound (on the right hand side)  is universal,  independent of both theory and spacetime dimension; equality is obtained for extremal black holes in the small cycle limit.

  \begin{figure}[htp]
\centering
\includegraphics[width=0.75\textwidth]{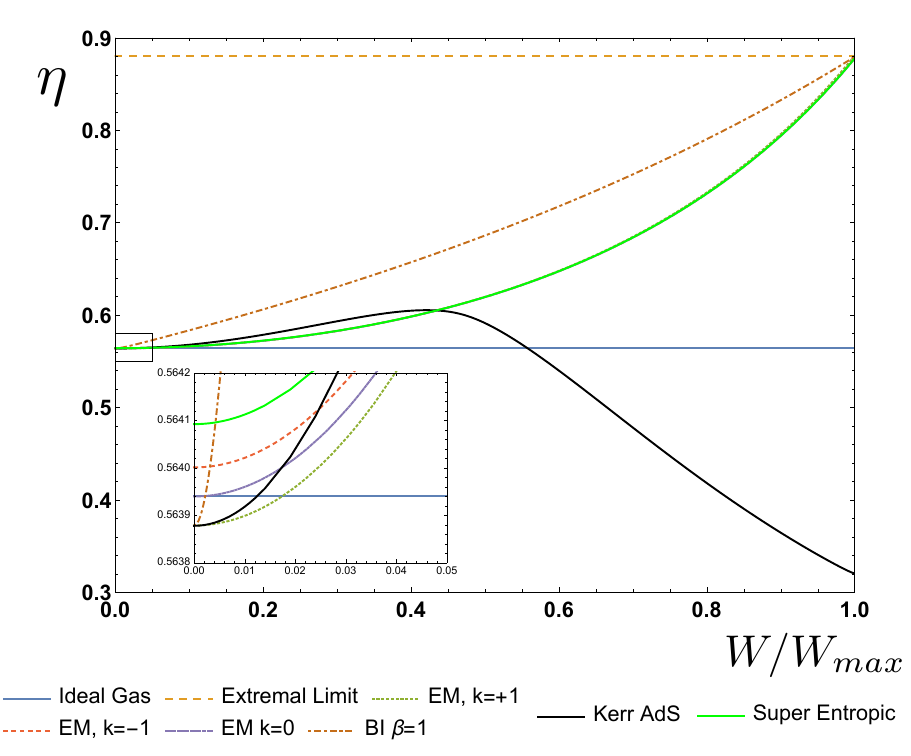}
\caption{{\bf Benchmarking curves} 
The quantity $W$ represents the work term (e.g. electric charge) and $W = W_{\rm max}$ gives an extremal black hole at one point on the cycle. In each case, the cycle was centered at $(V_0, P_0) = (200, 20)$. Here $k=-1,0,1$ corresponds to the horizon topology of the metric 
}
\label{fig:bench_marking}
\end{figure}

Fig.~\ref{fig:bench_marking} depicts plots of benchmarking curves for various black hole heat engines,
where EM denotes an Einstein-Maxwell-AdS black hole of mass $M$,  charge $Q$,  and topology $k$ in \eqref{Schw-Ads} and \eqref{dOmegak} with
\be
f = k - \frac{2m}{r^{D-3}} +\frac{q^2}{r^{2(D-3)}}  + \frac{r^2}{\ell^2}
\ee
in $D$ spacetime dimensions, and 
where BI denotes a Born-Infeld black hole \cite{Gunasekaran:2012dq},  
with metric function
\begin{align}
f &= 1 -  \frac{2m}{r^{D-3}} + \frac{r^2}{\ell^2}
 + \frac{4 \beta^2 r^2}{(D-1)(D-2)} \left[1 - \sqrt{1 + \frac{(D-2)(D-3)q^2}{2 \beta^2 r^{2D - 4}}} \right]
\nn\\
&+ \frac{2(D-2) q^2}{(D-1)r^{2D-6}}  {}_2F_{1} \left[\frac{D-3}{2D-4}, \frac{1}{2}, \frac{3D - 7}{2D - 4}, - \frac{(D-2)(D-3)q^2}{2 \beta^2 r^{2D-4}} \right]\, ,
\end{align}
in \eqref{Schw-Ads} with $k=1$, where ${}_2F_1$ is the hypergeometric function and $\beta$ is the Born-Infeld parameter.  The parameters $(m,q)$
are given by
\be
 m = \frac{16 \pi M}{\Omega_k (D-2)}. \qquad q = \frac{8 \pi Q}{\Omega_k \sqrt{2(D-2)(D-3)}}
\ee
with $\Omega_k$ given by \eqref{volsphere} for $k=1$.  The super-entropic black hole \cite{Hennigar:2014cfa} is a particular ultraspinning limit of a 
rotating black hole \eqref{KAdSmetric}. The ideal gas case and the extremal limit  are included for comparison, and
in cases where $C_V \neq 0$,  a numerical integration was carried out to obtain the heat \cite{Hennigar:2017apu};  the curves in Fig.~\ref{fig:bench_marking} are for $D=4$.
 
Normalizing the work $W$ (from charge or angular momentum) by the value $W_{\rm max}$ corresponding to an extremal black hole,  we see a
number of interesting results.  The efficiencies of all black holes  having $C_V = 0$  approach $\eta_\circ$ in the extremal limit.  The Kerr-AdS solution, with $C_V \neq 0$, attains a peak in its efficiency near $J/J_{\rm max} \approx 0.5$ and then becomes rapidly less efficient as the extremal limit is approached.  
The location and height of the maximum  depend on the details of the benchmarking cycle chosen.  The  super-entropic black hole \cite{Hennigar:2014cfa} also has $C_V \neq 0$, 
and has a benchmarking curve that closely follows  those of the  $C_V = 0$, and  in the extremal limit its efficiency approaches $\eta_\circ$.  The most efficient black hole is that of  non-linear   Born-Infeld electrodynamics, except for very small values of the charge (or $W$), shown in the inset, where the super-entropic black hole is most efficient.

 The efficiency of a black hole heat engine can approach the Carnot efficiency while maintaining finite power in the vicinity of a critical point~\cite{Johnson:2017hxu}.  This phenomenon can be  characterized to show  how the rate of approach to the Carnot efficiency is governed by the critical exponents~\cite{DiMarco:2022yhp}. In the case of the isolated critical discussed in section~\ref{5p3}, this approach can be used to show that  even-order Lovelock black holes with isolated critical points cannot exist, as this would constitute a violation of the second law of thermodynamics.

\subsection{Joule-Thompson Expansion}
\label{5p7}

The Joule-Thompson expansion is a chemical process in which gas at  high pressure passes through a porous plug to a region of  low pressure, during which the expansion enthalpy is constant \cite{Johnston_2014}. Often  used to liquefy gases, all real gases  at ordinary temperatures and pressures,  except hydrogen and helium,  cool in this process. As charged black holes thermodynamically resemble Van der Waals fluids \cite{Kubiznak:2012wp} it is natural to ask how black holes behave in a Joule-Thompson process \cite{Okcu:2016tgt}.

As the gas passes from the high pressure region to the low-pressure one (with constant enthalpy $H$) the  temperature changes with respect to the pressure, characterized by 
\begin{equation}\label{JT1}
\mu=\left(\frac{\partial T}{\partial P}\right)_{H} \ .
\end{equation}
where $\mu$ is called the Joule-Thomson coefficient.  The sign of $\mu$ determines whether the gas cools or heats during the expansion.  Since it is an expansion, the  pressure decreases (so $dP < 0$) but the temperature may either increase or decrease. If $\mu$ is negative (positive)
the  temperature increases (decreases)  and the gas warms (cools). From table~1, the first law in terms of enthalpy is 
\begin{equation}\label{firstLawGeneralH}
dH=TdS+VdP 
\end{equation}
and for fixed enthalpy ($dH=0$)  this  becomes
\begin{equation}\label{enthalph2}
0=T\left(\frac{\partial S}{\partial P}\right)_{H}+V \ .
\end{equation}
The differential $dS$ is given by
\begin{equation}\label{entrop}
dS=\left(\frac{\partial S}{\partial P}\right)_{T}dP+\left(\frac{\partial S}{\partial T}\right)_{P}dT \ 
\end{equation}
since entropy is a state function. This  can be rearranged, yielding
\begin{equation}\label{entrop2}
\left(\frac{\partial S}{\partial P}\right)_{H}=\left(\frac{\partial S}{\partial P}\right)_{T}+\left(\frac{\partial S}{\partial T}\right)_{P}\left(\frac{\partial T}{\partial P}\right)_{H} 
= -\left(\frac{\partial V}{\partial T}\right)_{P} + \frac{1}{T} C_{P}\left(\frac{\partial T}{\partial P}\right)_{H}
\end{equation}
using the Maxwell relation $\left(\frac{\partial S}{\partial P}\right)_{T}=-\left(\frac{\partial V}{\partial T}\right)_{P}$.
Inserting \eqref{entrop2} into \eqref{enthalph2} and solving for $\left(\frac{\partial T}{\partial P}\right)_{H}$ yields
\begin{equation}\label{JT2}
\mu=\left(\frac{\partial T}{\partial P}\right)_{H}=\frac{1}{C_{P}}\left[T\left(\frac{\partial V}{\partial T}\right)_{P}-V\right] 
\end{equation}
expressing the Joule-Thompson coefficient in terms of volume and heat capacity at constant pressure.

We see from \eqref{JT2} that  $\mu$ will vanish if
\begin{equation}\label{iT}
T_{i}=V\left(\frac{\partial T}{\partial V}\right)_{P} \ 
\end{equation}
where $T_{i}$ is the {\it inversion temperature}: it demarcates the boundaries between  the heating and cooling regions in the $T-P$ plane.

For a Van der Waals fluid,  the equation of state \eqref{VdWstate} inserted into \eqref{iT}
 yields (noting  $V=v/N$)
\begin{equation}\label{ie2}
T_{i}= \left(P_{i}v-\frac{a}{v^{2}}(v-2b)\right) 
\end{equation} 
for the  inversion temperature,  where $P_{i}$ is  the inversion pressure.  Inserting $(T_i,P_i)$ into \eqref{VdWstate}
and subtracting this from \eqref{ie2} gives 
\begin{equation}\label{roots}
v=\frac{a\pm\sqrt{a^{2}-3ab^{2}P_{i}}}{bP_{i}}  
\end{equation}
which in turn gives
\begin{equation}\label{uplow}
T^{\pm}_{i}=\frac{2\left(5a-3b^{2}P_{i}\pm4\sqrt{a^{2}-3ab^{2}P_{i}}\right)}{9bk} \ ,
\end{equation}
upon insertion into  (\ref{ie2}).  Equations \eqref{uplow} together give the inversion curve in the $T-P$ plane separating the cooling region 
 from the warming one for a Van der Waals fluid.

For the charged AdS black hole \eqref{HDRN} the procedure is similar.  Using \eqref{specvol} to replace $r_+$ with $V$ in the equation of state \eqref{RNstate} gives the inversion temperature
\begin{eqnarray}\label{iTRnAdS}
&T_{i}=\frac{1}{3}\left[\frac{Q^{2}}{V}-\left(\frac{6}{\pi}\right)^{\frac{2}{3}}\frac{1}{12V^{\frac{1}{3}}}+ \left(\frac{6V}{\pi}\right)^{\frac{1}{3}} P_{i}\right]\ =\frac{Q^{2}}{4\pi r_{+}^{3}}-\frac{1}{12\pi r_{+}}+\frac{2P_{i}r_{+}}{3}
\end{eqnarray}
from \eqref{JT2}. Inserting $(T_i,P_i)$ into \eqref{RNstate} and eliminating $r_+$ using \eqref{iTRnAdS} gives
\begin{equation}\label{iTRnAdS3}
T_{i}=\frac{\sqrt{P_{i}}}{\sqrt{2\pi}}\frac{\left(1+16\pi P_{i}Q^{2}-\sqrt{1+24\pi P_{i}Q^{2}}\right)}{\left(-1+\sqrt{1+24\pi P_{i}Q^{2}}\right)^{\frac{3}{2}}} 
\end{equation}
for the inversion curve in the $T-P$ plane.  The enthalpy of the black hole is its mass, given by \eqref{TDchgBH}, 
or alternatively  
\be
M = \frac{\pi v^2}{4} T + \frac{2 Q^2}{v} + \frac{\pi v^3}{6} P
\ee
using the Smarr relation \eqref{RNSmarr} and \eqref{specvol}, 
where $v$  is regarded as a function of $(T,P,Q)$ via the equation of state \eqref{RNstate}.

\begin{figure}[htp]
		\centering
		\includegraphics[width=0.49\textwidth]{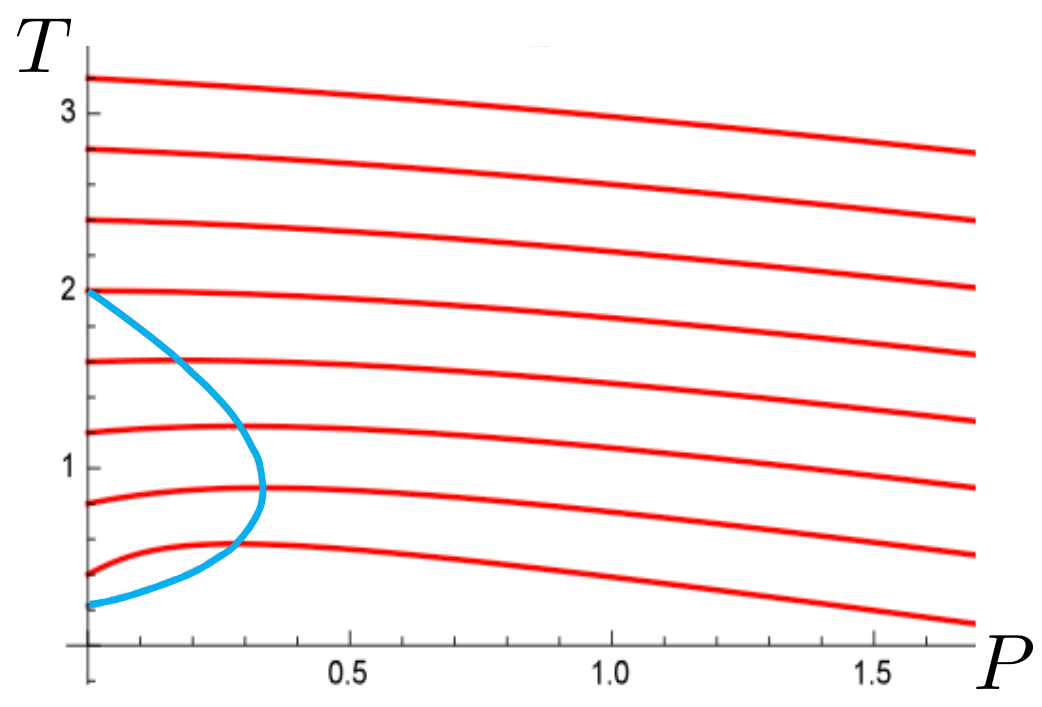}
		\includegraphics[width=0.49\textwidth]{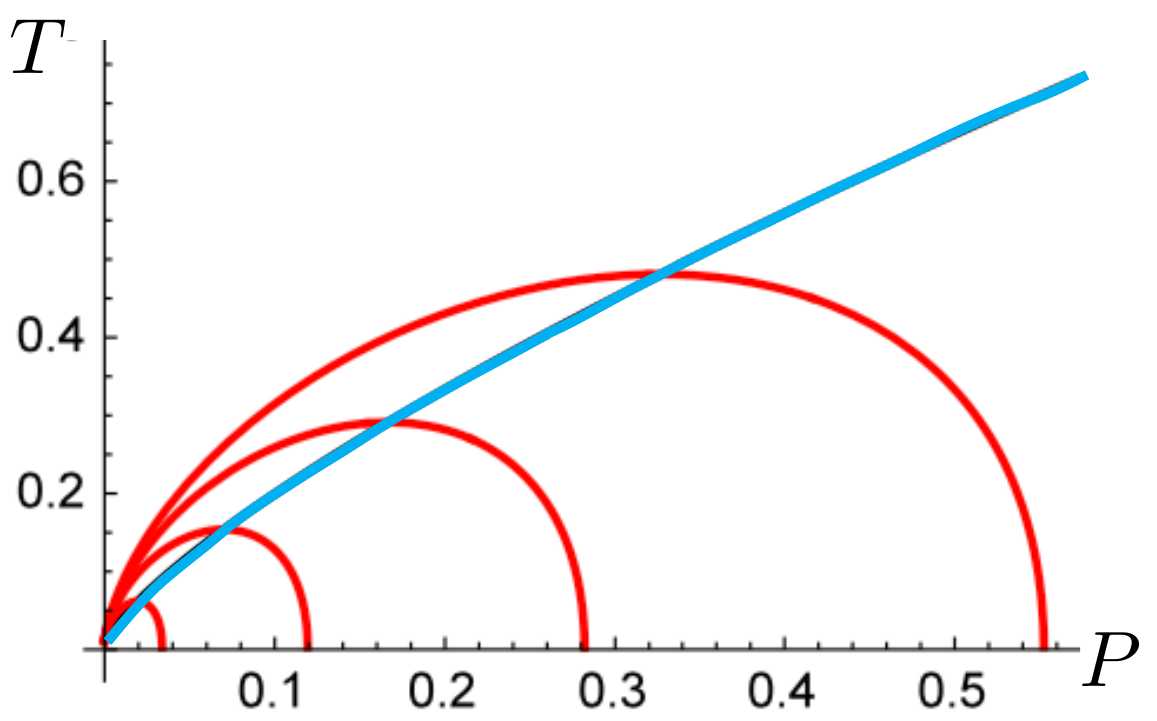}
	\caption{Inversion curves (blue) and isenthalpic curves (red) for a Van der Waals fluid (left) and a charged AdS black hole (right). 
In the left panel $a=b=1$ in \eqref{uplow} and \eqref{VdWenth}; from bottom to top the enthaplies are $H=1,2,3,4,5,6,7,8$. In the right panel
   $Q=2$ and the enthalpy (mass) is $M=2.5,3.0,3.5,4.0$ from lower-left to upper right.}
	\label{iE}
\end{figure}

A comparison of the inversion curves for the Van der Waals fluid and the charged AdS black hole is shown in Fig.~\ref{iE}.
We see a notable difference between   the Van der Waals fluid (left panel), whose
enthalpy is
\begin{equation}\label{VdWenth}
H(T,v)=\frac{3 T}{2}+\frac{Tv}{v-b}-\frac{2a}{v} 
\end{equation} 
(where $v=v(T,P)$ from \eqref{VdWstate})
and the charged AdS black hole (right panel).  In the Van der Waals case the inversion curve bounds a cooling region of positive $\mu$.
Isenthalpic curves have positive slopes inside this region that vanish at its boundary and become negative outside in the heating region.
Charged AdS black holes, however have an unbounded cooling region above the blue curve.  Note also for the Van der Waals fluid that
there is a largest pressure  $P=\frac{a}{3b}$, above which only a heating region exists and where the inversion curve is extremized; for
the charged black hole the inversion curve has no extremum, extending to arbitrarily large values of $P$.

The Joule-Thompson expansion has been applied to a broad range of black holes in a variety of contexts, including rotation, non-linear electrodynamics,
higher curvature corrections, and more \cite{Okcu:2017qgo,Mo:2018rgq,Chabab:2018zix,Mo:2018qkt,Lan:2018nnp,RizwanCL:2018cyb,Cisterna:2018jqg,Nam:2019zyk,Sadeghi:2020bon,Hegde:2020cdm,Bi:2020vcg,Guo:2020ysx,Zhang:2021kha,Cao:2022hmd,Barrientos:2022uit,Kruglov:2022bhx,Sekhmani:2023est,Kruglov:2023ogn}. The generic behaviour
is the same as shown in the right panel of Fig.~\ref{iE}: the cooling region is at higher temperatures and lower pressures, and the warming region
is at lower temperatures and higher pressures.

\subsection{Acceleration}
\label{5p8}

Although accelerating black hole solutions to the Einstein equations were obtained
not long after the advent of general relativity \cite{Weyl:1917gp}, their interpretation
 was a subject of discussion for decades afterward. Their  thermodynamics 
was particularly puzzling.  Conflicting results were obtained regarding the
relationship between the conserved mass and its thermodynamic counterpart, 
the relationship between the action and the free energy, and the nature of the conical deficits that
appear in the solution(s). It was not clear that 
a consistent formulation was even possible for these objects \cite{Appels:2016uha,Appels:2017xoe,Gregory:2017ogk,
Astorino:2016xiy,Astorino:2016ybm}.   These inconsistencies were only resolved fairly recently for  
accelerating AdS black holes  in a certain range of parameter space, and a consistent thermodynamics
for charged, rotating, and accelerating black holes has now been established
\cite{Anabalon:2018ydc,Anabalon:2018qfv}.

Accelerating black holes are described by the {\it C-metric}
\cite{Kinnersley:1970zw,Plebanski:1976gy,Dias:2002mi,Griffiths:2005qp}, an exact solution
to the Einstein-Maxwell-AdS equations (and variants, such as $f(R)$ gravity \cite{Zhang:2019vpf}), given by
\begin{align}
ds^2 &= \frac{1}{H^2}\bigg\{ -\frac{f(r)}{\Sigma}\Big[
\frac{dt}{\alpha}-a\sin^2\!\theta \frac{ d\varphi}{K} \Big]^2 + \frac{\Sigma}{f(r)}dr^2 
+ \frac{\Sigma r^2}{h(\theta)}d\theta^2 \nonumber\\
& \qquad \qquad \qquad + \frac{h(\theta) \sin^2\!\theta}{\Sigma r^2} \Big[\frac{adt}{\alpha}-(r^2+a^2)
\frac{d\varphi}{K}\Big]^2\bigg\}
\label{eq:Cmetric}
\end{align}
where the metric functions are  
\begin{eqnarray}
f(r)&=(1-A^2r^2)\bigg[1-\frac{2m}{r}+\frac{a^2+e^2}{r^2}\bigg]
+\frac{r^2+a^2}{\ell^2}\,,\\
{h(\theta)} &= 1+2mA\cos\theta+\bigg[A^2(a^2+e^2)
-\frac{a^2}{\ell^2}\bigg]\cos^2\!\theta\,,\\
\Sigma &=1+\frac{a^2}{r^2}\cos^2\!\theta\,, \qquad
H=1+Ar\cos\theta 
\end{eqnarray}
and $m,a,e,A$, and $K$ are the respective mass, rotation, charge, acceleration, and conical deficit parameters in the solution.
The range of $r$ is constrained by the conformal factor $H$, so $A\cos\theta \leq  1/r  \leq 1/r_+$; when $\cos\theta < 0$, $1/r$ crosses the origin and  the boundary is situated `beyond infinity'.  A black hole will be present provided  $f(r)$ has  at least one root in the range $r\in(0,1/A)$.
The gauge field one-form $\textsf{B}$ and its field strength $\textsf{F}$ are
\be
\textsf{F}=d\textsf{B}\,,\qquad \textsf{B}=-\frac{e}{\Sigma r}\Big[\frac{dt}{\alpha}
-a\sin^2\!\theta \frac{d\varphi}{K}\Big]  +\Phi_t dt\,,
\label{eq:elec}
\ee
with
\be\label{Bt}
\Phi_t=\frac{er_+}{(a^2+r_+^2)\alpha}
\ee
chosen 
so that the gauge potential, defined by $-\xi\cdot B$,
vanishes at the horizon, where $\xi=\partial_t+\Omega_H\partial_\varphi$
is the generator of the horizon whose angular velocity is 
 $\Omega_H$.

The metric \eqref{eq:Cmetric} has a string singularity along the polar axes, which can be seen by
expanding the  $\theta-\varphi$ section of the metric near $\theta=\theta_\pm=0,\pi$
respectively:
\be\label{nearpole}
ds_{\theta,\varphi}^2 \propto d\theta^2 +  h^2(\theta) \sin^2\theta
\frac{d\varphi^2}{K^2} \sim d\vartheta^2 + (\Xi\pm2mA)^2 \vartheta^2 d\varphi^2
\ee
where
\be
\Xi = 1 - \frac{a^2}{\ell^2} + A^2 (e^2+a^2)
\ee 
and  the  conical deficit is parameterized by $K$ so that the periodicity of $\varphi$ is  $2\pi$. Near the poles the coordinate  
$\vartheta_\pm = \pm(\theta-\theta_\pm)$ acts as a local radial coordinate.  From \eqref{nearpole} we see that the circumference of  a circle 
${\cal C_\pm}=\Delta \varphi \sqrt{g_{\varphi\varphi}}$ at fixed $\vartheta_\pm$ is not $2\pi \vartheta_\pm$ ($2\pi$ times the proper radius); consequently
there is a deficit angle $\delta_\pm = 2\pi -{\cal C_\pm}/{\vartheta_\pm}$ at each pole. Interpreting these angles as arising due to a cosmic string on each axis implies that the string tensions are
\be\label{eq:deficits}
\mu_\pm  = \delta_\pm/8\pi =\frac{1}{4}\bigg[1-\frac{\Xi \pm 2mA}{K}\bigg] 
\ee
and so the  acceleration is due to a mismatch $\mu_- -\mu_+ = mA/K$ of conical
deficits at the north and south poles.  The  overall deficit in the spacetime is
$\bar{\mu} = (\mu_++\mu_-)/2 = \frac14(1-\Xi/K)$.  A judicious choice of $K$ can remove one
of these tensions, but not both. These  conical defects  have been shown to be a form of true hair -- a new charge that the black hole can carry \cite{Gregory:2019dtq}.

The strings are thus a form of matter that provides an interpretation of
the force that accelerates the black holes. This in turn has been used in a number of scenarios  to demonstrate that the pair creation rate
of black holes is proportional to their entropy \cite{Gregory:1995hd,
Dowker:1993bt,Hawking:1994ii,
Emparan:1995je,Eardley:1995au,Mann:1995vb,Booth:1998pb}. However such metrics have two horizons, an acceleration horizon and a black hole horizon, 
each with their own temperature. This renders   their thermodynamic interpretation problematic, though a first law that takes the acceleration horizon into account has been formulated  \cite{Ball:2020vzo}.

However in the AdS case there exists a parameter regime where the acceleration horizon is absent, given by the condition that
$f(-1/A\cos\theta)$ has no roots.    An additional condition on the parameters is that
\be
mA<\begin{cases}
\frac{1}{2}\Xi\quad\mbox{for}\quad \Xi\in(0,2]\,,\\
\sqrt{\Xi-1}\quad  \mbox{for}\quad \Xi>2\,.
\end{cases}
\ee
ensuring that $h(\theta)>0$ on $[0,\pi]$ thereby preserving the metric signature.  With these constraints in mind
 the thermodynamic quantities and their conjugates are \cite{Anabalon:2018qfv}
\begin{align}
M&= \frac{m(\Xi+a^2/\ell^2)(1-A^2 \ell^2\Xi)}{K\Xi\alpha(1+a^2A^2)} \qquad 
T = \frac{f'_+ r_+^2}{4\pi\alpha(r_+^2+a^2)}  \qquad
S=\frac{\pi(r_+^2+a^2)}{K(1-A^2r_+^2)} \\
Q&= \frac{e}{K} \quad \Phi=\Phi_t=\frac{er_+}{(r_+^2+a^2)\alpha} \\
J& =\frac{ma}{K^2}   \quad \Omega=  \Omega_H-\Omega_\infty\,
=\left ( \frac{Ka}{\alpha(r_+^2+a^2)}\right ) -
\left ( -\frac{aK(1-A^2\ell^2\Xi)}{\ell^2\Xi \alpha(1+a^2A^2)}\right)  \\
P &= \frac{3}{8\pi \ell^2}   \quad
V = \frac{4\pi}{3K\alpha} \left [ \frac{r_+(r_+^2 + a^2)}{(1-A^2 r_+^2)^{2}}
+ \frac{m[a^2(1-A^2\ell^2 \Xi) + A^2 \ell^4 \Xi (\Xi+a^2/\ell^2)]}{(1+a^2 A^2) \Xi} \right] \\
\lambda_\pm &= \frac{r_+}{\alpha(1\pm Ar_+)} - \frac{m}{\alpha}
\frac{[\Xi + a^2/\ell^2 +   \frac{a^2}{\ell^2} (1-A^2\ell^2 \Xi)]}{(1+a^2 A^2)\Xi^2}
\mp \frac{A \ell^2 (\Xi +  a^2/\ell^2 )}{\alpha(1+a^2A^2)}
\label{AdSthermo} 
\end{align}
along with the tensions $\mu_\pm$ defined in \eqref{eq:deficits}, and where
\be\label{alp-eq}
\alpha=\frac{\sqrt{(\Xi+a^2/\ell^2)(1-A^2 \ell^2\Xi)}}{1+a^2A^2}\,.
\ee
 
These quantities can be shown to 
satisfy the Smarr relation \eqref{smarrBH} with $D=4$ and the 
first law \eqref{firstBH}
\be\label{flawacc}
\delta M=T\delta S+\Phi \delta Q+\Omega \delta J-\lambda_+\delta \mu_+
-\lambda_-\delta \mu_- +V\delta P
\ee
necessarily modified to include  the tensions $\mu_\pm$.    Since these latter quantities are dimensionless
they do not appear in the Smarr relation.  The parameter $\alpha$ is  a choice of gauge, chosen so that
$t$ corresponds to the “time” of an asymptotic observer \cite{Anabalon:2018ydc,Anabalon:2018qfv}. A similar situation occurs for  the non accelerating Kerr-AdS metric \cite{Gibbons:2004ai}.
  
\begin{figure}[htp]
		\centering
		\includegraphics[width=0.65\textwidth]{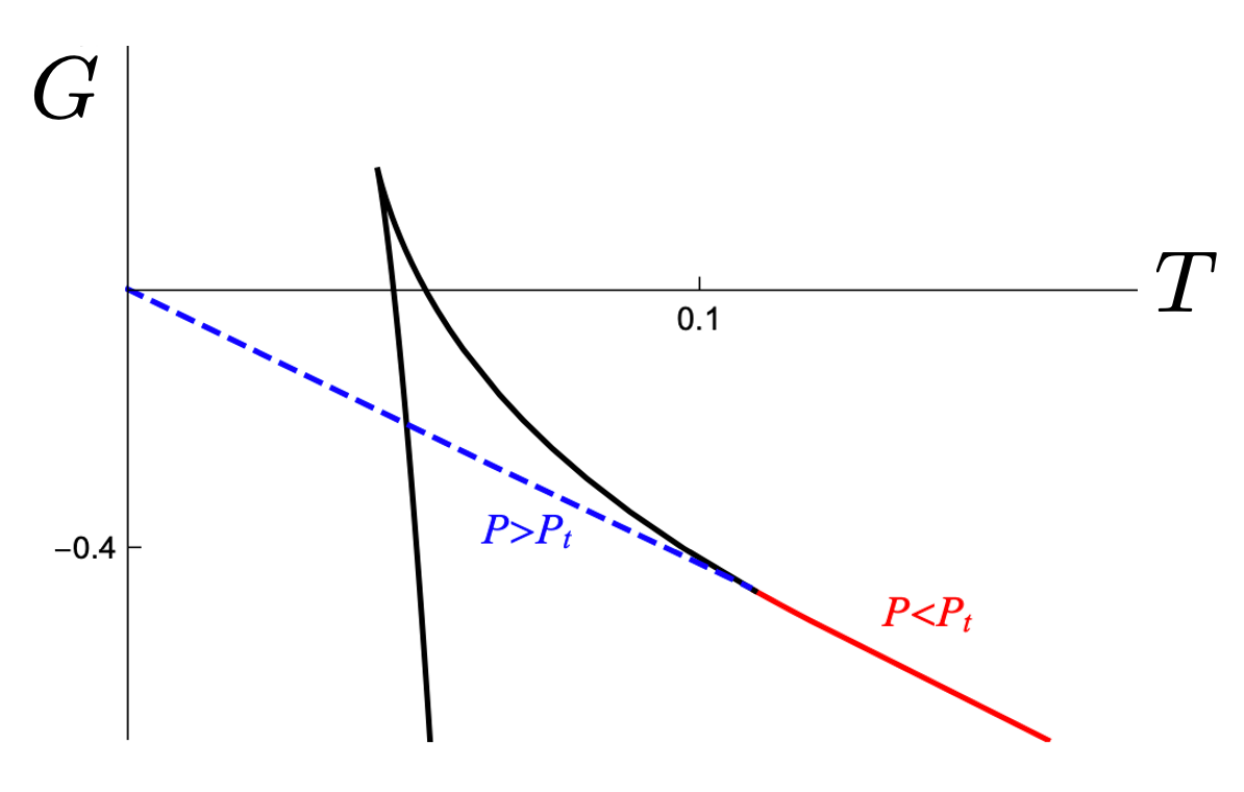}
			\caption{The snapping swallowtail phenomenon occurs for charged accelerating black holes at sufficiently low pressures $P<P_t$.
For $P>P_t$ (blue dashed curve and  solid black curve) a standard swallowtail, similar to that in Fig.~\ref{Fig:Swallow},  is present. However
as $P$ decreases, the swallowtail snaps once $P=P_t$, and  a new branch of black holes displayed by the red curve appears. The cusp point remains invariant at this transition. For $P < P_t$ the free energy is similar to that of Fig.~\ref{Fig:HP} for  the Schwarzschild-AdS black hole, given by the union of red and black curves.}
	\label{snap}
\end{figure}

The thermodynamic properties of these slowly accelerating black holes have some interesting features.  A new `snapping swallowtail'  phenomenon
appears \cite{Abbasvandi:2018vsh}, in which (simplifying to the case of zero rotation and $\mu_+=0$), there exists a transition pressure $P_t = \frac{3\mu^2}{8\pi Q^2}$ at 
which the standard swallowtail `snaps'. For $P<P_t$  the branch of low temperature black holes present in the left panel of Fig.~\ref{Fig:Swallow}
in the non-accelerating charged case disappears,  leading to a pressure induced zeroth order phase transition between small and large black holes.
This is illustrated in Fig.~\ref{snap}.

\begin{figure}[htp]
		\centering
		\includegraphics[width=0.32\textwidth]{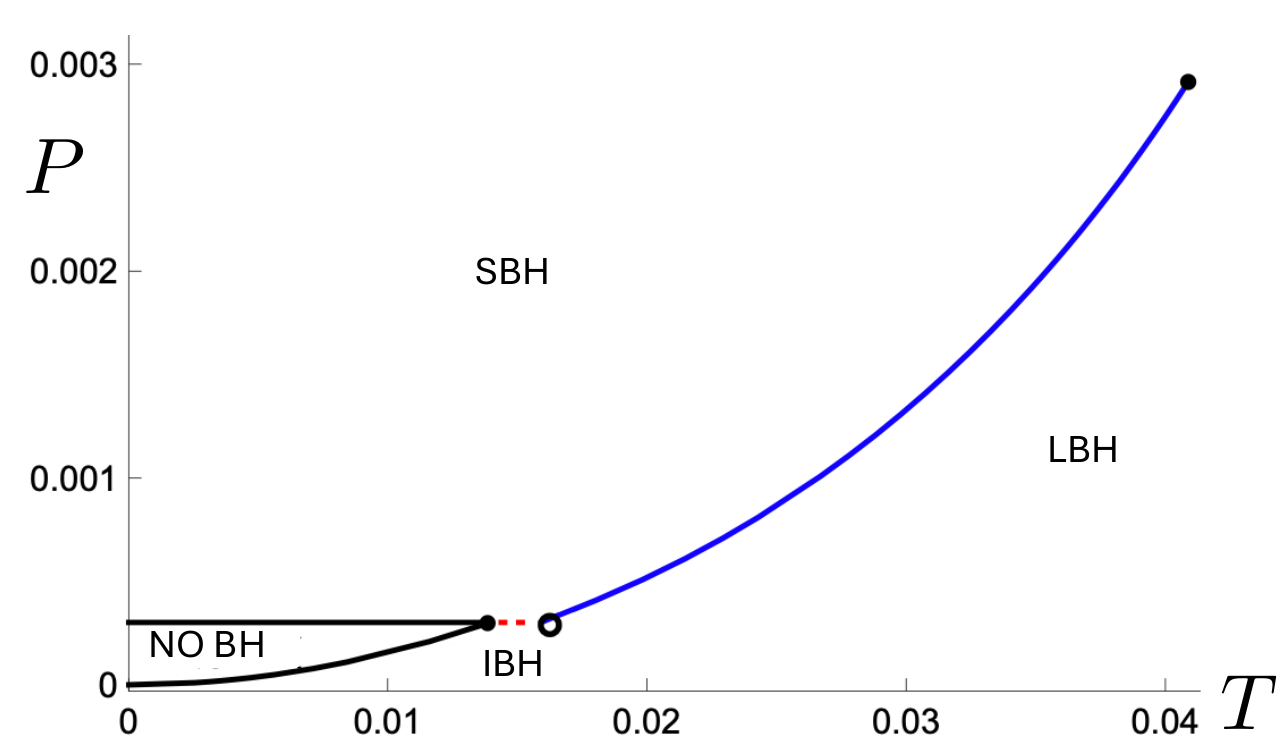}
		\includegraphics[width=0.32\textwidth]{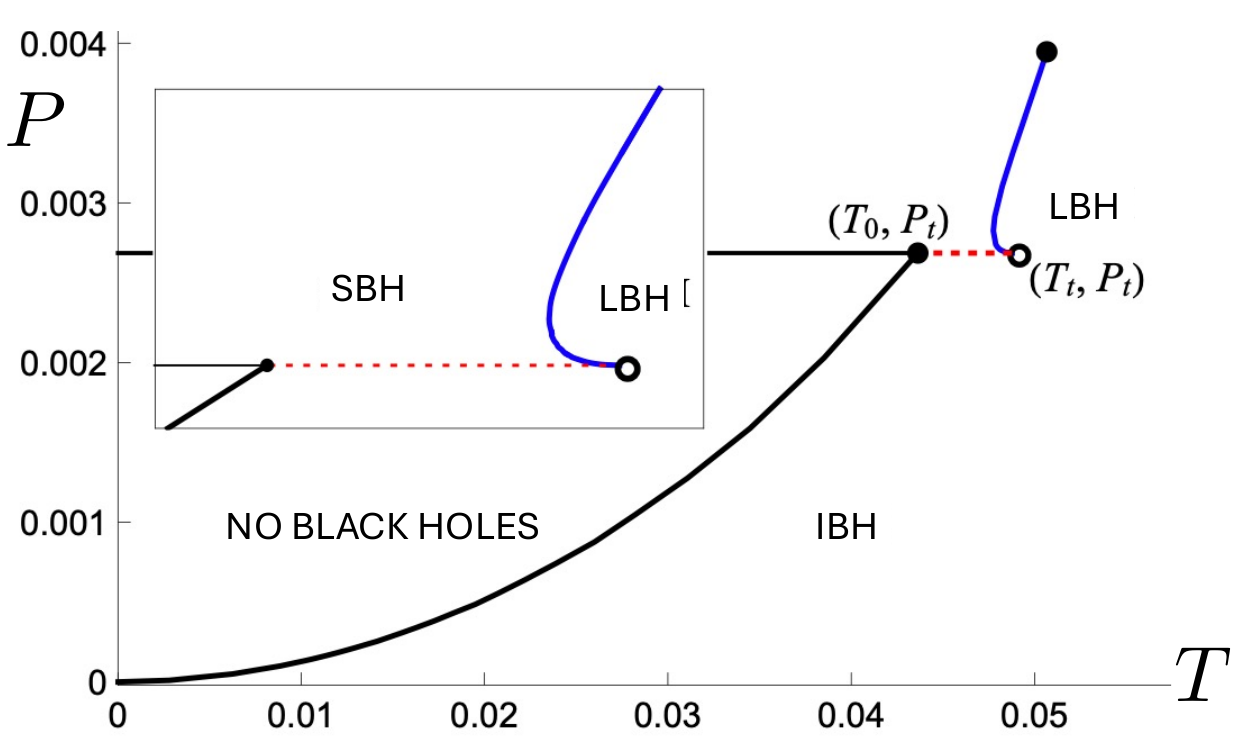}
		\includegraphics[width=0.32\textwidth]{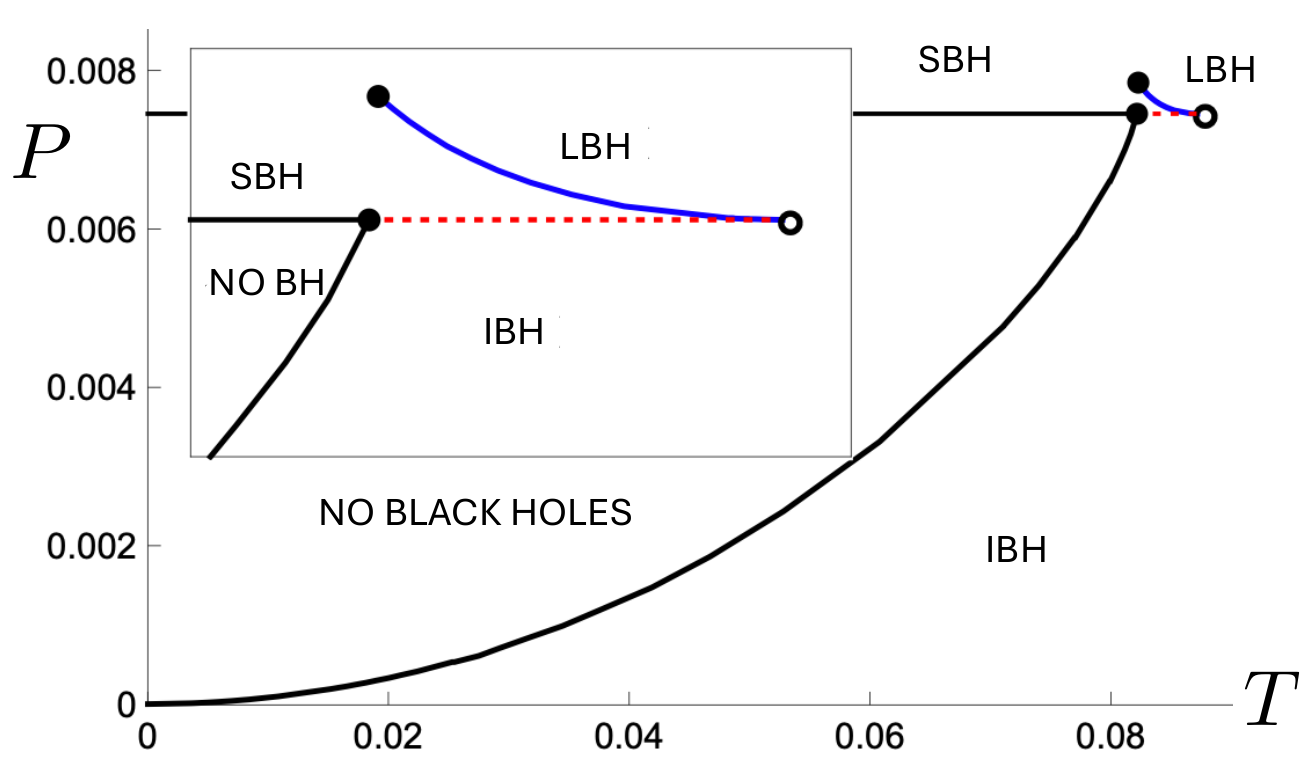}
			\caption{Phase diagrams are displayed for various string tensions, $\mu=0.05$ (left), $\mu=0.25$ (centre), and $\mu=0.50$ (right), illustrating the various types of phase transitions that can occur for a charged non-rotating slowly accelerating AdS black hole. Coexistence curves for first order phase transitions are indicated by solid blue lines; those for zeroth order transitions by dashed red lines. Black curves bound a region for which no slowly accelerating black holes exist, the ‘no black hole’ (NBH) region. The first order transitions terminate at high pressures at a critical point (black circles), whereas at low
pressures they terminate at a bicritical point  denoted by empty circle and characterized by ($T_t, P_t$).    The inset  in the centre diagram   illustrates the presence of a reentrant phase transition,  with pressure as the control parameter. The slope of the blue coexistence line in the right panel is negative,   opposite to that for the non-accelerated case.}
	\label{snapphase}
\end{figure}

The value of the string tension $\mu$ governs the type of phase diagram, shown in Fig.~\ref{snapphase}.  Considering first small
$\mu$ (left panel), as in the non-accelerating case
(section~\ref{4p2}), there is a coexistence line for a set of first order phase transitions  (blue curves) from large to small acclerating black holes
as the temperature decreases.  This coexistence line terminates at a critical point (solid circle), at which  the transition becomes second order. 
This behaviour is fully analogous to that of the charged AdS black hole in section~\ref{4p2}.  However a new  bicritical point (empty circle)
is also present, along with a  zeroth order phase transition (red dashed curve) from small to intermediate black holes as the pressure decreases.
There is also  a `no black hole' (NBH) region  since low temperature slowly accelerating black holes do not exist for $P < P_t$.  This situation is
notably different from the   Hawking–Page transition in section~\ref{4p1}, in which a radiation phase of lower free energy is present instead of
an NBH region.

As $\mu$ increases, a region of reentrant phase transitions emerges, shown in the central panel of Fig.~\ref{snapphase}.    There
is now a pressure driven reentrant phase transition, indicated by the double-valuedness of the
 coexistence curve for the first order phase transitions.  For temperatures slightly small than $T_t$, the system goes from being an
 intermediate black hole (IBH) to a large one, and then to a small one (SBH) as pressure increases.  The zeroth order transition from IBH to SBH
 remains. Once $\mu$ gets sufficiently large (right panel in  Fig.~\ref{snapphase}), the slope of the first-order coexistence line becomes negative:
  as pressure decreases from $P=P_c$ (black circle), the temperature increases, a situation notably different from that of the liquid/gas phase transition.

Rotating accelerating black holes also  have the ``snapping swallow tail" phenomenon.   However  the zeroth order phase transition now occurs over a range of pressures  and the "no black hole region" emerges continuously from a $T=0$ extremal black hole. The
no black hole region and the zeroth-order phase transition now appear at different pressures, the latter in turn   different  from the termination pressure of the first order phase transition. Reentrant phase transitions occur in two ways, being either pressure driven (as in the charged non-rotating case) or
temperature driven \cite{Abbasvandi:2019vfz}.
 
 Heat engines for accelerating black holes have also been studied
 \cite{Zhang:2018hms,Debnath:2020zdv,EslamPanah:2019szt,Ahmed:2019yci}. It has been shown that adding a 
 a conical deficit to a black hole heat engine increases its efficiency, whereas its efficiency decreases if it accelerates (provided the same average conical deficit is maintained). This effect is robust, being present if   other charges to the black hole are incorporated \cite{Ahmed:2019yci}.

\section{Holography}
\label{6s}

String theory emerged in the latter part of the 20th century as a prime candidate for a unified theory of all the forces in nature, gravity included.
From this theory a correspondence was posited between non-gravitational conformal field theories (CFTs) and gravitational theory with asymptotically Anti de Sitter boundary conditions \cite{Maldacena:1997re}.  The AdS/CFT duality claims  that strongly coupled 4-dimensional gauge theory is equivalent to gravitational theory in 5-dimensional asymptotically AdS spacetime. The proposal has since been broadened to arbitrary numbers of dimensions and for gauge theories that are not CFTs -- for this reason it has come to be known as gauge-gravity duality.  However the relationship is always between a gravitational theory in a given dimension and a non-gravitational gauge theory in one dimension less, so the concept is often referred to as holography.

AdS/CFT duality is a powerful tool for analyzing strongly-coupled gauge theories using classical gravitational theories and has been actively investigated in particle physics, particularly in the context of conformal field theories \cite{Maldacena:1997re, Witten:1998qj, Witten:1998zw}. It has since been expanded to applications in experimental quark-gluon plasmas  \cite{Kovtun:2004de},   nuclear physics \cite{Pahlavani:2014dma}, non-equilibrium physics \cite{Matsumoto:2018ukk}, and condensed-matter physics \cite{Hartnoll:2007ih, Hartnoll:2008vx}.

An important aspect of the duality is the holographic dictionary, which relates quantities on one side of the duality to those on the other.  In the context of black hole thermodynamics, it states that the thermodynamics of a CFT is completely equivalent  to the thermodynamics of an AdS black hole in its
gravitational dual.  This has the implication that black hole evaporation is a unitary process, since  a CFT is a standard unitary non-Abelian gauge theory
(perhaps with a large number $N$ of colours).     More generally, one expects  that holographic duality can be employed in  the dual field theory   to understand the perplexing features of black holes,  and vice-versa.

One of the first applications of holography to black hole chemistry was in terms interpreting the Hawking-Page transition \cite{Hawking:1982dh}
discussed in section~\ref{4p1} in terms of a liquid-solid transition \cite{Kubiznak:2014zwa}, as compared to the  the confinement/deconfinement phase transition of a quark gluon plasma in the context of AdS/CFT~\cite{Witten:1998zw}. This was the first hint that a chemical-type phase transition in the bulk had a CFT dual. Not long afterward black hole heat engines (discussed in section~\ref{5p6}) were shown to be motivated from a holographic perspective.
The engines are were referred to as  holographic because, for $\Lambda < 0$, the engine cycle corresponds to a process defined on the space of dual field theories in one dimension lower~\cite{Johnson:2014yja}.  A few years later   integration of the holographic stress tensor for slowly accelerating black holes (sec~\ref{5p8}) was shown to give the conserved mass, charge, and angular momentum  (equation~\eqref{AdSthermo}) of these objects \cite{Anabalon:2018qfv}.   An interpretation of variable $\Lambda$ originating from scalar hair in a string-theoretic context has also been proposed
\cite{Astefanesei:2019ehu,Astefanesei:2023sep}.

Despite this,   in the context of holographic duality, interpretation of a variable $\Lambda$ is problematic \cite{Johnson:2014yja, Dolan:2014cja, Kastor:2014dra, Zhang:2014uoa, Zhang:2015ova, Dolan:2016jjc, McCarthy:2017amh}.
From a cosmological perspective $\Lambda$ 
is the energy/pressure of the vacuum, and so could reasonably be expected to be a variable quantity, but in 
 the AdS/CFT correspondence \cite{Maldacena:1997re,Witten:1998qj} $\Lambda$ is regarded as fixed -- it sets the asymptotic structure of the bulk spacetime. The first attempts  \cite{Johnson:2014yja,Dolan:2014cja,Kastor:2014dra,Zhang:2014uoa, Dolan:2016jjc} to understand the meaning of a variable $\Lambda$ suggested that the $V\delta P$ term in the AdS bulk should be related  to a
$\mu \delta C $
term in the dual CFT, where $C$ is the central charge and $\mu$   its  thermodynamically conjugate chemical potential. Variations of $C$ then correspond to changing both
 the CFT volume ${\cal V}$ and the number of colours $N$ (or the central charge $C$)  \cite{Johnson:2014yja,Dolan:2014cja,Kastor:2014dra,Caceres:2015vsa,Couch:2016exn}; furthermore, electric charge and its conjugate potential  both rescale with the AdS length $\ell$.  The first law \eqref{firstBH} cannot therefore be straightforwardly related to the corresponding dual field theory 
 thermodynamics \cite{Karch:2015rpa,Sinamuli:2017rhp, Visser:2021eqk}.

Considerable progress  has been
made in understanding holographic black hole chemistry in the past few years and is the subject of a recent review \cite{Mann:2024sru}. The three major developments have been establishing a dual for the Smarr relation \eqref{smarrBH}, obtaining an understanding of CFT complexity in terms of the thermodynamic volume of a black hole, and the construction of an exact dictionary between the  laws of Black Hole Chemistry and their  CFT  counterparts.  This is now
yielding an emerging  understanding of CFT phase behaviour  and its bulk correspondents.

\subsection{Holographic Smarr Relation}
\label{6p1}

In its original formulation, the AdS/CFT correspondence \cite{Maldacena:1997re} posited that
Type IIB string theory on an $\mbox{AdS}_5\times S^5$ spacetime of AdS length $\ell$ was  dual to ${\cal{N}}=4$ $(3+1)$-dimensional $U(N)$  super-Yang-Mills theory. This conjecture quickly became generalized to arbitrary dimensions, and takes the form
\begin{equation}
\label{correspond1}
C= \textsf{k} \frac{\ell^{D-2}}{16\pi G_D}
\end{equation}
relating the central charge $C$ of the CFT (with $C\propto N^2$ for $SU(N)$ gauge theories with conformal symmetry) to the AdS length $\ell$, where the $D$-dimensional Newton constant of gravity has been restored and   $\textsf{k}$ is a constant that depends on the details of the particular holographic system.  Equation \eqref{correspond1} implicitly assumes that $\ell$ (and so by \eqref{eq:press} $\Lambda$) is fixed.   
 Essentially all   investigations of the AdS/CFT correspondence pivot on the assumption \eqref{correspond1}.
   
Confusion regarding the  holographic interpretation of black hole chemistry   \cite{Johnson:2014yja, Dolan:2014cja, Kastor:2014dra, Zhang:2014uoa, Zhang:2015ova, Dolan:2016jjc, McCarthy:2017amh} stems from \eqref{correspond1},  since variable pressure implies variable $\ell$,  which in turn implies variation of either $C$ (the dimension of the gauge group) or  the CFT volume ${\cal V}$, or both. Variation of $C$ corresponds to variation in the space of field theories in the boundary. Variation of ${\cal V}$ allows the field theory to remain fixed \cite{Karch:2015rpa}, admitting a holographic Smarr relation.

To obtain a holographic understanding of the Smarr relation  \eqref{smarrBH}, consider the free energy of the dual field theory  \cite{Karch:2015rpa}.  
This quantity  scales   as $N^{2}$, and so the grand canonical thermodynamic potential  $\tilde\Omega(N,\mu,T,\ell)$  scales as 
\begin{equation}\label{holosmarr}
\tilde\Omega(N,\tilde \mu ,T,\ell,\tilde Q,J)= N^{2}\tilde\Omega_{0}(\tilde \mu ,T,\ell,\tilde Q,J)
\end{equation}
for large $N$.  The CFT first law is \cite{Visser:2021eqk}
\be\label{FirstHol}
\delta E=T\delta S-p d{\cal V}+\tilde\phi \delta \tilde Q+\Omega \delta J+ \tilde \mu \delta C \,, 
\ee
for a CFT at energy $E$, pressure  $p$, and volume ${\cal V}={\cal V}_0 \ell^{D-2}$.  The quantity $\tilde \mu$ is the chemical potential for the central charge $C$,    $J$ and $\Omega$  are the respective angular momentum and conjugate angular velocity, and $ \tilde Q, \tilde\Phi$ are   the respective holographic electric charge and conjugate potential.

 These quantities obey the scaling relations
\ba
{}[S]&=&[\tilde Q]=[J]=[C]=L^0 \nonumber\\
{}[E]&=&[T]=[\Omega]=[\tilde \mu]= L^{-1} \qquad [{\cal V}]=L^{D-2} 
\label{units1}
\ea
yielding the {\em holographic Smarr}   relation 
\be\label{SmarrHol}
E=TS+\tilde \Phi \tilde Q+\Omega J+\tilde\mu C\,, 
\ee
from a standard  dimensional Euler scaling argument.   A $p-{\cal V}$ term does not appear in \eqref{SmarrHol}, but a similar scaling argument
gives 
\be\label{CFTEoS}
E = (D-2)  p {\cal V}  
\ee
which is the equation of state for the CFT.

Going beyond  leading order in $N$ \cite{Sinamuli:2017rhp},   the natural extension of \eqref{holosmarr} is 
 \begin{equation}\label{felov}
\tilde\Omega(N, T, \alpha_j,  R,  \tilde Q, J)=\sum_{k=0}g_k(N)\tilde\Omega^k(T, \alpha_j,  R,  \tilde Q, J)
\end{equation}
where  $\tilde\mu$ has been fixed,  the $g_k(N)$ are assumed to be polynomial functions  of $N$,  and $R$  is the radius of the sphere on which the field theory is formulated.
The quantities $\alpha_j$ are the coupling constants in Lovelock gravity (see~\ref{appB}), which scale as 
\begin{equation}
\label{extendedholo2}
\hat{\alpha}_j\sim L^{2(j-1)} 
\end{equation}
in turn implying that the functions $g_k(N)$ have the form
\begin{equation}
\label{extendedholo1}
g_k(N) = \beta_j(\hat{\alpha}_j)^{\frac{D-2}{2(j-1)}} 
\end{equation}
in \eqref{felov}. From \eqref{holosmarr}   the $j=0$ term is  
\begin{equation}
g_0(N) = \beta_0 \ell^{D-2}=N^2
\end{equation}
 recovering \eqref{correspond1}, with $\beta_0=\frac{k}{16\pi G_D}$.

Equation (\ref{extendedholo1}) implies for any arbitrary function $X = X(\hat{\alpha}_j)$ that 
\begin{equation}
\label{holosmarr1}
\hat{\alpha}_j\frac{\partial X}{\partial\hat{\alpha}_j}=\frac{D-2}{2(j-1)}g_j \frac{\partial X}{\partial g_j}\; .
\end{equation} 
Setting $X=\tilde\Omega$ in \eqref{felov}, 
multiplying both sides by $2(j-1)$ and summing over $j$ yields
\begin{equation}
\label{homo1}
\sum_{j=0}2(j-1)\hat{\alpha}_j\Psi^{(j)}=(D-2)\sum_{j=0}g_j\frac{\partial\tilde\Omega}{\partial g_j} =(D-2)\tilde\Omega
\end{equation}
where $\Psi^{(j)} = \frac{\partial\tilde\Omega}{\partial{\hat{\alpha}_j}}$ and, since $\Omega$ is   an homogeneous function of the $g_k$ of degree 1, the second equality  holds due to Eulerian scaling.  

For any function $f(\ell, Z) = f(\ell,Z_0 \ell^p)$, its derivative with respect to $\ell$ will be
\begin{equation}
\partial_\ell f(\ell, Z)|_{Z_b}=\partial_\ell f|_Z+ p \frac{Z}{\ell}\partial_Z f|_\ell 
\end{equation}
 for some constant $Z_0$.  Consequently 
 \begin{equation}
\label{homo5}
\ell\frac{\partial \tilde\Omega}{\partial \ell}+\sum_{j=1}2(j-1)\hat{\alpha}_j\frac{\partial \tilde\Omega}{\partial{\hat{\alpha}_j}}=(D-2)\sum_{j=0}g_j\frac{\partial \tilde\Omega}{\partial g_j}+R\frac{\partial \tilde\Omega}{\partial R}+Q\frac{\partial \tilde\Omega}{\partial Q}  
\end{equation}
since the radius  $R= R_0 \ell$ of the boundary CFT  and  the bulk charge
$Q = \tilde{Q}/\ell$. The first term on the left-hand side of \eqref{homo5} is zero, and so
 we obtain using \eqref{homo1} 
\begin{eqnarray}
&&\sum_{j=0}2(j-1)\hat{\alpha}_j\Psi^j\nonumber\\
&&=(D-2)\sum_{j=0}g_j\partial_{g_j}\tilde\Omega\big|_{\mu, T,R,Q,J}+R\partial_R\tilde\Omega\big|_{\mu, T,\hat{\alpha}_{j\geq 1},Q,J} + Q\partial_Q\tilde\Omega\big|_{\mu, T,\hat{\alpha}_j,R,J}  \nonumber\\
&&=(D-2)\tilde\Omega - M-\Phi Q\nonumber\\
&&=(D-3)M-(D-2)(TS+\Omega J)-(D-3)\Phi Q 
\label{smarr2}
\end{eqnarray}
which is the Smarr relation \eqref{SmarrLovelock} in Lovelock gravity, where $\Phi = \partial_Q\tilde\Omega$ and $M=\partial_R\tilde\Omega$.

\subsection{Complexity and Volume}
\label{6p2}

 Complexity is a concept in quantum information theory that quantifies  how difficult it is to prepare a particular target state $\ket{\psi_T}$ from a given reference state $\ket{\psi_R}$ (typically assumed to be a simple unentangled state) and an initial set of elementary gates $\mathcal{G}$
\begin{align}
V_n&\equiv g_n\dots g_1g_0
\end{align}
where $g_0,\dots g_n\in\mathcal{G}$. The complexity  is  defined as the  smallest number $n$ of elementary gates that can approximate   the target state $\ket{\psi_T}$ according to some norm
\begin{equation}
\mathcal{C}(\ket{\psi_T}) = \textrm{min}_{n}{||\ket{\psi_T}-V_n\ket{\psi_R}||}^2
\end{equation}
starting from a fixed reference state $\ket{\psi_R}$.  

From the holographic perspective, complexity was originally proposed to be   dual to the volume of the Einstein-Rosen (ER) bridge in eternal black holes \cite{Susskind:2014rva}. 
The eternal Schwarzschild-AdS black hole is dual to two copies of the CFT prepared in the thermofield double state (TFD)~\cite{Maldacena:2001kr}.   The volume of the ER bridge continues to grow in time even after thermalization, and it was conjectured that this growth captures some notion of complexity for the CFT state,  since this quantity likewise  evolves after equilibrium is reached~\cite{Hartman:2013qma}.  

\begin{figure}
\centering
\includegraphics[width=\textwidth]{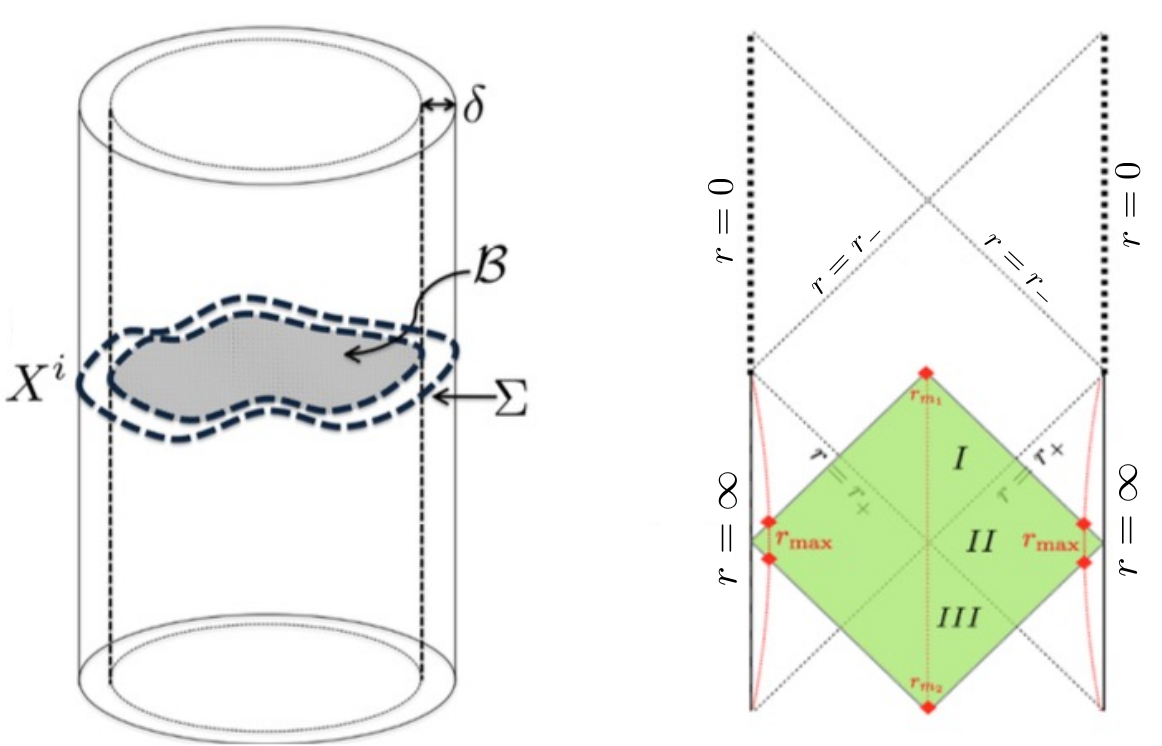}
\caption{{\it Left}: Depiction of  the relevant parts of the AdS geometry for the CV conjecture.  {\it Right}: 
Depiction of the Penrose diagram of 
a charged-AdS black hole; the WDW path is in shaded green.
}
\label{figcomp}
\end{figure}

Two proposals for implementing this idea emerged, expressed by the equations
\begin{equation}\label{CV}
\mathcal{C}_\mathcal{V}(\Upsilon)=\max\limits_{\Upsilon=\partial\mathcal{B}}\left[\frac{\mathcal{V}(\mathcal{B})}{G_D R}\right]
\end{equation}
known as the `Complexity equals Volume"  (CV) conjecture and 
\begin{equation}\label{CA}
\mathcal{C}_\mathcal{A}(\Upsilon)=\frac{I_{\text{WDW}}}{\pi\hbar}  
\end{equation}
known as the ``Complexity equals Action" (CA) conjecture.  The CV conjecture asserts states  that  the volume an the extremal/maximal spacelike 
slice $\mathcal{B}$ anchored at $t_L$ and $t_R$ at the boundaries  \cite{Stanford:2014jda} of  the boundary section $\Upsilon$ in the AdS spacetime is equal to the complexity of the TFD state.  The quantity  $R$, typically taken to be the AdS length $\ell$, 
 is an arbitrary length scale chosen to make the complexity $\mathcal{C}_\mathcal{V}(\Upsilon)$ dimensionless.  The CA conjecture posits that 
 the complexity  of the CFT state is given by the numerical value of the whole domain of dependence of $\mathcal{B}$, a region of
 spacetime called the  Wheeler-DeWitt (WDW) patch. The two proposals are illustrated in Fig.~\ref{figcomp}.

The quantities in \eqref{CV} and \eqref{CA} are in general divergent. It is more useful to compute
\be\label{CFvol}
\Delta \mathcal{C}_\mathcal{V} = \lim_{r_{\rm max} \to \infty} \frac{\left[ \mathcal{V} - 2 \mathcal{V}_{AdS} \right]}{G_D R} \, . 
\ee
instead of \eqref{CV},  where 
\be 
\mathcal{V}_{\rm AdS} = \Omega_{D-2} \int_{0}^{r_{\rm max}} dr \frac{r^{D-2} }{\sqrt{1 + r^2/\ell^2}} \, ,
\ee 
and  the limit $r_{\rm max} \to \infty$ is taken. Similarly  one calculates
\begin{equation}
\Delta\mathcal{C}_\mathcal{A}(\Upsilon)=\frac{1}{\pi\hbar}\left[I_{\text{WDW}}(\text{BH})-2I_{\text{WDW}}(\text{AdS})\right]
\label{formationC}
\end{equation}
instead of \eqref{CA}.  The quantities in \eqref{CFvol} and \eqref{CA} measure the additional complexity present in preparing the TFD state compared to two copies of the AdS vacuum alone, and are called  the complexities of formation \cite{Chapman:2017rqy}.

Various calculations for spherically symmetric AdS black holes \cite{Chapman:2017rqy,Carmi:2017jqz,Chapman:2019clq}  indicated that  the complexity of formation
grew  linearly with entropy in the high-temperature (or large black hole) limit. However this   general expectation proved  not to be correct.  Instead
for both proposals it has been shown that~\cite{Andrews:2019hvq,AlBalushi:2020rqe,AlBalushi:2020heq}
\be \label{DCbehaveprop} 
\Delta\mathcal{C} \propto V^{(D-2)/(D-1)}
\ee
in the high temperature limit, where $V$ is the thermodynamic volume of the black hole. This was shown for solitons~\cite{Andrews:2019hvq}, higher-dimensional multiply rotating black holes of equal angular momenta~\cite{AlBalushi:2020rqe,AlBalushi:2020heq}, and  for $D=4$ Kerr-AdS black holes~\cite{Bernamonti:2021jyu}.

The rationale for considering rotating solutions is that the thermodynamic volume $V$ and entropy $S$ are
quantities that independently depend on the horizon radius $r_+$ and the rotation parameters, unlike the situation for spherical symmetry 
in which $S \propto V^{(D-2)/(D-1)}$. However rotating black holes have a more complicated causal structure, in which null hypersurfaces
depend on the polar angle  \cite{AlBalushi:2019obu,Imseis:2020vsw}, in contrast to the situation shown at the right in Fig.~\ref{figcomp}. However in
Kerr-AdS spacetimes in odd dimensions with equal angular momenta in each orthogonal rotation plane the null hypersurfaces do not have such dependence. Their causal structure is the same as shown in Fig.~\ref{figcomp} and has some similarities with the charged case \cite{Sinamuli:2019utz,Chapman:2019clq}.

Setting all rotation parameters equal, the Kerr-AdS black hole solutions \eqref{KAdSmetric}  in $D = 2N + 3$ odd dimensions are~\cite{Kunduri:2006qa}
\begin{align}\label{multrot}
ds^2 =& -f(r)^2 dt^2 + g(r)^2 dr^2 + h(r)^2 \left[ d\psi +  A - \Omega(r) dt\right]^2 
+ r^2 \hat{g}_{ab} dx^a dx^b
\end{align}
where $\psi$ is periodically identified so that $\psi \sim \psi + 2\pi$ and
\begin{align}
g(r) ^2 =& \left( 1+ \frac{r^2}{\ell^2} - \frac{2m \Xi}{r^{2N}} + \frac{2m a^2}{r^{2N + 2}}\right)^{-1} 
\qquad h(r)^2 = r^2 \left( 1 + \frac{2m a^2}{r^{2N+2}}\right)  \nonumber\\
 f(r) =& \frac{r}{g(r) h(r)} \qquad \Omega(r) = \frac{2ma}{r^{2N} h^2} \qquad \Xi = 1 - \frac{a^2}{\ell^2}
\end{align} 
with 
$A$  a 1-form on $\mathbb{CP}^N$ that satisfies $dA =2J$, where $J$ is the K\"ahler form. The constant $(t,r,\psi)$ section is $\mathbb{CP}^N$ 
with  Fubini-Study metric $\hat{g}$ and curvature normalized so that $\hat{R}_{ij} = 2 (N+1) \hat{g}_{ij}$.   For example if $N=1$ ($D=5$)
\begin{equation}
 A = \frac{1}{2} \cos \theta d\phi  \qquad \hat{g} = \frac{1}{4} \left( d\theta^2 + \sin^2 \theta d\phi^2 \right) 
\end{equation}
 corresponding to $\mathbb{CP}^1 \cong S^2$. 
The asymptotic region is obtained in the limit $r \to \infty$, where  the  usual AdS$_{2N+3}$ metric is recovered.   The horizon $r_+$ is given by the largest root of $1/g^{2}(r_+) = 0$, and is a smooth null hypersurface with generator 
\begin{equation}
\xi = \frac{\partial}{\partial t} + \Omega_H \frac{\partial}{\partial \psi} , \qquad \Omega_H = \frac{2 m a}{r_+^{2N + 2} + 2m a^2} \, .
\end{equation}  
There is also an inner Cauchy horizon at $r = r_-$ which is the smaller of the two positive real roots of $1/g^{2}(r)$. 

 The black hole's temperature, entropy, volume, are \cite{Altamirano:2014tva}
\begin{align}
T &=\frac{1}{2\pi h(r_+)} \left[(N+1) \left(1 + \frac{r_+^2}{\ell^2}\right) - \frac{\ell^2 r_+^2}{(r_+^2-a^2)\ell^2 - r_+^2 a^2} \right] \label{tempmult}\\
S &=\frac{\Omega_{2N+1}h(r_+)r_+^{2N}}{4G_N} \label{entropymult} \\
V &= \frac{r_+^{2(N+1)} \Omega_{2N+1}}{2(N+1)} + \frac{4 \pi a J}{(2N+1)(N+1)}   \label{volmult}
\end{align}
from which it is clear that  the entropy \eqref{entropymult} and thermodynamic volume \eqref{volmult} are independent functions of   $m$ and $a$ (or $r_+$ and $r_-$).

Evaluating  \eqref{CFvol} is a tedious calculation, whose result is~\cite{AlBalushi:2020rqe,AlBalushi:2020heq}
\be \label{DCbehave} 
\Delta \mathcal{C} = {\Sigma}_{\rm g} C_T \left(\frac{\mathcal{V}}{\mathcal{V}_{\rm AdS}}\right)^{\frac{D-2}{D-1}} 
\ee
where
\begin{align}\label{extVol}
\mathcal{V}&=2\Omega_{D-2}\int_{r_+}^{r_{\rm max}}dr \, r^{(D-3)}h(r) g(r) \, ,
\end{align}
with $C_T \sim \ell^{D-2}/G_D$   the central charge of the CFT,  $V_{\rm AdS} = \ell^{D-1}$, and $\Sigma_{\rm g}$  a factor that depends on the specific metric, dimension, etc. but not on the size of the black hole.  The relationship 
\eqref{DCbehave} can be explicitly checked   by considering the ratio
\be 
R(\beta) = \frac{ R G_N \Delta \mathcal{C}_\mathcal{V}}{(r_+/\ell)^\beta}  
\ee
and   numerically determining the value of $\beta$ so that $R(\beta)$ exhibits no dependence on $r_+/\ell$ when $r_+/\ell$ is large.  Explicit
checks have been carried out~\cite{AlBalushi:2020heq} for all  $D \leq 27$. 
\begin{figure}
\centering
\includegraphics[width=\textwidth]{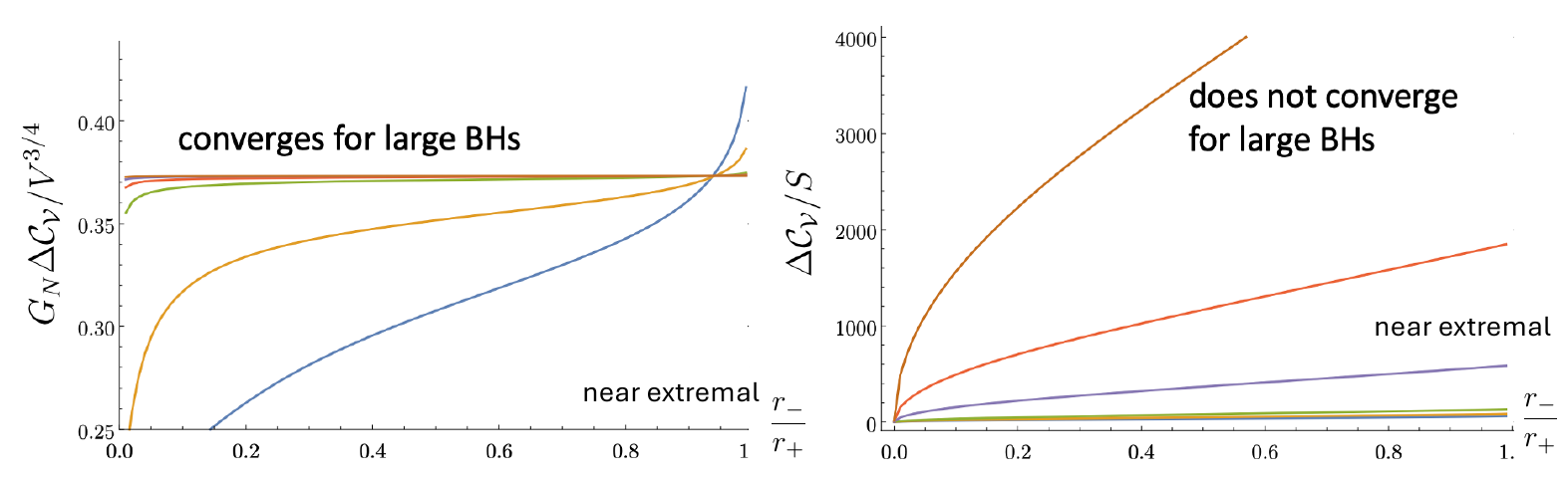}
\caption{Left: A plot of the CV complexity of formation \eqref{DCbehave} for $D=5$ as a function of the ratio $r_-/r_+$, normalized by $V^{3/4}$. 
Moving from bottom to top the curves correspond to fixed $r_+/\ell = 10, 10^2, 10^3, 10^4, 10^5, 10^6$ ; for $r_+/\ell \geq 1000$ the curves are visually indistinguishable. Right: A plot of the CV complexity of formation normalized by the entropy $S$.  Again, from bottom to top, the curves correspond to fixed $r_+/\ell = 10, 10^2, 10^3, 10^4, 10^5, 10^6$.  The scaling relation for spherically symmetric black holes is clearly not obeyed.
}
\label{CVplot}
\end{figure}
Computing \eqref{formationC} for the CA conjecture is even more difficult, 
and results have been obtained numerically~\cite{AlBalushi:2020heq} for $D=5,7$.  These are commensurate with \eqref{DCbehave}. An illustration
of this scaling is shown in Fig.~\ref{CVplot} for  $D=5$ 
 
The relationship \eqref{DCbehave} is the first time that thermodynamic volume has explicitly appeared in a holographic context.  However whereas $V$ generally
governs the behaviour of the complexity of formation, the entropy $S$ can be shown to provide a lower bound~\cite{AlBalushi:2020heq}:
\be\label{isoBound} 
\Delta \mathcal{C} \ge \beta_D S
\ee
where $\beta_D$ is a positive constant.  

More recently the ambiguities inherent in both the CV and CA proposals (in terms of boundary terms on null slices in the CA proposal or
length scales in the CV proposal) have prompted the question: ``does complexity equal anything?'' \cite{Belin:2021bga}. Insofar as there is an infinite class of diffeomorphism-invariant gravitational observables that display universal features of complexity, the answer appears to be yes. Consequently there is an infinite class of equally viable generalized holographic complexity proposals. The constraints on all such proposals is that (i) the complexity must grow linearly with time at late times (commensurate with  evolution of the entangled CFT thermofield double state (TFD)) and (ii)   growth of complexity is delayed by effects of the far-past shock wave geometry  dual to the perturbations of the TFD \cite{Susskind:2014jwa,Roberts:2014isa}. 

The constraints for the  ``complexity equals (almost) anything'' proposal \cite{Belin:2022xmt} have been checked  for a range of spherically symmetric black holes \cite{Wang:2023eep,Jiang:2023jti} and recently for the multiply rotating Kerr-AdS black holes \eqref{multrot} \cite{Zhang:2024mxb}. The volume complexity \eqref{extVol} generalizes to include higher curvature terms. While the scaling relation \eqref{DCbehave} is maintained,   the generalized volume complexity undergoes several types of phase transitions at early times, and obeys a generalization of the Maxwell  area law.  Multi-critical points for complexity phase transition, where two first-order phase transition points intersect at a single point, can also exist, reminiscent of the multicritical behaviour discussed in section~\ref{5p2}.
Perhaps there is a `complexity chemistry' that awaits discovery.

\subsection{Central Charge Criticality} 
\label{6p3}

The holographic dictionary pivots on the relationship \eqref{correspond1} relating the central charge $C$ to the
AdS length $\ell$.
%
%
However it is also standard to identify the curvature radius of the spatial geometry on which the CFT is formulated with  $\ell$   \cite{Karch:2015rpa}.
Conformal completion of the bulk AdS spacetime yields  the boundary metric  \cite{Gubser:1998bc,Witten:1998qj}
\be
ds^2= \omega^2 \Bigl(-dt^2+\ell^2 d\Omega_{k,d-2}^2\Bigr)\,,
\ee
 of the dual CFT, where   $\omega$ is an `arbitrary'  dimensionless  conformal factor, a function of boundary coordinates, that reflects the conformal symmetry of the boundary theory.  
The volume of the CFT 
\be \label{volume}
{\cal V} \propto (\omega \ell)^{D-2} 
\ee
so if  $\omega=1$,  a variation of the cosmological constant in the bulk induces a variation of the CFT volume ${\cal V}$, indicating that a pressure-volume work term, $-p\delta {\cal V}$,   should be present on the CFT side.  This in turn implies that  the $\mu \delta C$ and $-p \delta {\cal V}$ terms are not truly independent, rendering the corresponding CFT first law \eqref{FirstHol} {\em degenerate}, 
leaving the holographic interpretation of  black hole chemistry obscure at best. 

Retaining independence in the variations of $C$ and $\cal V$   is therefore preferable.  This can be done by varying Newton's constant \cite{Karch:2015rpa,Cong:2021fnf} or by treating $\omega$ as a (dimensionless) thermodynamic parameter   (similar to the horizon radius or AdS radius), instead of a function of the boundary coordinates~\cite{Ahmed:2023snm}, regarding the two  quantities on the right hand side of \eqref{volume}  
as being independent.  Variation of the central charge~$C$,  from \eqref{correspond1},  is therefore purely induced by  variations 
of~$\ell$, with $G_N$ remaining fixed.  Previous treatments set   $\omega=R/\ell$ (as in section~\ref{6p1}), with $R$   a constant  boundary curvature radius \cite{Visser:2021eqk,Cong:2021jgb}.  However this is not necessary: the  curvature radius~$\omega \ell$ can be arbitrary, distinct  from the AdS radius $\ell$, and the  CFT is formulated on this geometry, with     the central charge $C$ being a thermodynamic variable.

For the Einstein-Maxwell   Lagrangian density \eqref{EinMaxact},  
this results  in the following generalized dictionary 
\ba \label{extendeddictionary}
\tilde S&=& S=\frac{A}{4 G_N}\,,\quad 
\tilde E=\frac{M}{\omega}\,,\quad \tilde T=\frac{T}{\omega}\,,\quad \tilde\Omega=\frac{\Omega}{\omega}\,,\nonumber\\
\tilde J &=& J, \quad \tilde \Phi=\frac{\Phi\sqrt{G_N}}{\omega \ell}\,,\quad \tilde Q=\frac{Q\ell}{\sqrt{G_N}} 
\ea
between  bulk (without tildes) and dual CFT (with tildes) thermodynamic quantities.  The bulk first law  \eqref{firstBH} can then be rewritten as follows:
\begin{align}
&\delta \Bigl(\frac{M}{\omega}\Bigr)= \frac{T}{\omega}\delta \Bigl(\frac{A}{4 G_N}\Bigr)+\frac{\Omega}{\omega}\delta J+\frac{\Phi\sqrt{G_N}}{\omega \ell}\delta \Bigl(\frac{Q\ell}{\sqrt{G_N}}\Bigl) -\frac{M}{\omega(D-2)}\frac{\delta (\omega \ell)^{D-2}}{(\omega \ell)^{d-2}} \nonumber\\
&\qquad\qquad +   \Bigl(\frac{M}{\omega}-\frac{TS}{\omega}-\frac{\Omega J }{\omega}-\frac{\Phi Q}{\omega}\Bigr) \!\frac{\delta(\ell^{D-2}/G_N)}{\ell^{D-2}/G_N}
\label{Eq8}
\end{align}
using \eqref{smarrBH}.  The dictionary \eqref{extendeddictionary} then implies that \eqref{Eq8} becomes \cite{Ahmed:2023snm}
\be
\delta \tilde E=\tilde T\delta S+\tilde \Omega \delta J+\tilde \Phi \delta \tilde Q+\mu \delta C-p\delta {\cal V}\,,\label{j1}
\ee
which is the CFT first law \eqref{FirstHol}, with
\ba
p&=& \frac{\tilde E}{(d-2){\cal V}} \label{j3}
\ea 
recovering the CFT equation of state \eqref{CFTEoS}
and
\ba
\mu&=&\frac{1}{C}( \tilde E - \tilde TS-\tilde \Omega J-\tilde \Phi \tilde Q)\,, \label{j2}
\ea
the chemical potential conjugate to $C$. 

We see that the CFT first law \eqref{FirstHol} is exactly dual to the bulk first law \eqref{firstBH} of black hole chemistry.  Note that the variation of $\ell$
(or bulk pressure $P$)  enters not only in the variation of the central charge, but also in the dictionary for the  spatial volume  and electric charge. The   $V\delta P$ term in \eqref{firstBH}  thus splits into several pieces   related  to  variation of the  electric charge, volume, and central charge of the CFT. 
 The Euler relation for holographic CFTs is given by \eqref{j2} and does not contain a $p\mathcal V$ term, reflecting the fact that  the internal energy is not an extensive variable on compact spaces at finite temperature in the deconfined phase, a feature of holographic CFTs.    In the  large-volume (or high-temperature) regime, where $\omega L \tilde T  \gg 1$, the $\mu C$ term becomes equal to $ - p \mathcal V$, and   the energy becomes extensive.   

The dimension dependent factors in the bulk Smarr relation \eqref{smarrBH} stand in notable contrast to the dimensionless ones in the
CFT Euler relation \eqref{j2}. We can understand this by noting that
\be
\label{PVterm}
- 2 P V = -2 P \left ( \frac{\partial M}{\partial P}\right)_{A,J,Q,G_N}\!\!\!\! = 
\ell \left ( \frac{\partial M}{\partial \ell}\right)_{A,J,Q,G_N}\!\!\!\! = \ell \omega \left (\frac{\partial \tilde E}{\partial \ell}\right)_{A,J,Q,G_N}  
\ee  
using the dictionary \eqref{extendeddictionary} to obtain the last equality.  However $\tilde E$ is a function $\tilde E = \tilde E (S(A,G_N),J,\tilde Q(Q,L,G_N),C(L,G_N),V(L,\omega))$ of  bulk quantities, and so  
\begin{align}
\label{partial}
&\left ( \frac{\partial \tilde E}{\partial \ell}\right)_{A,J,Q,G_N} \!\!\!\!\!\!=\frac{1}{\ell} ( \tilde \Phi \tilde Q + (D-2)\mu C - (D-2) p {\cal V})  \\
&\qquad \qquad \,\, =\frac{1}{\ell} ( (D-3) (\tilde E - \tilde \Phi \tilde Q) - (D-2) (\tilde \Omega J + \tilde T S))  \nonumber
 \end{align}
 using \eqref{volume}, \eqref{extendeddictionary}, the  Euler relation, and the equation of state.  Inserting  \eqref{partial} into   \eqref{PVterm}  and using the holographic dictionary  \eqref{extendeddictionary} yields the bulk Smarr relation  \eqref{smarrBH}.

A significant body of literature has emerged discussing holographic black hole chemistry for a variety of 
scenarios \cite{Rafiee:2021hyj,Alfaia:2021cnk,Kumar:2022fyq,Dutta:2022wbh,Lobo:2022eyr,Kumar:2022afq,Qu:2022nrt,Bai:2022vmx,Bai:2023wjm,Zhang:2023uay,Bai:2023woh,Sadeghi:2023tuj,Chen:2023pgs,Ladghami:2024wkv,Paul:2024rto,Cui:2024cnj,Sadeghi:2024dnw,Zheng:2024glr,Baruah:2024yzw,TranNHung:2024tyg,Zhang:2024jlp,Sadeghi:2024ish}.
An illustration of how this works can be done for charged AdS black holes \cite{Cong:2021fnf,Cong:2021jgb}, whose metric is \eqref{metfunction}. From this, we can write the entries in the dictionary \eqref{extendeddictionary} as
 \begin{eqnarray}
E &=& \frac{D-1}{R} C x^{D-2} \left ( 1 + x^2 + \frac{y^2}{x^{2D-4}} \right) \label{energycft} \\
T &=& \frac{D-2}{4 \pi R} \frac{1}{x} \left ( 1 + \frac{D}{D-2} x^2 - \frac{y^2}{x^{2D-4}} \right)  \label{energytempcft} \\
\mu &=& \frac{x^{D-2}}{R} \left ( 1 - x^2 - \frac{y^2}{x^{2D -4}}\right)
 \label{mu}
 \end{eqnarray}
 for the energy, temperature, and chemical potential, 
where 
\begin{equation}
 	x \equiv \frac{r_h}{\ell} \,,\qquad   y \equiv \frac{q}{\ell^{D-2}}\,.
 \end{equation}
 are dimensionless parameters.  Similarly, we can write 
 \begin{equation}
 \label{SQphi}
 	S = 4 \pi C x^{D-1}, \qquad \tilde Q = 2 \alpha (D-1)  C  y , \qquad \tilde \Phi = \frac{1}{\alpha R} \frac{y}{  x^{D-2}} 
 \end{equation}
for the entropy, electric charge,  and conjugate potential, setting $\textsf{k}=\Omega_{D-2}$ in \eqref{correspond1} and $\omega\ell = R$
is the boundary curvature radius.

In addition to   $(T,S)$ there are three pairs  $(\tilde \Phi, \tilde Q)$, $(p,{\cal V})$ and $(\mu, C)$ of conjugate thermodynamic variables 
in the CFT description of charged AdS black holes, for a total of  $2^3=8$ (grand) canonical ensembles.  Three of these ensembles
     \begin{equation}
     \begin{aligned}  \label{freeenergies2}
    &\text{fixed} \quad  (\tilde Q, {\cal V}, C): \qquad &&F \equiv E - TS  = \tilde \Phi \tilde Q +\mu C \,,\\
      &\text{fixed} \quad  (\tilde \Phi, {\cal V},C)\,: \qquad &&W \equiv E- TS - \tilde \Phi \tilde Q =\mu C\,,\\
       &\text{fixed} \quad  (\tilde Q, {\cal V}, \mu)\,: \qquad &&G \equiv E - TS-     \mu C=\tilde \Phi \tilde Q
     \end{aligned}
 \end{equation}
 exhibit interesting phase behaviour.  
 
\begin{figure}
    \centering
    \includegraphics[width=0.48\textwidth]{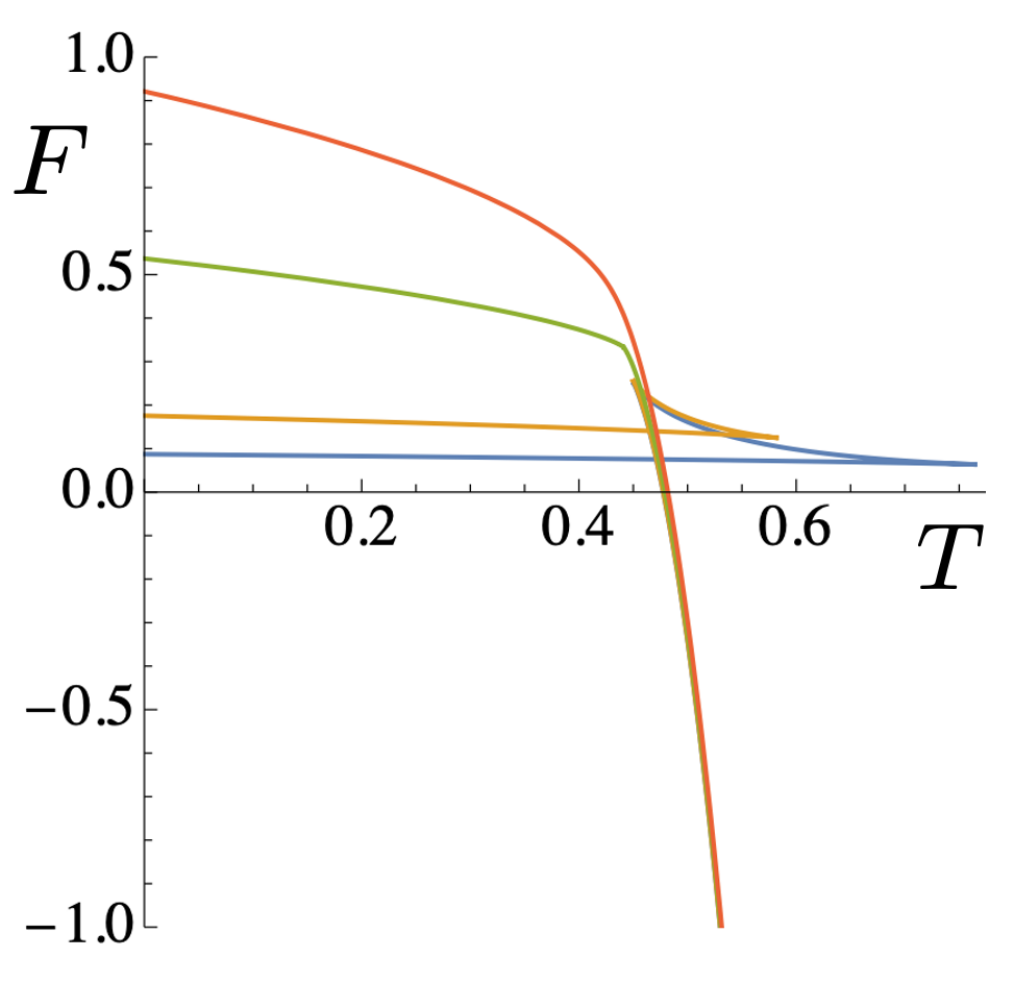} \hspace{0.3 cm}
    \includegraphics[width=0.48\textwidth]{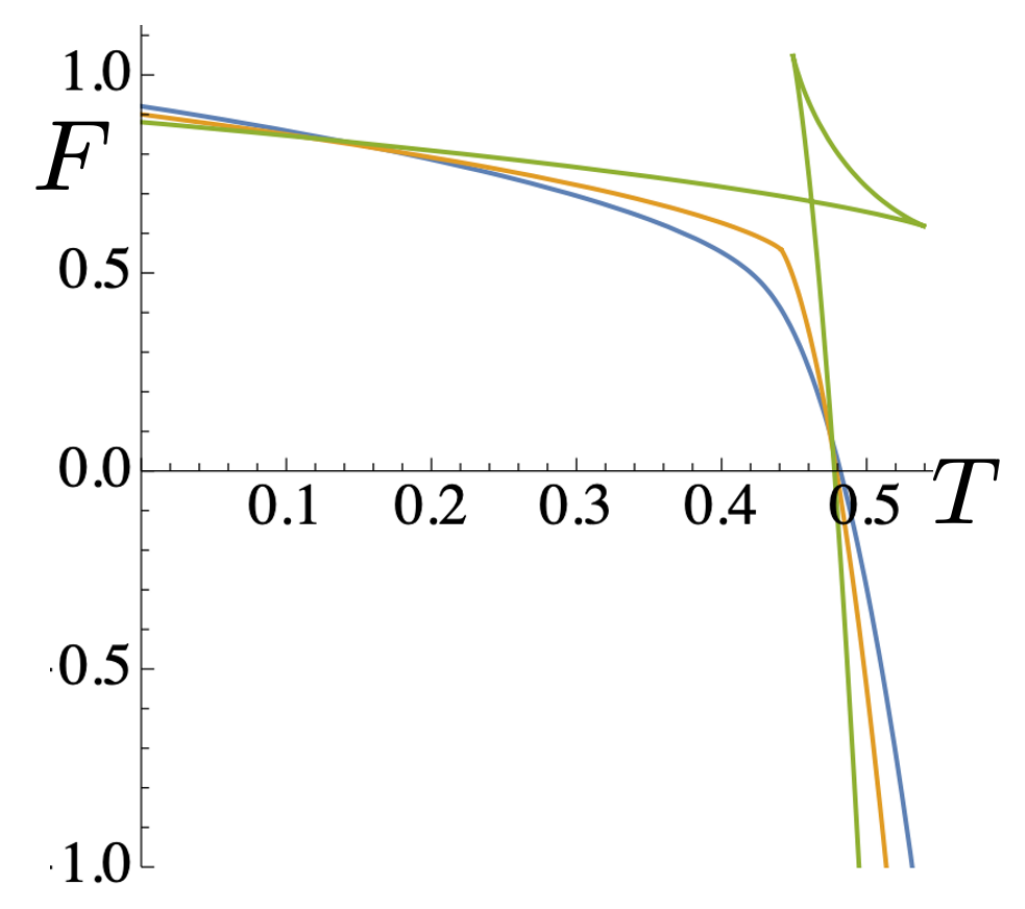}
    \caption{{\bf Plots of the free energy $F$ vs. temperature $T$  in $D=4$ dimensions for the fixed $(\tilde Q,{\cal V},C)$ ensemble.} \textbf{Left}:  The
    different curves correspond to $\tilde Q=0.1,0.2, 4/3 \sqrt{5}, 1$ (blue, orange, green, red) for fixed $R=1$, $C=1$.  We see  for  $Q< Q_{crit}$ (blue, orange)  that   ``swallowtail'' behaviour is present.  A first-order phase transition occurs between a low entropy (`horizontal') branch 
    and a high entropy (`vertical')  branch; both branches are stable.  The intermediate branch connecting them has negative heat capacity and is thus unstable.  
    For $Q=Q_{crit}$ (green) there is a second-order phase transition,   and for $Q > Q_{crit}$ (red) there are no phase transitions. 
    \textbf{Right}: The curves correspond to $C=1, 3\sqrt{5}/4, 4$ (blue, orange, green) for fixed  $R=1$ and $\tilde Q=1$. Again,    different behaviours 
   appear      below and above  a critical value of the  central charge. First-order phase transitions (with the accompanying swallowtail) occur for $C>C_{crit}$ (green), and   becomes of second order at $C=C_{crit}$ (orange). For $C<C_{crit}$ only a single phase exists, as implied by the smooth single-valued curve (blue).  No triple intersections are present; any that appear to be there are due to  plotting resolution.  
    }
    \label{fig:qvcFT}
\end{figure}

 Consider, for example  the fixed ($\tilde Q, {\cal V}, C$) ensemble with Helmholtz free energy  $F$, given by
  \begin{align}
 \label{F1}
 	F  \equiv E - T S 
 	 = C \frac{x^{D-2}}{R} \left( 1 - x^2 + (2D-3) \frac{y^2}{x^{2D-4}}\right)\,.  
 \end{align} 
 and plotted in Fig.~\ref{fig:qvcFT} for fixed $C$ and various values of $\tilde Q$ (left panel), and fixed $\tilde Q$ and various values of $C$  (right panel),
 both for fixed $R$. 
 The differential of $F$ is
 \begin{equation}
     dF = dE-TdS-SdT = -SdT+\tilde \Phi d\tilde Q-pd{\cal V}+\mu dC
 \end{equation}
 which is clearly stationary at   fixed   $(T,\tilde Q,{\cal V},C)$. This ensemble is equivalent to the fixed charge ensemble \cite{Kubiznak:2012wp} discussed
 in section~\ref{4p2}, but   implicitly  ${\cal V}$ and   $C$ are also kept fixed in the dual CFT description.

 We observe in Fig.~\ref{fig:qvcFT}  ``swallowtail'' behaviour dual to that in Fig.~\ref{Fig:Swallow} for the bulk.   Fixing   $C$ and $R$ (left panel)
  a ``swallowtail'' shape (blue curve) is present for $\tilde Q<\tilde Q_{crit}$. This becomes a kink at $\tilde Q=\tilde Q_{crit}$ (green curve), and then a smooth monotonic (orange) curve for $\tilde Q>\tilde Q_{crit}$. The size of the black hole (the value of $x$) increases along the curves increases as $T$ increases, beginning at $T=0$.
 Black holes with small $x\equiv r_h /\ell$  are dual to CFT thermal states with small $S/C = 4 \pi x^{C-1} $, which are states with  low entropy per degree of freedom. The states initially have the lowest free energy $F$ and are   the only available ones near $T=0$.  As $T$ increases, eventually   the self-intersection point is reached, where a first-order phase transition to a  CFT state with  high entropy per degree of freedom (large $x$) takes place.
 These high entropy states lie on the near-vertical  branch of the curve, and are the states  with lowest free energy $F$ for larger values of $T$, thereby  dominating the canonical ensemble.  This  first-order phase transition takes place for all $\tilde Q < \tilde Q_{crit}$, with the transition temperature 
 increasing with  increasing $\tilde Q$. The swallowtail shrinks  in size, becoming just a kink in the curve once $\tilde Q=\tilde Q_{crit}$, at which point the phase transition is second order.  This scenario depicted in the left panel  is commensurate with the canonical ensemble for AdS black holes at fixed charge~\cite{Kubiznak:2012wp}.

The values of $(\tilde Q_{crit}, T_{crit})$ critical point depend on the value of $C$ and can be found by noting that  the temperature has an inflection point as a function of $x$  at this critical value \cite{Chamblin:1999tk}, yielding
 \begin{equation}
 \label{critpt}
 	x_{crit}  =	\frac{(D-2) }{\sqrt{D(D-1)}} \qquad   y_{crit}  = \frac{1}{\sqrt{(D-1)(2D-3)}} x_{crit}^{D-2 }\,.
 \end{equation}
 and thereby implying
 \begin{equation}
 \label{Ccrit}
 	\frac{C_{crit}}{\tilde Q_{crit}} =\frac{1}{2 \alpha (D-1) y_{crit}  }   = \sqrt{\frac{ (2D-3)}{8(D-2)  }}\frac{1}{x_{crit}^{D-2}} 
 \end{equation}
  for the ratio of the central charge and the electric charge at criticality. There is no critical point ( and hence no critical value of the central charge)
 if $\tilde Q=0$.  

The right panel in   Fig.~\ref{fig:qvcFT}  illustrates the situation for various values of   $C$ at fixed $\tilde Q$, with $C<C_{crit}$ (blue), $C=C_{crit}$ (orange) and $C>C_{crit}$ (green). The phase behaviour is similar to that in the left panel, but with the role of $C_{crit}$ reversed: there is a single phase for $C<C_{crit}$, a second-order phase transition at $C_{crit}$, and a first-order phase transition between states with low- and high-entropy per degree of freedom for $C>C_{crit}$. As $C$ decreases, the value of the free energy at which the first-order transition occurs decreases, whereas from the left panel we see that  the free energy of the first-order phase   transition  increases  as $\tilde Q$ decreases  \cite{Cong:2021fnf}. 

These results indicate that   $\tilde Q$ and $1/C$  play  a role  analogous to the bulk pressure $P$ of the Van der Waals fluid in driving the system to its critical point, with their respective conjugates  $\tilde \Phi$ (electric potential) and  $\mu$ (chemical potential) analogous to the bulk volume $V$.  
The critical exponents of the CFT critical point can be shown to be the same as those of  the Van der Waals fluid \cite{Cong:2021jgb}.  These  analogies do  not identify the same physical quantities, and so CFT states dual to charged AdS black holes are in the same universality class
as Van der Waals fluids, but are not identical to them \cite{Cong:2021fnf}. 

Notably there is no critical behaviour in the $p-{\cal V}$ plane in the CFT,  in stark contrast to the $P-V$ criticality for charged AdS black holes discussed in
section~\ref{4p2} \cite{Kubiznak:2012wp}. A plot of the CFT pressure $p$ as a function of ${\cal V}$  show that for any temperature $p$ first decreases to a global minimum before increasing with the volume $\cal{V}$ at any given $(\tilde Q, C)$. In particular, there is no critical temperature at which the $p(\cal{V})$ plot displays an inflection point. The CFT fluid dual to a charged black holes is \emph{not} a standard Van der Waals fluid \cite{Cong:2021fnf}.

Holographic black hole chemistry is now an active area of research, having been extended to include rotation~\cite{Ahmed:2023dnh,Gong:2023ywu},
non-linear electrodynamics~\cite{Bai:2022vmx}, higher curvatue effects~\cite{Paul:2024rto}, and more 
\cite{Zeyuan:2021uol,Alfaia:2021cnk,Dutta:2022wbh,Lobo:2022eyr,Kumar:2022afq,TranNHung:2024tyg,Sadeghi:2024ish,Bousder:2023zhl,Zhang:2024jlp,Cui:2024cnj,Ladghami:2024wkv,Feng:2024uia,Climent:2024nuj}.   A long-term aim of this research is to find   holographic duals of the phenomena described in section~\ref{5s}. 
  
\subsection{Holographic Origins of Black Hole Chemistry}

Very recently a mechanism for a holographic origin of a dynamical cosmological constant was  proposed 
 \cite{Frassino:2022zaz}. In this approach,  classical, asymptotically AdS black holes are mapped to a brane of one less dimension. The resultant object on the brane is called a  quantum black hole \cite{Emparan:2020znc}, and has a conformal matter sector that back-reacts on the brane geometry. Variation of the cosmological constant (the pressure) on the brane corresponds to varying the tension of the brane. In this way  standard thermodynamics in the bulk, including a work term coming from the brane, induces black hole chemistry on the brane, exactly, to all orders in the back-reaction.  
 
A third description of the system is possible insofar as the induced gravity on the brane -- which is also asymptotic to 3-dimensional AdS -- is dual to a 2-dimensional defect CFT on its boundary.   Variations in the brane tension then correspond to variations in the central charge of this 2-dimensional CFT, and in this sense the system is `doubly holographic'  \cite{Frassino:2022zaz}. 

This approach to holographic black hole chemistry is now an active area of research
\cite{Panella:2023lsi,Zhang:2023uay,Gong:2023ywu,Punia:2023ilo,Johnson:2023dtf,Frassino:2023wpc,HosseiniMansoori:2024bfi,
Wu:2024txe,Feng:2024uia,Frassino:2024bjg}.  Extensions more general brane-world models merit investigation, as does 
 a more fundamental string theoretic description of the holographic origin of black hole chemistry.   
 
\section{Microstructure}

Ludwig Boltzmann, who advocated for atomic theory at a time when it was highly controversial, is alleged to have said ``If you can heat it, it has atoms".
Atoms -- or more generally microscopic degrees of freedom --  leave imprints on natural phenomena at large scales  in the form of temperature and heat. 
One can see this from the ideal gas law
\be\label{idgas}
PV = N k_{\rm B} T
\ee
which provides information on the number of microscopic degrees of freedom $N$ in terms of pressure, volume, and temperature, all of which are
 macroscopic variables that can be empirically measured.  In other words, without any ability to probe matter at microscopic scales, we can infer and quantify  the existence of microstructure simply because we can make something hot \cite{Padmanabhan:2015zmr}. It is reasonable to infer that, since a black hole can get hot (it has nonzero temperature), it must therefore possess its own microscopic structure. 

Today we can use statistical mechanics to construct the macroscopic thermodynamic quantities of a fluid from  its microscopic molecular constituents and their interactions.  However we don't know what these constituents are for a black hole.  Early investigations of this problem for  a charged AdS black hole began with the notion of a  {\it black hole molecular density}
\cite{Wei:2015iwa} 
\be
n=\frac{1}{v}=\frac{1}{2l_p^2 r_+}
\ee
similar to \eqref{idgas}.   As we cross the coexistence line shown in Fig.~\ref{Fig:Swallow}, the number densities of large and small black holes
experience a discontinuous jump accompanied by a latent heat 
\be
\textsf{L}=T\Delta S = T\Delta v \frac{\partial P}{\partial T} = T \left(\frac{1}{n_{\textrm{LBH}}} -  \frac{1}{n_{\textrm{SBH}}}\right) 
\frac{\partial P}{\partial T} 
\ee
reminiscent of the magnetization/temperature behaviour of an Ising ferromagnet. 
If the system passes the critical point the  latent heat vanishes due to a continuous change of the number density $n$.   

To gain further insight into black hole microstructure, we can begin with the Boltzmann entropy formula 
\begin{equation}
 S=k_{\rm B}\ln\Omega,
\end{equation}
where  $\Omega$ is the number of the microscopic states of the corresponding thermodynamic system. Inverting this yields
\begin{equation}
 \Omega=e^{\frac{S}{k_{\rm B}}},\label{Os}
\end{equation}
which is the starting point of thermodynamic fluctuation theory. For a system of $N+1$ independent variables $z^{\mu}$ with $\mu$=0, 1, ..., $N$, the probability of finding its state to be between $(z^{0}, ..., z^{N})$  and $( z^{0}+ dz^{0}, ..., z^{\rm N} + dz^{\rm N})$ is proportional to the number of microstates
\begin{equation}\label{probOm}
 P(z^{0}, ..., z^{\rm N})dz^{0}\cdot \cdot \cdot dz^{\rm N}=
 C\Omega (z^{0}, ..., z^{\rm N})dz^{0}\cdot \cdot \cdot dz^{\rm N} = C e^{\frac{S}{k_{\rm B}}}dz^{0}\cdot \cdot \cdot dz^{\rm N}
\end{equation}
where $C$ is a normalization constant.

Consider partitioning a thermodynamic system  into a small sub-system $\textsf{S}$ plus its remainder $\textsf{E}$, regarded as the environment. The total entropy can be written as 
\begin{align}
S(z^{0}, ..., z^{\rm N}) &= S_{\textsf{S}}(z^{0}, ..., z^{\rm N})+S_{\textsf{E}}(z^{0}, ..., z^{\rm N}) \nonumber \\
&=S(z_{0}^{0}, ..., z_{0}^{\rm N}) + \left. \frac{\partial S_\textsf{S}}{\partial z^{\mu}}  \right|_{z^{\mu} = z^{\mu}_0} \Delta z^{\mu}_{\textsf{S}}
+ \left.\frac{\partial S_\textsf{E}}{\partial z^{\mu}}   \right|_{z^{\mu} = z^{\mu}_0}   \Delta z^{\mu}_{\textsf{E}}
\label{qwq}\\
&+ \left. \frac{1}{2}\frac{\partial^{2}S_{\textsf{S}}}{\partial z^{\mu}\partial z^{\nu}}
        \right|_{z^{\mu} = z^{\mu}_0}  \Delta z^{\mu}_{\textsf{S}}\Delta z^{\nu}_{\textsf{S}}
       + \left. \frac{1}{2}\frac{\partial^{2}S_{\textsf{E}}}{\partial z^{\mu}\partial z^{\nu}}
        \right|_{z^{\mu} = z^{\mu}_0}  \Delta z^{\mu}_{\textsf{E}}\Delta z^{\nu}_{\textsf{E}}
   +\cdots \nonumber
\end{align}
where $S_{\textsf{S}}\ll S_{\textsf{E}}\sim S$. Taking the ezpansion point $z^{\mu} = z^{\mu}_0$ to be the local mazimum of the entropy
yields
\begin{eqnarray} \label{entfluc}
 \Delta S=S-S_{0} = \left. \frac{1}{2}\frac{\partial^{2}S_{\textsf{S}}}{\partial z^{\mu}\partial z^{\nu}}
        \right|_{0}  \Delta z^{\mu}_{\textsf{S}}\Delta z^{\nu}_{\textsf{S}}
        +\cdots  
\end{eqnarray}
since  $\frac{\partial S_{\textsf{S}}}{\partial z^{\mu}} |_{0} \Delta z^{\mu}_{\textsf{S}}=-\frac{\partial S_{\textsf{E}}}{\partial z^{\mu}} |_{0} \Delta z^{\mu}_{\textsf{E}}$
(the fluctuating parameters $z^\mu$ are conservative and additive).  The last term of (\ref{qwq}) is much smaller than the fourth term since
$S_{\textsf{E}}\sim S$ remains nearly constant. Inserting \eqref{entfluc} into \eqref{probOm} gives
\begin{equation}\label{probDl}
 P(z^{0}, ..., z^{\rm N}) \propto e^{-\frac{1}{2}\Delta l^{2}},
\end{equation}
upon absorbing $S_{0}$ into the normalization constant, 
where
\begin{eqnarray}
 \Delta l^{2}&=&-\frac{1}{k_{\rm B}}\frac{\partial^{2}S_{\textsf{S}}}{\partial z^{\mu}\partial z^{\nu}} \Delta z^{\mu}\Delta z^{\nu}
 = \textsf{g}_{\mu\nu} \Delta z^{\mu}\Delta z^{\nu}
 \label{Ds}
\end{eqnarray}
is the distance  between two neighboring fluctuation states \cite{Ruppeiner}.    

The relation \eqref{probDl} shows that thermodynamic fluctuations can be understood geometrically: the greater the geometric
distance \label{Ds} between  two thermodynamic states, the less likely there will be a fluctuation between them. 
The line element \eqref{Ds} thus encodes information about the effective interaction between two microscopic fluctuation states.  The scalar curvature
of the  information metric $\textsf{g}_{\mu\nu}$ in \eqref{Ds} provides an indicator of its
microstructure interactions \cite{Ruppeiner:1995zz,HiroshiOshima_1999}:  a repulsive/attractive interaction dominates for 
positive/negative scalar curvature;  vanishing curvature indicates   repulsive and attractive interactions are in balance. 
It is reasonable to posit further  that the value of the scalar curvature measures the strength of the interactions.   The thermodynamic potential is the entropy of this information geometry,  known as   Ruppeiner geometry. 

It is useful to separate the thermodynamic variables $(z^{0}, ..., z^{N}) = (x^{0}, ..., x^{N},y^{0}, ..., y^{N})$ where the 
$x^{i}$ are the extensive thermodynamic variables,  and $y_{i}$ their corresponding conjugate potentials. Setting $k_{\rm B}=1$, the first law
of thermodynamics can be written as  
\begin{equation}
 dS=\frac{1}{T}dU-\sum_{i}\frac{y_{i}}{T}dx^{i} = \frac{\partial S}{\partial x^{\mu}} \Delta x^{\mu} 
 \label{firstenlaw}
\end{equation}
where $\mu$=0,1,2,$\cdots$ and $i$=1,2,3,$\cdots$. The line element \eqref{Ds} is then
\begin{align}
 \Delta l^{2} & =-\Delta \left(\frac{\partial S}{\partial x^{\mu}} \right) \Delta x^{\mu}
 =  \frac{\Delta T}{T^{2}} \Delta U -  (\frac{y_{i}\Delta T}{T^{2}}-\frac{\Delta y_{i}}{T})\Delta x^{i} 
\nonumber\\  
&= \frac{1}{T}\Delta T\Delta S+\frac{1}{T}\Delta y_{i}\Delta x^{i} 
 \label{dl2}
\end{align}
using \eqref{firstenlaw}.

If the fluctuation coordinate variables are chosen to be ($T$, $y^{i}$), then
using the thermodynamic potential $W=U-TS-y_{i}x^{i}$ the line element
\eqref{dl2} becomes
\begin{equation}
 \Delta l^{2}=-\frac{1}{T}\left(\frac{\partial^{2}W}{\partial p_{\mu}\partial p_{\nu}}\right) \Delta p_{\mu}\Delta p_{\nu},\quad p_{\mu}=(T, y_{i}).
\end{equation}
where $S=-\partial_{T}W$ and $x_{i}=-\partial_{y^{i}}W$.   This form of the  information metric $\textsf{g}_{\mu\nu}$ is a bit more awkward to
work with since it has off-diagonal terms.  However if 
($T$, $x^{i}$) are chosen as the fluctuation variables then the relation
\eqref{dl2} becomes
\begin{equation}
 \Delta l^{2}=-\frac{1}{T}\left(\frac{\partial^{2}F}{\partial T^{2}}\right)\Delta T^{2}
   +\frac{1}{T}\left(\frac{\partial^{2}F}{\partial x^{i}\partial x^{j}}\right)\Delta x^{i}\Delta x^{j}. \label{ddl2}
\end{equation}
using the free energy $F=U-TS$ and 
noting that $S=-\partial_{T}F$ and $y_{i}=\partial_{x^{i}}F$.  

Restricting attention to fluctuations of two variables $(T,x)$,  the metric is  two-dimensional 
\begin{equation}
 \textsf{g}_{\mu\nu}=\frac{1}{T}\left(
  \begin{array}{cc}
     -\left(\frac{\partial^{2}F}{\partial T^{2}}\right)_{x} & 0\\
    0 & \left(\frac{\partial^{2}F}{\partial x^{2}}\right)_{T}\\
  \end{array}
\right)
=
\frac{1}{T}\left(
  \begin{array}{cc}
     \left(\frac{\partial S}{\partial T}\right)_{x} & 0\\
    0 & -\left(\frac{\partial y}{\partial x}\right)_{T}\\
  \end{array}
\right) 
\end{equation}
where 
\begin{equation}
 C_{x}=T\left(\frac{\partial S}{\partial T}\right)_{x},
\end{equation}
is   the heat capacity at constant $x$. The Riemann scalar curvature of this metric is  \begin{eqnarray}
 \textsf{R}&=&\frac{1}{2C_{x}^{2}(\partial_{x}y)^{2}}
 \bigg\{
 T(\partial_{x}y)\bigg[(\partial_{x}C_{x})^{2}+(\partial_{T}C_{x})(\partial_{x}y-T\partial_{T,x}y)\bigg]\nonumber\\
 &+&C_{x}\bigg[(\partial_{x}y)^{2}+T\left((\partial_{x}C_{x})(\partial_{x,x}y)-T(\partial_{T,x}y)^{2}\right)
 +2T(\partial_{x}y)(-(\partial_{x,x}C_{x})+T(\partial_{T,T,x}y))\bigg]
 \bigg\} \nonumber\\\label{RR}
\end{eqnarray}
which can potentially diverge at $C_{x}=0$ or $(\partial_{x}y)_{T} =0$. The latter relation is one of the two conditions
\begin{equation}\label{critpt1}
 \left(\partial_{x}y\right)_{T}=\left(\partial_{x,x}y\right)_{T}=0
\end{equation}
determining the critical point of a VdW-like phase transition. Hence the scalar curvature $\textsf{R}$ of the Ruppeiner geometry has  divergent behaviour at the critical point of the phase transition. This property provides a possible link  between $\textsf{R}$ and the correlation length, which approaches infinity at the critical point.

Consider a Van der Waals fluid, whose Helmholtz free energy is \cite{Landau:1980mil}
\begin{equation}
  F=-\frac{3}{2}T\ln T-\xi T+\epsilon-T\ln(e(v-b))-\frac{a}{v},\label{fffd}
\end{equation}
where $e$, $\xi$ and $\epsilon$ are constants and  $v=V/N > b$ is the specific volume.  The entropy, energy, and heat capacity are
\begin{eqnarray}
 S&=&-\left(\frac{\partial F}{\partial T}\right)_{v}
  =\frac{3}{2}\left(1 +\frac{2}{3}\xi+\ln T\right) +\ln(e(v-b)),\\
 U&=&F+TS
  =\frac{3}{2}T-\frac{a}{v} +\epsilon \\
  C_{v}&=&\frac{3}{2}k_{\rm B}
  \label{cvVdW}
\end{eqnarray}
temporarily restoring the Boltzmann constant $k_{\rm B}$ to emphasize the small value of $C_{v}$.  The equation of state \eqref{VdWstate}
follows from  $P=-\left(\frac{\partial F}{\partial v}\right)_{T}$
and has a critical point at
\begin{equation}\label{redspace}
 P_{\rm c}=\frac{a}{27b^{2}},\quad v_{\rm c}=3b,\quad T_{\rm c}=\frac{8a}{27b}
\end{equation}
which can be obtained from \eqref{critpt1} upon setting $y=P$ and $x=v$.  Using this, one can rewrite \eqref{VdWstate} as
\begin{equation}\label{VdWeos}
 \tilde{P}=\frac{8\tilde{T}}{3\tilde{v}-1}-\frac{3}{\tilde{v}^{2}},
\end{equation}
where the reduced pressure, temperature, and specific volume are defined by $\tilde{P}=P/P_{\rm c}$, $\tilde{T}=T/T_{\rm c}$, and $\tilde{v}=v/v_{\rm c}$. 

\begin{figure*}
\begin{center}
{\label{ISOa}
\includegraphics[width=0.48\textwidth]{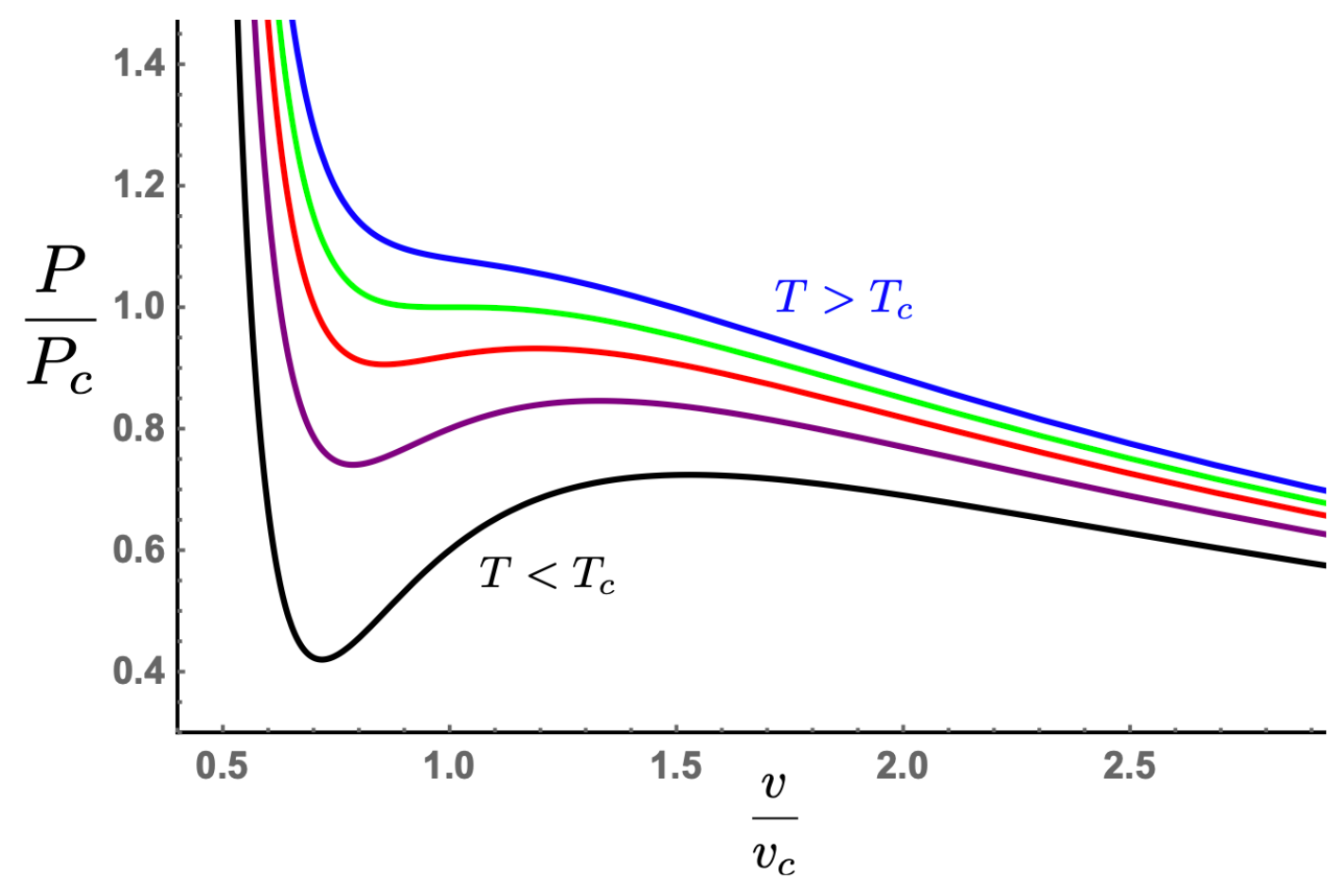}}
{\label{ISPb}
\includegraphics[width=0.48\textwidth]{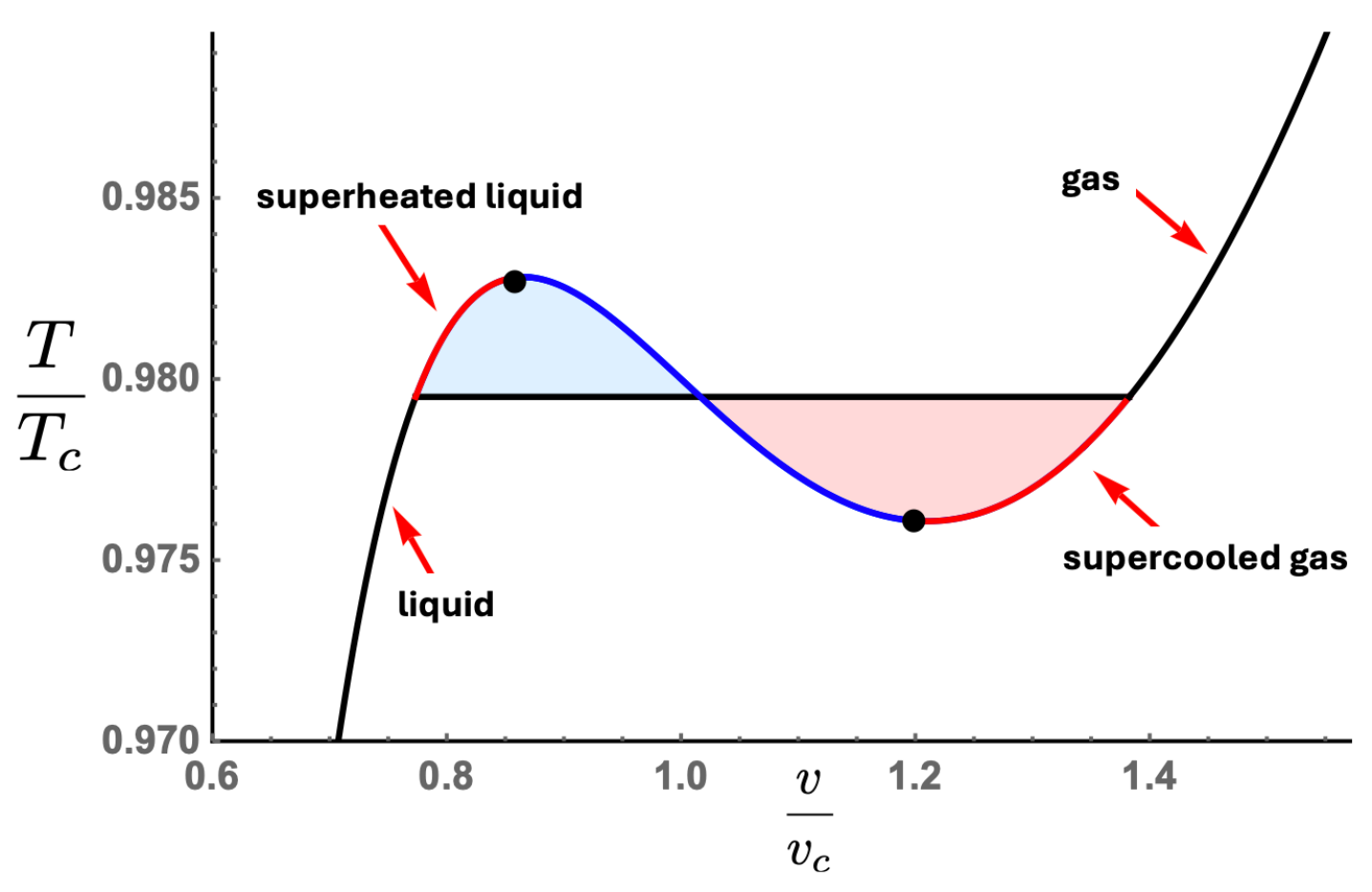}}
\end{center}
\caption{{\bf Left}  $\tilde{P}$ as a function of $\tilde{v}$ for a VdW fluid  near the critical point, with isotherms $\tilde{T}=$0.9, 0.95, 0.98, 1.00, and 1.02 from bottom to top. {\bf Right} $\tilde{T}$ as a function of $\tilde{v}$ for a VdW fluid  near the critical point, for fixed  $\tilde{P}=0.92$. The black horizontal line of $T/T_{\rm c}=0.98$, corresponds to the phase transition temperature calculated from the free energy $F$. The stable gas and liquid branches are given by
the  black curves, and the red curves are the two metastable  superheated liquid and supercooled gas branches. The blue curve of negative slope is an unstable branch; at the phase transition  it is replaced by the horizontal line. The spinodal points separate the metastable branches from the unstable one
and are depicted by the two black dots. }
\label{pISPb}
\end{figure*}

Isothermal and isobaric curves for the VdW fluid are shown in Fig. \ref{pISPb}. The isotherms (left panel) are for  $\tilde{T}=$0.9, 0.95, 0.98, 1.00, and 1.02 from bottom to top and have two extremal points for $\tilde{T}<1$; for $\tilde{T}>1$ (blue curve) no extremal point exists. The parts of the curves with negative slopes are the stable liquid and gas phases, whereas the part with positive slope is an unstable phase.  For a given $\tilde{P}$, the phase transition point on an isothermal curve can be obtained via the equal area law. An isobaric curve with $\tilde{P}=0.92$ is shown in the right panel; the phase transition temperature is at $T/T_{\rm c}=0.98$.  The equal area law does not hold for a $\tilde{T}$--$\tilde{v}$ diagram and the  light blue/red shaded regions are not of equal area.  Stable phases are shown in black and metastable phases in red;  the unstable branch is the blue curve of negative slope.  The extremal points (black dots) separate the two metastable branches from the unstable branch, and are called spinodal points.

\begin{figure}
\begin{center} 
\includegraphics[width=0.48\textwidth]{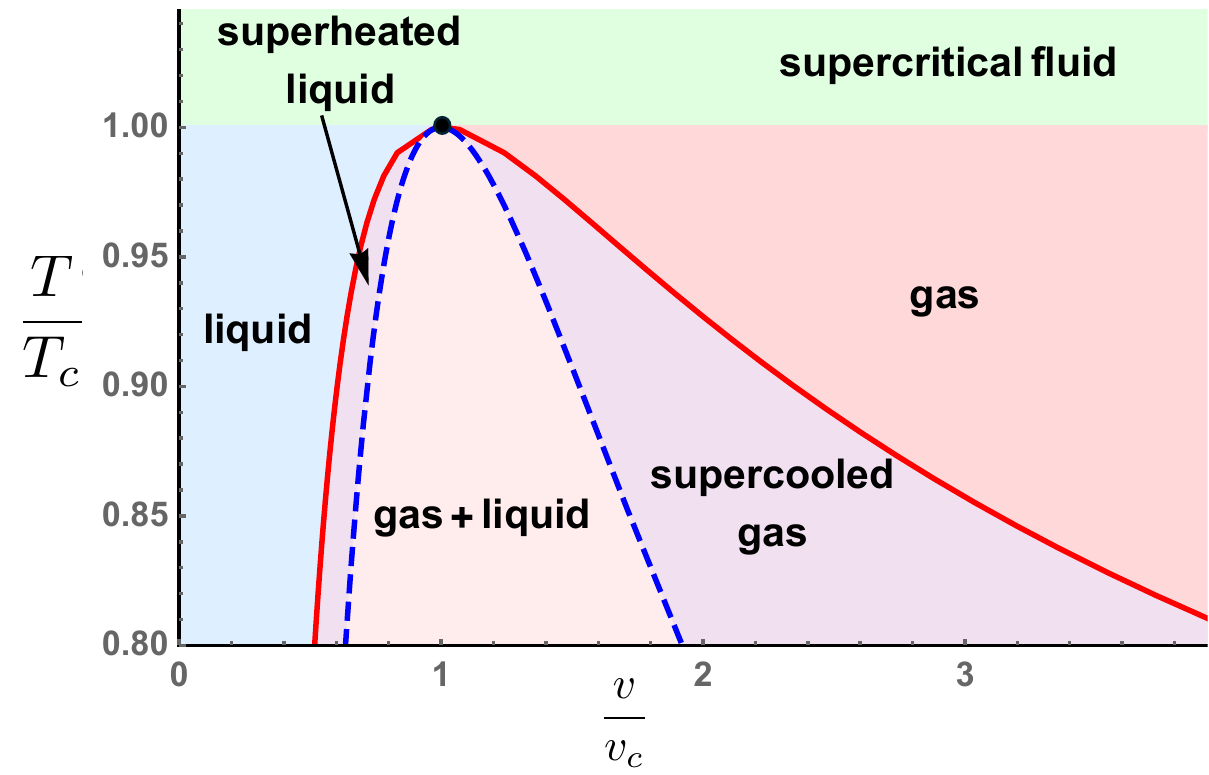}
\includegraphics[width=0.48\textwidth]{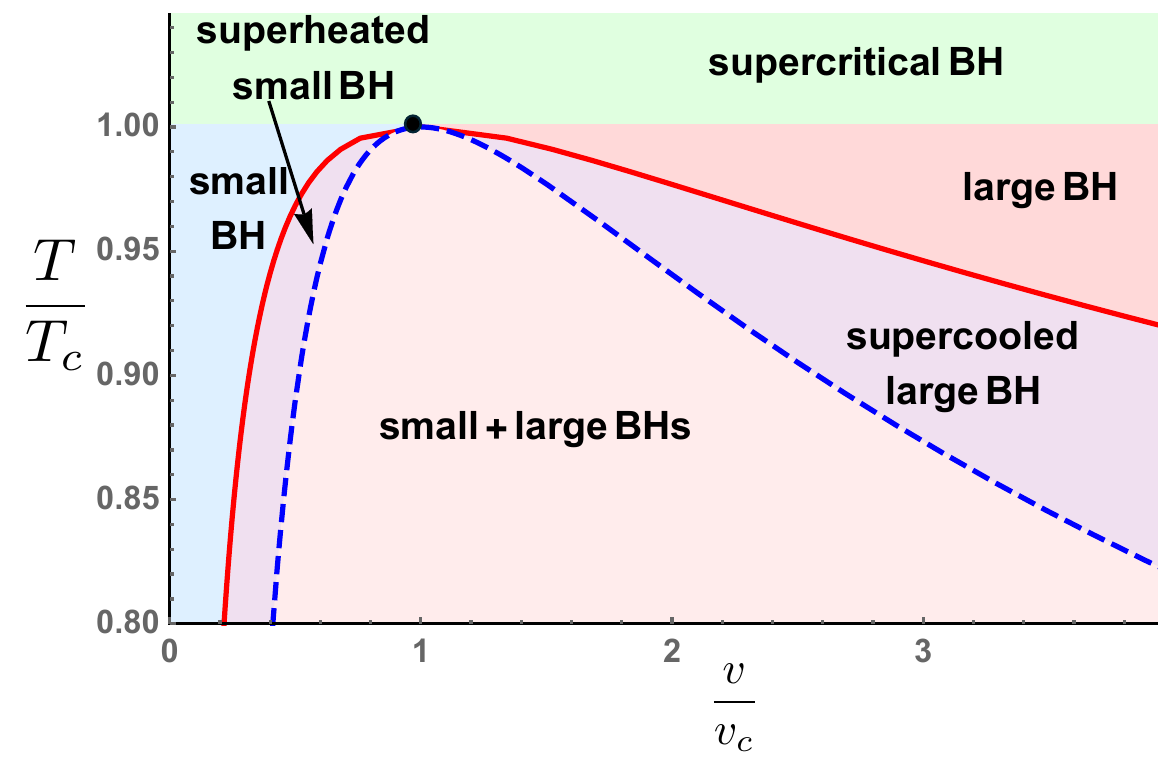}
\end{center}
\caption{ Phase diagrams for a VdW fluid (left) and charged AdS black hole (right). Coexistence curves (red, solid),  spinodal curves (blue dashed)
and critical point (black dot) are shown.  Regions for the gas, liquid, metastable superheated, metastable supercooled,  and supercritical  phases are also displayed. For the small black hole the spinodal curve has a reduced volume $\tilde{V}=1/3\sqrt{3}$ when $\tilde{T}=0$, whereas for the large black hole spinodal curve, $\tilde{V}$ approaches infinity as $\tilde{T}\to 0$.
} 
\label{VdWPTPb}
\end{figure}

The locus of spinodal points as $\tilde{P}$ varies is a spinodal curve, determined by the condition \cite{Wei:2019yvs}
\begin{equation}
 (\partial_{v}P)_{T}=0,\quad \text{or}\quad
 (\partial_{v}T)_{P}=0
\end{equation}
which yields the relation 
\begin{equation}
 \tilde{T}_{\rm sp}=\frac{(3\tilde{v}-1)^{2}}{4\tilde{v}^{3}}
 \label{spcurvevdw}
\end{equation}
where $1/3<\tilde{v}<1$ is for the liquid spinodal curve and $\tilde{v}>1$ is for the gas spinodal curve, shown as  blue dashed curves 
in the left panel of Fig.~\ref{VdWPTPb}.  The red coexistence curve separating the liquid and gas phases does not have an analytic form, but can
be computed numerically from a fitting   formula \cite{Johnston_2014}. 
The black dot denotes the critical point and the light green region is the supercritical fluid phase, in which gas and liquid cannot be clearly distinguished. 
 In the  coexistence phase region  the  equation of state (\ref{VdWeos}) is invalid.

The line element of the VdW Ruppeiner geometry is
\begin{eqnarray}
 dl^{2}&=&\frac{C_{v}}{T^{2}}dT^{2}+\frac{(\partial_{v}P)_{T}}{T}dv^{2} 
= \frac{3}{2T^{2}}dT^{2}+\frac{Tv^{3}-2a(v-b)^{2}}{Tv^{3}(v-b)^{2}}dv^{2}
\end{eqnarray}
and has scalar curvature
\begin{eqnarray}
\textsf{R}=\frac{(3\tilde{v}-1)^{2}\big((3\tilde{v}-1)^{2}-8\tilde{T}\tilde{v}^{3}\big)}{3\big((3\tilde{v}-1)^{2}-4\tilde{T}\tilde{v}^{3}\big)^{2}},
\end{eqnarray}
in the reduced parameter space \eqref{redspace}. There is no explicit dependence on the parameters $a$ and $b$ of the VdW fluid, indicating that  the scalar curvature  expresses universal properties of different VdW fluid systems.
It is also clear that $\textsf{R}$ vanishes when $\tilde{v}=1/3$, or $v=b$. At this point all the volume is occupied by the fluid molecules and the fluid becomes a rigid body, No interaction  between the fluid molecules can exist, consistent with $\textsf{R}=0$ corresponding to vanishing interaction \cite{HiroshiOshima_1999}. 
 
The denominator of $\textsf{R}$ vanishes at the critical point and along the spinodal curves \eqref{spcurvevdw}. It changes sign at
\begin{equation}
 T_{0}=\frac{\tilde{T}_{\rm sp}}{2}=\frac{(3\tilde{v}-1)^{2}}{8\tilde{v}^{3}}
 \label{to}
\end{equation}
which suggests a repulsive microstructure interaction. However the region where $\textsf{R}$ changes sign is in the coexistence phase, where
the equation of state is invalid; the divergent behaviour also occurs in this region.  Hence both the 
positive and divergent behaviours of $\textsf{R}$ are excluded, implying only an attractive interaction between the   molecules of the VdW fluid in the liquid and gas phases.

 Black hole chemistry offers the possibility of comparing the microstructure of a VdW fluid to that of a charged AdS black hole using these tools
\cite{Wei:2015iwa,Wei:2019uqg}, taking $(T,v)$ to be the fluctuating variables, with $v=V/N$ the specific thermodynamic volume of the black hole. 
Rescaling parameters in the equation of state \eqref{RNstate} in  terms of values at the critical point \eqref{critRNAdS} yields 
\begin{equation}
 \tilde{P}=\frac{8 \tilde{T}}{3\tilde{v}}-\frac{2}{\tilde{v}^{2}}+\frac{1}{3\tilde{v}^{4}}
 \label{crstate}
\end{equation}
for the reduced equation of state, which is the black hole analogue of \eqref{VdWeos}; note that the charge $Q$ does not explicitly appear.
The reduced temperature and Gibbs free energy \eqref{GibbsRN} in terms of $\tilde{P}$ and $\tilde{V}$ are \cite{Wei:2015ana}
\begin{eqnarray}
 \tilde{T}&=&\frac{3 \tilde{P} \tilde{V}^{4/3}+6
   \tilde{V}^{2/3}-1}{8 \tilde{V}},\\
 \tilde{G}&=&\frac{1}{8} \left(-\tilde{P}\tilde{V}+6
   \sqrt[3]{\tilde{V}}+\frac{3}{\sqrt[3]{\tilde{V}}}\right),
\end{eqnarray}
where $\tilde{G}=G/G_{\rm c}$ with $G_{\rm c}=\frac{\sqrt{6}}{3}Q$.  The swallowtail behaviour in Fig.~\ref{Fig:Swallow} occurs for   $\tilde{P}<1$. 

The spinodal curves for the charged AdS black hole can be straightforwardly obtained by solving $(\partial_{\tilde{v}}\tilde{P})=0=(\partial_{\tilde{V}}\tilde{P})$ from (\ref{crstate}). This gives
\begin{equation}
 \tilde{T}_{\rm sp}=\frac{3\tilde{V}^{\frac{2}{3}}-1}{2\tilde{V}},\label{spcurve}
\end{equation}
where $1/3\sqrt{3}<\tilde{V}<1$ holds for the small black hole spinodal curve, and $\tilde{V}>1$ for the large black hole spinodal curve.
An analytic form of the large/small coexistence curve in  Fig.~\ref{Fig:Swallow} 
for small/large black hole phases can also be obtained  \cite{Spallucci:2013osa} and reads
\begin{eqnarray}
 \tilde{T}^{2}=\tilde{P}(3-\sqrt{\tilde{P}})/2,\label{tt}
\end{eqnarray}
in terms of the reduced quantities.  These curves are plotted in the right panel of Fig.~\ref{VdWPTPb}.  The qualitative resemblance to a VdW fluid
is quite striking, with only the sizes of the regions differing, and the coexistence phase extending to arbitrarily large values of $\tilde{T}$.

The  line element of the Ruppeiner geometry for the charged AdS black hole is \cite{Wei:2019uqg}
\begin{equation}
\label{RNRupmet}
 dl^{2}=\frac{C_{V}}{T^{2}}dT^{2}+\frac{(\partial_{V}P)_{T}}{T}dV^{2}
\end{equation}
which requires some care to deal with since 
\begin{equation}
 C_{V}=T\left(\frac{\partial S}{\partial T}\right)_{V} = 0
\end{equation}
since constant volume means $dr_{r_+}=0$ since $V\sim r_{r_+}^{3}$, in turn implying $dS=0$  since $S=\pi r_{r_+}^{2}$.
This can be dealt with by noting from \eqref{cvVdW} that $C_V$ for a VdW fluid is very small; hence in computing the Ruppeiner
curvature, we can use the metric \eqref{RNRupmet}, treating   $C_{V}$ as a constant with its value infinitessimally close to zero.  
Normalizing the scalar curvature $\textsf{R}$ such that
\begin{equation}
 \textsf{R}_{\rm N}= C_{V} \textsf{R} 
\end{equation}
then gives
\begin{equation}
\textsf{R}_{\rm N}=\frac{(3\tilde{V}^{\frac{2}{3}}-1)(3\tilde{V}^{\frac{2}{3}}-4\tilde{T}\tilde{V}-1)}{2(3\tilde{V}^{\frac{2}{3}}-2\tilde{T}\tilde{V}-1)^{2}},\label{crnn}
\end{equation}
in terms of the reduced volume and temperature. As with the VdW fluid,  $R_{\rm N}$ has no explicit dependence on the charge $Q$, indicating
that it captures universal properties of charged AdS black holes.
  
Like the VdW fluid, $\textsf{R}_{\rm N}$ diverges on the spinodal curves, vanishes at a finite value
\begin{equation}
 \tilde{V}=\frac{1}{3\sqrt{3}}
\end{equation}
of $\tilde{V}$,  and also vanishes at 
\begin{equation}\label{schg}
 T_{0}=\frac{\tilde{T}_{\rm sp}}{2}=\frac{3\tilde{V}^{\frac{2}{3}}-1}{4\tilde{V}}
\end{equation}
which is half of the spinodal curve temperature, similar to (\ref{to}). However unlike the VdW fluid, the sign-changing curve
\eqref{schg} has values in regions where the equation of state \eqref{crstate} is valid.

\begin{figure}
\begin{center} 
\includegraphics[width=0.48\textwidth]{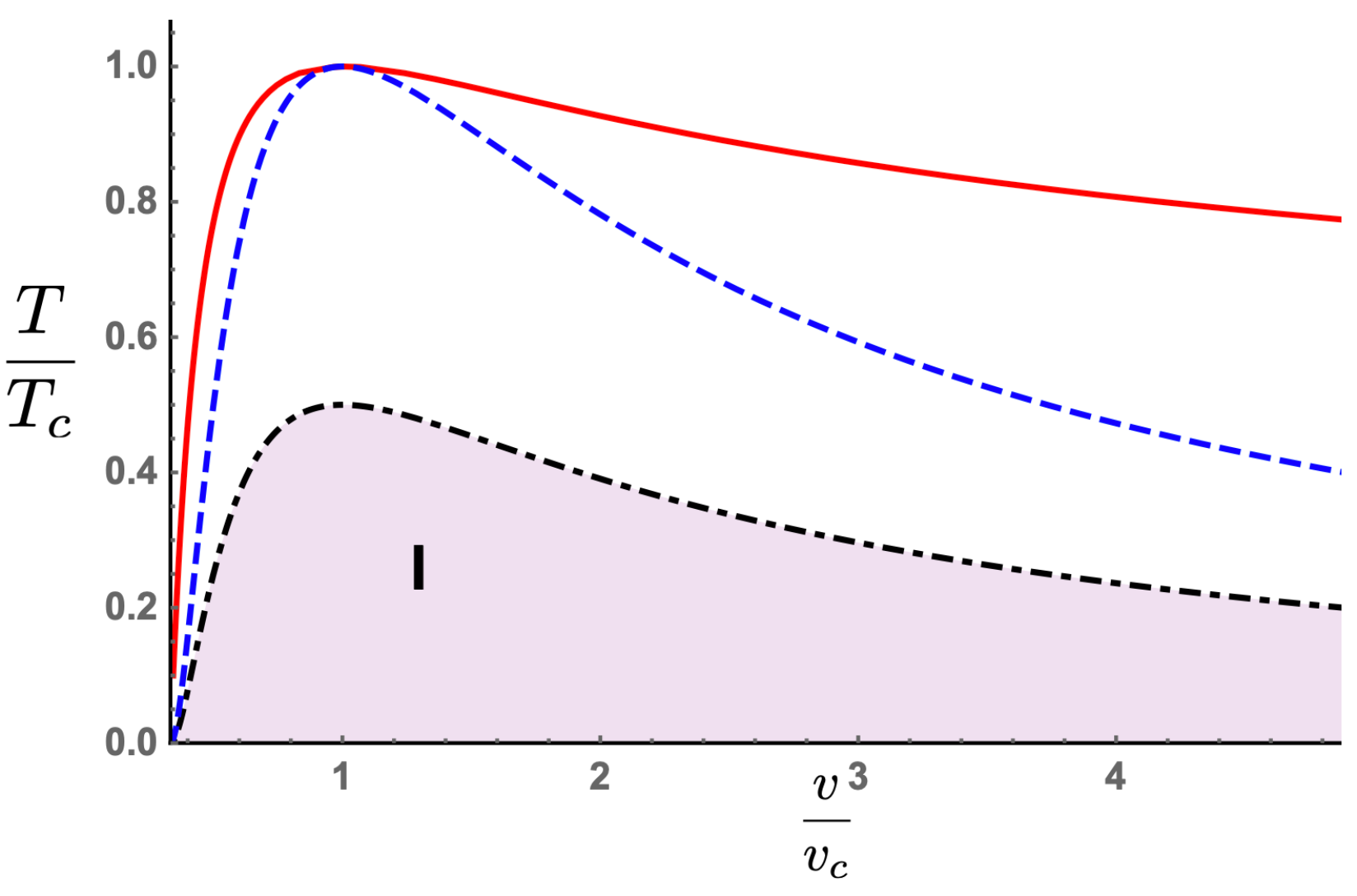}
\includegraphics[width=0.48\textwidth]{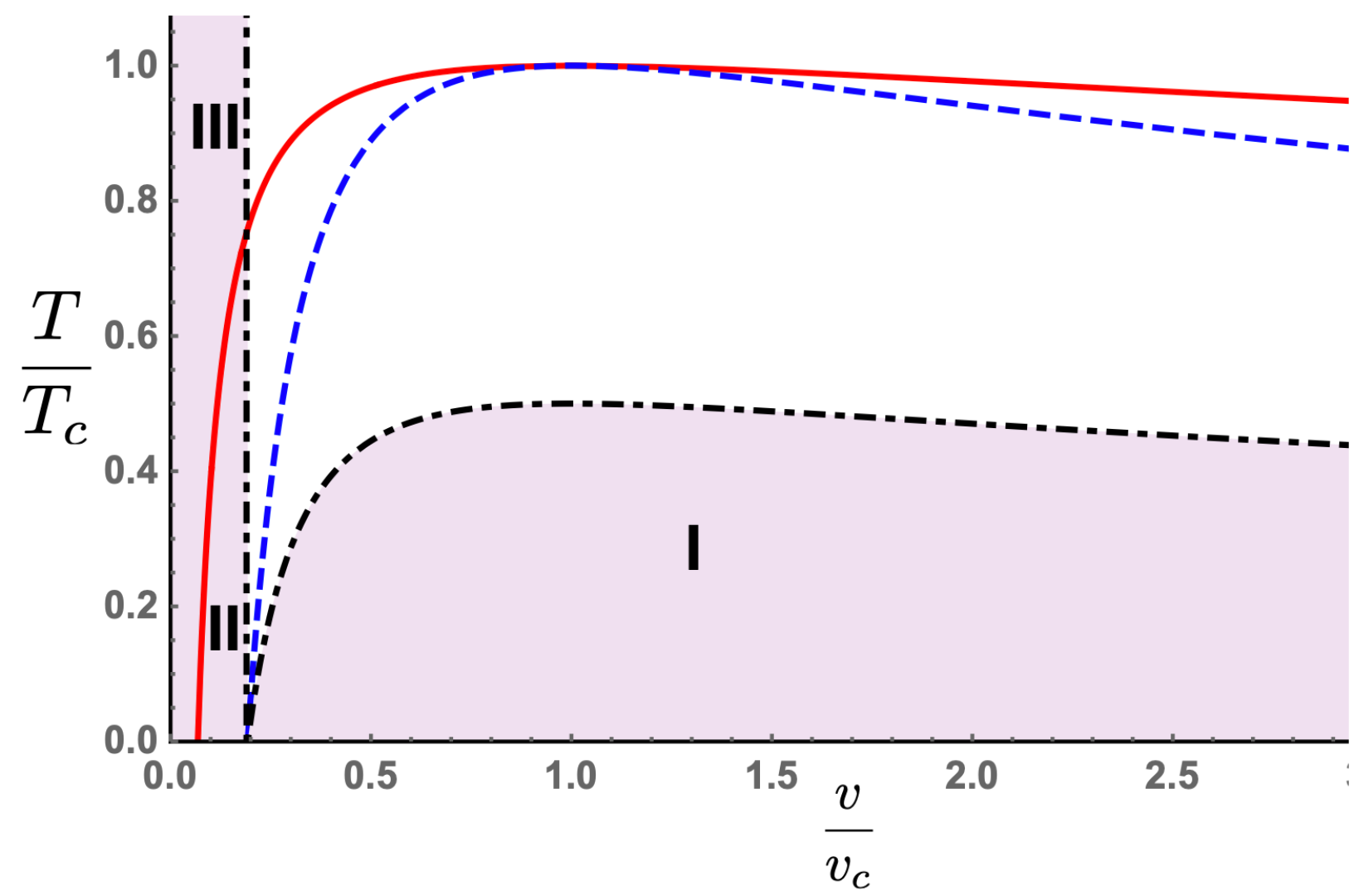}
\end{center}
\caption{Characteristic curves for the VdW fluid (left) and charged AdS black hole (right), with coexistence (red solid) and  sign-changing (black dot-dash) curves  displayed. The blue dashed line corresponds to the temperature $\tilde{T}_{\rm div}$ at which $\textsf{R} \to -\infty$. In the shaded regions I, II, and III $\textsf{R} > 0$;  otherwise,  $\textsf{R}<0$.
} 
\label{F22}
\end{figure}

A comparison of the coexistence  (red solid line) and   sign-changing (black dot dashed line) curves  between the VdW 
and charged AdS black hole is shown in Fig.~\ref{F22}.  While the phase structure for both is qualitatively similar at large $\tilde{v}$,
the situation is markedly different  for small $\tilde{v}$. The black hole has three regions with positive $\textsf{R}_{\rm N}$, whereas the VdW fluid
has only one region.  For each, the equation of state is inapplicable below the coexistence curve, and so no conclusions can be drawn with regards
to the repulsive/attractive character of the microstructure interactions  in regions  I and II. However  region III for the black hole exists 
above the coexistence curve, indicating that  repulsive interactions dominate among the microstructures of small charged AdS black holes  at sufficiently high temperature.  Since region III is far from the $\tilde{T}_{\rm div}$ curve, $R_{\rm N}$ is small and so the repulsive microstructure interaction, while dominant, is weak. A weak attractive interaction dominates in other parameter regions above the coexistence curve.

The   `molecules' of the charged AdS black hole system thus can have either repulsive or attractive interactions, depending on the size of the black hole, unlike those of a VdW fluid, whose interactions are only attractive.

It is possible to draw some further conclusions about the thermodynamic characteristics using the Ruppeiner curvature.  Notwithstanding the inapplicability of the equation of state in the coexistence region, $\textsf{R}_{\rm N}$  diverges at the critical point.  Expanding this quantity near the critical point
along the saturated small (SBH) and large (LBH) black hole curves yields
\begin{eqnarray}
 R_{\rm N}({\rm SBH})&=&-\frac{1}{8}t^{-2}+\frac{1}{2\sqrt{2}}t^{-\frac{3}{2}}+\mathcal{O}(t^{-1}),\label{rnsb1}\\
 R_{\rm N}({\rm LBH})&=&-\frac{1}{8}t^{-2}-\frac{1}{2\sqrt{2}}t^{-\frac{3}{2}}+\mathcal{O}(t^{-1}),\label{rnsb2}
\end{eqnarray}
where $t=1-\tilde{T}$. The scalar curvature $\textsf{R}_{\rm N} \to -\infty$ at the critical point with a universal critical exponent of $2$.  Relating this to
the correlation length $\xi\sim t^{-\nu}$ implies
\begin{equation}\label{RNnu}
R_{\rm N}  \sim-\xi^{\frac{2}{\nu}}   \sim   -\xi^{4}
\end{equation}
or $\nu=1/2$, where the latter relation follows from mean field theory.  Finally using 
(\ref{rnsb1}) and (\ref{rnsb2}) yields the relation
\begin{equation} \label{RNeighth}
 \lim_{t\to 1} R_{\rm N}t^{2}=-\frac{1}{8}
\end{equation}
indicating another dimensionless universal constant of $-1/8$.  The quantity $\nu$ in \eqref{RNnu} and the coefficient in \eqref{RNeighth}
can be shown to be the same for a VdW fluid \cite{Wei:2019yvs}.

The methods of Ruppeiner geometry have been applied to gain insight into the microstructure of black holes in a number of contexts, including
higher-curvature gravity \cite{Wei:2019ctz,NaveenaKumara:2020hov,Zhou:2020vzf,Wei:2021krr,Wei:2021pql}, non-linear electrodynamics \cite{NaveenaKumara:2020biu,Ye:2022uuj}, comparison with condensed matter systems \cite{Mahish:2020gwg}, cosmological settings
\cite{Saavedra:2023lds,Abdusattar:2023pck,Feng:2024zor}
and many other scenarios
\cite{Xu:2019nnp,Guo:2019hxa,Xu:2020gud,Singh:2020tkf,Ghosh:2020kba,NaveenaKumara:2020lgq,Xu:2020iic,Dehyadegari:2020ebz,Yerra:2020oph,Rizwan:2020bhp,Wu:2020fij,Yerra:2020tzg,Wang:2022bqu,Dutta:2021whz,Ruppeiner:2023wkq,Liu:2022nzo}.   However  what the microstates actually are is still unknown. It is possible they could   counted by the D-brane states \cite{Wei:2015iwa}, or perhaps consist of sphere of strings similar to the fuzzball proposal \cite{Mathur:2005zp}.  Regardless, the molecular-like structure of the microconstituents of a black hole seem to be a robust feature of black hole chemistry.

\section{Other Topics}

There are a number of other topics in Black Hole Chemistry that are currently undergoing active research.  Here is a brief summary.

\subsection{Chemistry of de Sitter Black Holes}

One of the outstanding problems in the subject is that of understanding thermodynamics for $\Lambda > 0$, for which the pressure in
\eqref{eq:press} become negative and thus becomes tension.   This problem is primarily motivated by 
the physical relevance of de Sitter black holes in cosmology, but also by the  formulation of the de Sitter/conformal field theory correspondence (dS/CFT) \cite{Strominger:2001pn}.   It is possible to extend the first law to incorporate variable $\Lambda > 0$ \cite{Dolan:2013ft}, but
new issues arise because, in addition to the black hole horizon, a  cosmological horizon is necessarily present
\cite{Sekiwa:2006qj}. The existence of two horizons precludes the assignment of a single equilibrium temperature to the entire spacetime. Furthermore, isolated de Sitter black holes evaporate due to Hawking radiation, unlike AdS black holes where reflecting boundary conditions at infinity ensure thermal stability.  Responses to this situation include fine-tuning the system to impose thermodynamic equilibrium \cite{Romans1992}, studying soliton solutions
that have a single cosmological horizon \cite{Mbarek:2018bau}, 
considering each horizon as a separate thermodynamic system  \cite{Sekiwa:2006qj,Kubiznak:2015bya}, adopting   an {\it effective temperature} approach \cite{Urano:2009xn}, or placing the black hole in a cavity \cite{Carlip2003,Simovic:2018tdy}.

This last approach is quite fruitful, and has been used to properly formulate the thermodynamics of asymptotically flat black holes \cite{York:1986it}.
The idea is that the (spherical) cavity can be held at a  fixed temperature corresponding to the redshifted black hole temperature at the radius of the cavity.
A grand canonical ensemble  in which the cavity acts as a reservoir can be defined, which in turn allows thermodynamically stable black holes to exist. 
This ensemble can be shown to be stable  \cite{Braden:1990hw}.  A Hawking-Page-like phase transition in both the asymptotically flat and de Sitter cases was somewhat later shown to exist \cite{Carlip2003}, but exploration of black hole chemistry in the de Sitter case is fairly recent. 

The first considerations were for neutral and charged de Sitter black holes \cite{Simovic:2018tdy}.  These were found to exhibit behaviour analogous to that of their AdS counterparts, with the former having a Hawking-Page phase transition, and the latter a small/large black hole phase transition. However   the equation of state has a non-linear  dependence on the temperature, and does not exhibit behaviour characteristic of a van der Waals fluid.  Quite unlike the AdS case, charged dS black holes can undergo  a new type of  `compact' first-order phase transition  that exists within a finite  range of $P$. 
These studies have been extended to include non-linear curvature corrections \cite{Haroon:2020vpr,Marks:2021fpe}, scalar hair \cite{Simovic:2020dke,Li:2023sig}, and non-linear electrodynamics \cite{Simovic:2019zgb,Du:2021dxp,ElMoumni:2021woq}.  Further investigations have included extensions to Renyi statistics \cite{Tannukij:2020njz}, microstructure \cite{Du:2022jcb}, and considerations of an entropic force between
the two horizons \cite{Zhang:2020odg,Ma:2020aab,Ma:2022vwt}.
 
Methods from black hole chemistry have been used to study the evolution and thermodynamics of a slow-roll transition between early and late time de Sitter phases, with and without the presence of a black hole \cite{Gregory:2017sor}.  The evolution is driven by a scalar field rolling slowly between a maximum and minimum of its potential, both  assumed to be positive.  The late time de Sitter phase has finite cosmological tension when the scalar field oscillation around its minimum is underdamped. In the  overdamped case the cosmological tension diverges. The black hole temperature dynamically evolves, and the  first law of thermodynamics in de Sitter space \cite{Dolan:2013ft} can be generalized to variation of the effective cosmological constant at the potential minima of this model. It  is satisfied between the initial and final states. Extensions to a ``mass first law" and  ``Schwarzschild-de Sitter patch first law" can be shown to be satisfied throughout the evolution \cite{Gregory:2018ghc,Beyen:2023rca}.

\subsection{Phase Dynamics}

To the extent that black holes can be understood as chemical systems, it should be possible to understand the dynamics of their various
phase transitions.  Phase transition dynamics has recently emerged as another interesting aspect of black hole chemistry \cite{Li:2020khm,Li:2020nsy,Li:2021vdp,Li:2020spm}, in which
the Smoluchowski equation  (a special case of the Fokker-Planck equation) is used to study how a black hole system in one particular phase dynamically evolves to another phase \cite{Wei:2020rcd}. 

The Smoluchowski equation models the diffusion process of a system that has one or more potential barriers. For black hole phase transitions, these barriers emerge from the maxima of the off-shell Gibbs free energy, which is a continuous function of the black hole horizon radius at a given ensemble temperature. The minima correspond to the various stable  black hole states, whereas the maxima  correspond to unstable states. This approach provides a comprehensive way to visualize how black hole phase transitions occur.

The  distribution function $\rho(r_+,t)$  quantifies the probability that a system stays in a specific black hole state with horizon radius $r_+$ at time $t$. 
A given black hole will experience thermal fluctuations that result in a change to a new state, governed by the order parameter $r$. The Smoluchowski equation \cite{Zwanzig}
\begin{eqnarray}
 \frac{\partial \rho(r_+,0)}{\partial t}=D
 \frac{\partial}{\partial r_+}\left(e^{-\beta_E G_{\rm L}(r_+)}
 \frac{\partial}{\partial r_+}\left(e^{{\beta_e G_{\rm L}(r_+)}}\rho(r_+,t)\right)\right), \label{FPE}
\end{eqnarray}
describes these dynamics,
 where $G_{\rm L}(r_+)$ represents the off-shell Gibbs free energy acting as the potential, $\beta_E$  the inverse ensemble  temperature, 
$D=k_{\rm B}T_{\rm E}/\zeta$   the diffusion coefficient,  $\zeta$ the dissipation coefficient, and   $k_{\rm B}$  Boltzmann's constant.  These latter two constants can be set to unity  without loss of generality.  

To observe phase transition dynamics, one begins with an initial distribution for the probability density, for example
\begin{eqnarray}\label{rhoinit}
 \rho(r_+,0)=\frac{1}{\sigma\sqrt{\pi}}e^{-\frac{(r-r_{{+} })^2}{\sigma^2}}
\end{eqnarray}
where $\sigma$ is related to the standard deviation of the distribution, 
and then numerically solves \eqref{FPE} for $\rho(r_+,t)$.  To maintain the conservation of total probability over time, reflective boundary conditions \begin{eqnarray}
e^{-\beta G(r_+)}
 \frac{\partial}{\partial r}\left(e^{{\beta_E G(r_+)}}\rho(r_+,t)\right)\bigg|_{r=r_{bdy}}=0\label{FPE2}.
\end{eqnarray}
are imposed at $r_+=0$ and $r_+=\infty$, where extremely high potential barriers are present.
\begin{figure}
\begin{center} 
\includegraphics[width=0.48\textwidth]{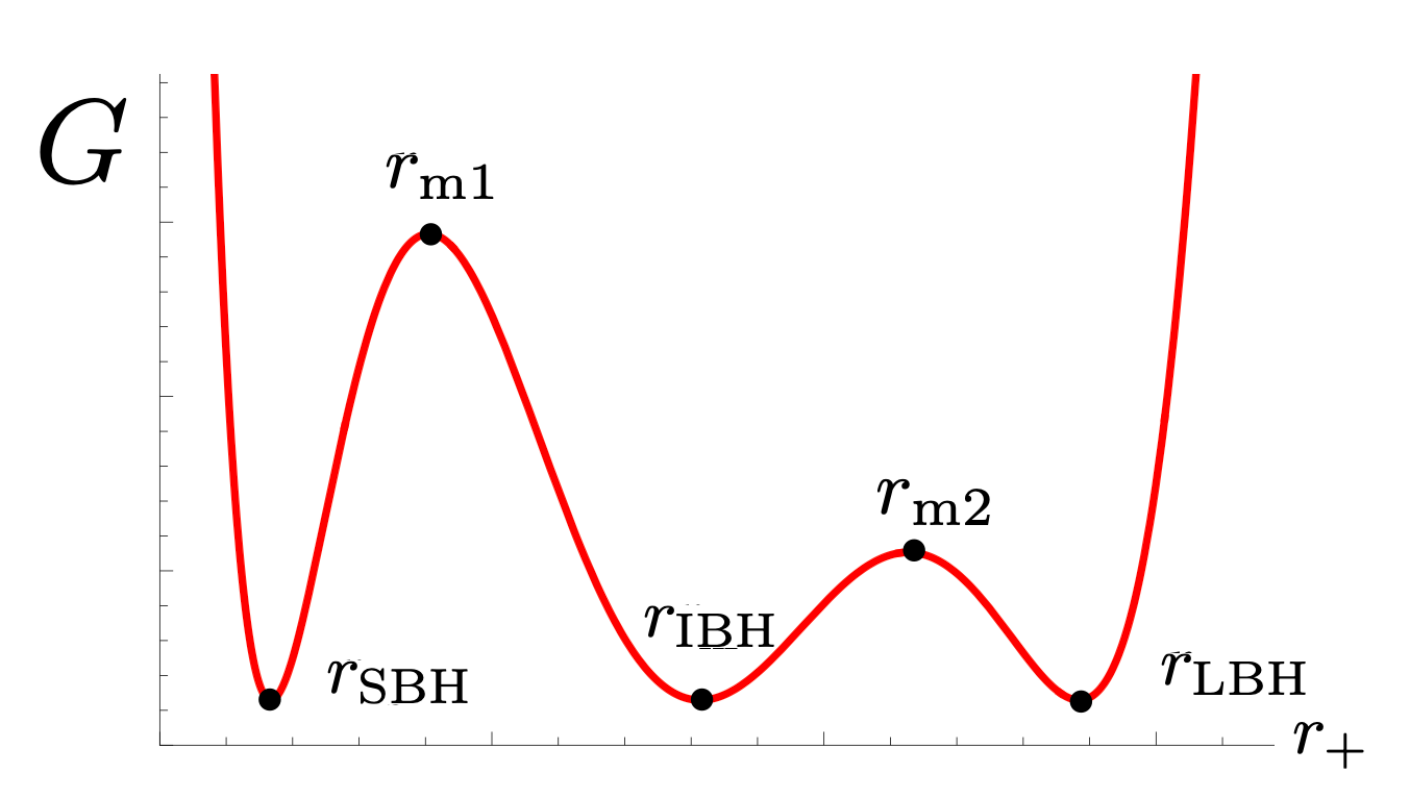}
\includegraphics[width=0.48\textwidth]{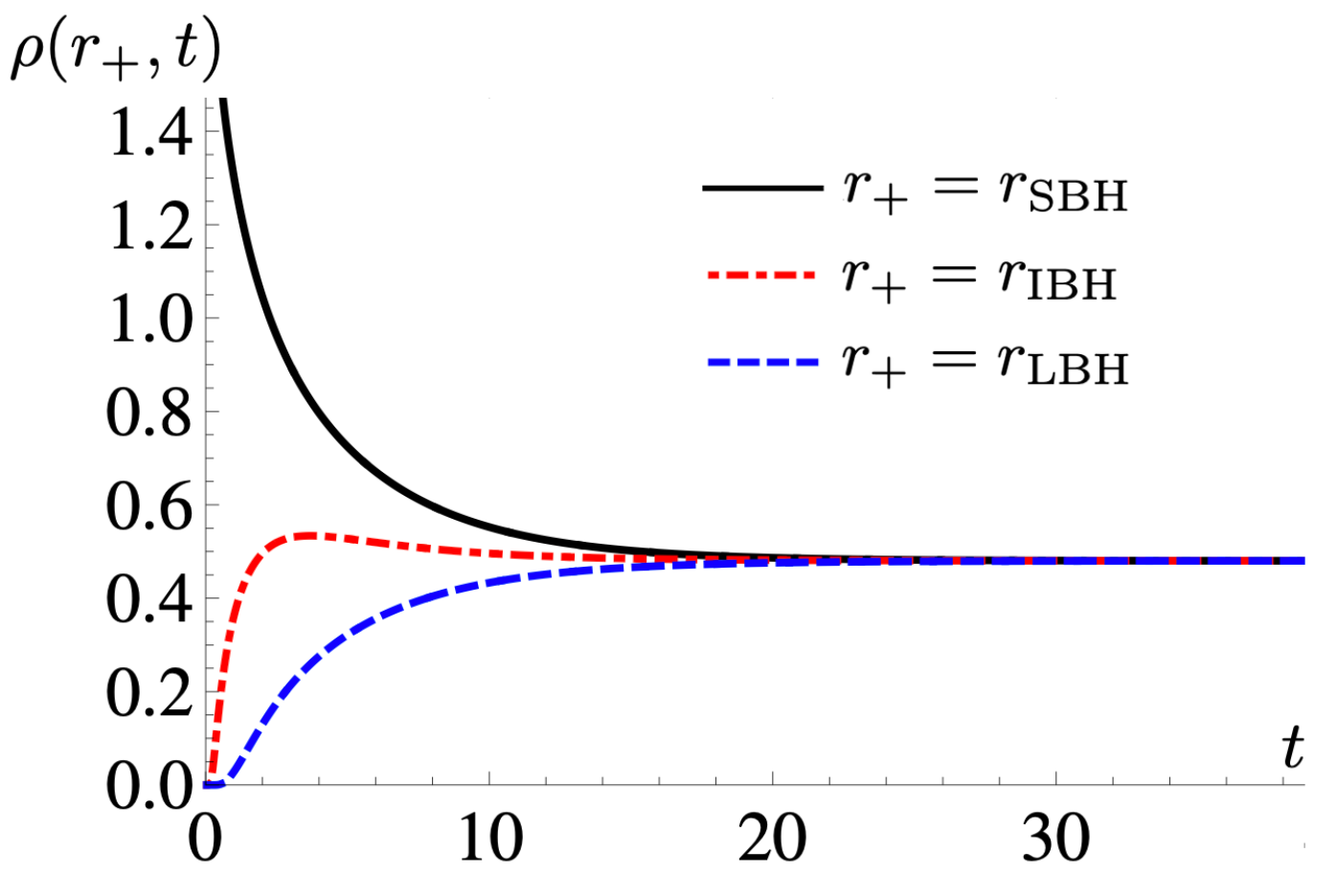}
\end{center} 
\caption{Left: behaviour of the Gibbs free energy via landscape at the triple point of a charged AdS black hole in Einstein-Gauss-Bonnet gravity at the equilibrium temperature. Right: Evolution of the probability $\rho(r_+, t)$ for each of the SBH, IBH, and LBH  states, when the initial Gaussian wave packet is peaked at the coexistent SBH state.}
\label{F23}
\end{figure}

\begin{figure}
\begin{center} 
\includegraphics[width=0.68\textwidth]{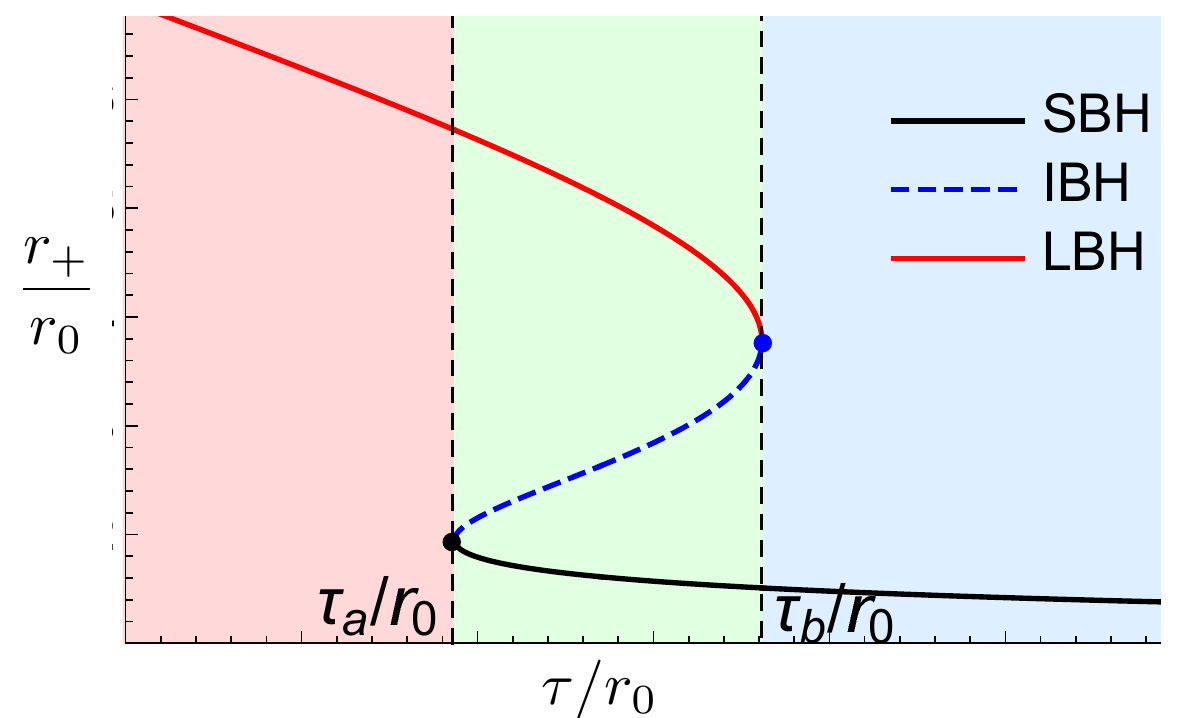} 
\end{center}
\caption{Zero points of $\phi^{r_{\text{h}}}$ shown in the $r_{\text{h}}$-$\tau$ plane for the charged-AdS black hole. The scales on the axes depend on the particular choice of $P r^2_0$  and $Q/r_0$, where $r_0$ is an arbitrary length scale set by the size of a cavity surrounding the black hole. 
The red solid, blue dashed, and black solid lines are for the large  (LBH), intermediate  (IBH), and small (SBH)  black holes respectively. The annihilation and generation points are respectively given by the black and blue dots The green region has three black hole branches as compared to the single branch red and blue regions, but for each region  $W=1$.}
\label{F24}
\end{figure}

The basic ideas are easily seen for the  triple point of a charged AdS black hole in Einstein-Gauss-Bonnet gravity \cite{Wei:2021bwy,Cai:2021sag} , for which the equation of state is
\cite{Wei:2014hba}
\begin{equation}
 P=\frac{T}{r_{+}}-\frac{3}{4\pi r_{+}^2}+\frac{2\alpha T}{r_{+}^3}-\frac{\alpha}{4\pi r_{+}^4}+\frac{Q^2}{8\pi r_{+}^8}
\end{equation}
and whose free energy is 
\begin{align}
G(r_+) & =\frac{2\pi r^5_{+}}{15}\left(4\pi P - \frac{5\pi T_{\rm E}}
   {r_{+}}+ \frac{5}{r_{+}^2} - \frac{20\pi\alpha T_{\rm E}}{r_{+}^3} +\frac{5\alpha}{r_{+}^4}\right)+\frac{\pi Q^2}{9r_{+}^3} \label{glll}
\end{align}
with $T_{\rm E}$ the ensemble temperature; this is in general not the temperature of any of the phases of the black hole.
In the left panel of Fig.~\ref{F24}, the landscape of the free energy \eqref{glll} is presented, with the ensemble temperature chosen to
be the equilibrium temperature at the triple point.  The minima respectively correspond to the black hole system being in the 
SBH (small) state, the IBH (intermediate) state, and large (LBH) state as $r_+$ increases; there are two potential barriers of differing height,
whose maxima are $r_{\textrm{m1}}$ and $r_{\textrm{m2}}$.  If the ensemble temperature is chosen to differ from the equilibrium temperature then  the black hole system will migrate to (or remain at) the well of  greatest depth.  The differing heights of the peaks govern the transit rate between the various black hole states.  An example of the evolution of $\rho(r_+,t)$ is shown in the right panel of Fig.~\ref{F24}.  We see that at the equilibrium temperature,  the system
at late times becomes equally populated for all three states. This late-time behaviour holds regardless of which initial  coexistent black holes state is chosen. 
Studies of the dynamics at black hole quadruple points have also been carried out \cite{Yang:2023xzv}.

The dynamics of black hole phase transitions is now quite a lively avenue of research. Incorporation of various effects such as dilatons \cite{Mo:2021jff,Li:2023sig},  dark energy \cite{Lan:2021crt}, non-linear electrodynamics \cite{Du:2021cxs,Dai:2022mko,Ali:2023wkq}, massive gravity \cite{Liu:2022sot,Wu:2024zig}, monopoles \cite{Luo:2022gss}, dyons \cite{Liu:2023sbf}, and higher curvature effects \cite{Wang:2024zbp,Ma:2024ysf}
have all been considered. Extensions to the Langevin equation have been carried out in
order to the stochastic dynamics of the black hole phase transition 
\cite{Li:2021vdp,Li:2021zep,Yang:2021ljn,Kumara:2021hlt,Li:2021tpu,Li:2022yti,Li:2022ylz,Li:2023ppc,Li:2024hje}.

\subsection{Thermodynamic Topology}

A recent development in Black Hole Chemistry has been the employment of topological concepts into understanding black hole thermodynamics \cite{Wei:2021vdx,Wei:2022dzw}. The basic idea is that black hole solutions can be regarded as defects in the thermodynamic parameter space, with the positive/negative sign of the winding number of a curve around the defect indicative of its stability/instability.  This perspective does not require the presence of pressure, and is applicable for black hole thermodynamics with all possible values/signs of $\Lambda$.

This approach employs the generalized free energy
\begin{eqnarray}
 \mathcal{F}=E-\frac{S}{\tau} \label{grf}
\end{eqnarray}
for a black hole system with energy $E$ and entropy $S$.  The parameter $\tau$  can be thought of as the inverse temperature of a cavity enclosing the black hole, and is taken to vary freely.  For $\tau=T^{-1}$ the free energy \eqref{grf} corresponds to a  
black hole solution that satisfies the gravitational field equations; otherwise it is `off-shell', as per the previous subsection.  From this one then 
introduces a parameter $0\leq\Theta\leq\pi$ (similar to what has been done for studying light rings of black holes \cite{Cunha:2020azh})
 and then defines 
\begin{eqnarray}
 \phi=\left(\frac{\partial \mathcal{F}}{\partial r_{\text{h}}}, -\cot\Theta \csc\Theta\right) \label{vectorField}
\end{eqnarray}
whose zero points are at $\Theta=\pi/2$ and $\tau=T^{-1}$, which is the actual black hole solution;  at
 $\Theta=0$, $\pi$, the component $\phi^{\Theta}$ diverges, indicating the direction of the vector is outwards. Each black hole can thus be assigned a topological charge by using $\phi$.
 
 A topological current 
\begin{eqnarray}
 j^{\mu} &=& \frac{1}{2\pi}\epsilon^{\mu\nu\rho}\epsilon_{ab}\partial_{\nu}n^{a}\partial_{\rho}n^{b},
 \quad \mu,\nu,\rho=0,1,2,
 \label{topcur}
\end{eqnarray}
can then be constructed \cite{Duan:1979ucg,Duan:1984ws},   
where $\partial_{\nu}=\frac{\partial}{\partial x^{\nu}}$ with $x^{\nu}=(\tau,~r_{\text{h}},~\Theta)$ and  $n^a=\frac{\phi^a}{||\phi||}$ ($a=r, \Theta$).
 The topological current \eqref{topcur} can be reexpressed as \cite{Wei:2020rbh}
\begin{equation}
 j^{\mu}=\delta^{2}(\phi)J^{\mu}\left(\frac{\phi}{x}\right)\label{juu}
\end{equation}
and is conserved, i.e., $\partial_{\mu}j^{\mu}=0$.

The current $j^{\mu}$ is nonvanishing only  at the zeroes of $\phi^a(x^i)$, and so must take the form  \cite{Schouton:1951}
\begin{equation}
 j^{0}=\sum_{i=1}^{N}\beta_{i}\eta_{i}\delta^{2}(\vec{x}-\vec{z}_{i}).
\end{equation}
where the $i$-th solution is denoted $\vec{x}=\vec{z}_{i}$.   The number of  loops that $\phi^a$ makes in the vector $\phi$ space as $x^{\mu}$ goes around a zero point $z_i$ is counted by the positive Hopf index $\beta_i$, and the Brouwer degree $\eta_{i}=\text{sign}(J^{0}({\phi}/{x})_{z_i})=\pm1$. 
Assuming the Jacobian $J_{0}(\phi/x)\neq0$, the corresponding topological number is
\begin{eqnarray}
 W=\int_{\Sigma}j^{0}d^2x
 =\sum_{i=1}^{N}\beta_{i}\eta_{i}=\sum_{i=1}^{N}w_{i}
\end{eqnarray}
for a parameter region $\Sigma$,  where $w_{i}$ is the winding number for the $i$-th zero point of $\phi \in \Sigma$. Any two loops  enclosing the same zero point will have the same winding number;   if there is no zero point  in the enclosed region, then $W=0$. Local topological properties are discerned by taking  $\Sigma$ to be a neighbourhood of a zero point of $\phi$; the global topological  number $W$ is obtained by taking  $\Sigma$  to be the entire parameter space. If the Jacobian $J_{0}(\phi/x)= 0$ the defect bifurcates \cite{Fu:2000pb}.

As an example, for the charged-AdS black hole \eqref{metfunction}  the generalized free energy is
\begin{eqnarray}
 \mathcal{F}=\frac{8 \pi  P r_{+}^4+3 r_{\text{h}}^2+3 Q^2}{6 r_{+}}-\frac{\pi r_{+}^2}{\tau}
\end{eqnarray}
in $D=4$. It is straightforward to show that the zero points of $\phi^{r_{\text{h}}}$, shown as the winding curve in Fig. \ref{F23} in the $r_{+}-\tau$ plane,
delineate three regions: one with  three distinct three black hole branches for $\tau_{a}<\tau<\tau_{b}$, one with a single small black hole branch for $\tau<\tau_{a}$  and one with a single large black hole branch for $\tau>\tau_{b}$. The winding number for each of the stable small and large black hole branches can be shown \cite{Wei:2022dzw} to be $1$, whereas the unstable intermediate black hole branch has $w=-1$. The global topological number is always $W=1$; this is clear for the LBH and SBH regions, and for the three-branch region  $W=1-1+1=1$,  independent of $\tau$. Since the pressure $P$ is positive for the RN-AdS black hole, it does not affect the asymptotic behaviour of $\tau$ at small and large $r_{\text{h}}$.  These results hold  for the black hole 
\eqref{metfunction} in any spacetime dimension \cite{Wei:2022dzw}. 

Many studies have been carried out on the thermodynamic topology of black holes \cite{Wang:2024fhf,Chen:2024kmy,Bhattacharya:2024bjp,Zhu:2024jhw,Wu:2024rmv,Zhang:2023svu,Sadeghi:2023dsg,Chen:2023pqk,Tong:2023kob,Shahzad:2023cis,Chen:2023elp,Wu:2023meo,Wang:2023qxw,Sadeghi:2023aii,Wu:2023fcw,Zhang:2023tlq,Gogoi:2023qku,Alipour:2023uzo,Gogoi:2023xzy,Li:2023men,Du:2023wwg,Du:2023nkr,Wu:2023xpq,Wu:2023sue,Ahmed:2022kyv,Wu:2022whe,Fang:2022rsb,Fan:2022bsq,Bai:2022klw,Yerra:2022coh}, and the subject has become a lively area of research. Several different topological categories have been discovered
\cite{Wei:2024gfz,Wu:2024asq}.  Their relevance for black hole thermodynamics is not yet fully understood.

\section{Conclusions}

The wealth of new chemical-like phenomena revealed in the Black Hole Chemistry program warrants a measure of scientific reflection
and assessment.  What has been learned from all of these endeavours?  What might they ultimately mean?

One thing we have learned is that the subject has opened new frontiers for understanding black holes as thermodynamic systems.
The lessons learned from existing discoveries should prove to be useful in investigating black holes in asymptotically de Sitter spacetime,
a setting commensurate with observation.  Likewise, the holography of black hole chemistry is just emerging as a new research frontier,
and I expect many interesting discoveries await.  

More generally, if the original considerations of Bekenstein  \cite{Bekenstein:1973ur} and Hawking \cite{Hawking:1974rv} indeed prove to be correct --
and it is difficult to see how things could be otherwise -- then the considerations and results from Black Hole Chemistry are necessarily correct,
since they straightforwardly follow from the generalized laws of black hole mechanics  \cite{Bekenstein:1973ur} interpreted as the laws of
thermodynamics \cite{Kubiznak:2016qmn}.  This has significant implications for any quantum theory of gravity.  If any such theory predicts the absence
of these phenomena -- in whole or in part -- there will have to be an extraordinarily outstanding reason as to why.  Much more likely is that --
much in the same way that a theory of atomic structure had to give rise to chemical thermodynamics --  a theory of quantum gravity must contain within it the phenomena of Black Hole Chemistry, at least in an emergent way.

\appendix

\section{The generalized first law of black hole mechanics}\label{appA}

The original derivation of the first law of black hole mechanics \cite{Bardeen:1973gs} was extended to include a cosmological constant \cite{Kastor:2009wy},
providing the foundations of Black Hole Chemistry.  Here this derivation is reproduced.

Consider a  $D$-dimensional  spacetime  that describes a black hole with a Killing field.  The metric can be written as
\begin{equation}\label{metricsplit}
g_{ab} = h_{ab}-n_a n_b\,,
\end{equation}
where $h_{ab}$ is the induced metric on a hypersurface $\Sigma$ with unit timelike normal 
 $n^a$ ($n\cdot n=-1$), and so $h_a{}^b n_b =0$.  The system can
be taken to evolve along the vector field
\begin{equation}\label{xidecomp}
\xi^a=N n^a +N^a 
\end{equation}
where the spacetime is foliated by a family of such hypersurfaces.  The quantities
 $N=-\xi\cdot n$  and $N^a$ are, respectively,  the lapse function and shift vector, the latter being   tangential to $ {\Sigma}$.

Considering Einstein gravity, the  dynamical variables in the phase space are $(h_{ab},\pi^{ab})$, where $\pi ^{ab} =-\sqrt{h} (K^{ab}-K h ^{ab} )$
is the   conjugate momentum to the induced metric, and   $K_{ab} =h_a{}^c \nabla _c n_b$ is the extrinsic curvature  of ${\Sigma}$.   The full gravitational  Hamiltonian  is 
\be
{\cal H} =NH+N^a H_a
\ee
where
\ba
H&\equiv& -2G_{ab} n^a n^b =  -\, R^{(D-1)}  + {1 \over |h|}  \Bigl({\pi ^2 \over D-2 } - \pi^{ab} \pi_{ab} \Bigr)\,,\label{hamconstraint}\nonumber\\
H_b&\equiv& -2G_{ac} n^a h^c _b=-2\,  D_a (|h|^{-{1 \over 2}} \pi^{ab} )\,.\label{momconstraint}
\ea
and  $K\equiv K^a{}_a$ and $\pi \equiv \pi^a{}_a$ are the traces
of these respective tensors. The quantity $D_a$ is the covariant derivative operator with respect to $h_{ab}$  on $ {\Sigma}$, with $h$ is the determinant of $h_{ab}$ and $R^{(D-1)}$ its scalar curvature. 

Consider a solution  $g_{ab}$  of the field equations with Killing vector $\xi^a$
and cosmological constant $\Lambda$.  Since the only `matter' is that of a cosmological constant, then 
\begin{equation}\label{Hconstraint}
H =- {2} \Lambda\,,  \quad H_b =0 
\end{equation}
for the constraint equations upon setting ${8\pi} T^a _b =-\Lambda g^a _b $.

Perturbing the metric to an `infinitesimally close' solution (not necessarily admitting any Killing vector),  such that $\tilde g_{ab} =g_{ab} +\delta g_{ab}$ and $\tilde\Lambda = \Lambda +\delta\Lambda$, we have
$$
\tilde h_{ab}= h_{ab} + \gamma_{ab} \qquad \tilde \pi_{ab}= \pi_{ab} + p_{ab}
$$
 where  $h_{ab}$ and $\pi^{ab}$ are regarded as  initial data for the original (background) solution $g_{ab}$, and $\gamma_{ab} = \delta h_{ab}$\,, $p_{ab} = \delta\pi_{ab}$.  Incorporating this into
\eqref{Hconstraint} yields \cite{Traschen:1985,SudarskyWald:1992,TraschenFox:2004,Kastor:2009wy}
\be\label{altgauss}
 D_a B^a = N\delta H + N^a\delta H_a = - {2}N \delta\Lambda \Rightarrow D_a ( B^a   -{2}\delta \Lambda \omega ^{ab} n_b ) =0\,,
 \ee
where  
\be\label{gaussvector}
 B^a [\xi ] =    N(D^a \gamma^c_c \!-\! D_b \gamma^{ab})  - \gamma^c_c D^a N + \gamma^{ab} D_b N
+ |h|^{-\frac{1}{2}} N^b\bigl(\pi^{cd} \gamma_{cd} h^a{}_b \!-\! 2 \pi^{ac} \gamma_{bc} \!-\!2p^a{}_b \bigr)
\ee
with $N= - \xi^a n_a=-D_c (\omega ^{cb} n_b )$. The quantity
  $\omega^{ab} = -\omega^{ba} $ is the {\it Killing co-potential}, which by definition satisfies
\cite{Kastor:2008xb,Kastor:2009wy}:
 \begin{equation}\label{omegadef}
 \nabla _c \omega ^{cb} = \xi ^b\,,
 \end{equation}
and is  defined only up to a divergence-less term. In other words, if  $ \omega_{ab}$ solves \eqref{omegadef}
  then   $\omega^\prime_{ab}=  \omega_{ab} + \zeta_{ab}$ likewise solves this equation, where
$ \nabla ^a \zeta_{ab} =0$.

The left-hand side of   \eqref{altgauss} is  a total divergence, which yields 
\be \label{gaussint}
\int_{\partial \hat V_{out}} d\mathcal{S} r_c \left( B^c [\xi ]  - {2}\delta \Lambda \omega ^{cb} n_b \right)
 = \int_{\partial \hat V_{in}} d\mathcal{S} r_c \left( B^c [\xi ]   - {2} \delta \Lambda \omega ^{cb} n_b \right)\,,
\ee
upon integration  over a volume $\hat V$ having inner and outer boundaries   $\partial \hat V_{in, out}$ contained in $\Sigma$, with 
unit normal $r^c$ respectively pointing into and out of these  boundaries.  
 
 Writing
$\omega ^{cb}= \omega ^{cb} - \omega_{AdS}^{cb} + \omega_{AdS}^{cb}$ for the $\partial \hat V_{out}$ integral
yields
\ba \label{gaussint2}
\int_{\partial \hat V_{out}} d\mathcal{S} r_c \left( B^c [\xi ]  - {2}\delta \Lambda  \omega_{AdS}^{cb}  n_b \right)
&=& \int_{\partial \hat V_{out}} d\mathcal{S} r_c \left({2}\delta \Lambda  (\omega ^{cb} - \omega_{AdS}^{cb}) n_b \right)\nonumber\\
&&\qquad + \int_{\partial \hat V_{in}} d\mathcal{S} r_c \left( B^c [\xi ]   - {2} \delta \Lambda \omega ^{cb} n_b \right) \qquad
\ea
where $\omega_{AdS}^{ab}$ satisfies \eqref{omegadef} for AdS spacetime; it 
 is the Killing co-potential of the `background AdS spacetime'.  Note that only the difference of the values of the integrals on the outer and inner boundaries is meaningful due to the ambiguity in  $ \omega_{ab}$.   
 
Setting the outer boundary at spatial infinity,
the respective variations in the total mass  $M$ and angular momentum $J$ of the space-time  
obtained by respectively setting $\xi^a = (\partial_t)^a$ (time translations) and $ \xi^a = (\partial_\varphi)^a$
(rotations), yield
\begin{eqnarray}
 16\pi \delta M  &=&
 -\int_\infty d\mathcal{S} r_c \left(B^c  [{\partial_t}]  - {2}\delta \Lambda \omega_{AdS}^{cb} n_b\right)\,,
   \label{flatdm} \\
  16\pi \delta J  &=&  \int_\infty d\mathcal{S} r_c  B^c  [{\partial_\varphi} ]\,, \label{flatdj}
 \end{eqnarray}
where the $ \omega_{AdS}^{cb}$ term ensures  $\delta M$  is finite \cite{Kastor:2009wy}.

If the event horizon $H$ of the black hole is a  bifurcate Killing horizon of area $A$ on which the Killing vector
$\xi^a = ({\partial_t+ \Omega \partial_\varphi})^a$ vanishes, then
\be
   2\kappa \delta A  =  -\int_{H} d\mathcal{S} r_c  B^c  [ {\partial_t+ \Omega \partial_\varphi} ]\,,  
   \label{bhhor}
 \ee
upon taking the  inner boundary to be the horizon, whose surface gravity is $\kappa=\sqrt{ -\nabla^a\xi^b \nabla_a\xi_b /2}\;\bigr\vert_{r=r_+}$.

Since $\delta \Lambda$ is spacetime-independent, the remaining terms on the right hand side of \eqref{gaussint2} become
\begin{equation}\label{vdef}
2\delta \Lambda \left( \int _{\infty} d\mathcal{S} r_c n_b \left(\omega ^{cb} - \omega_{AdS}^{cb}\right) -\int _{H} d\mathcal{S} r_c n_b \omega ^{cb} \right)
= -16\pi V \delta P
\end{equation}
thereby  defining the thermodynamic volume, since from \eqref{eq:press} $\delta \Lambda = - 8\pi \delta P$.
The $\omega_{AdS}^{cb}$ term in \eqref{vdef} ensures that $V$ is finite.
 
Together  \eqref{flatdm}, \eqref{flatdj}, \eqref{bhhor} and \eqref{vdef} inserted into \eqref{gaussint2} yield
\be
\delta M =T\delta S  + V \delta P+\Omega \delta J\,, 
 \ee
 which  becomes the first law \eqref{firstBH} for zero charge and one angular momentum.  Inclusion of charges, all possible angular momenta
 in $D$-dimensions, and a positive cosmological constant are straightforward extensions of this argument \cite{Dolan:2013ft}
and a covariant treatment in a more general setting of variable background fields has also been considered  \cite{Wu:2016auq}.

\section{Higher-Curvature Gravity}
\label{appB}

Einstein's theory of gravity equates stress-energy with a particular linear combination of contractions of the Riemann tensor, and so
is linear in the curvature.  There is no logical reason to require this -- one could imagine taking an arbitrary, generally covariant, function 
of the Riemann curvature and equating it to the stress-energy.  Such higher-curvature theories, however, in general suffer from pathologies due to the fact that their field equations contain derivatives of the metric that are greater than second order.

In recent years investigations have been undertaken to find the most general class of higher-curvature theories that ameliorate these pathologies. This class is known as `Generalized Quasi-Topological Gravity' (GQTG)  \cite{Bueno:2019ycr} and satisfies the criterion that for metrics
of the form
\be\label{GQTGmet}
ds^2  = - N(r) f(r) dt^2+\frac{dr^2}{f(r)}+r^2  d\Omega_k^2 
\ee
where $d\Omega_k^2$ is given by \eqref{dOmegak}, the variation of the action with respect to the metric function $f(r)$ vanishes identically. This in turn forces $N(r)$ to be a constant, implying that the  metric can be written in the form \eqref{Schw-Ads}. GQTGs have a number of phenomenologically interesting properties. They exist in spacetime dimensions $D\geq 4$ and their field equations are second-order equations of motion identical to those of Einstein gravity (up to a redefinition of Newton's constant) when linearized around maximally symmetric backgrounds.  Black hole solutions to GQTGs are fully characterized by their ADM mass/energy and their thermodynamic properties can be obtained from an algebraic system of equations
\cite{Bueno:2016xff,Bueno:2016lrh,Hennigar:2015esa,Ahmed:2017jod}. These theories have a well-defined and continuous Einstein gravity limit, corresponding to setting all higher-curvature couplings to zero.  It has been shown that every gravitational effective action constructed from arbitrary contractions of the metric and the Riemann tensor is equivalent, through a metric redefinition, to some GQTG  \cite{Bueno:2019ycr}.   GQTGs for which the equation of the metric function $f$ is algebraic are called `Quasi-topological'   \cite{Oliva:2010eb,Myers:2010jv,Myers:2010ru} and only exist for $D\geq 5$. 

GQTGs are of considerable interest in the context of both quantum gravity and phenomenology.  In quantum gravity  it is expected that the Einstein--Hilbert action is only an effective gravitational action valid for small curvature or low energies, and so will be modified by higher-curvature terms.  To this end they have been of considerable important in black hole chemistry \cite{Hennigar:2015esa,Mir:2019ecg,Mir:2019rik,Lu:2023hgu}. Phenomenologically they form interesting foils against which Einstein gravity can be tested \cite{Hennigar:2018hza,Poshteh:2018wqy,Khodabakhshi:2020hny,Khodabakhshi:2020ddv}.

\subsection{Lovelock Gravity}\label{appB1}

Lovelock gravity is a special subclass of GQTGs  whose field equations are always of second order   \cite{Lovelock:1971yv}. The interest in these theories stems from this unique feature,  whereas a generic higher-curvature theory will have higher derivative terms in the field equations. The only GQTG
that is second-order in the curvature tensor is the simplest Lovelock theory \cite{Bueno:2016lrh,Hennigar:2015esa}, known as Gauss-Bonnet gravity, and is non-trivial in $D\geq 5$.

In $D$ spacetime dimensions, the action of a Lovelock gravity theory  is  \cite{Lovelock:1971yv}
 \begin{equation}
S = \int d^Dx \sqrt{-g} \mathcal{L}=  \int d^Dx \sqrt{-g} \left[\frac{1}{16\pi}\sum_{k=0}^{K}{\alpha}_k\mathcal{L}^{\left(k\right)}+{\cal L}_m \right]
\label{eq:actLove}
\end{equation}
where the  ${\alpha}_k$ are the  Lovelock coupling constants. The quantities $\mathcal{L}^{\text{\ensuremath{\left(k\right)}}}$ are given by the contraction of $k$ powers of the Riemann tensor
\begin{equation}
\mathcal{L}^{\left(k\right)}=\frac{1}{2^{k}}\,\delta_{c_{1}d_{1}\ldots c_{k}d_{k}}^{a_{1}b_{1}\ldots a_{k}b_{k}}R_{a_{1}b_{1}}^{\quad c_{1}d_{1}}\ldots R_{a_{k}b_{k}}^{\quad c_{k}d_{k}}\,,
\end{equation}
and  are the $2k$-dimensional Euler densities, where the `generalized Kronecker delta function' is totally antisymmetric in both sets of indices. 

By definition  $\mathcal{L}^{\left(0\right)} = 1$; it corresponds to  the cosmological constant term, with $\alpha_0=- 2 \Lambda = 16\pi P$. 
The Ricci scalar is $\mathcal{L}^{\left(1\right)}$; its coupling $\alpha_1=1$,  and its action is the Einstein-Hilbert action. The quantity
\begin{equation}
\mathcal{L}^{\left(2\right)}  = \frac{1}{4} \left(R^{abcd}R_{abcd} - 4 R^{ab}R_{ab} + R^2 \right)
\end{equation}
 corresponds to the quadratic Gauss--Bonnet term.   Coupling to matter is given by   ${\cal L}_m$. 
 
 It is common to rescale the coupling constants
 so that
 \be
\hat \alpha_0=\frac{\alpha_0}{(D-1)(D-2)}\,,\quad
\hat \alpha_1=\alpha_1\,,\quad \hat \alpha_k=\alpha_k\prod_{n=3}^{2k}(D-n)\quad \mbox{for} \quad k\geq 2
\ee
The quantity $K$ has an upper bound of $ K=\lfloor{\frac{D-1}{2}}\rfloor$,  reflecting the fact that a given
 term $\mathcal{L}^{\left(k\right)}$  identically vanishes for $D< 2k$, and is a purely topological object in $D=2k$. In order for this term to contribute to the equations of motion, $D>2k$.  Einstein gravity (general relativity)  is recovered by  setting ${\alpha}_k=0$ for $k\geq 2$.
  
Two special subclasses of Lovelock theories have been of interest in Black Hole Chemistry. In one of them 
\be\label{SpecialCouplings}
\hat \alpha_k=\hat \alpha_K\bigl(K\hat \alpha_K\bigr)^{-\frac{K-k}{K-1}}\left({K\atop k}\right)\quad \mbox{for}\quad 2\leq k<K\,,
\ee
 where $\hat \alpha_K\neq 0$, while $\hat \alpha_1=1$ and $\hat \alpha_0$ is arbitrary. This particular choice led to the discovery of 
 an {\it isolated critical point} \cite{Dolan:2014vba}, modifying the standard mean-field exponents as in \eqref{Widrush}, 
 discussed  in Sec.~\ref{5p3}.   The other special subclass
 arises in odd dimensions for the choice
\be\label{CS-couplings}
\hat \alpha_p =\frac{\ell^{2p-2n+1}}{2n-2p-1}\left( {n-1\atop p}  \right)  \qquad p=1,2,\ldots, n-1 = \frac{D-1}{2}
\ee
and is known as{\it Chern--Simons gravity} \cite{Zanelli:2005sa}, where  $\ell$ is the AdS radius. It is 
a non-trivial function of  the `bare' cosmological constant $\Lambda=-\alpha_0/2$ and the higher-order Lovelock couplings. In this particular case   the local Lorentz invariance of the Lovelock action is enhanced to a local (A)dS symmetry.

 Taking the variation of \eqref{eq:actLove} with respect to the metric yields the field equations
\begin{equation}
\mathcal{G}_{\, b}^{a}=\sum_{k=0}^{K}{\alpha}_k\mathcal{G}_{\,\,\quad b}^{\left(k\right)\, a}=8\pi T^a{}_b\,,
\end{equation}
where the  tensors $\mathcal{G}_{\,\,\quad b}^{\left(k\right)\, a}$ 
\begin{equation}
\mathcal{G}_{\,\,\quad b}^{\left(k\right)\, a}=-\frac{1}{2^{\left(k+1\right)}}\delta_{b\, e_{1}f_{1}\ldots e_{k}f_{k}}^{a\, c_{1}d_{1}\ldots c_{k}d_{k}}R_{c_{1}d_{1}}^{\quad e_{1}f_{1}}\ldots R_{c_{k}d_{k}}^{\quad e_{k}f_{k}}\label{eq:G}\,,
\end{equation}
generalize the Einstein tensor.  Each $\mathcal{G}_{\,\,\quad b}^{\left(k\right)\, a}$ 
 independently satisfies a conservation law $\nabla_{a}\mathcal{G}_{\,\,\quad b}^{\left(k\right)\, a}=0\,$ and so any choice of 
 couplings $\{\alpha_k\}$ yields a consistent theory of gravity coupled to matter, though some choices (such as $\alpha_k = 0$ for $k\geq 1$) yield trivial theories.

The arguments in~\ref{appA} generalize to the Lovelock case, yielding
\be\label{flawLove}
\delta M = T\delta S+V\delta P+\sum_{k=2}^K\Psi^k\delta\alpha_k+\sum_i\Omega^i\delta J^i+\sum_j\Phi^j\delta Q^j
\ee
for the first  law of black hole thermodynamics  \cite{Jacobson:1993xs,Kastor:2010gq}, where the Smarr relation becomes
\be\label{SmarrLovelock}
\frac{d-3}{d-2}M=TS-\frac{2}{d-2}PV+\sum_{k=2}^K\frac{2(k-1)}{d-2}\Psi^k\alpha_k+\sum_i\Omega^iJ^i+\frac{d-3}{d-2}\sum_j\Phi^jQ^j
\ee
using the Euler scaling argument. The Smarr formula  can be expressed in terms of a Noether charge surface integral plus  a suitable volume integral \cite{Liberati:2015xcp}.

An important change from $k=1$ Einstein gravity is that  the entropy is  no longer proportional to the area of the horizon. Instead it is given by the expression
\begin{equation}\label{S}
S= \frac{1}{4} \sum_k {\alpha}_{k} {\cal A}^{(k)}\,,\quad   {\cal A}^{(k)}  = k\int_{\mathcal{H}} \sqrt{\sigma}{\mathcal{L}}^{(k-1)}\,,
\end{equation}
where  $\sigma$ denotes the determinant of  the induced metric $\sigma_{ab}$ on the black hole horizon ${\mathcal{H}}$, and the Lovelock terms ${\mathcal{L}}^{(k-1)}$ are evaluated on that surface.  For $k=1$ the quantity $ {\cal A}^{(1)}$ equals the horizon area, and the usual entropy/area relation is recovered.  
 
 The Lovelock   coupling constants ${\alpha}_k$ are regarded  as thermodynamic variables in \eqref{flawLove}, a situation that
 also holds for coupling constants in non-linear electrodynamics \cite{Gunasekaran:2012dq}.  Both of these properties have recently been
 exploited to observe multicriticality \cite{Tavakoli:2022kmo,Wu:2022plw}, as discussed in section~\ref{5p2}. The quantity $\alpha_2$ plays a particularly important role in multiphase behaviour for asymptotically flat black holes \cite{Wu:2022xmp}. The conjugate potentials $\Psi^k$ can be 
 be explicitly computed  in the case of spherical symmetry \cite{Frassino:2014pha}.

\section*{Acknowledgments}
This work was supported in part by the Natural Sciences and Engineering Research Council of Canada. I am grateful to Niloofar Abbasvandi, Jamil Ahmed,  Moaathe Belhaj Ahmed, Rahim Al Balushi, Andres Anabalon, Natacha Altimirano, Michael Appels, Alvaro Ballon, Wilson Brenna, Ivan Booth, J. David Brown, Wan Cong, Jolien Creighton, Saurya Das, Ali Dehghani, Hossein Dehghani, Terence Delsate, Maria DiMarco, Brian Dolan, Hanna Dykaar, Antonia Frassino, David Garkfinkle, Sharmila Gunasekaran, Finnian Gray, Ruth Gregory, Seyed Hendi, Robie Hennigar, Brayden Hull, Michael Imseis, Sierra Jess, David Kastor, Anna Kostouki, David Kubiznak, Hari Kunduri, Mengqi Lu, Seyed Mansoori,  Fiona McCarthy, Mozghan Mir, Jonas Mureika, Julio Oliva, Ali Ovgun, Shahram Panahiyan, Leo Pando-Zayas, Miok Park, Constanza Quijada,  Morteza Rafiee, Aruna Rajagopal, Simon Ross, Zeinab Sherkatgnad, Fil Simovic, Cedric Sinamuli, Cris Stelea, Jialong Sun, Masoumeh Tavakoli, Mae Teo, Jennie Traschen, Erickson Tjoa, Manus Visser, Yong-Qiang Wang, Shao Wen Wei, Jerry Wu, Yu-Xiao Liu, Jiayue Yang, and Ming Zhang for their collaborations that led to the results presented here.

\bibliography{GR110BHChem-15yrs}

\providecommand{\href}[2]{#2}\begingroup\raggedright\begin{thebibliography}{100}

\bibitem{Kubiznak:2016qmn}
D.~Kubiznak, R.~B. Mann, and M.~Teo, {\it {Black hole chemistry: thermodynamics
  with Lambda}},  {\em Class. Quant. Grav.} {\bf 34} (2017), no.~6 063001,
  [\href{http://xxx.arxiv.org/abs/1608.06147}{{\tt arXiv:1608.06147}}].

\bibitem{Michell:1783}
J.~Michell, {\it {On the means of discovering the distance, magnitude etc. of
  the fixed stars}},  {\em Phil. Trans. R. Soc. London} {\bf 74} (1783) 35.

\bibitem{Bekenstein:1973ur}
J.~D. Bekenstein, {\it {Black holes and entropy}},  {\em Phys.Rev.} {\bf D7}
  (1973) 2333--2346.

\bibitem{Hawking:1974rv}
S.~W. Hawking, {\it {Black hole explosions}},  {\em Nature} {\bf 248} (1974)
  30--31.

\bibitem{Dolan:2010ha}
B.~P. Dolan, {\it {The cosmological constant and the black hole equation of
  state}},  {\em Class. Quant. Grav.} {\bf 28} (2011) 125020,
  [\href{http://xxx.arxiv.org/abs/1008.5023}{{\tt arXiv:1008.5023}}].

\bibitem{Kastor:2009wy}
D.~Kastor, S.~Ray, and J.~Traschen, {\it {Enthalpy and the Mechanics of AdS
  Black Holes}},  {\em Class. Quant. Grav.} {\bf 26} (2009) 195011,
  [\href{http://xxx.arxiv.org/abs/0904.2765}{{\tt arXiv:0904.2765}}].

\bibitem{1985Guggenheim}
E.~A. {Guggenheim}, {\em {Thermodynamics - An advanced treatment for chemists
  and physicists (7th edition)}}.
\newblock 1985.

\bibitem{Maxwell_2011}
J.~C. Maxwell, {\em Theory of Heat}.
\newblock Cambridge Library Collection - Physical  Sciences. Cambridge
  University Press, 2011.

\bibitem{fowler1965statistical}
R.~Fowler and E.~Guggenheim, {\em Statistical Thermodynamics}.
\newblock Cambridge University Press, 1965.

\bibitem{carnot1890}
N.~Carnot, {\em Reflections on the Motive Power of Fire}.
\newblock MacMillan and Company, 1890.

\bibitem{Nernst1926}
W.~Nernst, {\em The New Heat Theorem}.
\newblock Methuen and Company, Ltd.; Reprinted in 1969 by Dover, 1926.

\bibitem{Schwarzschild:1916}
K.~Schwarzschild, {\it Uber das gravitationsfeld eines massenpunktes nach der
  einsteinschen theorie},  {\em Deutsch. Akad. Wiss. Berlin, Kl. Math. Phys.
  Tech.} {\bf 1916} (1916) 189--196.

\bibitem{Oppenheimer:1939ue}
J.~R. Oppenheimer and H.~Snyder, {\it {On Continued gravitational
  contraction}},  {\em Phys. Rev.} {\bf 56} (1939) 455--459.

\bibitem{WheelerFord1998}
J.~Wheeler and K.~Ford, {\em Geons, Black Holes, and Quantum Foam: A Life in
  Physics}.
\newblock New York: W.W. Norton \& Co.

\bibitem{Oppenheim:2015}
J.~Oppenheim {\em Nature Phys.} {\bf 11} (2015) 805.

\bibitem{Hawking:1971tu}
S.~W. Hawking, {\it {Gravitational radiation from colliding black holes}},
  {\em Phys. Rev. Lett.} {\bf 26} (1971) 1344--1346.

\bibitem{Bardeen:1973gs}
J.~M. Bardeen, B.~Carter, and S.~W. Hawking, {\it {The Four laws of black hole
  mechanics}},  {\em Commun. Math. Phys.} {\bf 31} (1973) 161--170.

\bibitem{Israel:1986gqz}
W.~Israel, {\it {Third Law of Black-Hole Dynamics: A Formulation and Proof}},
  {\em Phys. Rev. Lett.} {\bf 57} (1986), no.~4 397.

\bibitem{Sorce:2017dst}
J.~Sorce and R.~M. Wald, {\it {Gedanken experiments to destroy a black hole.
  II. Kerr-Newman black holes cannot be overcharged or overspun}},  {\em Phys.
  Rev. D} {\bf 96} (2017), no.~10 104014,
  [\href{http://xxx.arxiv.org/abs/1707.05862}{{\tt arXiv:1707.05862}}].

\bibitem{Penrose:1969pc}
R.~Penrose, {\it {Gravitational collapse: The role of general relativity}},
  {\em Riv. Nuovo Cim.} {\bf 1} (1969) 252--276.

\bibitem{Wald:1997wa}
R.~M. Wald, {\it {Gravitational collapse and cosmic censorship}},  pp.~69--85,
  10, 1997.
\newblock \href{http://xxx.arxiv.org/abs/gr-qc/9710068}{{\tt gr-qc/9710068}}.

\bibitem{Smarr:1972kt}
L.~Smarr, {\it {Mass formula for Kerr black holes}},  {\em Phys. Rev. Lett.}
  {\bf 30} (1973) 71--73. [Erratum: Phys. Rev. Lett.30,521(1973)].

\bibitem{Wald:1993nt}
R.~M. Wald, {\it {Black hole entropy is the Noether charge}},  {\em Phys. Rev.}
  {\bf D48} (1993), no.~8 R3427--R3431,
  [\href{http://xxx.arxiv.org/abs/gr-qc/9307038}{{\tt gr-qc/9307038}}].

\bibitem{Jacobson:1995ab}
T.~Jacobson, {\it {Thermodynamics of space-time: The Einstein equation of
  state}},  {\em Phys. Rev. Lett.} {\bf 75} (1995) 1260--1263,
  [\href{http://xxx.arxiv.org/abs/gr-qc/9504004}{{\tt gr-qc/9504004}}].

\bibitem{Padmanabhan:2009vy}
T.~Padmanabhan, {\it {Thermodynamical Aspects of Gravity: New insights}},  {\em
  Rept. Prog. Phys.} {\bf 73} (2010) 046901,
  [\href{http://xxx.arxiv.org/abs/0911.5004}{{\tt arXiv:0911.5004}}].

\bibitem{Bianchi:2012ev}
E.~Bianchi and R.~C. Myers, {\it {On the Architecture of Spacetime Geometry}},
  {\em Class. Quant. Grav.} {\bf 31} (2014) 214002,
  [\href{http://xxx.arxiv.org/abs/1212.5183}{{\tt arXiv:1212.5183}}].

\bibitem{Gruber:2016mqb}
C.~Gruber, O.~Luongo, and H.~Quevedo, {\it {Geometric approaches to the
  thermodynamics of black holes}},  in {\em {14th Marcel Grossmann Meeting on
  Recent Developments in Theoretical and Experimental General Relativity,
  Astrophysics, and Relativistic Field Theories (MG14) Rome, Italy, July 12-18,
  2015}}, 2016.
\newblock \href{http://xxx.arxiv.org/abs/1603.09443}{{\tt arXiv:1603.09443}}.

\bibitem{Ryu:2006bv}
S.~Ryu and T.~Takayanagi, {\it {Holographic derivation of entanglement entropy
  from AdS/CFT}},  {\em Phys.Rev.Lett.} {\bf 96} (2006) 181602,
  [\href{http://xxx.arxiv.org/abs/hep-th/0603001}{{\tt hep-th/0603001}}].

\bibitem{Faulkner:2013ica}
T.~Faulkner, M.~Guica, T.~Hartman, R.~C. Myers, and M.~Van~Raamsdonk, {\it
  {Gravitation from Entanglement in Holographic CFTs}},  {\em JHEP} {\bf 03}
  (2014) 051, [\href{http://xxx.arxiv.org/abs/1312.7856}{{\tt
  arXiv:1312.7856}}].

\bibitem{Giddings:1995gd}
S.~B. Giddings, {\it {The Black hole information paradox}},  in {\em
  {Particles, strings and cosmology. Proceedings, 19th Johns Hopkins Workshop
  and 5th PASCOS Interdisciplinary Symposium, Baltimore, USA, March 22-25,
  1995}}, pp.~415--428, 1995.
\newblock \href{http://xxx.arxiv.org/abs/hep-th/9508151}{{\tt hep-th/9508151}}.

\bibitem{Mathur:2009hf}
S.~D. Mathur, {\it {The Information paradox: A Pedagogical introduction}},
  {\em Class.Quant.Grav.} {\bf 26} (2009) 224001,
  [\href{http://xxx.arxiv.org/abs/0909.1038}{{\tt arXiv:0909.1038}}].

\bibitem{Mathur:2005zp}
S.~D. Mathur, {\it {The Fuzzball proposal for black holes: An Elementary
  review}},  {\em Fortsch.Phys.} {\bf 53} (2005) 793--827,
  [\href{http://xxx.arxiv.org/abs/hep-th/0502050}{{\tt hep-th/0502050}}].

\bibitem{Almheiri:2012rt}
A.~Almheiri, D.~Marolf, J.~Polchinski, and J.~Sully, {\it {Black Holes:
  Complementarity or Firewalls?}},  {\em JHEP} {\bf 1302} (2013) 062,
  [\href{http://xxx.arxiv.org/abs/1207.3123}{{\tt arXiv:1207.3123}}].

\bibitem{Hawking:2016msc}
S.~W. Hawking, M.~J. Perry, and A.~Strominger, {\it {Soft Hair on Black
  Holes}},  {\em Phys. Rev. Lett.} {\bf 116} (2016), no.~23 231301,
  [\href{http://xxx.arxiv.org/abs/1601.00921}{{\tt arXiv:1601.00921}}].

\bibitem{Almheiri:2020cfm}
A.~Almheiri, T.~Hartman, J.~Maldacena, E.~Shaghoulian, and A.~Tajdini, {\it
  {The entropy of Hawking radiation}},  {\em Rev. Mod. Phys.} {\bf 93} (2021),
  no.~3 035002, [\href{http://xxx.arxiv.org/abs/2006.06872}{{\tt
  arXiv:2006.06872}}].

\bibitem{Wald:1999vt}
R.~M. Wald, {\it {The thermodynamics of black holes}},  {\em Living Rev. Rel.}
  {\bf 4} (2001) 6, [\href{http://xxx.arxiv.org/abs/gr-qc/9912119}{{\tt
  gr-qc/9912119}}].

\bibitem{Traschen:1999zr}
J.~H. Traschen, {\it {An Introduction to black hole evaporation}},  in {\em
  {Mathematical methods in physics. Proceedings, Winter School, Londrina,
  Brazil, August 17-26, 1999}}, 1999.
\newblock \href{http://xxx.arxiv.org/abs/gr-qc/0010055}{{\tt gr-qc/0010055}}.

\bibitem{Grumiller:2014qma}
D.~Grumiller, R.~McNees, and J.~Salzer, {\it {Black holes and thermodynamics -
  The first half century}},  {\em Fundam. Theor. Phys.} {\bf 178} (2015)
  27--70, [\href{http://xxx.arxiv.org/abs/1402.5127}{{\tt arXiv:1402.5127}}].

\bibitem{Carlip:2014pma}
S.~Carlip, {\it {Black Hole Thermodynamics}},  {\em Int. J. Mod. Phys.} {\bf
  D23} (2014) 1430023, [\href{http://xxx.arxiv.org/abs/1410.1486}{{\tt
  arXiv:1410.1486}}].

\bibitem{Hawking:1982dh}
S.~W. Hawking and D.~N. Page, {\it {Thermodynamics of Black Holes in anti-De
  Sitter Space}},  {\em Commun. Math. Phys.} {\bf 87} (1983) 577.

\bibitem{DeWitt1979}
B.~S. DeWitt, S.~Hawking, and W.~Israel, {\em General Relativity: An Einstein
  Centenary Survey}.
\newblock Cambridge University Press Cambridge, 1979.

\bibitem{DeSabbata:1992cb}
R.~Wald, {\it {Black hole physics. Proceedings, NATO Advanced Study Institute,
  12th Course of the International School of Cosmology and Gravitation, Erice,
  Italy, (1991) eds. V. De Sabbata and Z. Zhang}}, .

\bibitem{York:1986it}
J.~W. York, Jr., {\it {Black hole thermodynamics and the Euclidean Einstein
  action}},  {\em Phys. Rev.} {\bf D33} (1986) 2092--2099.

\bibitem{Whiting:1988qr}
B.~F. Whiting and J.~W. York, Jr., {\it {Action Principle and Partition
  Function for the Gravitational Field in Black Hole Topologies}},  {\em Phys.
  Rev. Lett.} {\bf 61} (1988) 1336.

\bibitem{Brown:1992bq}
J.~D. Brown and J.~W. York, Jr., {\it {The Microcanonical functional integral.
  1. The Gravitational field}},  {\em Phys. Rev. D} {\bf 47} (1993) 1420--1431,
  [\href{http://xxx.arxiv.org/abs/gr-qc/9209014}{{\tt gr-qc/9209014}}].

\bibitem{Brown:1992br}
J.~D. Brown and J.~W. York, {\it Quasilocal energy and conserved charges
  derived from the gravitational action},  {\em Phys. Rev. D} {\bf 47} (Feb,
  1993) 1407--1419.

\bibitem{Brown:1994gs}
J.~D. Brown, J.~Creighton, and R.~B. Mann, {\it {Temperature, energy and heat
  capacity of asymptotically anti-de Sitter black holes}},  {\em Phys. Rev. D}
  {\bf 50} (1994) 6394--6403,
  [\href{http://xxx.arxiv.org/abs/gr-qc/9405007}{{\tt gr-qc/9405007}}].

\bibitem{BanadosEtal:1992}
M.~Banados, C.~Teitelboim, and J.~Zanelli, {\it {The Black hole in
  three-dimensional space-time}},  {\em Phys. Rev. Lett.} {\bf 69} (1992)
  1849--1851, [\href{http://xxx.arxiv.org/abs/hep-th/9204099}{{\tt
  hep-th/9204099}}].

\bibitem{Creighton:1995au}
J.~D.~E. Creighton and R.~B. Mann, {\it {Quasilocal thermodynamics of dilaton
  gravity coupled to gauge fields}},  {\em Phys. Rev. D} {\bf 52} (1995)
  4569--4587, [\href{http://xxx.arxiv.org/abs/gr-qc/9505007}{{\tt
  gr-qc/9505007}}].

\bibitem{Caldarelli:1999xj}
M.~M. Caldarelli, G.~Cognola, and D.~Klemm, {\it {Thermodynamics of
  Kerr-Newman-AdS black holes and conformal field theories}},  {\em Class.
  Quant. Grav.} {\bf 17} (2000) 399--420,
  [\href{http://xxx.arxiv.org/abs/hep-th/9908022}{{\tt hep-th/9908022}}].

\bibitem{Myers:1986un}
R.~C. Myers and M.~J. Perry, {\it {Black Holes in Higher Dimensional
  Space-Times}},  {\em Annals Phys.} {\bf 172} (1986) 304.

\bibitem{Frassino:2015oca}
A.~M. Frassino, R.~B. Mann, and J.~R. Mureika, {\it {Lower-Dimensional Black
  Hole Chemistry}},  {\em Phys. Rev.} {\bf D92} (2015), no.~12 124069,
  [\href{http://xxx.arxiv.org/abs/1509.05481}{{\tt arXiv:1509.05481}}].

\bibitem{Dolan:2013ft}
B.~P. Dolan, D.~Kastor, D.~Kubiznak, R.~B. Mann, and J.~Traschen, {\it
  {Thermodynamic Volumes and Isoperimetric Inequalities for de Sitter Black
  Holes}},  {\em Phys. Rev.} {\bf D87} (2013), no.~10 104017,
  [\href{http://xxx.arxiv.org/abs/1301.5926}{{\tt arXiv:1301.5926}}].

\bibitem{Gunasekaran:2012dq}
S.~Gunasekaran, R.~B. Mann, and D.~Kubiznak, {\it {Extended phase space
  thermodynamics for charged and rotating black holes and Born-Infeld vacuum
  polarization}},  {\em JHEP} {\bf 1211} (2012) 110,
  [\href{http://xxx.arxiv.org/abs/1208.6251}{{\tt arXiv:1208.6251}}].

\bibitem{Kastor:2010gq}
D.~Kastor, S.~Ray, and J.~Traschen, {\it {Smarr Formula and an Extended First
  Law for Lovelock Gravity}},  {\em Class. Quant. Grav.} {\bf 27} (2010)
  235014, [\href{http://xxx.arxiv.org/abs/1005.5053}{{\tt arXiv:1005.5053}}].

\bibitem{Brenna:2015pqa}
W.~G. Brenna, R.~B. Mann, and M.~Park, {\it {Mass and Thermodynamic Volume in
  Lifshitz Spacetimes}},  {\em Phys. Rev.} {\bf D92} (2015), no.~4 044015,
  [\href{http://xxx.arxiv.org/abs/1505.06331}{{\tt arXiv:1505.06331}}].

\bibitem{Caldarelli:2008pz}
M.~M. Caldarelli, R.~Emparan, and M.~J. Rodriguez, {\it {Black Rings in
  (Anti)-deSitter space}},  {\em JHEP} {\bf 11} (2008) 011,
  [\href{http://xxx.arxiv.org/abs/0806.1954}{{\tt arXiv:0806.1954}}].

\bibitem{Altamirano:2014tva}
N.~Altamirano, D.~Kubiznak, R.~B. Mann, and Z.~Sherkatghanad, {\it
  {Thermodynamics of rotating black holes and black rings: phase transitions
  and thermodynamic volume}},  {\em Galaxies} {\bf 2} (2014) 89--159,
  [\href{http://xxx.arxiv.org/abs/1401.2586}{{\tt arXiv:1401.2586}}].

\bibitem{Hennigar:2014cfa}
R.~A. Hennigar, D.~Kubiznak, and R.~B. Mann, {\it {Entropy Inequality
  Violations from Ultraspinning Black Holes}},  {\em Phys. Rev. Lett.} {\bf
  115} (2015), no.~3 031101, [\href{http://xxx.arxiv.org/abs/1411.4309}{{\tt
  arXiv:1411.4309}}].

\bibitem{Bertoldi:2009dt}
G.~Bertoldi, B.~A. Burrington, and A.~W. Peet, {\it {Thermodynamics of black
  branes in asymptotically Lifshitz spacetimes}},  {\em Phys. Rev.} {\bf D80}
  (2009) 126004, [\href{http://xxx.arxiv.org/abs/0907.4755}{{\tt
  arXiv:0907.4755}}].

\bibitem{Bertoldi:2009vn}
G.~Bertoldi, B.~A. Burrington, and A.~Peet, {\it {Black Holes in asymptotically
  Lifshitz spacetimes with arbitrary critical exponent}},  {\em Phys. Rev.}
  {\bf D80} (2009) 126003, [\href{http://xxx.arxiv.org/abs/0905.3183}{{\tt
  arXiv:0905.3183}}].

\bibitem{Dehghani:2010kd}
M.~H. Dehghani and R.~B. Mann, {\it {Lovelock-Lifshitz Black Holes}},  {\em
  JHEP} {\bf 07} (2010) 019, [\href{http://xxx.arxiv.org/abs/1004.4397}{{\tt
  arXiv:1004.4397}}].

\bibitem{Liu:2014dva}
H.-S. Liu and H.~Lü, {\it {Thermodynamics of Lifshitz Black Holes}},  {\em
  JHEP} {\bf 12} (2014) 071, [\href{http://xxx.arxiv.org/abs/1410.6181}{{\tt
  arXiv:1410.6181}}].

\bibitem{Berglund:2011cp}
P.~Berglund, J.~Bhattacharyya, and D.~Mattingly, {\it {Charged Dilatonic AdS
  Black Branes in Arbitrary Dimensions}},  {\em JHEP} {\bf 08} (2012) 042,
  [\href{http://xxx.arxiv.org/abs/1107.3096}{{\tt arXiv:1107.3096}}].

\bibitem{Dehghani:2011hf}
M.~H. Dehghani and S.~Asnafi, {\it {Thermodynamics of Rotating
  Lovelock-Lifshitz Black Branes}},  {\em Phys. Rev.} {\bf D84} (2011) 064038,
  [\href{http://xxx.arxiv.org/abs/1107.3354}{{\tt arXiv:1107.3354}}].

\bibitem{Dehghani:2013mba}
M.~H. Dehghani, C.~Shakuri, and M.~H. Vahidinia, {\it {Lifshitz black brane
  thermodynamics in the presence of a nonlinear electromagnetic field}},  {\em
  Phys. Rev.} {\bf D87} (2013), no.~8 084013,
  [\href{http://xxx.arxiv.org/abs/1306.4501}{{\tt arXiv:1306.4501}}].

\bibitem{Way:2012gr}
B.~Way, {\it {Holographic Confinement/Deconfinement Transitions in
  Asymptotically Lifshitz Spacetimes}},  {\em Phys. Rev.} {\bf D86} (2012)
  086007, [\href{http://xxx.arxiv.org/abs/1207.4205}{{\tt arXiv:1207.4205}}].

\bibitem{Gibbons:2004ai}
G.~W. Gibbons, M.~J. Perry, and C.~N. Pope, {\it {The First law of
  thermodynamics for Kerr-anti-de Sitter black holes}},  {\em Class. Quant.
  Grav.} {\bf 22} (2005) 1503--1526,
  [\href{http://xxx.arxiv.org/abs/hep-th/0408217}{{\tt hep-th/0408217}}].

\bibitem{CveticEtal:2010}
M.~Cvetic, G.~Gibbons, D.~Kubiznak, and C.~Pope, {\it {Black Hole Enthalpy and
  an Entropy Inequality for the Thermodynamic Volume}},  {\em Phys.Rev.} {\bf
  D84} (2011) 024037, [\href{http://xxx.arxiv.org/abs/1012.2888}{{\tt
  arXiv:1012.2888}}].

\bibitem{Aminneborg:1996iz}
S.~Aminneborg, I.~Bengtsson, S.~Holst, and P.~Peldan, {\it {Making anti-de
  Sitter black holes}},  {\em Class. Quant. Grav.} {\bf 13} (1996) 2707--2714,
  [\href{http://xxx.arxiv.org/abs/gr-qc/9604005}{{\tt gr-qc/9604005}}].

\bibitem{Mann:1996gj}
R.~Mann, {\it {Pair production of topological anti-de Sitter black holes}},
  {\em Class.Quant.Grav.} {\bf 14} (1997) L109--L114,
  [\href{http://xxx.arxiv.org/abs/gr-qc/9607071}{{\tt gr-qc/9607071}}].

\bibitem{Maldacena:1997re}
J.~M. Maldacena, {\it {The Large N limit of superconformal field theories and
  supergravity}},  {\em Int.J.Theor.Phys.} {\bf 38} (1999) 1113--1133,
  [\href{http://xxx.arxiv.org/abs/hep-th/9711200}{{\tt hep-th/9711200}}].

\bibitem{Witten:1998zw}
E.~Witten, {\it {Anti-de Sitter space, thermal phase transition, and
  confinement in gauge theories}},  {\em Adv. Theor. Math. Phys.} {\bf 2}
  (1998) 505--532, [\href{http://xxx.arxiv.org/abs/hep-th/9803131}{{\tt
  hep-th/9803131}}].

\bibitem{Kubiznak:2014zwa}
D.~Kubiznak and R.~B. Mann, {\it {Black hole chemistry}},  {\em Can. J. Phys.}
  {\bf 93} (2015), no.~9 999--1002,
  [\href{http://xxx.arxiv.org/abs/1404.2126}{{\tt arXiv:1404.2126}}].

\bibitem{Kubiznak:2012wp}
D.~Kubiznak and R.~B. Mann, {\it {P-V criticality of charged AdS black holes}},
   {\em JHEP} {\bf 07} (2012) 033,
  [\href{http://xxx.arxiv.org/abs/1205.0559}{{\tt arXiv:1205.0559}}].

\bibitem{Chamblin:1999tk}
A.~Chamblin, R.~Emparan, C.~V. Johnson, and R.~C. Myers, {\it {Charged AdS
  black holes and catastrophic holography}},  {\em Phys. Rev.} {\bf D60} (1999)
  064018, [\href{http://xxx.arxiv.org/abs/hep-th/9902170}{{\tt
  hep-th/9902170}}].

\bibitem{Chamblin:1999hg}
A.~Chamblin, R.~Emparan, C.~V. Johnson, and R.~C. Myers, {\it {Holography,
  thermodynamics and fluctuations of charged AdS black holes}},  {\em Phys.
  Rev.} {\bf D60} (1999) 104026,
  [\href{http://xxx.arxiv.org/abs/hep-th/9904197}{{\tt hep-th/9904197}}].

\bibitem{Cvetic:1999ne}
M.~Cvetic and S.~S. Gubser, {\it {Phases of R charged black holes, spinning
  branes and strongly coupled gauge theories}},  {\em JHEP} {\bf 04} (1999)
  024, [\href{http://xxx.arxiv.org/abs/hep-th/9902195}{{\tt hep-th/9902195}}].

\bibitem{Cvetic:1999rb}
M.~Cvetic and S.~S. Gubser, {\it {Thermodynamic stability and phases of general
  spinning branes}},  {\em JHEP} {\bf 07} (1999) 010,
  [\href{http://xxx.arxiv.org/abs/hep-th/9903132}{{\tt hep-th/9903132}}].

\bibitem{Dolan:2011xt}
B.~P. Dolan, {\it {Pressure and volume in the first law of black hole
  thermodynamics}},  {\em Class. Quant. Grav.} {\bf 28} (2011) 235017,
  [\href{http://xxx.arxiv.org/abs/1106.6260}{{\tt arXiv:1106.6260}}].

\bibitem{Rajagopal:2014ewa}
A.~Rajagopal, D.~Kubiznak, and R.~B. Mann, {\it {Van der Waals black hole}},
  {\em Phys. Lett.} {\bf B737} (2014) 277--279,
  [\href{http://xxx.arxiv.org/abs/1408.1105}{{\tt arXiv:1408.1105}}].

\bibitem{Delsate:2014zma}
T.~Delsate and R.~Mann, {\it {Van Der Waals Black Holes in $d$ dimensions}},
  {\em JHEP} {\bf 02} (2015) 070,
  [\href{http://xxx.arxiv.org/abs/1411.7850}{{\tt arXiv:1411.7850}}].

\bibitem{Mo:2016sel}
J.-X. Mo and G.-Q. Li, {\it {Coexistence curves and molecule number densities
  of AdS black holes in the reduced parameter space}},  {\em Phys. Rev.} {\bf
  D92} (2015), no.~2 024055, [\href{http://xxx.arxiv.org/abs/1604.07931}{{\tt
  arXiv:1604.07931}}].

\bibitem{Wei:2014qwa}
S.-W. Wei and Y.-X. Liu, {\it {Clapeyron equations and fitting formula of the
  coexistence curve in the extended phase space of charged AdS black holes}},
  {\em Phys. Rev.} {\bf D91} (2015), no.~4 044018,
  [\href{http://xxx.arxiv.org/abs/1411.5749}{{\tt arXiv:1411.5749}}].

\bibitem{Spallucci:2013osa}
E.~Spallucci and A.~Smailagic, {\it {Maxwell's equal area law for charged
  Anti-deSitter black holes}},  {\em Phys. Lett.} {\bf B723} (2013) 436--441,
  [\href{http://xxx.arxiv.org/abs/1305.3379}{{\tt arXiv:1305.3379}}].

\bibitem{Lan:2015bia}
S.-Q. Lan, J.-X. Mo, and W.-B. Liu, {\it {A note on Maxwell's equal area law
  for black hole phase transition}},  {\em Eur. Phys. J.} {\bf C75} (2015),
  no.~9 419, [\href{http://xxx.arxiv.org/abs/1503.07658}{{\tt
  arXiv:1503.07658}}].

\bibitem{Xu:2015hba}
H.~Xu and Z.-M. Xu, {\it {Maxwell's equal area law for Lovelock
  Thermodynamics}},  \href{http://xxx.arxiv.org/abs/1510.06557}{{\tt
  arXiv:1510.06557}}.

\bibitem{Wei:2015ana}
S.-W. Wei, P.~Cheng, and Y.-X. Liu, {\it {Analytical and exact critical
  phenomena of $d$-dimensional singly spinning Kerr-AdS black holes}},  {\em
  Phys. Rev.} {\bf D93} (2016), no.~8 084015,
  [\href{http://xxx.arxiv.org/abs/1510.00085}{{\tt arXiv:1510.00085}}].

\bibitem{Cheng:2016bpx}
P.~Cheng, S.-W. Wei, and Y.-X. Liu, {\it {Critical phenomena in the extended
  phase space of Kerr-Newman-AdS black holes}},  {\em Phys. Rev.} {\bf D94}
  (2016) 024025, [\href{http://xxx.arxiv.org/abs/1603.08694}{{\tt
  arXiv:1603.08694}}].

\bibitem{Wei:2015iwa}
S.-W. Wei and Y.-X. Liu, {\it {Insight into the Microscopic Structure of an AdS
  Black Hole from a Thermodynamical Phase Transition}},  {\em Phys. Rev. Lett.}
  {\bf 115} (2015), no.~11 111302,
  [\href{http://xxx.arxiv.org/abs/1502.00386}{{\tt arXiv:1502.00386}}].
  [Erratum: Phys. Rev. Lett.116,no.16,169903(2016)].

\bibitem{Mo:2013ela}
J.-X. Mo and W.-B. Liu, {\it {Ehrenfest scheme for P-V criticality in the
  extended phase space of black holes}},  {\em Phys. Lett.} {\bf B727} (2013)
  336--339.

\bibitem{Mo:2014wca}
J.-X. Mo, G.-Q. Li, and W.-B. Liu, {\it {Another novel Ehrenfest scheme for P-V
  criticality of RN-AdS black holes}},  {\em Phys. Lett.} {\bf B730} (2014)
  111--114.

\bibitem{Mo:2014mba}
J.-X. Mo and W.-B. Liu, {\it {Ehrenfest scheme for $P-V$ criticality of higher
  dimensional charged black holes, rotating black holes and Gauss-Bonnet AdS
  black holes}},  {\em Phys. Rev.} {\bf D89} (2014), no.~8 084057,
  [\href{http://xxx.arxiv.org/abs/1404.3872}{{\tt arXiv:1404.3872}}].

\bibitem{Zhao:2014fea}
H.-H. Zhao, L.-C. Zhang, M.-S. Ma, and R.~Zhao, {\it {Phase transition and
  Clapeyron equation of black holes in higher dimensional AdS spacetime}},
  {\em Class. Quant. Grav.} {\bf 32} (2015), no.~14 145007,
  [\href{http://xxx.arxiv.org/abs/1411.3554}{{\tt arXiv:1411.3554}}].

\bibitem{Li:2020nsy}
R.~Li, K.~Zhang, and J.~Wang, {\it {Thermal dynamic phase transition of
  Reissner-Nordstr\"om Anti-de Sitter black holes on free energy landscape}},
  {\em JHEP} {\bf 10} (2020) 090,
  [\href{http://xxx.arxiv.org/abs/2008.00495}{{\tt arXiv:2008.00495}}].

\bibitem{Wei:2020rcd}
S.-W. Wei, Y.-X. Liu, and Y.-Q. Wang, {\it {Dynamic properties of thermodynamic
  phase transition for five-dimensional neutral Gauss-Bonnet AdS black hole on
  free energy landscape}},  {\em Nucl. Phys. B} {\bf 976} (2022) 115692,
  [\href{http://xxx.arxiv.org/abs/2009.05215}{{\tt arXiv:2009.05215}}].

\bibitem{Li:2020spm}
R.~Li and J.~Wang, {\it {Energy and entropy compensation, phase transition and
  kinetics of four dimensional charged Gauss-Bonnet Anti-de Sitter black holes
  on the underlying free energy landscape}},  {\em Nucl. Phys. B} {\bf 976}
  (2022) 115714, [\href{http://xxx.arxiv.org/abs/2012.05424}{{\tt
  arXiv:2012.05424}}].

\bibitem{Li:2021vdp}
R.~Li, K.~Zhang, and J.~Wang, {\it {Probing black hole microstructure with the
  kinetic turnover of phase transition}},  {\em Phys. Rev. D} {\bf 104} (2021),
  no.~8 084076, [\href{http://xxx.arxiv.org/abs/2102.09439}{{\tt
  arXiv:2102.09439}}].

\bibitem{Lan:2021crt}
S.-Q. Lan, J.-X. Mo, G.-Q. Li, and X.-B. Xu, {\it {Effects of dark energy on
  dynamic phase transition of charged AdS black holes}},  {\em Phys. Rev. D}
  {\bf 104} (2021), no.~10 104032,
  [\href{http://xxx.arxiv.org/abs/2104.11553}{{\tt arXiv:2104.11553}}].

\bibitem{Li:2021zep}
R.~Li, K.~Zhang, and J.~Wang, {\it {Kinetics and its turnover of Hawking-Page
  phase transition under the black hole evaporation}},  {\em Phys. Rev. D} {\bf
  104} (2021), no.~8 084060, [\href{http://xxx.arxiv.org/abs/2105.00229}{{\tt
  arXiv:2105.00229}}].

\bibitem{Yang:2021ljn}
S.-J. Yang, R.~Zhou, S.-W. Wei, and Y.-X. Liu, {\it {Kinetics of a phase
  transition for a Kerr-AdS black hole on the free-energy landscape}},  {\em
  Phys. Rev. D} {\bf 105} (2022), no.~8 084030,
  [\href{http://xxx.arxiv.org/abs/2105.00491}{{\tt arXiv:2105.00491}}].

\bibitem{Mo:2021jff}
J.-X. Mo and S.-Q. Lan, {\it {Dynamic phase transition of charged dilaton black
  holes}},  {\em Chin. Phys. C} {\bf 45} (2021), no.~10 105106,
  [\href{http://xxx.arxiv.org/abs/2105.00868}{{\tt arXiv:2105.00868}}].

\bibitem{Kumara:2021hlt}
A.~N. Kumara, S.~Punacha, K.~Hegde, C.~L.~A. Rizwan, K.~M. Ajith, and M.~S.
  Ali, {\it {Dynamics and kinetics of phase transition for regular AdS black
  holes in general relativity coupled to nonlinear electrodynamics}},  {\em
  Int. J. Mod. Phys. A} {\bf 38} (2023), no.~29n30 2350151,
  [\href{http://xxx.arxiv.org/abs/2106.11095}{{\tt arXiv:2106.11095}}].

\bibitem{Li:2021tpu}
R.~Li and J.~Wang, {\it {Free energy landscape and kinetics of phase transition
  in two coupled SYK models and the corresponding wormhole-two black hole
  switching}},  {\em JHEP} {\bf 12} (2021) 208,
  [\href{http://xxx.arxiv.org/abs/2109.07635}{{\tt arXiv:2109.07635}}].

\bibitem{Xu:2021usl}
Z.-M. Xu, {\it {Fokker-Planck equation for black holes in thermal potential}},
  {\em Phys. Rev. D} {\bf 104} (2021), no.~10 104022,
  [\href{http://xxx.arxiv.org/abs/2111.05856}{{\tt arXiv:2111.05856}}].

\bibitem{Du:2021cxs}
Y.-Z. Du, H.-F. Li, F.~Liu, and L.-C. Zhang, {\it {Dynamic property of phase
  transition for non-linear charged anti-de Sitter black holes *}},  {\em Chin.
  Phys. C} {\bf 46} (2022), no.~5 055104,
  [\href{http://xxx.arxiv.org/abs/2112.10398}{{\tt arXiv:2112.10398}}].

\bibitem{Li:2022ylz}
R.~Li and J.~Wang, {\it {Kinetics of Hawking-Page phase transition with the
  non-Markovian effects}},  {\em JHEP} {\bf 05} (2022) 128,
  [\href{http://xxx.arxiv.org/abs/2201.06138}{{\tt arXiv:2201.06138}}].

\bibitem{Li:2022oup}
R.~Li and J.~wang, {\it {Generalized free energy landscape of a black hole
  phase transition}},  {\em Phys. Rev. D} {\bf 106} (2022), no.~10 106015,
  [\href{http://xxx.arxiv.org/abs/2206.02623}{{\tt arXiv:2206.02623}}].

\bibitem{Li:2023ppc}
R.~Li, C.~Liu, K.~Zhang, and J.~Wang, {\it {Topology of the landscape and
  dominant kinetic path for the thermodynamic phase transition of the charged
  Gauss-Bonnet-AdS black holes}},  {\em Phys. Rev. D} {\bf 108} (2023), no.~4
  044003, [\href{http://xxx.arxiv.org/abs/2302.06201}{{\tt arXiv:2302.06201}}].

\bibitem{Safir:2023thg}
T.~K. Safir, A.~N. Kumara, S.~Punacha, C.~L. Ahmed~Rizwan, C.~Fairoos, and
  D.~Vaid, {\it {Dynamic phase transition of black holes in massive gravity}},
  {\em Annals Phys.} {\bf 458} (2023) 169480,
  [\href{http://xxx.arxiv.org/abs/2306.10383}{{\tt arXiv:2306.10383}}].

\bibitem{Liu:2023sbf}
C.~Liu, R.~Li, K.~Zhang, and J.~Wang, {\it {Generalized free energy and
  dynamical state transition of the dyonic AdS black hole in the grand
  canonical ensemble}},  {\em JHEP} {\bf 11} (2023) 068,
  [\href{http://xxx.arxiv.org/abs/2309.13931}{{\tt arXiv:2309.13931}}].

\bibitem{Wang:2024zbp}
Y.-S. Wang, Z.-M. Xu, and B.~Wu, {\it {Thermodynamic phase transition rate for
  the third-order Lovelock black hole in diverse dimensions}},  {\em Phys.
  Lett. B} {\bf 853} (2024) 138690,
  [\href{http://xxx.arxiv.org/abs/2402.08887}{{\tt arXiv:2402.08887}}].

\bibitem{Wu:2024zig}
W.-Y. Wu, Z.~Luo, and J.~Li, {\it {Thermodynamic Phase Transition of AdS Black
  Holes in Massive Gravity on Free Energy Landscape}},  {\em Int. J. Theor.
  Phys.} {\bf 63} (2024), no.~7 177.

\bibitem{Li:2024tyk}
Q.-Q. Li, Y.~Zhang, Q.~Sun, C.-H. Xie, and Y.-L. Lou, {\it {Phase structure of
  quantum corrected charged AdS black hole surrounded by perfect fluid dark
  matter}},  {\em Chin. J. Phys.} {\bf 92} (2024) 1--9.

\bibitem{narayanan1994reentrant}
T.~Narayanan and A.~Kumar, {\it Reentrant phase transitions in multicomponent
  liquid mixtures},  {\em Physics Reports} {\bf 249} (1994), no.~3 135--218.

\bibitem{Hudson:1904}
C.~Hudson, {\it {Die gegenseitige loslichkeit von nikotin in wasser}},  {\em
  Z. Phys. Chem.} {\bf 47} (1904) 113.

\bibitem{Altamirano:2013ane}
N.~Altamirano, D.~Kubiznak, and R.~B. Mann, {\it {Reentrant phase transitions
  in rotating anti-de Sitter black holes}},  {\em Phys. Rev.} {\bf D88} (2013),
  no.~10 101502, [\href{http://xxx.arxiv.org/abs/1306.5756}{{\tt
  arXiv:1306.5756}}].

\bibitem{Altamirano:2013uqa}
N.~Altamirano, D.~Kubiznak, R.~B. Mann, and Z.~Sherkatghanad, {\it {Kerr-AdS
  analogue of triple point and solid/liquid/gas phase transition}},  {\em
  Class. Quant. Grav.} {\bf 31} (2014) 042001,
  [\href{http://xxx.arxiv.org/abs/1308.2672}{{\tt arXiv:1308.2672}}].

\bibitem{Frassino:2014pha}
A.~M. Frassino, D.~Kubiznak, R.~B. Mann, and F.~Simovic, {\it {Multiple
  Reentrant Phase Transitions and Triple Points in Lovelock Thermodynamics}},
  {\em JHEP} {\bf 1409} (2014) 080,
  [\href{http://xxx.arxiv.org/abs/1406.7015}{{\tt arXiv:1406.7015}}].

\bibitem{Wei:2014hba}
S.-W. Wei and Y.-X. Liu, {\it {Triple points and phase diagrams in the extended
  phase space of charged Gauss-Bonnet black holes in AdS space}},  {\em Phys.
  Rev.} {\bf D90} (2014), no.~4 044057,
  [\href{http://xxx.arxiv.org/abs/1402.2837}{{\tt arXiv:1402.2837}}].

\bibitem{Hennigar:2015esa}
R.~A. Hennigar, W.~G. Brenna, and R.~B. Mann, {\it {$P$--$v$ criticality in
  quasitopological gravity}},  {\em JHEP} {\bf 07} (2015) 077,
  [\href{http://xxx.arxiv.org/abs/1505.05517}{{\tt arXiv:1505.05517}}].

\bibitem{Sherkatghanad:2014hda}
Z.~Sherkatghanad, B.~Mirza, Z.~Mirzaeyan, and S.~A.~H. Mansoori, {\it {Critical
  behaviors and phase transitions of black holes in higher order gravities and
  extended phase spaces}},  \href{http://xxx.arxiv.org/abs/1412.5028}{{\tt
  arXiv:1412.5028}}.

\bibitem{Hennigar:2015wxa}
R.~A. Hennigar and R.~B. Mann, {\it {Reentrant phase transitions and van der
  Waals behaviour for hairy black holes}},  {\em Entropy} {\bf 17} (2015),
  no.~12 8056--8072, [\href{http://xxx.arxiv.org/abs/1509.06798}{{\tt
  arXiv:1509.06798}}].

\bibitem{Astefanesei:2021vcp}
D.~Astefanesei, P.~Cabrera, R.~B. Mann, and R.~Rojas, {\it {Reentrant phase
  transitions in Einstein-Maxwell-scalar black holes}},  {\em Phys. Rev. D}
  {\bf 105} (2022), no.~4 046021,
  [\href{http://xxx.arxiv.org/abs/2110.12005}{{\tt arXiv:2110.12005}}].

\bibitem{GibbonsEtal:2005}
G.~W. Gibbons, H.~L{\"u}, D.~N. Page, and C.~N. Pope, {\it {The general Kerr-de
  Sitter metrics in all dimensions}},  {\em J. Geom. Phys.} {\bf 53} (2005)
  49--73, [\href{http://xxx.arxiv.org/abs/hep-th/0404008}{{\tt
  hep-th/0404008}}].

\bibitem{Das:2000cu}
S.~Das and R.~B. Mann, {\it {Conserved quantities in Kerr-anti-de Sitter
  space-times in various dimensions}},  {\em JHEP} {\bf 08} (2000) 033,
  [\href{http://xxx.arxiv.org/abs/hep-th/0008028}{{\tt hep-th/0008028}}].

\bibitem{maslov2004zeroth}
V.~P. Maslov, {\it Zeroth-order phase transitions},  {\em Mathematical Notes}
  {\bf 76} (2004), no.~5-6 697--710.

\bibitem{Wei:2021krr}
S.-W. Wei and Y.-X. Liu, {\it {The microstructure and Ruppeiner geometry of
  charged anti-de Sitter black holes in Gauss\textendash{}Bonnet gravity: from
  the critical point to the triple point}},  {\em Commun. Theor. Phys.} {\bf
  74} (2022), no.~9 095402, [\href{http://xxx.arxiv.org/abs/2107.14523}{{\tt
  arXiv:2107.14523}}].

\bibitem{Dehghani:2020blz}
A.~Dehghani, S.~H. Hendi, and R.~B. Mann, {\it {Range of novel black hole phase
  transitions via massive gravity: Triple points and $N$-fold reentrant phase
  transitions}},  {\em Phys. Rev. D} {\bf 101} (2020), no.~8 084026,
  [\href{http://xxx.arxiv.org/abs/2009.07980}{{\tt arXiv:2009.07980}}].

\bibitem{Li:2022vcd}
M.-D. Li, H.-M. Wang, and S.-W. Wei, {\it {Triple points and novel phase
  transitions of dyonic AdS black holes with quasitopological
  electromagnetism}},  {\em Phys. Rev. D} {\bf 105} (2022), no.~10 104048,
  [\href{http://xxx.arxiv.org/abs/2201.09026}{{\tt arXiv:2201.09026}}].

\bibitem{Mou:2023nrx}
P.-H. Mou, Q.-Q. Jiang, K.-J. He, and G.-P. Li, {\it {Triple points and phase
  transitions of D-dimensional dyonic AdS black holes with quasitopological
  electromagnetism in Einstein\textendash{}Gauss\textendash{}Bonnet gravity}},
  {\em Chin. Phys. B} {\bf 33} (2024), no.~6 060401,
  [\href{http://xxx.arxiv.org/abs/2310.08010}{{\tt arXiv:2310.08010}}].

\bibitem{Hull:2022xew}
B.~R. Hull and F.~Simovic, {\it {Exotic black hole thermodynamics in
  third-order Lovelock gravity}},  {\em Class. Quant. Grav.} {\bf 40} (2023),
  no.~14 145016, [\href{http://xxx.arxiv.org/abs/2208.05500}{{\tt
  arXiv:2208.05500}}].

\bibitem{Hull:2021bry}
B.~R. Hull and R.~B. Mann, {\it {Thermodynamics of exotic black holes in
  Lovelock gravity}},  {\em Phys. Rev. D} {\bf 104} (2021), no.~8 084032,
  [\href{http://xxx.arxiv.org/abs/2102.05282}{{\tt arXiv:2102.05282}}].

\bibitem{Quijada:2023fkc}
C.~Quijada, A.~Anabal\'on, R.~B. Mann, and J.~Oliva, {\it {Triple Points of
  Gravitational AdS Solitons and Black Holes}},
  \href{http://xxx.arxiv.org/abs/2308.16341}{{\tt arXiv:2308.16341}}.

\bibitem{Wei:2021bwy}
S.-W. Wei, Y.-Q. Wang, Y.-X. Liu, and R.~B. Mann, {\it {Observing dynamic
  oscillatory behavior of triple points among black hole thermodynamic phase
  transitions}},  {\em Sci. China Phys. Mech. Astron.} {\bf 64} (2021), no.~7
  270411, [\href{http://xxx.arxiv.org/abs/2102.00799}{{\tt arXiv:2102.00799}}].

\bibitem{Cai:2021sag}
R.-G. Cai, {\it {Oscillatory behaviors near a black hole triple point}},  {\em
  Sci. China Phys. Mech. Astron.} {\bf 64} (2021), no.~9 290432.

\bibitem{Akahane2016}
K.~{Akahane}, J.~{Russo}, and H.~{Tanaka}, {\it {A possible four-phase
  coexistence in a single-component system}},  {\em Nature Communications} {\bf
  7} (Aug., 2016) 12599, [\href{http://xxx.arxiv.org/abs/1611.07148}{{\tt
  arXiv:1611.07148}}].

\bibitem{Garcia2017}
{\'A}.~{Gonz{\'a}lez Garc{\'\i}a}, H.~H. {Wensink}, H.~N.~W. {Lekkerkerker},
  and R.~{Tuinier}, {\it {Entropic patchiness drives multi-phase coexistence in
  discotic colloid-depletant mixtures}},  {\em Scientific Reports} {\bf 7}
  (Dec., 2017) 17058, [\href{http://xxx.arxiv.org/abs/1711.04143}{{\tt
  arXiv:1711.04143}}].

\bibitem{Sun:2021gpr}
W.~{Sun}, M.~J. {Powell-Palm}, and J.~{Chen}, {\it {The geometry of
  high-dimensional phase diagrams: I. Generalized Gibbs Phase Rule}},  {\em
  arXiv e-prints} (May, 2021) arXiv:2105.01337,
  [\href{http://xxx.arxiv.org/abs/2105.01337}{{\tt arXiv:2105.01337}}].

\bibitem{Tavakoli:2022kmo}
M.~Tavakoli, J.~Wu, and R.~B. Mann, {\it {Multi-critical points in black hole
  phase transitions}},  {\em JHEP} {\bf 12} (2022) 117,
  [\href{http://xxx.arxiv.org/abs/2207.03505}{{\tt arXiv:2207.03505}}].

\bibitem{Wu:2022plw}
J.~Wu and R.~B. Mann, {\it {Multicritical phase transitions in Lovelock AdS
  black holes}},  {\em Phys. Rev. D} {\bf 107} (2023), no.~8 084035,
  [\href{http://xxx.arxiv.org/abs/2212.08087}{{\tt arXiv:2212.08087}}].

\bibitem{Wu:2022xmp}
J.~Wu and R.~B. Mann, {\it {Thermodynamically stable phases of asymptotically
  flat Lovelock black holes}},  {\em Class. Quant. Grav.} {\bf 40} (2023),
  no.~14 145009, [\href{http://xxx.arxiv.org/abs/2212.08673}{{\tt
  arXiv:2212.08673}}].

\bibitem{Wu:2022bdk}
J.~Wu and R.~B. Mann, {\it {Multicritical Phase Transitions in Multiply
  Rotating Black Holes}},  \href{http://xxx.arxiv.org/abs/2208.00012}{{\tt
  arXiv:2208.00012}}.

\bibitem{Gibbons:2004js}
G.~W. Gibbons, H.~Lu, D.~N. Page, and C.~N. Pope, {\it {Rotating black holes in
  higher dimensions with a cosmological constant}},  {\em Phys. Rev. Lett.}
  {\bf 93} (2004) 171102, [\href{http://xxx.arxiv.org/abs/hep-th/0409155}{{\tt
  hep-th/0409155}}].

\bibitem{Gibbons:2004uw}
G.~W. Gibbons, H.~Lu, D.~N. Page, and C.~N. Pope, {\it {The General Kerr-de
  Sitter metrics in all dimensions}},  {\em J. Geom. Phys.} {\bf 53} (2005)
  49--73, [\href{http://xxx.arxiv.org/abs/hep-th/0404008}{{\tt
  hep-th/0404008}}].

\bibitem{Lu:2023hgu}
M.~Lu and R.~B. Mann, {\it {Maxwell construction and multi-criticality in
  uncharged generalized quasi-topological black holes}},  {\em Class. Quant.
  Grav.} {\bf 41} (2024), no.~1 015016,
  [\href{http://xxx.arxiv.org/abs/2306.06733}{{\tt arXiv:2306.06733}}].

\bibitem{Mo:2018rks}
J.-X. Mo and S.-Q. Lan, {\it {Criticality associated with the variation of the
  Lovelock parameter}},  {\em Phys. Lett. B} {\bf 783} (2018) 368--374.

\bibitem{Yang:2023xzv}
J.~Yang and R.~B. Mann, {\it {Dynamic behaviours of black hole phase
  transitions near quadruple points}},  {\em JHEP} {\bf 08} (2023) 028,
  [\href{http://xxx.arxiv.org/abs/2304.08969}{{\tt arXiv:2304.08969}}].

\bibitem{Dolan:2014vba}
B.~P. Dolan, A.~Kostouki, D.~Kubiznak, and R.~B. Mann, {\it {Isolated critical
  point from Lovelock gravity}},  {\em Class. Quant. Grav.} {\bf 31} (2014),
  no.~24 242001, [\href{http://xxx.arxiv.org/abs/1407.4783}{{\tt
  arXiv:1407.4783}}].

\bibitem{Oliva:2010eb}
J.~Oliva and S.~Ray, {\it {A new cubic theory of gravity in five dimensions:
  Black hole, Birkhoff's theorem and C-function}},  {\em Class. Quant. Grav.}
  {\bf 27} (2010) 225002, [\href{http://xxx.arxiv.org/abs/1003.4773}{{\tt
  arXiv:1003.4773}}].

\bibitem{Myers:2010ru}
R.~C. Myers and B.~Robinson, {\it {Black Holes in Quasi-topological Gravity}},
  {\em JHEP} {\bf 1008} (2010) 067,
  [\href{http://xxx.arxiv.org/abs/1003.5357}{{\tt arXiv:1003.5357}}].

\bibitem{prigogine1974chem}
I.~Prigogine and R.~Defay, {\it Chemical thermodynamics},  1974.

\bibitem{gupta1976prigogine}
P.~K. Gupta and C.~T. Moynihan, {\it Prigogine--defay ratio for systems with
  more than one order parameter},  {\em The Journal of Chemical Physics} {\bf
  65} (1976), no.~10 4136--4140.

\bibitem{gundermann2011predicting}
D.~Gundermann and et~all., {\it Predicting the density-scaling exponent of a
  glass-forming liquid from prigogine-defay ratio measurements},  {\em Nature
  Physics} {\bf 7} (2011), no.~10 816--821.

\bibitem{Dykaar:2017mba}
H.~Dykaar, R.~A. Hennigar, and R.~B. Mann, {\it {Hairy black holes in cubic
  quasi-topological gravity}},  {\em JHEP} {\bf 05} (2017) 045,
  [\href{http://xxx.arxiv.org/abs/1703.01633}{{\tt arXiv:1703.01633}}].

\bibitem{RevModPhys.71.S318}
A.~J. Leggett, {\it Superfluidity},  {\em Rev. Mod. Phys.} {\bf 71} (Mar, 1999)
  S318--S323.

\bibitem{Hennigar:2016xwd}
R.~A. Hennigar, R.~B. Mann, and E.~Tjoa, {\it {Superfluid Black Holes}},  {\em
  Phys. Rev. Lett.} {\bf 118} (2017), no.~2 021301,
  [\href{http://xxx.arxiv.org/abs/1609.02564}{{\tt arXiv:1609.02564}}].

\bibitem{Oliva:2011np}
J.~Oliva and S.~Ray, {\it {Conformal couplings of a scalar field to higher
  curvature terms}},  {\em Class. Quant. Grav.} {\bf 29} (2012) 205008,
  [\href{http://xxx.arxiv.org/abs/1112.4112}{{\tt arXiv:1112.4112}}].

\bibitem{Giribet:2014bva}
G.~Giribet, M.~Leoni, J.~Oliva, and S.~Ray, {\it {Hairy black holes sourced by
  a conformally coupled scalar field in D dimensions}},  {\em Phys. Rev.} {\bf
  D89} (2014), no.~8 085040, [\href{http://xxx.arxiv.org/abs/1401.4987}{{\tt
  arXiv:1401.4987}}].

\bibitem{nogo_hairy}
C.~Martinez, {\it Black holes with a conformally coupled scalar field},  in
  {\em Quantum Mechanics of Fundamental Systems: The Quest for Beauty and
  Simplicity}.

\bibitem{Hartnoll:2008vx}
S.~A. Hartnoll, C.~P. Herzog, and G.~T. Horowitz, {\it {Building a Holographic
  Superconductor}},  {\em Phys. Rev. Lett.} {\bf 101} (2008) 031601,
  [\href{http://xxx.arxiv.org/abs/0803.3295}{{\tt arXiv:0803.3295}}].

\bibitem{Nie:2015zia}
Z.-Y. Nie and H.~Zeng, {\it {P-T phase diagram of a holographic s+p model from
  Gauss-Bonnet gravity}},  {\em JHEP} {\bf 10} (2015) 047,
  [\href{http://xxx.arxiv.org/abs/1505.02289}{{\tt arXiv:1505.02289}}].

\bibitem{Giribet:2014fla}
G.~Giribet, A.~Goya, and J.~Oliva, {\it {Different phases of hairy black holes
  in AdS$_5$ space}},  {\em Phys. Rev.} {\bf D91} (2015), no.~4 045031,
  [\href{http://xxx.arxiv.org/abs/1501.00184}{{\tt arXiv:1501.00184}}].

\bibitem{Galante:2015voa}
M.~Galante, G.~Giribet, A.~Goya, and J.~Oliva, {\it {Chemical potential driven
  phase transition of black holes in anti de Sitter space}},  {\em Phys. Rev.}
  {\bf D92} (2015), no.~10 104039,
  [\href{http://xxx.arxiv.org/abs/1508.03780}{{\tt arXiv:1508.03780}}].

\bibitem{Chernicoff:2016jsu}
M.~Chernicoff, M.~Galante, G.~Giribet, A.~Goya, M.~Leoni, J.~Oliva, and
  G.~Perez-Nadal, {\it {Black hole thermodynamics, conformal couplings, and
  R$^{2}$ terms}},  {\em JHEP} {\bf 06} (2016) 159,
  [\href{http://xxx.arxiv.org/abs/1604.08203}{{\tt arXiv:1604.08203}}].

\bibitem{RevModPhys.67.279}
D.~M. Ceperley, {\it Path integrals in the theory of condensed helium},  {\em
  Rev. Mod. Phys.} {\bf 67} (Apr, 1995) 279--355.

\bibitem{Chong:2005hr}
Z.~W. Chong, M.~Cvetic, H.~Lu, and C.~N. Pope, {\it {General non-extremal
  rotating black holes in minimal five- dimensional gauged supergravity}},
  {\em Phys. Rev. Lett.} {\bf 95} (2005) 161301,
  [\href{http://xxx.arxiv.org/abs/hep-th/0506029}{{\tt hep-th/0506029}}].

\bibitem{Bai:2023woh}
N.-C. Bai, L.~Li, and J.~Tao, {\it {Superfluid \ensuremath{\lambda} transition
  in charged AdS black holes}},  {\em Sci. China Phys. Mech. Astron.} {\bf 66}
  (2023), no.~12 120411, [\href{http://xxx.arxiv.org/abs/2305.15258}{{\tt
  arXiv:2305.15258}}].

\bibitem{Taub:1950ez}
A.~H. Taub, {\it {Empty space-times admitting a three parameter group of
  motions}},  {\em Annals Math.} {\bf 53} (1951) 472--490.

\bibitem{Newman:1963yy}
E.~Newman, L.~Tamburino, and T.~Unti, {\it {Empty space generalization of the
  Schwarzschild metric}},  {\em J. Math. Phys.} {\bf 4} (1963) 915.

\bibitem{Miller:1971em}
J.~G. Miller, M.~D. Kruskal, and B.~B. Godfrey, {\it {Taub-NUT (Newman, Unti,
  Tamburino) Metric and Incompatible Extensions}},  {\em Phys. Rev. D} {\bf 4}
  (1971) 2945--2948.

\bibitem{Misner:1963fr}
C.~W. Misner, {\it {The Flatter regions of Newman, Unti and Tamburino's
  generalized Schwarzschild space}},  {\em J. Math. Phys.} {\bf 4} (1963)
  924--938.

\bibitem{HawkingEllis:book}
S.~W. Hawking and G.~F.~R. Ellis, {\em The Large Scale Structure of
  Space-Time}.
\newblock Cambridge University Press, Cambridge, England, 1973.

\bibitem{Hajicek:1971}
P.~Hajicek, {\it {Causality in non-Hausdorff space-times}},  {\em Comm. Math.
  Phys.} {\bf 21} (1971) 75--84.

\bibitem{Clement:2015cxa}
G.~Clement, D.~Gal'tsov, and M.~Guenouche, {\it {Rehabilitating space-times
  with NUTs}},  {\em Phys. Lett.} {\bf B750} (2015) 591--594,
  [\href{http://xxx.arxiv.org/abs/1508.07622}{{\tt arXiv:1508.07622}}].

\bibitem{Clement:2015aka}
G.~Cl\'ement, D.~Gal'tsov, and M.~Guenouche, {\it {NUT wormholes}},  {\em Phys.
  Rev. D} {\bf 93} (2016), no.~2 024048,
  [\href{http://xxx.arxiv.org/abs/1509.07854}{{\tt arXiv:1509.07854}}].

\bibitem{Hennigar:2019ive}
R.~A. Hennigar, D.~Kubiz\v{n}\'ak, and R.~B. Mann, {\it {Thermodynamics of
  Lorentzian Taub-NUT spacetimes}},  {\em Phys. Rev. D} {\bf 100} (2019), no.~6
  064055, [\href{http://xxx.arxiv.org/abs/1903.08668}{{\tt arXiv:1903.08668}}].

\bibitem{Bordo:2019slw}
A.~B. Bordo, F.~Gray, and D.~Kubiz\v{n}\'ak, {\it {Thermodynamics and Phase
  Transitions of NUTty Dyons}},  {\em JHEP} {\bf 07} (2019) 119,
  [\href{http://xxx.arxiv.org/abs/1904.00030}{{\tt arXiv:1904.00030}}].

\bibitem{BallonBordo:2019vrn}
A.~Ballon~Bordo, F.~Gray, R.~A. Hennigar, and D.~Kubiz\v{n}\'ak, {\it {The
  First Law for Rotating NUTs}},  {\em Phys. Lett. B} {\bf 798} (2019) 134972,
  [\href{http://xxx.arxiv.org/abs/1905.06350}{{\tt arXiv:1905.06350}}].

\bibitem{Hawking:1998ct}
S.~W. Hawking, C.~J. Hunter, and D.~N. Page, {\it {Nut charge, anti-de Sitter
  space and entropy}},  {\em Phys. Rev.} {\bf D59} (1999) 044033,
  [\href{http://xxx.arxiv.org/abs/hep-th/9809035}{{\tt hep-th/9809035}}].

\bibitem{Ashtekar:1999jx}
A.~Ashtekar and S.~Das, {\it {Asymptotically Anti-de Sitter space-times:
  Conserved quantities}},  {\em Class. Quant. Grav.} {\bf 17} (2000) L17--L30,
  [\href{http://xxx.arxiv.org/abs/hep-th/9911230}{{\tt hep-th/9911230}}].

\bibitem{Wu:2019pzr}
S.-Q. Wu and D.~Wu, {\it {Thermodynamical hairs of the four-dimensional
  Taub-Newman-Unti-Tamburino spacetimes}},  {\em Phys. Rev. D} {\bf 100}
  (2019), no.~10 101501, [\href{http://xxx.arxiv.org/abs/1909.07776}{{\tt
  arXiv:1909.07776}}].

\bibitem{Wu:2022xpp}
D.~Wu and S.-Q. Wu, {\it {Revisiting mass formulas of the four-dimensional
  Reissner-Nordstr\"om-NUT-AdS solutions in a different metric form}},  {\em
  Phys. Lett. B} {\bf 846} (2023) 138227,
  [\href{http://xxx.arxiv.org/abs/2210.17504}{{\tt arXiv:2210.17504}}].

\bibitem{Awad:2022jgn}
A.~Awad and S.~Eissa, {\it {Lorentzian Taub-NUT spacetimes: Misner string
  charges and the first law}},  {\em Phys. Rev. D} {\bf 105} (2022), no.~12
  124034, [\href{http://xxx.arxiv.org/abs/2206.09124}{{\tt arXiv:2206.09124}}].

\bibitem{Awad:2023lyt}
A.~Awad and E.~Elkhateeb, {\it {Dyonic Taub-NUT-AdS: Unconstrained
  thermodynamics and phase structure}},  {\em Phys. Rev. D} {\bf 108} (2023),
  no.~6 064022, [\href{http://xxx.arxiv.org/abs/2304.06705}{{\tt
  arXiv:2304.06705}}].

\bibitem{Durka:2019ajz}
R.~Durka, {\it {The first law of black hole thermodynamics for
  Taub\textendash{}NUT spacetime}},  {\em Int. J. Mod. Phys. D} {\bf 31}
  (2022), no.~04 2250021, [\href{http://xxx.arxiv.org/abs/1908.04238}{{\tt
  arXiv:1908.04238}}].

\bibitem{Rodriguez:2021hks}
N.~H. Rodr\'\i{}guez and M.~J. Rodriguez, {\it {First law for Kerr Taub-NUT AdS
  black holes}},  {\em JHEP} {\bf 10} (2022) 044,
  [\href{http://xxx.arxiv.org/abs/2112.00780}{{\tt arXiv:2112.00780}}].

\bibitem{Mann1999}
R.~B. Mann, {\it Misner string entropy},  {\em Phys.Rev. D} {\bf 60} (1999)
  104047, [\href{http://xxx.arxiv.org/abs/hep-th/9903229}{{\tt
  hep-th/9903229}}].

\bibitem{Frodden:2021ces}
E.~Frodden and D.~Hidalgo, {\it {The first law for the Kerr-NUT spacetime}},
  {\em Phys. Lett. B} {\bf 832} (2022) 137264,
  [\href{http://xxx.arxiv.org/abs/2109.07715}{{\tt arXiv:2109.07715}}].

\bibitem{Garfinkle:2000ms}
D.~Garfinkle and R.~B. Mann, {\it {Generalized entropy and Noether charge}},
  {\em Class. Quant. Grav.} {\bf 17} (2000) 3317--3324,
  [\href{http://xxx.arxiv.org/abs/gr-qc/0004056}{{\tt gr-qc/0004056}}].

\bibitem{Bordo:2019tyh}
A.~B. Bordo, F.~Gray, R.~A. Hennigar, and D.~Kubiz\v{n}\'ak, {\it {Misner
  Gravitational Charges and Variable String Strengths}},  {\em Class. Quant.
  Grav.} {\bf 36} (2019), no.~19 194001,
  [\href{http://xxx.arxiv.org/abs/1905.03785}{{\tt arXiv:1905.03785}}].

\bibitem{Mann:2004mi}
R.~B. Mann and C.~Stelea, {\it {On the thermodynamics of NUT charged spaces}},
  {\em Phys. Rev.} {\bf D72} (2005) 084032,
  [\href{http://xxx.arxiv.org/abs/hep-th/0408234}{{\tt hep-th/0408234}}].

\bibitem{Chen:2019uhp}
Z.~Chen and J.~Jiang, {\it {General Smarr relation and first law of a NUT
  dyonic black hole}},  {\em Phys. Rev. D} {\bf 100} (2019), no.~10 104016,
  [\href{http://xxx.arxiv.org/abs/1910.10107}{{\tt arXiv:1910.10107}}].

\bibitem{Abbasvandi:2021nyv}
N.~Abbasvandi, M.~Tavakoli, and R.~B. Mann, {\it {Thermodynamics of Dyonic NUT
  Charged Black Holes with entropy as Noether charge}},  {\em JHEP} {\bf 08}
  (2021) 152, [\href{http://xxx.arxiv.org/abs/2107.00182}{{\tt
  arXiv:2107.00182}}].

\bibitem{Mann:2020wad}
R.~B. Mann, L.~A. Pando~Zayas, and M.~Park, {\it {Complement to thermodynamics
  of dyonic Taub-NUT-AdS spacetime}},  {\em JHEP} {\bf 03} (2021) 039,
  [\href{http://xxx.arxiv.org/abs/2012.13506}{{\tt arXiv:2012.13506}}].

\bibitem{Liu:2022wku}
H.-S. Liu, H.~Lu, and L.~Ma, {\it {Thermodynamics of Taub-NUT and Plebanski
  solutions}},  {\em JHEP} {\bf 10} (2022) 174,
  [\href{http://xxx.arxiv.org/abs/2208.05494}{{\tt arXiv:2208.05494}}].

\bibitem{Wu:2022mlz}
D.~Wu and S.-Q. Wu, {\it {Consistent mass formulas for higher even-dimensional
  Taub-NUT spacetimes and their AdS counterparts}},  {\em Phys. Rev. D} {\bf
  108} (2023), no.~6 064034, [\href{http://xxx.arxiv.org/abs/2209.01757}{{\tt
  arXiv:2209.01757}}].

\bibitem{Chen:2024knw}
Y.-Q. Chen, H.-S. Liu, and H.~Lu, {\it {Taub-NUT black hole in higher
  derivative gravity and its thermodynamics}},  {\em Phys. Rev. D} {\bf 110}
  (2024), no.~10 104068, [\href{http://xxx.arxiv.org/abs/2409.07692}{{\tt
  arXiv:2409.07692}}].

\bibitem{Siahaan:2022ecb}
H.~M. Siahaan, {\it {Charged Taub-NUT-de Sitter spacetime in DGP braneworld and
  its thermodynamics*}},  {\em Chin. Phys. C} {\bf 47} (2023), no.~3 035105,
  [\href{http://xxx.arxiv.org/abs/2212.03051}{{\tt arXiv:2212.03051}}].

\bibitem{BallonBordo:2020jtw}
A.~Ballon~Bordo, D.~Kubiz\v{n}\'ak, and T.~R. Perche, {\it {Taub-NUT solutions
  in conformal electrodynamics}},  {\em Phys. Lett. B} {\bf 817} (2021) 136312,
  [\href{http://xxx.arxiv.org/abs/2011.13398}{{\tt arXiv:2011.13398}}].

\bibitem{Zhang:2021qga}
M.~Zhang and J.~Jiang, {\it {Conformal scalar NUT-like dyons in conformal
  electrodynamics}},  {\em Phys. Rev. D} {\bf 104} (2021), no.~8 084094,
  [\href{http://xxx.arxiv.org/abs/2110.04757}{{\tt arXiv:2110.04757}}].

\bibitem{Yang:2023hll}
S.-J. Yang, W.-D. Guo, S.-W. Wei, and Y.-X. Liu, {\it {First law of black hole
  thermodynamics and the weak cosmic censorship conjecture for
  Kerr\textendash{}Newman Taub\textendash{}NUT black holes}},  {\em Eur. Phys.
  J. C} {\bf 83} (2023), no.~12 1111,
  [\href{http://xxx.arxiv.org/abs/2306.05266}{{\tt arXiv:2306.05266}}].

\bibitem{Yang:2020iat}
S.-J. Yang, J.~Chen, J.-J. Wan, S.-W. Wei, and Y.-X. Liu, {\it {Weak cosmic
  censorship conjecture for a Kerr-Taub-NUT black hole with a test scalar field
  and particle}},  {\em Phys. Rev. D} {\bf 101} (2020), no.~6 064048,
  [\href{http://xxx.arxiv.org/abs/2001.03106}{{\tt arXiv:2001.03106}}].

\bibitem{Ciambelli:2020qny}
L.~Ciambelli, C.~Corral, J.~Figueroa, G.~Giribet, and R.~Olea, {\it
  {Topological Terms and the Misner String Entropy}},  {\em Phys. Rev. D} {\bf
  103} (2021), no.~2 024052, [\href{http://xxx.arxiv.org/abs/2011.11044}{{\tt
  arXiv:2011.11044}}].

\bibitem{Cano:2021qzp}
P.~A. Cano and D.~Pere\~niguez, {\it {Quasinormal modes of NUT-charged black
  branes in the AdS/CFT correspondence}},  {\em Class. Quant. Grav.} {\bf 39}
  (2022), no.~16 165003, [\href{http://xxx.arxiv.org/abs/2101.10652}{{\tt
  arXiv:2101.10652}}].

\bibitem{Ghezelbash:2021lcf}
M.~Ghezelbash and H.~M. Siahaan, {\it {Magnetized
  Kerr\textendash{}Newman\textendash{}Taub-NUT spacetimes}},  {\em Eur. Phys.
  J. C} {\bf 81} (2021), no.~7 621,
  [\href{http://xxx.arxiv.org/abs/2103.04865}{{\tt arXiv:2103.04865}}].

\bibitem{Siahaan:2021uqo}
H.~M. Siahaan, {\it {Magnetized Kerr-Taub-NUT spacetime and Kerr/CFT
  correspondence}},  {\em Phys. Lett. B} {\bf 820} (2021) 136568,
  [\href{http://xxx.arxiv.org/abs/2102.04345}{{\tt arXiv:2102.04345}}].

\bibitem{Liu:2024bzh}
H.-S. Liu and L.~Zhang, {\it {Scalarization of Taub-NUT black holes in extended
  scalar-tensor-Gauss-Bonnet theory}},  {\em JHEP} {\bf 10} (2024) 067,
  [\href{http://xxx.arxiv.org/abs/2407.08208}{{\tt arXiv:2407.08208}}].

\bibitem{Johnson:2014yja}
C.~V. Johnson, {\it {Holographic Heat Engines}},  {\em Class.Quant.Grav.} {\bf
  31} (2014) 205002, [\href{http://xxx.arxiv.org/abs/1404.5982}{{\tt
  arXiv:1404.5982}}].

\bibitem{Johnson:2015ekr}
C.~V. Johnson, {\it {Gauss-Bonnet Black Holes and Holographic Heat Engines
  Beyond Large N}},  \href{http://xxx.arxiv.org/abs/1511.08782}{{\tt
  arXiv:1511.08782}}.

\bibitem{Johnson:2015fva}
C.~V. Johnson, {\it {Born-Infeld AdS black holes as heat engines}},  {\em
  Class. Quant. Grav.} {\bf 33} (2016), no.~13 135001,
  [\href{http://xxx.arxiv.org/abs/1512.01746}{{\tt arXiv:1512.01746}}].

\bibitem{Balart:2021glm}
L.~Balart and S.~Fernando, {\it {Thermodynamics and heat engines of black holes
  with Born\textendash{}Infeld-type electrodynamics}},  {\em Mod. Phys. Lett.
  A} {\bf 36} (2021), no.~15 2150102,
  [\href{http://xxx.arxiv.org/abs/2103.15040}{{\tt arXiv:2103.15040}}].

\bibitem{Bhamidipati:2016gel}
C.~Bhamidipati and P.~K. Yerra, {\it {Heat Engines for Dilatonic Born-Infeld
  Black Holes}},  \href{http://xxx.arxiv.org/abs/1606.03223}{{\tt
  arXiv:1606.03223}}.

\bibitem{Sadeghi:2016xal}
J.~Sadeghi and K.~Jafarzade, {\it {The correction of Horava-Lifshitz black hole
  from holographic engine}},  \href{http://xxx.arxiv.org/abs/1604.02973}{{\tt
  arXiv:1604.02973}}.

\bibitem{EslamPanah:2019szt}
B.~Eslam~Panah and K.~Jafarzade, {\it {Thermal stability, $P{-}V$ criticality
  and heat engine of charged rotating accelerating black holes}},  {\em Gen.
  Rel. Grav.} {\bf 54} (2022), no.~2 19,
  [\href{http://xxx.arxiv.org/abs/1906.09478}{{\tt arXiv:1906.09478}}].

\bibitem{Debnath:2020zdv}
U.~Debnath, {\it {The general class of accelerating, rotating and charged
  Plebanski-Demianski black holes as heat engine}},  {\em Nucl. Phys. B} {\bf
  982} (2022) 115883, [\href{http://xxx.arxiv.org/abs/2006.02920}{{\tt
  arXiv:2006.02920}}].

\bibitem{Roy:2023qqy}
T.~Roy, A.~Sardar, and U.~Debnath, {\it {Thermodynamic overview and heat engine
  efficiency of Kerr\textendash{}Sen\textendash{}AdS black hole}},  {\em Int.
  J. Geom. Meth. Mod. Phys.} {\bf 20} (2023), no.~08 2350136.

\bibitem{Zhong:2023lhc}
Y.~Zhong and Y.-Z. Du, {\it {Heat engines of the Kerr-AdS black hole}},  {\em
  Commun. Theor. Phys.} {\bf 75} (2023), no.~12 125405.

\bibitem{Zhang:2018hms}
J.~Zhang, Y.~Li, and H.~Yu, {\it {Thermodynamics of charged accelerating AdS
  black holes and holographic heat engines}},  {\em JHEP} {\bf 02} (2019) 144,
  [\href{http://xxx.arxiv.org/abs/1808.10299}{{\tt arXiv:1808.10299}}].

\bibitem{Zhang:2016wek}
M.~Zhang and W.-B. Liu, {\it {f(R) Black Holes as Heat Engines}},  {\em Int. J.
  Theor. Phys.} {\bf 55} (2016), no.~12 5136--5145.

\bibitem{Wei:2016hkm}
S.-W. Wei and Y.-X. Liu, {\it {Implementing black hole as efficient power
  plant}},  \href{http://xxx.arxiv.org/abs/1605.04629}{{\tt arXiv:1605.04629}}.

\bibitem{Setare:2015yra}
M.~R. Setare and H.~Adami, {\it {Polytropic black hole as a heat engine}},
  {\em Gen. Rel. Grav.} {\bf 47} (2015), no.~11 133.

\bibitem{Sadeghi:2015ksa}
J.~Sadeghi and K.~Jafarzade, {\it {Heat Engine of black holes}},
  \href{http://xxx.arxiv.org/abs/1504.07744}{{\tt arXiv:1504.07744}}.

\bibitem{Mo:2017nhw}
J.-X. Mo, F.~Liang, and G.-Q. Li, {\it {Heat engine in the three-dimensional
  spacetime}},  {\em JHEP} {\bf 03} (2017) 010,
  [\href{http://xxx.arxiv.org/abs/1701.00883}{{\tt arXiv:1701.00883}}].

\bibitem{Liu:2017baz}
H.~Liu and X.-H. Meng, {\it {Effects of dark energy on the efficiency of
  charged AdS black holes as heat engines}},  {\em Eur. Phys. J. C} {\bf 77}
  (2017), no.~8 556, [\href{http://xxx.arxiv.org/abs/1704.04363}{{\tt
  arXiv:1704.04363}}].

\bibitem{Johnson:2017ood}
C.~V. Johnson, {\it {Taub\textendash{}Bolt heat engines}},  {\em Class. Quant.
  Grav.} {\bf 35} (2018), no.~4 045001,
  [\href{http://xxx.arxiv.org/abs/1705.04855}{{\tt arXiv:1705.04855}}].

\bibitem{Mo:2017nes}
J.-X. Mo and G.-Q. Li, {\it {Holographic Heat engine within the framework of
  massive gravity}},  {\em JHEP} {\bf 05} (2018) 122,
  [\href{http://xxx.arxiv.org/abs/1707.01235}{{\tt arXiv:1707.01235}}].

\bibitem{Hendi:2017bys}
S.~H. Hendi, B.~Eslam~Panah, S.~Panahiyan, H.~Liu, and X.~H. Meng, {\it {Black
  holes in massive gravity as heat engines}},  {\em Phys. Lett. B} {\bf 781}
  (2018) 40--47, [\href{http://xxx.arxiv.org/abs/1707.02231}{{\tt
  arXiv:1707.02231}}].

\bibitem{Wei:2017vqs}
S.-W. Wei and Y.-X. Liu, {\it {Charged AdS black hole heat engines}},  {\em
  Nucl. Phys.} {\bf B} (2019) 114700,
  [\href{http://xxx.arxiv.org/abs/1708.08176}{{\tt arXiv:1708.08176}}].

\bibitem{Zhang:2018vqs}
J.~Zhang, Y.~Li, and H.~Yu, {\it {Accelerating AdS black holes as the
  holographic heat engines in a benchmarking scheme}},  {\em Eur. Phys. J. C}
  {\bf 78} (2018), no.~8 645, [\href{http://xxx.arxiv.org/abs/1801.06811}{{\tt
  arXiv:1801.06811}}].

\bibitem{Johnson:2018amj}
C.~V. Johnson and F.~Rosso, {\it {Holographic Heat Engines, Entanglement
  Entropy, and Renormalization Group Flow}},  {\em Class. Quant. Grav.} {\bf
  36} (2019), no.~1 015019, [\href{http://xxx.arxiv.org/abs/1806.05170}{{\tt
  arXiv:1806.05170}}].

\bibitem{Johnson:2019olt}
C.~V. Johnson, {\it {Holographic Heat Engines as Quantum Heat Engines}},  {\em
  Class. Quant. Grav.} {\bf 37} (2020), no.~3 034001,
  [\href{http://xxx.arxiv.org/abs/1905.09399}{{\tt arXiv:1905.09399}}].

\bibitem{Chakraborty:2016ssb}
A.~Chakraborty and C.~V. Johnson, {\it {Benchmarking black hole heat engines,
  I}},  {\em Int. J. Mod. Phys. D} {\bf 27} (2018), no.~16 1950012,
  [\href{http://xxx.arxiv.org/abs/1612.09272}{{\tt arXiv:1612.09272}}].

\bibitem{Hennigar:2017apu}
R.~A. Hennigar, F.~McCarthy, A.~Ballon, and R.~B. Mann, {\it {Holographic heat
  engines: general considerations and rotating black holes}},  {\em Class.
  Quant. Grav.} {\bf 34} (2017), no.~17 175005,
  [\href{http://xxx.arxiv.org/abs/1704.02314}{{\tt arXiv:1704.02314}}].

\bibitem{Mann:1997jb}
R.~Mann, {\it {Black holes of negative mass}},  {\em Class.Quant.Grav.} {\bf
  14} (1997) 2927--2930, [\href{http://xxx.arxiv.org/abs/gr-qc/9705007}{{\tt
  gr-qc/9705007}}].

\bibitem{Johnson:2017hxu}
C.~V. Johnson, {\it {Exact model of the power-to-efficiency trade-off while
  approaching the Carnot limit}},  {\em Phys. Rev. D} {\bf 98} (2018), no.~2
  026008, [\href{http://xxx.arxiv.org/abs/1703.06119}{{\tt arXiv:1703.06119}}].

\bibitem{DiMarco:2022yhp}
M.~C. DiMarco, S.~L. Jess, R.~A. Hennigar, and R.~B. Mann, {\it {Universality
  for black hole heat engines near critical points}},  {\em Phys. Rev. D} {\bf
  107} (2023), no.~4 044001, [\href{http://xxx.arxiv.org/abs/2211.14856}{{\tt
  arXiv:2211.14856}}].

\bibitem{Johnston_2014}
D.~C. Johnston, {\em Advances in Thermodynamics of the van der Waals Fluid}.
\newblock Morgan and Claypool Publishers, Sept., 2014.

\bibitem{Okcu:2016tgt}
O.~\"Okc\"u and E.~Ayd\i{}ner, {\it {Joule\textendash{}Thomson expansion of the
  charged AdS black holes}},  {\em Eur. Phys. J. C} {\bf 77} (2017), no.~1 24,
  [\href{http://xxx.arxiv.org/abs/1611.06327}{{\tt arXiv:1611.06327}}].

\bibitem{Okcu:2017qgo}
O.~\"Okc\"u and E.~Ayd\i{}ner, {\it {Joule\textendash{}Thomson expansion of
  Kerr\textendash{}AdS black holes}},  {\em Eur. Phys. J. C} {\bf 78} (2018),
  no.~2 123, [\href{http://xxx.arxiv.org/abs/1709.06426}{{\tt
  arXiv:1709.06426}}].

\bibitem{Mo:2018rgq}
J.-X. Mo, G.-Q. Li, S.-Q. Lan, and X.-B. Xu, {\it {Joule-Thomson expansion of
  $d$-dimensional charged AdS black holes}},  {\em Phys. Rev. D} {\bf 98}
  (2018), no.~12 124032, [\href{http://xxx.arxiv.org/abs/1804.02650}{{\tt
  arXiv:1804.02650}}].

\bibitem{Chabab:2018zix}
M.~Chabab, H.~El~Moumni, S.~Iraoui, K.~Masmar, and S.~Zhizeh, {\it
  {Joule-Thomson Expansion of RN-AdS Black Holes in $f(R)$ gravity}},  {\em
  LHEP} {\bf 02} (2018) 05, [\href{http://xxx.arxiv.org/abs/1804.10042}{{\tt
  arXiv:1804.10042}}].

\bibitem{Mo:2018qkt}
J.-X. Mo and G.-Q. Li, {\it {Effects of Lovelock gravity on the
  Joule\textendash{}Thomson expansion}},  {\em Class. Quant. Grav.} {\bf 37}
  (2020), no.~4 045009, [\href{http://xxx.arxiv.org/abs/1805.04327}{{\tt
  arXiv:1805.04327}}].

\bibitem{Lan:2018nnp}
S.-Q. Lan, {\it {Joule-Thomson expansion of charged Gauss-Bonnet black holes in
  AdS space}},  {\em Phys. Rev. D} {\bf 98} (2018), no.~8 084014,
  [\href{http://xxx.arxiv.org/abs/1805.05817}{{\tt arXiv:1805.05817}}].

\bibitem{RizwanCL:2018cyb}
A.~Rizwan C.~L., N.~Kumara~A., D.~Vaid, and K.~M. Ajith, {\it {Joule-Thomson
  expansion in AdS black hole with a global monopole}},  {\em Int. J. Mod.
  Phys. A} {\bf 33} (2019), no.~35 1850210,
  [\href{http://xxx.arxiv.org/abs/1805.11053}{{\tt arXiv:1805.11053}}].

\bibitem{Cisterna:2018jqg}
A.~Cisterna, S.-Q. Hu, and X.-M. Kuang, {\it {Joule-Thomson expansion in AdS
  black holes with momentum relaxation}},  {\em Phys. Lett. B} {\bf 797} (2019)
  134883, [\href{http://xxx.arxiv.org/abs/1808.07392}{{\tt arXiv:1808.07392}}].

\bibitem{Nam:2019zyk}
C.~H. Nam, {\it {Heat engine efficiency and Joule\textendash{}Thomson expansion
  of nonlinear charged AdS black hole in massive gravity}},  {\em Gen. Rel.
  Grav.} {\bf 53} (2021), no.~3 30,
  [\href{http://xxx.arxiv.org/abs/1906.05557}{{\tt arXiv:1906.05557}}].

\bibitem{Sadeghi:2020bon}
J.~Sadeghi and R.~Toorandaz, {\it {Joule-Thomson expansion of hyperscaling
  violating black holes with spherical and hyperbolic horizons}},  {\em Nucl.
  Phys. B} {\bf 951} (2020) 114902.

\bibitem{Hegde:2020cdm}
K.~Hegde, A.~Naveena~Kumara, C.~L. Ahmed~Rizwan, A.~K. M., M.~S. Ali, and
  S.~Punacha, {\it {Thermodynamics, phase transition and
  Joule\textendash{}Thomson expansion of 4-D Gauss\textendash{}Bonnet AdS black
  hole}},  {\em Int. J. Mod. Phys. A} {\bf 39} (2024), no.~21 2450080,
  [\href{http://xxx.arxiv.org/abs/2003.08778}{{\tt arXiv:2003.08778}}].

\bibitem{Bi:2020vcg}
S.~Bi, M.~Du, J.~Tao, and F.~Yao, {\it {Joule-Thomson expansion of Born-Infeld
  AdS black holes}},  {\em Chin. Phys. C} {\bf 45} (2021), no.~2 025109,
  [\href{http://xxx.arxiv.org/abs/2006.08920}{{\tt arXiv:2006.08920}}].

\bibitem{Guo:2020ysx}
S.~Guo, Y.-L. Huang, and G.-P. Li, {\it {Cooling-heating phase transition and
  critical behavior of the charged accelerating AdS black hole}},
  \href{http://xxx.arxiv.org/abs/2009.09401}{{\tt arXiv:2009.09401}}.

\bibitem{Zhang:2021kha}
C.-M. Zhang, M.~Zhang, and D.-C. Zou, {\it {Joule\textendash{}Thomson expansion
  of Born\textendash{}Infeld AdS black holes in consistent 4D
  Einstein\textendash{}Gauss\textendash{}Bonnet gravity}},  {\em Mod. Phys.
  Lett. A} {\bf 37} (2022), no.~11 2250063,
  [\href{http://xxx.arxiv.org/abs/2106.00183}{{\tt arXiv:2106.00183}}].

\bibitem{Cao:2022hmd}
Y.~Cao, H.~Feng, J.~Tao, and Y.~Xue, {\it {Black holes in a cavity: Heat engine
  and Joule-Thomson expansion}},  {\em Gen. Rel. Grav.} {\bf 54} (2022), no.~9
  105, [\href{http://xxx.arxiv.org/abs/2201.07584}{{\tt arXiv:2201.07584}}].

\bibitem{Barrientos:2022uit}
J.~Barrientos and J.~Mena, {\it {Joule-Thomson expansion of AdS black holes in
  quasitopological electromagnetism}},  {\em Phys. Rev. D} {\bf 106} (2022),
  no.~4 044064, [\href{http://xxx.arxiv.org/abs/2206.06018}{{\tt
  arXiv:2206.06018}}].

\bibitem{Kruglov:2022bhx}
S.~I. Kruglov, {\it {AdS Black Holes in the Framework of Nonlinear
  Electrodynamics, Thermodynamics, and Joule\textendash{}Thomson Expansion}},
  {\em Symmetry} {\bf 14} (2022), no.~8 1597,
  [\href{http://xxx.arxiv.org/abs/2209.05394}{{\tt arXiv:2209.05394}}].

\bibitem{Sekhmani:2023est}
Y.~Sekhmani, R.~Myrzakulov, and R.~Ali, {\it {Joule\textendash{}Thomson
  expansion of black holes in STU supergravity}},  {\em Int. J. Mod. Phys. A}
  {\bf 38} (2023), no.~33n34 2350176.

\bibitem{Kruglov:2023ogn}
S.~I. Kruglov, {\it {Magnetic black holes within Einstein\textendash{}AdS
  gravity coupled to nonlinear electrodynamics, extended phase space
  thermodynamics and Joule\textendash{}Thomson expansion}},  {\em Can. J.
  Phys.} {\bf 101} (2023), no.~12 739--748,
  [\href{http://xxx.arxiv.org/abs/2401.15115}{{\tt arXiv:2401.15115}}].

\bibitem{Weyl:1917gp}
H.~Weyl, {\it {The theory of gravitation}},  {\em Annalen Phys.} {\bf 54}
  (1917) 117--145.

\bibitem{Appels:2016uha}
M.~Appels, R.~Gregory, and D.~Kubiznak, {\it {Thermodynamics of Accelerating
  Black Holes}},  \href{http://xxx.arxiv.org/abs/1604.08812}{{\tt
  arXiv:1604.08812}}.

\bibitem{Appels:2017xoe}
M.~Appels, R.~Gregory, and D.~Kubiznak, {\it {Black Hole Thermodynamics with
  Conical Defects}},  {\em JHEP} {\bf 05} (2017) 116,
  [\href{http://xxx.arxiv.org/abs/1702.00490}{{\tt arXiv:1702.00490}}].

\bibitem{Gregory:2017ogk}
R.~Gregory, {\it {Accelerating Black Holes}},  {\em J. Phys. Conf. Ser.} {\bf
  942} (2017), no.~1 012002, [\href{http://xxx.arxiv.org/abs/1712.04992}{{\tt
  arXiv:1712.04992}}].

\bibitem{Astorino:2016xiy}
M.~Astorino, {\it {CFT Duals for Accelerating Black Holes}},  {\em Phys. Lett.
  B} {\bf 760} (2016) 393--405,
  [\href{http://xxx.arxiv.org/abs/1605.06131}{{\tt arXiv:1605.06131}}].

\bibitem{Astorino:2016ybm}
M.~Astorino, {\it {Thermodynamics of Regular Accelerating Black Holes}},  {\em
  Phys. Rev. D} {\bf 95} (2017), no.~6 064007,
  [\href{http://xxx.arxiv.org/abs/1612.04387}{{\tt arXiv:1612.04387}}].

\bibitem{Anabalon:2018ydc}
A.~Anabal\'on, M.~Appels, R.~Gregory, D.~Kubiz\v{n}\'ak, R.~B. Mann, and
  A.~Ovg\"un, {\it {Holographic Thermodynamics of Accelerating Black Holes}},
  {\em Phys. Rev. D} {\bf 98} (2018), no.~10 104038,
  [\href{http://xxx.arxiv.org/abs/1805.02687}{{\tt arXiv:1805.02687}}].

\bibitem{Anabalon:2018qfv}
A.~Anabal\'on, F.~Gray, R.~Gregory, D.~Kubiz\v{n}\'ak, and R.~B. Mann, {\it
  {Thermodynamics of Charged, Rotating, and Accelerating Black Holes}},  {\em
  JHEP} {\bf 04} (2019) 096, [\href{http://xxx.arxiv.org/abs/1811.04936}{{\tt
  arXiv:1811.04936}}].

\bibitem{Kinnersley:1970zw}
W.~Kinnersley and M.~Walker, {\it {Uniformly accelerating charged mass in
  general relativity}},  {\em Phys. Rev.} {\bf D2} (1970) 1359--1370.

\bibitem{Plebanski:1976gy}
J.~F. Plebanski and M.~Demianski, {\it {Rotating, charged, and uniformly
  accelerating mass in general relativity}},  {\em Annals Phys.} {\bf 98}
  (1976) 98--127.

\bibitem{Dias:2002mi}
O.~J.~C. Dias and J.~P.~S. Lemos, {\it {Pair of accelerated black holes in
  anti-de Sitter background: AdS C metric}},  {\em Phys. Rev.} {\bf D67} (2003)
  064001, [\href{http://xxx.arxiv.org/abs/hep-th/0210065}{{\tt
  hep-th/0210065}}].

\bibitem{Griffiths:2005qp}
J.~B. Griffiths and J.~Podolsky, {\it {A New look at the Plebanski-Demianski
  family of solutions}},  {\em Int. J. Mod. Phys.} {\bf D15} (2006) 335--370,
  [\href{http://xxx.arxiv.org/abs/gr-qc/0511091}{{\tt gr-qc/0511091}}].

\bibitem{Zhang:2019vpf}
M.~Zhang and R.~B. Mann, {\it {Charged accelerating black hole in $f(R)$
  gravity}},  {\em Phys. Rev. D} {\bf 100} (2019), no.~8 084061,
  [\href{http://xxx.arxiv.org/abs/1908.05118}{{\tt arXiv:1908.05118}}].

\bibitem{Gregory:2019dtq}
R.~Gregory and A.~Scoins, {\it {Accelerating Black Hole Chemistry}},  {\em
  Phys. Lett. B} {\bf 796} (2019) 191--195,
  [\href{http://xxx.arxiv.org/abs/1904.09660}{{\tt arXiv:1904.09660}}].

\bibitem{Gregory:1995hd}
R.~Gregory and M.~Hindmarsh, {\it {Smooth metrics for snapping strings}},  {\em
  Phys. Rev.} {\bf D52} (1995) 5598--5605,
  [\href{http://xxx.arxiv.org/abs/gr-qc/9506054}{{\tt gr-qc/9506054}}].

\bibitem{Dowker:1993bt}
F.~Dowker, J.~P. Gauntlett, D.~A. Kastor, and J.~H. Traschen, {\it {Pair
  creation of dilaton black holes}},  {\em Phys. Rev.} {\bf D49} (1994)
  2909--2917, [\href{http://xxx.arxiv.org/abs/hep-th/9309075}{{\tt
  hep-th/9309075}}].

\bibitem{Hawking:1994ii}
S.~W. Hawking, G.~T. Horowitz, and S.~F. Ross, {\it {Entropy, Area, and black
  hole pairs}},  {\em Phys. Rev.} {\bf D51} (1995) 4302--4314,
  [\href{http://xxx.arxiv.org/abs/gr-qc/9409013}{{\tt gr-qc/9409013}}].

\bibitem{Emparan:1995je}
R.~Emparan, {\it {Pair creation of black holes joined by cosmic strings}},
  {\em Phys. Rev. Lett.} {\bf 75} (1995) 3386--3389,
  [\href{http://xxx.arxiv.org/abs/gr-qc/9506025}{{\tt gr-qc/9506025}}].

\bibitem{Eardley:1995au}
D.~M. Eardley, G.~T. Horowitz, D.~A. Kastor, and J.~H. Traschen, {\it {Breaking
  cosmic strings without monopoles}},  {\em Phys. Rev. Lett.} {\bf 75} (1995)
  3390--3393, [\href{http://xxx.arxiv.org/abs/gr-qc/9506041}{{\tt
  gr-qc/9506041}}].

\bibitem{Mann:1995vb}
R.~B. Mann and S.~F. Ross, {\it {Cosmological production of charged black hole
  pairs}},  {\em Phys. Rev.} {\bf D52} (1995) 2254--2265,
  [\href{http://xxx.arxiv.org/abs/gr-qc/9504015}{{\tt gr-qc/9504015}}].

\bibitem{Booth:1998pb}
I.~S. Booth and R.~B. Mann, {\it {Complex instantons and charged rotating black
  hole pair creation}},  {\em Phys. Rev. Lett.} {\bf 81} (1998) 5052--5055,
  [\href{http://xxx.arxiv.org/abs/gr-qc/9806015}{{\tt gr-qc/9806015}}].

\bibitem{Ball:2020vzo}
A.~Ball and N.~Miller, {\it {Accelerating black hole thermodynamics with boost
  time}},  {\em Class. Quant. Grav.} {\bf 38} (2021), no.~14 145031,
  [\href{http://xxx.arxiv.org/abs/2008.03682}{{\tt arXiv:2008.03682}}].

\bibitem{Abbasvandi:2018vsh}
N.~Abbasvandi, W.~Cong, D.~Kubiznak, and R.~B. Mann, {\it {Snapping
  swallowtails in accelerating black hole thermodynamics}},  {\em Class. Quant.
  Grav.} {\bf 36} (2019), no.~10 104001,
  [\href{http://xxx.arxiv.org/abs/1812.00384}{{\tt arXiv:1812.00384}}].

\bibitem{Abbasvandi:2019vfz}
N.~Abbasvandi, W.~Ahmed, W.~Cong, D.~Kubiz\v{n}\'ak, and R.~B. Mann, {\it
  {Finely Split Phase Transitions of Rotating and Accelerating Black Holes}},
  {\em Phys. Rev. D} {\bf 100} (2019), no.~6 064027,
  [\href{http://xxx.arxiv.org/abs/1906.03379}{{\tt arXiv:1906.03379}}].

\bibitem{Ahmed:2019yci}
W.~Ahmed, H.~Z. Chen, E.~Gesteau, R.~Gregory, and A.~Scoins, {\it {Conical
  Holographic Heat Engines}},  {\em Class. Quant. Grav.} {\bf 36} (2019),
  no.~21 214001, [\href{http://xxx.arxiv.org/abs/1906.10289}{{\tt
  arXiv:1906.10289}}].

\bibitem{Witten:1998qj}
E.~Witten, {\it {Anti-de Sitter space and holography}},  {\em Adv. Theor. Math.
  Phys.} {\bf 2} (1998) 253--291,
  [\href{http://xxx.arxiv.org/abs/hep-th/9802150}{{\tt hep-th/9802150}}].

\bibitem{Kovtun:2004de}
P.~Kovtun, D.~T. Son, and A.~O. Starinets, {\it {Viscosity in strongly
  interacting quantum field theories from black hole physics}},  {\em Phys.
  Rev. Lett.} {\bf 94} (2005) 111601,
  [\href{http://xxx.arxiv.org/abs/hep-th/0405231}{{\tt hep-th/0405231}}].

\bibitem{Pahlavani:2014dma}
M.~R. Pahlavani and R.~Morad, {\it {Application of AdS/CFT in Nuclear
  Physics}},  {\em Adv. High Energy Phys.} {\bf 2014} (2014) 863268,
  [\href{http://xxx.arxiv.org/abs/1403.2501}{{\tt arXiv:1403.2501}}].

\bibitem{Matsumoto:2018ukk}
M.~Matsumoto and S.~Nakamura, {\it {Critical Exponents of Nonequilibrium Phase
  Transitions in AdS/CFT Correspondence}},  {\em Phys. Rev. D} {\bf 98} (2018),
  no.~10 106027, [\href{http://xxx.arxiv.org/abs/1804.10124}{{\tt
  arXiv:1804.10124}}].

\bibitem{Hartnoll:2007ih}
S.~A. Hartnoll, P.~K. Kovtun, M.~Muller, and S.~Sachdev, {\it {Theory of the
  Nernst effect near quantum phase transitions in condensed matter, and in
  dyonic black holes}},  {\em Phys. Rev.} {\bf B76} (2007) 144502,
  [\href{http://xxx.arxiv.org/abs/0706.3215}{{\tt arXiv:0706.3215}}].

\bibitem{Astefanesei:2019ehu}
D.~Astefanesei, R.~B. Mann, and R.~Rojas, {\it {Hairy Black Hole Chemistry}},
  {\em JHEP} {\bf 11} (2019) 043,
  [\href{http://xxx.arxiv.org/abs/1907.08636}{{\tt arXiv:1907.08636}}].

\bibitem{Astefanesei:2023sep}
D.~Astefanesei, P.~Cabrera, R.~B. Mann, and R.~Rojas, {\it {Extended phase
  space thermodynamics for hairy black holes}},  {\em Phys. Rev. D} {\bf 108}
  (2023), no.~10 104047, [\href{http://xxx.arxiv.org/abs/2304.09203}{{\tt
  arXiv:2304.09203}}].

\bibitem{Dolan:2014cja}
B.~P. Dolan, {\it {Bose condensation and branes}},  {\em JHEP} {\bf 10} (2014)
  179, [\href{http://xxx.arxiv.org/abs/1406.7267}{{\tt arXiv:1406.7267}}].

\bibitem{Kastor:2014dra}
D.~Kastor, S.~Ray, and J.~Traschen, {\it {Chemical Potential in the First Law
  for Holographic Entanglement Entropy}},  {\em JHEP} {\bf 11} (2014) 120,
  [\href{http://xxx.arxiv.org/abs/1409.3521}{{\tt arXiv:1409.3521}}].

\bibitem{Zhang:2014uoa}
J.-L. Zhang, R.-G. Cai, and H.~Yu, {\it {Phase transition and thermodynamical
  geometry for Schwarzschild AdS black hole in AdS$_{5}$ × S$^{5}$
  spacetime}},  {\em JHEP} {\bf 02} (2015) 143,
  [\href{http://xxx.arxiv.org/abs/1409.5305}{{\tt arXiv:1409.5305}}].

\bibitem{Zhang:2015ova}
J.-L. Zhang, R.-G. Cai, and H.~Yu, {\it {Phase transition and thermodynamical
  geometry of Reissner-Nordstr\"om-AdS black holes in extended phase space}},
  {\em Phys.Rev.} {\bf D91} (2015), no.~4 044028,
  [\href{http://xxx.arxiv.org/abs/1502.01428}{{\tt arXiv:1502.01428}}].

\bibitem{Dolan:2016jjc}
B.~P. Dolan, {\it {Pressure and compressibility of conformal field theories
  from the AdS/CFT correspondence}},  {\em Entropy} {\bf 18} (2016) 169,
  [\href{http://xxx.arxiv.org/abs/1603.06279}{{\tt arXiv:1603.06279}}].

\bibitem{McCarthy:2017amh}
F.~McCarthy, D.~Kubiz\v{n}\'ak, and R.~B. Mann, {\it {Breakdown of the equal
  area law for holographic entanglement entropy}},  {\em JHEP} {\bf 11} (2017)
  165, [\href{http://xxx.arxiv.org/abs/1708.07982}{{\tt arXiv:1708.07982}}].

\bibitem{Caceres:2015vsa}
E.~Caceres, P.~H. Nguyen, and J.~F. Pedraza, {\it {Holographic entanglement
  entropy and the extended phase structure of STU black holes}},  {\em JHEP}
  {\bf 09} (2015) 184, [\href{http://xxx.arxiv.org/abs/1507.06069}{{\tt
  arXiv:1507.06069}}].

\bibitem{Couch:2016exn}
J.~Couch, W.~Fischler, and P.~H. Nguyen, {\it {Noether charge, black hole
  volume, and complexity}},  {\em JHEP} {\bf 03} (2017) 119,
  [\href{http://xxx.arxiv.org/abs/1610.02038}{{\tt arXiv:1610.02038}}].

\bibitem{Karch:2015rpa}
A.~Karch and B.~Robinson, {\it {Holographic Black Hole Chemistry}},  {\em JHEP}
  {\bf 12} (2015) 073, [\href{http://xxx.arxiv.org/abs/1510.02472}{{\tt
  arXiv:1510.02472}}].

\bibitem{Sinamuli:2017rhp}
M.~Sinamuli and R.~B. Mann, {\it {Higher Order Corrections to Holographic Black
  Hole Chemistry}},  {\em Phys. Rev. D} {\bf 96} (2017), no.~8 086008,
  [\href{http://xxx.arxiv.org/abs/1706.04259}{{\tt arXiv:1706.04259}}].

\bibitem{Visser:2021eqk}
M.~R. Visser, {\it {Holographic thermodynamics requires a chemical potential
  for color}},  {\em Phys. Rev. D} {\bf 105} (2022), no.~10 106014,
  [\href{http://xxx.arxiv.org/abs/2101.04145}{{\tt arXiv:2101.04145}}].

\bibitem{Mann:2024sru}
R.~B. Mann, {\it {Recent Developments in Holographic Black Hole Chemistry}},
  {\em JHAP} {\bf 4} (2024), no.~1 1--26,
  [\href{http://xxx.arxiv.org/abs/2403.02864}{{\tt arXiv:2403.02864}}].

\bibitem{Susskind:2014rva}
L.~Susskind, {\it {Computational Complexity and Black Hole Horizons}},  {\em
  Fortsch. Phys.} {\bf 64} (2016) 24--43,
  [\href{http://xxx.arxiv.org/abs/1403.5695}{{\tt arXiv:1403.5695}}].

\bibitem{Maldacena:2001kr}
J.~M. Maldacena, {\it {Eternal black holes in anti-de Sitter}},  {\em JHEP}
  {\bf 04} (2003) 021, [\href{http://xxx.arxiv.org/abs/hep-th/0106112}{{\tt
  hep-th/0106112}}].

\bibitem{Hartman:2013qma}
T.~Hartman and J.~Maldacena, {\it {Time Evolution of Entanglement Entropy from
  Black Hole Interiors}},  {\em JHEP} {\bf 05} (2013) 014,
  [\href{http://xxx.arxiv.org/abs/1303.1080}{{\tt arXiv:1303.1080}}].

\bibitem{Stanford:2014jda}
D.~Stanford and L.~Susskind, {\it {Complexity and Shock Wave Geometries}},
  {\em Phys. Rev.} {\bf D90} (2014), no.~12 126007,
  [\href{http://xxx.arxiv.org/abs/1406.2678}{{\tt arXiv:1406.2678}}].

\bibitem{Chapman:2017rqy}
S.~Chapman, M.~P. Heller, H.~Marrochio, and F.~Pastawski, {\it {Toward a
  Definition of Complexity for Quantum Field Theory States}},  {\em Phys. Rev.
  Lett.} {\bf 120} (2018), no.~12 121602,
  [\href{http://xxx.arxiv.org/abs/1707.08582}{{\tt arXiv:1707.08582}}].

\bibitem{Carmi:2017jqz}
D.~Carmi, S.~Chapman, H.~Marrochio, R.~C. Myers, and S.~Sugishita, {\it {On the
  Time Dependence of Holographic Complexity}},  {\em JHEP} {\bf 11} (2017) 188,
  [\href{http://xxx.arxiv.org/abs/1709.10184}{{\tt arXiv:1709.10184}}].

\bibitem{Chapman:2019clq}
S.~Chapman and H.~Z. Chen, {\it {Charged Complexity and the Thermofield Double
  State}},  {\em JHEP} {\bf 02} (2021) 187,
  [\href{http://xxx.arxiv.org/abs/1910.07508}{{\tt arXiv:1910.07508}}].

\bibitem{Andrews:2019hvq}
S.~Andrews, R.~A. Hennigar, and H.~K. Kunduri, {\it {Chemistry and complexity
  for solitons in AdS$_5$}},  {\em Class. Quant. Grav.} {\bf 37} (2020), no.~20
  204002, [\href{http://xxx.arxiv.org/abs/1912.07637}{{\tt arXiv:1912.07637}}].

\bibitem{AlBalushi:2020rqe}
A.~Al~Balushi, R.~A. Hennigar, H.~K. Kunduri, and R.~B. Mann, {\it {Holographic
  Complexity and Thermodynamic Volume}},  {\em Phys. Rev. Lett.} {\bf 126}
  (2021), no.~10 101601, [\href{http://xxx.arxiv.org/abs/2008.09138}{{\tt
  arXiv:2008.09138}}].

\bibitem{AlBalushi:2020heq}
A.~Al~Balushi, R.~A. Hennigar, H.~K. Kunduri, and R.~B. Mann, {\it {Holographic
  complexity of rotating black holes}},  {\em JHEP} {\bf 05} (2021) 226,
  [\href{http://xxx.arxiv.org/abs/2010.11203}{{\tt arXiv:2010.11203}}].

\bibitem{Bernamonti:2021jyu}
A.~Bernamonti, F.~Bigazzi, D.~Billo, L.~Faggi, and F.~Galli, {\it {Holographic
  and QFT complexity with angular momentum}},  {\em JHEP} {\bf 11} (2021) 037,
  [\href{http://xxx.arxiv.org/abs/2108.09281}{{\tt arXiv:2108.09281}}].

\bibitem{AlBalushi:2019obu}
A.~Al~Balushi and R.~B. Mann, {\it {Null hypersurfaces in
  Kerr\textendash{}(A)dS spacetimes}},  {\em Class. Quant. Grav.} {\bf 36}
  (2019), no.~24 245017, [\href{http://xxx.arxiv.org/abs/1909.06419}{{\tt
  arXiv:1909.06419}}].

\bibitem{Imseis:2020vsw}
M.~T.~N. Imseis, A.~Al~Balushi, and R.~B. Mann, {\it {Null hypersurfaces in
  Kerr\textendash{}Newman\textendash{}AdS black hole and super-entropic black
  hole spacetimes}},  {\em Class. Quant. Grav.} {\bf 38} (2021), no.~4 045018,
  [\href{http://xxx.arxiv.org/abs/2007.04354}{{\tt arXiv:2007.04354}}].

\bibitem{Sinamuli:2019utz}
M.~Sinamuli and R.~B. Mann, {\it {Holographic Complexity and Charged Scalar
  Fields}},  {\em Phys. Rev. D} {\bf 99} (2019), no.~10 106013,
  [\href{http://xxx.arxiv.org/abs/1902.01912}{{\tt arXiv:1902.01912}}].

\bibitem{Kunduri:2006qa}
H.~K. Kunduri, J.~Lucietti, and H.~S. Reall, {\it {Gravitational perturbations
  of higher dimensional rotating black holes: Tensor perturbations}},  {\em
  Phys. Rev. D} {\bf 74} (2006) 084021,
  [\href{http://xxx.arxiv.org/abs/hep-th/0606076}{{\tt hep-th/0606076}}].

\bibitem{Belin:2021bga}
A.~Belin, R.~C. Myers, S.-M. Ruan, G.~S\'arosi, and A.~J. Speranza, {\it {Does
  Complexity Equal Anything?}},  {\em Phys. Rev. Lett.} {\bf 128} (2022), no.~8
  081602, [\href{http://xxx.arxiv.org/abs/2111.02429}{{\tt arXiv:2111.02429}}].

\bibitem{Susskind:2014jwa}
L.~Susskind and Y.~Zhao, {\it {Switchbacks and the Bridge to Nowhere}},
  \href{http://xxx.arxiv.org/abs/1408.2823}{{\tt arXiv:1408.2823}}.

\bibitem{Roberts:2014isa}
D.~A. Roberts, D.~Stanford, and L.~Susskind, {\it {Localized shocks}},  {\em
  JHEP} {\bf 03} (2015) 051, [\href{http://xxx.arxiv.org/abs/1409.8180}{{\tt
  arXiv:1409.8180}}].

\bibitem{Belin:2022xmt}
A.~Belin, R.~C. Myers, S.-M. Ruan, G.~S\'arosi, and A.~J. Speranza, {\it
  {Complexity equals anything II}},  {\em JHEP} {\bf 01} (2023) 154,
  [\href{http://xxx.arxiv.org/abs/2210.09647}{{\tt arXiv:2210.09647}}].

\bibitem{Wang:2023eep}
M.-T. Wang, H.-Y. Jiang, and Y.-X. Liu, {\it {Generalized volume-complexity for
  RN-AdS black hole}},  {\em JHEP} {\bf 07} (2023) 178,
  [\href{http://xxx.arxiv.org/abs/2304.05751}{{\tt arXiv:2304.05751}}].

\bibitem{Jiang:2023jti}
H.-Y. Jiang, M.-T. Wang, and Y.-X. Liua, {\it {Holographic complexity and phase
  transition for AdS black holes}},  {\em Phys. Rev. D} {\bf 110} (2024), no.~4
  046013, [\href{http://xxx.arxiv.org/abs/2307.09223}{{\tt arXiv:2307.09223}}].

\bibitem{Zhang:2024mxb}
M.~Zhang, J.~Sun, and R.~B. Mann, {\it {Generalized holographic complexity of
  rotating black holes}},  {\em JHEP} {\bf 09} (2024) 050,
  [\href{http://xxx.arxiv.org/abs/2401.08571}{{\tt arXiv:2401.08571}}].

\bibitem{Gubser:1998bc}
S.~S. Gubser, I.~R. Klebanov, and A.~M. Polyakov, {\it {Gauge theory
  correlators from noncritical string theory}},  {\em Phys. Lett. B} {\bf 428}
  (1998) 105--114, [\href{http://xxx.arxiv.org/abs/hep-th/9802109}{{\tt
  hep-th/9802109}}].

\bibitem{Cong:2021fnf}
W.~Cong, D.~Kubiznak, and R.~B. Mann, {\it {Thermodynamics of AdS Black Holes:
  Critical Behavior of the Central Charge}},  {\em Phys. Rev. Lett.} {\bf 127}
  (2021), no.~9 091301, [\href{http://xxx.arxiv.org/abs/2105.02223}{{\tt
  arXiv:2105.02223}}].

\bibitem{Ahmed:2023snm}
M.~B. Ahmed, W.~Cong, D.~Kubiz\v{n}\'ak, R.~B. Mann, and M.~R. Visser, {\it
  {Holographic Dual of Extended Black Hole Thermodynamics}},  {\em Phys. Rev.
  Lett.} {\bf 130} (2023), no.~18 181401,
  [\href{http://xxx.arxiv.org/abs/2302.08163}{{\tt arXiv:2302.08163}}].

\bibitem{Cong:2021jgb}
W.~Cong, D.~Kubiznak, R.~B. Mann, and M.~R. Visser, {\it {Holographic CFT phase
  transitions and criticality for charged AdS black holes}},  {\em JHEP} {\bf
  08} (2022) 174, [\href{http://xxx.arxiv.org/abs/2112.14848}{{\tt
  arXiv:2112.14848}}].

\bibitem{Rafiee:2021hyj}
M.~Rafiee, S.~A.~H. Mansoori, S.-W. Wei, and R.~B. Mann, {\it {Universal
  criticality of thermodynamic geometry for boundary conformal field theories
  in gauge/gravity duality}},  \href{http://xxx.arxiv.org/abs/2107.08883}{{\tt
  arXiv:2107.08883}}.

\bibitem{Alfaia:2021cnk}
R.~B. Alfaia, I.~P. Lobo, and L.~C.~T. Brito, {\it {Central charge criticality
  of charged AdS black hole surrounded by different fluids}},  {\em Eur. Phys.
  J. Plus} {\bf 137} (2022), no.~3 402,
  [\href{http://xxx.arxiv.org/abs/2109.06599}{{\tt arXiv:2109.06599}}].

\bibitem{Kumar:2022fyq}
N.~Kumar, S.~Sen, and S.~Gangopadhyay, {\it {Phase transition structure and
  breaking of universal nature of central charge criticality in a Born-Infeld
  AdS black hole}},  {\em Phys. Rev. D} {\bf 106} (2022), no.~2 026005,
  [\href{http://xxx.arxiv.org/abs/2206.00440}{{\tt arXiv:2206.00440}}].

\bibitem{Dutta:2022wbh}
S.~Dutta and G.~S. Punia, {\it {String theory corrections to holographic black
  hole chemistry}},  {\em Phys. Rev. D} {\bf 106} (2022), no.~2 026003,
  [\href{http://xxx.arxiv.org/abs/2205.15593}{{\tt arXiv:2205.15593}}].

\bibitem{Lobo:2022eyr}
I.~P. Lobo, J.~a.~P. Morais~Gra\c{c}a, E.~Folco~Capossoli, and H.~Boschi-Filho,
  {\it {A varying gravitational constant map in asymptotically AdS black hole
  thermodynamics}},  {\em Phys. Lett. B} {\bf 835} (2022) 137559,
  [\href{http://xxx.arxiv.org/abs/2206.13664}{{\tt arXiv:2206.13664}}].

\bibitem{Kumar:2022afq}
N.~Kumar, S.~Sen, and S.~Gangopadhyay, {\it {Breaking of the universal nature
  of the central charge criticality in AdS black holes in Gauss-Bonnet
  gravity}},  {\em Phys. Rev. D} {\bf 107} (2023), no.~4 046005,
  [\href{http://xxx.arxiv.org/abs/2211.00925}{{\tt arXiv:2211.00925}}].

\bibitem{Qu:2022nrt}
Y.~Qu, J.~Tao, and H.~Yang, {\it {Thermodynamics and phase transition in
  central charge criticality of charged Gauss-Bonnet AdS black holes}},  {\em
  Nucl. Phys. B} {\bf 992} (2023) 116234,
  [\href{http://xxx.arxiv.org/abs/2211.08127}{{\tt arXiv:2211.08127}}].

\bibitem{Bai:2022vmx}
N.-C. Bai, L.~Song, and J.~Tao, {\it {Reentrant phase transition in holographic
  thermodynamicsof Born\textendash{}Infeld AdS black hole}},  {\em Eur. Phys.
  J. C} {\bf 84} (2024), no.~1 43,
  [\href{http://xxx.arxiv.org/abs/2212.04341}{{\tt arXiv:2212.04341}}].

\bibitem{Bai:2023wjm}
Y.-Y. Bai, X.-R. Chen, Z.-M. Xu, and B.~Wu, {\it {Revisiting the thermodynamics
  of the BTZ black hole with a variable gravitational constant}},  {\em Chin.
  Phys. C} {\bf 47} (2023), no.~11 115105.

\bibitem{Zhang:2023uay}
M.~Zhang and J.~Jiang, {\it {Bulk-boundary thermodynamic equivalence: a
  topology viewpoint}},  {\em JHEP} {\bf 06} (2023) 115,
  [\href{http://xxx.arxiv.org/abs/2303.17515}{{\tt arXiv:2303.17515}}].

\bibitem{Sadeghi:2023tuj}
J.~Sadeghi, M.~R. Alipour, S.~Noori~Gashti, and M.~A.~S. Afshar, {\it
  {Bulk-boundary and RPS thermodynamics from topology perspective}},  {\em
  Chin. Phys. C} {\bf 48} (2024), no.~9 095106,
  [\href{http://xxx.arxiv.org/abs/2306.16117}{{\tt arXiv:2306.16117}}].

\bibitem{Chen:2023pgs}
X.-R. Chen, B.~Wu, and Z.-M. Xu, {\it {Thermodynamics of black holes in massive
  gravity with holography}},  {\em Phys. Dark Univ.} {\bf 42} (2023) 101317.

\bibitem{Ladghami:2024wkv}
Y.~Ladghami and T.~Ouali, {\it {Black holes thermodynamics with CFT
  re-scaling}},  {\em Phys. Dark Univ.} {\bf 44} (2024) 101471,
  [\href{http://xxx.arxiv.org/abs/2402.15913}{{\tt arXiv:2402.15913}}].

\bibitem{Paul:2024rto}
S.~Paul, S.~Gangopadhyay, and A.~Saha, {\it {Gauss\textendash{}Bonnet AdS
  planar and spherical black hole thermodynamics and holography}},  {\em Class.
  Quant. Grav.} {\bf 41} (2024), no.~23 235010,
  [\href{http://xxx.arxiv.org/abs/2403.07543}{{\tt arXiv:2403.07543}}].

\bibitem{Cui:2024cnj}
H.-M. Cui and Z.-Y. Fan, {\it {Criticality of central charges for
  Gauss\textendash{}Bonnet black holes}},  {\em Eur. Phys. J. C} {\bf 84}
  (2024), no.~7 758, [\href{http://xxx.arxiv.org/abs/2404.05945}{{\tt
  arXiv:2404.05945}}].

\bibitem{Sadeghi:2024dnw}
J.~Sadeghi, S.~Noori~Gashti, M.~R. Alipour, and M.~A.~S. Afshar, {\it {Weak
  cosmic censorship and weak gravity conjectures in CFT thermodynamics}},  {\em
  JHEAp} {\bf 44} (2024) 482--493,
  [\href{http://xxx.arxiv.org/abs/2404.15981}{{\tt arXiv:2404.15981}}].

\bibitem{Zheng:2024glr}
H.~Zheng, Y.~Chen, and J.~Tang, {\it {Holographic thermodynamics of a charged
  AdS black hole with a global monopole}},  {\em Commun. Theor. Phys.} {\bf 77}
  (2025), no.~2 025403, [\href{http://xxx.arxiv.org/abs/2405.12227}{{\tt
  arXiv:2405.12227}}].

\bibitem{Baruah:2024yzw}
A.~Baruah and P.~Phukon, {\it {Holographic CFT thermodynamics of charged,
  rotating black holes in $D=4$ dimension}},
  \href{http://xxx.arxiv.org/abs/2407.02997}{{\tt arXiv:2407.02997}}.

\bibitem{TranNHung:2024tyg}
T.~N. Hung and C.~H. Nam, {\it {Topological equivalence and phase transition
  rate in holographic thermodynamics of regularized Maxwell theory}},  {\em
  Eur. Phys. J. C} {\bf 84} (2024), no.~8 870,
  [\href{http://xxx.arxiv.org/abs/2407.09122}{{\tt arXiv:2407.09122}}].

\bibitem{Zhang:2024jlp}
J.~Zhang, X.~Chen, and J.~Tao, {\it {Phase transition and central charge
  criticality of RN AdS black hole immersed in perfect fluid dark matter}},
  {\em Phys. Dark Univ.} {\bf 46} (2024) 101594.

\bibitem{Sadeghi:2024ish}
J.~Sadeghi, M.~R. Alipour, M.~A.~S. Afshar, and S.~Noori~Gashti, {\it
  {Exploring the phase transition in charged Gauss\textendash{}Bonnet black
  holes: a holographic thermodynamics perspectives}},  {\em Gen. Rel. Grav.}
  {\bf 56} (2024), no.~8 93, [\href{http://xxx.arxiv.org/abs/2408.03126}{{\tt
  arXiv:2408.03126}}].

\bibitem{Ahmed:2023dnh}
M.~B. Ahmed, W.~Cong, D.~Kubiznak, R.~B. Mann, and M.~R. Visser, {\it
  {Holographic CFT phase transitions and criticality for rotating AdS black
  holes}},  {\em JHEP} {\bf 08} (2023) 142,
  [\href{http://xxx.arxiv.org/abs/2305.03161}{{\tt arXiv:2305.03161}}].

\bibitem{Gong:2023ywu}
T.-F. Gong, J.~Jiang, and M.~Zhang, {\it {Holographic thermodynamics of
  rotating black holes}},  {\em JHEP} {\bf 06} (2023) 105,
  [\href{http://xxx.arxiv.org/abs/2305.00267}{{\tt arXiv:2305.00267}}].

\bibitem{Zeyuan:2021uol}
G.~Zeyuan and L.~Zhao, {\it {Restricted phase space thermodynamics for AdS
  black holes via holography}},  {\em Class. Quant. Grav.} {\bf 39} (2022),
  no.~7 075019, [\href{http://xxx.arxiv.org/abs/2112.02386}{{\tt
  arXiv:2112.02386}}].

\bibitem{Bousder:2023zhl}
M.~Bousder, E.~Salmani, and H.~Ez-Zahraouy, {\it {Entropy as logarithmic term
  of the central charge and modified Friedmann equation in AdS/CFT
  correspondence}},  {\em JHEAp} {\bf 38} (2023) 49--57,
  [\href{http://xxx.arxiv.org/abs/2306.16426}{{\tt arXiv:2306.16426}}].

\bibitem{Feng:2024uia}
Y.~Feng, H.~Ma, R.~B. Mann, Y.~Xue, and M.~Zhang, {\it {Quantum charged black
  holes}},  {\em JHEP} {\bf 08} (2024) 184,
  [\href{http://xxx.arxiv.org/abs/2404.07192}{{\tt arXiv:2404.07192}}].

\bibitem{Climent:2024nuj}
A.~Climent, R.~Emparan, and R.~A. Hennigar, {\it {Chemical potential and charge
  in quantum black holes}},  {\em JHEP} {\bf 08} (2024) 150,
  [\href{http://xxx.arxiv.org/abs/2404.15148}{{\tt arXiv:2404.15148}}].

\bibitem{Frassino:2022zaz}
A.~M. Frassino, J.~F. Pedraza, A.~Svesko, and M.~R. Visser, {\it
  {Higher-Dimensional Origin of Extended Black Hole Thermodynamics}},  {\em
  Phys. Rev. Lett.} {\bf 130} (2023), no.~16 161501,
  [\href{http://xxx.arxiv.org/abs/2212.14055}{{\tt arXiv:2212.14055}}].

\bibitem{Emparan:2020znc}
R.~Emparan, A.~M. Frassino, and B.~Way, {\it {Quantum BTZ black hole}},  {\em
  JHEP} {\bf 11} (2020) 137, [\href{http://xxx.arxiv.org/abs/2007.15999}{{\tt
  arXiv:2007.15999}}].

\bibitem{Panella:2023lsi}
E.~Panella and A.~Svesko, {\it {Quantum Kerr-de Sitter black holes in three
  dimensions}},  {\em JHEP} {\bf 06} (2023) 127,
  [\href{http://xxx.arxiv.org/abs/2303.08845}{{\tt arXiv:2303.08845}}].

\bibitem{Punia:2023ilo}
G.~S. Punia, {\it {Bulk-boundary thermodynamics of charged black holes in
  higher-derivative theory}},  {\em Eur. Phys. J. C} {\bf 84} (2024), no.~5
  467, [\href{http://xxx.arxiv.org/abs/2305.06552}{{\tt arXiv:2305.06552}}].

\bibitem{Johnson:2023dtf}
C.~V. Johnson and R.~Nazario, {\it {Specific heats for quantum BTZ black holes
  in extended thermodynamics}},  {\em Phys. Rev. D} {\bf 110} (2024), no.~10
  106004, [\href{http://xxx.arxiv.org/abs/2310.12212}{{\tt arXiv:2310.12212}}].

\bibitem{Frassino:2023wpc}
A.~M. Frassino, J.~F. Pedraza, A.~Svesko, and M.~R. Visser, {\it {Reentrant
  phase transitions of quantum black holes}},  {\em Phys. Rev. D} {\bf 109}
  (2024), no.~12 124040, [\href{http://xxx.arxiv.org/abs/2310.12220}{{\tt
  arXiv:2310.12220}}].

\bibitem{HosseiniMansoori:2024bfi}
S.~A. Hosseini~Mansoori, J.~F. Pedraza, and M.~Rafiee, {\it {Criticality and
  thermodynamic geometry of quantum BTZ black holes}},
  \href{http://xxx.arxiv.org/abs/2403.13063}{{\tt arXiv:2403.13063}}.

\bibitem{Wu:2024txe}
S.-P. Wu and S.-W. Wei, {\it {Thermodynamical topology of quantum BTZ black
  hole}},  {\em Phys. Rev. D} {\bf 110} (2024), no.~2 024054,
  [\href{http://xxx.arxiv.org/abs/2403.14167}{{\tt arXiv:2403.14167}}].

\bibitem{Frassino:2024bjg}
A.~M. Frassino, R.~A. Hennigar, J.~F. Pedraza, and A.~Svesko, {\it {Quantum
  Inequalities for Quantum Black Holes}},  {\em Phys. Rev. Lett.} {\bf 133}
  (2024), no.~18 181501, [\href{http://xxx.arxiv.org/abs/2406.17860}{{\tt
  arXiv:2406.17860}}].

\bibitem{Padmanabhan:2015zmr}
T.~Padmanabhan, {\it {Gravity and/is Thermodynamics}},  {\em Curr. Sci.} {\bf
  109} (2015) 2236--2242, [\href{http://xxx.arxiv.org/abs/1512.06546}{{\tt
  arXiv:1512.06546}}].

\bibitem{Ruppeiner}
G.~Ruppeiner, {\it Thermodynamics: A riemannian geometric model},  {\em Phys.
  Rev. A} {\bf 20} (Oct, 1979) 1608--1613.

\bibitem{Ruppeiner:1995zz}
G.~Ruppeiner, {\it {Riemannian geometry in thermodynamic fluctuation theory}},
  {\em Rev. Mod. Phys.} {\bf 67} (1995) 605--659. [Erratum: Rev.Mod.Phys. 68,
  313--313 (1996)].

\bibitem{HiroshiOshima_1999}
H.~Oshima, T.~Obata, and H.~Hara, {\it Riemann scalar curvature of ideal
  quantum gases obeying gentile's statistics},  {\em Journal of Physics A:
  Mathematical and General} {\bf 32} (sep, 1999) 6373.

\bibitem{Landau:1980mil}
L.~D. Landau and E.~M. Lifshitz, {\em {Statistical Physics, Part 1}}, vol.~5 of
  {\em Course of Theoretical Physics}.
\newblock Butterworth-Heinemann, Oxford, 1980.

\bibitem{Wei:2019yvs}
S.-W. Wei, Y.-X. Liu, and R.~B. Mann, {\it {Ruppeiner Geometry, Phase
  Transitions, and the Microstructure of Charged AdS Black Holes}},  {\em Phys.
  Rev. D} {\bf 100} (2019), no.~12 124033,
  [\href{http://xxx.arxiv.org/abs/1909.03887}{{\tt arXiv:1909.03887}}].

\bibitem{Wei:2019uqg}
S.-W. Wei, Y.-X. Liu, and R.~B. Mann, {\it {Repulsive Interactions and
  Universal Properties of Charged Anti\textendash{}de Sitter Black Hole
  Microstructures}},  {\em Phys. Rev. Lett.} {\bf 123} (2019), no.~7 071103,
  [\href{http://xxx.arxiv.org/abs/1906.10840}{{\tt arXiv:1906.10840}}].

\bibitem{Wei:2019ctz}
S.-W. Wei and Y.-X. Liu, {\it {Intriguing microstructures of five-dimensional
  neutral Gauss-Bonnet AdS black hole}},  {\em Phys. Lett. B} {\bf 803} (2020)
  135287, [\href{http://xxx.arxiv.org/abs/1910.04528}{{\tt arXiv:1910.04528}}].

\bibitem{NaveenaKumara:2020hov}
A.~Naveena~Kumara, C.~L.~A. Rizwan, K.~Hegde, M.~S. Ali, and A.~K. M, {\it
  {Microstructure of five-dimensional neutral Gauss\textendash{}Bonnet black
  hole in anti-de Sitter spacetime via $P-V$ criticality}},  {\em Gen. Rel.
  Grav.} {\bf 55} (2023), no.~1 4,
  [\href{http://xxx.arxiv.org/abs/2006.13907}{{\tt arXiv:2006.13907}}].

\bibitem{Zhou:2020vzf}
R.~Zhou, Y.-X. Liu, and S.-W. Wei, {\it {Phase transition and microstructures
  of five-dimensional charged Gauss-Bonnet-AdS black holes in the grand
  canonical ensemble}},  {\em Phys. Rev. D} {\bf 102} (2020), no.~12 124015,
  [\href{http://xxx.arxiv.org/abs/2008.08301}{{\tt arXiv:2008.08301}}].

\bibitem{Wei:2021pql}
S.-W. Wei and Y.-X. Liu, {\it {Testing the microstructure of d-dimensional
  charged Gauss-Bonnet anti\textendash{}de Sitter black holes}},  {\em Phys.
  Rev. D} {\bf 104} (2021), no.~2 024062,
  [\href{http://xxx.arxiv.org/abs/2105.01295}{{\tt arXiv:2105.01295}}].

\bibitem{NaveenaKumara:2020biu}
A.~Naveena~Kumara, C.~L. Ahmed~Rizwan, K.~Hegde, M.~S. Ali, and K.~M. Ajith,
  {\it {Ruppeiner geometry, reentrant phase transition, and microstructure of
  Born-Infeld AdS black hole}},  {\em Phys. Rev. D} {\bf 103} (2021), no.~4
  044025, [\href{http://xxx.arxiv.org/abs/2007.07861}{{\tt arXiv:2007.07861}}].

\bibitem{Ye:2022uuj}
X.~Ye, Z.-Q. Chen, M.-D. Li, and S.-W. Wei, {\it {QED effects on phase
  transition and Ruppeiner geometry of Euler-Heisenberg-AdS black holes*}},
  {\em Chin. Phys. C} {\bf 46} (2022), no.~11 115102,
  [\href{http://xxx.arxiv.org/abs/2202.09053}{{\tt arXiv:2202.09053}}].

\bibitem{Mahish:2020gwg}
S.~Mahish, A.~Ghosh, and C.~Bhamidipati, {\it {Thermodynamic curvature of the
  Schwarzschild-AdS black hole and Bose condensation}},  {\em Phys. Lett. B}
  {\bf 811} (2020) 135958, [\href{http://xxx.arxiv.org/abs/2006.02943}{{\tt
  arXiv:2006.02943}}].

\bibitem{Saavedra:2023lds}
J.~F. Saavedra and F.~Tello-Ortiz, {\it {Einstein-Gauss-Bonnet gravity: Phase
  transitions and micro-structure in an FLRW background}},
  \href{http://xxx.arxiv.org/abs/2311.14047}{{\tt arXiv:2311.14047}}.

\bibitem{Abdusattar:2023pck}
H.~Abdusattar, {\it {Insight into the Microstructure of FRW Universe from a P-V
  Phase Transition}},  {\em JHEP} {\bf 09} (2023) 147,
  [\href{http://xxx.arxiv.org/abs/2304.08348}{{\tt arXiv:2304.08348}}].

\bibitem{Feng:2024zor}
Z.-W. Feng, S.-Y. Li, X.~Zhou, and H.~Abdusattar, {\it {Phase transitions,
  critical behavior and microstructure of the FRW universe in the framework of
  higher order GUP}},  {\em Phys. Dark Univ.} {\bf 46} (2024) 101719,
  [\href{http://xxx.arxiv.org/abs/2404.17624}{{\tt arXiv:2404.17624}}].

\bibitem{Xu:2019nnp}
Z.-M. Xu, B.~Wu, and W.-L. Yang, {\it {Fine micro-thermal structures for
  Reissner-Nordstr\"om black hole}},  {\em Chin. Phys. C} {\bf 44} (2020),
  no.~9 095106, [\href{http://xxx.arxiv.org/abs/1910.03378}{{\tt
  arXiv:1910.03378}}].

\bibitem{Guo:2019hxa}
X.-Y. Guo, H.-F. Li, L.-C. Zhang, and R.~Zhao, {\it {Continuous Phase
  Transition and Microstructure of Charged AdS Black Hole with Quintessence}},
  {\em Eur. Phys. J. C} {\bf 80} (2020), no.~2 168,
  [\href{http://xxx.arxiv.org/abs/1911.09902}{{\tt arXiv:1911.09902}}].

\bibitem{Xu:2020gud}
Z.-M. Xu, B.~Wu, and W.-L. Yang, {\it {Ruppeiner thermodynamic geometry for the
  Schwarzschild-AdS black hole}},  {\em Phys. Rev. D} {\bf 101} (2020), no.~2
  024018, [\href{http://xxx.arxiv.org/abs/1910.12182}{{\tt arXiv:1910.12182}}].

\bibitem{Singh:2020tkf}
A.~Singh, A.~Ghosh, and C.~Bhamidipati, {\it {Thermodynamic curvature of AdS
  black holes with dark energy}},  {\em Front. in Phys.} {\bf 9} (2021) 65,
  [\href{http://xxx.arxiv.org/abs/2002.08787}{{\tt arXiv:2002.08787}}].

\bibitem{Ghosh:2020kba}
A.~Ghosh and C.~Bhamidipati, {\it {Thermodynamic geometry and interacting
  microstructures of BTZ black holes}},  {\em Phys. Rev. D} {\bf 101} (2020),
  no.~10 106007, [\href{http://xxx.arxiv.org/abs/2001.10510}{{\tt
  arXiv:2001.10510}}].

\bibitem{NaveenaKumara:2020lgq}
A.~Naveena~Kumara, C.~L.~A. Rizwan, K.~Hegde, M.~S. Ali, and A.~K. M., {\it
  {Microstructure and continuous phase transition of a regular Hayward black
  hole in anti-de Sitter spacetime}},  {\em PTEP} {\bf 2021} (2021), no.~7
  073E01, [\href{http://xxx.arxiv.org/abs/2003.00889}{{\tt arXiv:2003.00889}}].

\bibitem{Xu:2020iic}
Z.-M. Xu, B.~Wu, T.~Yang, and W.-L. Yang, {\it {A new measure of thermal
  micro-behavior for the AdS black hole}},  {\em Chin. Phys. C} {\bf 45}
  (2021), no.~1 015106, [\href{http://xxx.arxiv.org/abs/2005.04219}{{\tt
  arXiv:2005.04219}}].

\bibitem{Dehyadegari:2020ebz}
A.~Dehyadegari, A.~Sheykhi, and S.-W. Wei, {\it {Microstructure of charged AdS
  black hole via $P-V$ criticality}},  {\em Phys. Rev. D} {\bf 102} (2020),
  no.~10 104013, [\href{http://xxx.arxiv.org/abs/2006.12265}{{\tt
  arXiv:2006.12265}}].

\bibitem{Yerra:2020oph}
P.~K. Yerra and C.~Bhamidipati, {\it {Ruppeiner Geometry, Phase Transitions and
  Microstructures of Black Holes in Massive Gravity}},  {\em Int. J. Mod. Phys.
  A} {\bf 35} (2020), no.~22 2050120,
  [\href{http://xxx.arxiv.org/abs/2006.07775}{{\tt arXiv:2006.07775}}].

\bibitem{Rizwan:2020bhp}
C.~L.~A. Rizwan, A.~Naveena~Kumara, K.~Hegde, and D.~Vaid, {\it {Coexistent
  Physics and Microstructure of the Regular Bardeen Black Hole in Anti-de
  Sitter Spacetime}},  {\em Annals Phys.} {\bf 422} (2020) 168320,
  [\href{http://xxx.arxiv.org/abs/2008.06472}{{\tt arXiv:2008.06472}}].

\bibitem{Wu:2020fij}
B.~Wu, C.~Wang, Z.-M. Xu, and W.-L. Yang, {\it {Ruppeiner geometry and
  thermodynamic phase transition of the black hole in massive gravity}},  {\em
  Eur. Phys. J. C} {\bf 81} (2021), no.~7 626,
  [\href{http://xxx.arxiv.org/abs/2006.09021}{{\tt arXiv:2006.09021}}].

\bibitem{Yerra:2020tzg}
P.~K. Yerra and C.~Bhamidipati, {\it {Ruppeiner curvature along a
  renormalization group flow}},  {\em Phys. Lett. B} {\bf 819} (2021) 136450,
  [\href{http://xxx.arxiv.org/abs/2007.11515}{{\tt arXiv:2007.11515}}].

\bibitem{Wang:2022bqu}
C.~Wang, S.~P. Yin, Z.~M. Xu, B.~Wu, and W.~L. Yang, {\it {Ruppeiner geometry
  and the fluctuation of the RN-AdS black hole in framework of the extensive
  thermodynamics}},  {\em Nucl. Phys. B} {\bf 998} (2024) 116426,
  [\href{http://xxx.arxiv.org/abs/2210.08822}{{\tt arXiv:2210.08822}}].

\bibitem{Dutta:2021whz}
S.~Dutta and G.~S. Punia, {\it {Interactions between AdS black hole
  molecules}},  {\em Phys. Rev. D} {\bf 104} (2021), no.~12 126009,
  [\href{http://xxx.arxiv.org/abs/2108.06135}{{\tt arXiv:2108.06135}}].

\bibitem{Ruppeiner:2023wkq}
G.~Ruppeiner and A.-M. Sturzu, {\it {Black hole microstructures in the extremal
  limit}},  {\em Phys. Rev. D} {\bf 108} (2023), no.~8 086004,
  [\href{http://xxx.arxiv.org/abs/2304.06187}{{\tt arXiv:2304.06187}}].

\bibitem{Liu:2022nzo}
C.~Liu and J.~Wang, {\it {The radial distribution function reveals the
  underlying mesostructure of the AdS black hole}},  {\em JHEP} {\bf 10} (2022)
  171, [\href{http://xxx.arxiv.org/abs/2207.02011}{{\tt arXiv:2207.02011}}].

\bibitem{Strominger:2001pn}
A.~Strominger, {\it {The dS / CFT correspondence}},  {\em JHEP} {\bf 10} (2001)
  034, [\href{http://xxx.arxiv.org/abs/hep-th/0106113}{{\tt hep-th/0106113}}].

\bibitem{Sekiwa:2006qj}
Y.~Sekiwa, {\it {Thermodynamics of de Sitter black holes: Thermal cosmological
  constant}},  {\em Phys. Rev. D} {\bf 73} (2006) 084009,
  [\href{http://xxx.arxiv.org/abs/hep-th/0602269}{{\tt hep-th/0602269}}].

\bibitem{Romans1992}
L.~J. Romans, {\it Supersymmetric, cold and lukewarm black holes in
  cosmological einstein-maxwell theory},  {\em Nucl.Phys. B} {\bf 383} (1992)
  395--415, [\href{http://xxx.arxiv.org/abs/hep-th/9203018}{{\tt
  hep-th/9203018}}].

\bibitem{Mbarek:2018bau}
S.~Mbarek and R.~B. Mann, {\it {Reverse Hawking-Page Phase Transition in de
  Sitter Black Holes}},  {\em JHEP} {\bf 02} (2019) 103,
  [\href{http://xxx.arxiv.org/abs/1808.03349}{{\tt arXiv:1808.03349}}].

\bibitem{Kubiznak:2015bya}
D.~Kubiznak and F.~Simovic, {\it {Thermodynamics of horizons: de Sitter black
  holes}},  \href{http://xxx.arxiv.org/abs/1507.08630}{{\tt arXiv:1507.08630}}.

\bibitem{Urano:2009xn}
M.~Urano, A.~Tomimatsu, and H.~Saida, {\it {Mechanical First Law of Black Hole
  Spacetimes with Cosmological Constant and Its Application to Schwarzschild-de
  Sitter Spacetime}},  {\em Class. Quant. Grav.} {\bf 26} (2009) 105010,
  [\href{http://xxx.arxiv.org/abs/0903.4230}{{\tt arXiv:0903.4230}}].

\bibitem{Carlip2003}
S.~Carlip and S.~Vaidya, {\it Phase transitions and critical behavior for
  charged black holes},  {\em Class.Quant.Grav.} {\bf 20} (2003) 3827--3838,
  [\href{http://xxx.arxiv.org/abs/gr-qc/0306054}{{\tt gr-qc/0306054}}].

\bibitem{Simovic:2018tdy}
F.~Simovic and R.~B. Mann, {\it {Critical Phenomena of Charged de Sitter Black
  Holes in Cavities}},  {\em Class. Quant. Grav.} {\bf 36} (2019), no.~1
  014002, [\href{http://xxx.arxiv.org/abs/1807.11875}{{\tt arXiv:1807.11875}}].

\bibitem{Braden:1990hw}
H.~W. Braden, J.~D. Brown, B.~F. Whiting, and J.~W. York, Jr., {\it {Charged
  black hole in a grand canonical ensemble}},  {\em Phys. Rev. D} {\bf 42}
  (1990) 3376--3385.

\bibitem{Haroon:2020vpr}
S.~Haroon, R.~A. Hennigar, R.~B. Mann, and F.~Simovic, {\it {Thermodynamics of
  Gauss-Bonnet-de Sitter Black Holes}},  {\em Phys. Rev. D} {\bf 101} (2020)
  084051, [\href{http://xxx.arxiv.org/abs/2002.01567}{{\tt arXiv:2002.01567}}].

\bibitem{Marks:2021fpe}
G.~A. Marks, F.~Simovic, and R.~B. Mann, {\it {Phase transitions in 4D
  Gauss\textendash{}Bonnet\textendash{}de Sitter black holes}},  {\em Phys.
  Rev. D} {\bf 104} (2021), no.~10 104056,
  [\href{http://xxx.arxiv.org/abs/2107.11352}{{\tt arXiv:2107.11352}}].

\bibitem{Simovic:2020dke}
F.~Simovic, D.~Fusco, and R.~B. Mann, {\it {Thermodynamics of de Sitter Black
  Holes with Conformally Coupled Scalar Fields}},
  \href{http://xxx.arxiv.org/abs/2008.07593}{{\tt arXiv:2008.07593}}.

\bibitem{Li:2023sig}
X.-P. Li, L.-C. Zhang, Y.-B. Ma, and H.-F. Li, {\it {Thermodynamic quantities
  and phase transitions of five-dimensional de Sitter hairy spacetime}},  {\em
  Chin. Phys. C} {\bf 47} (2023), no.~10 105102.

\bibitem{Simovic:2019zgb}
F.~Simovic and R.~B. Mann, {\it {Critical Phenomena of Born-Infeld-de Sitter
  Black Holes in Cavities}},  {\em JHEP} {\bf 05} (2019) 136,
  [\href{http://xxx.arxiv.org/abs/1904.04871}{{\tt arXiv:1904.04871}}].

\bibitem{Du:2021dxp}
Y.-Z. Du, H.-F. Li, and L.-C. Zhang, {\it {Continuous phase transition of
  higher-dimensional de-Sitter spacetime with non-linear source}},  {\em Eur.
  Phys. J. C} {\bf 82} (2022), no.~4 370,
  [\href{http://xxx.arxiv.org/abs/2104.10309}{{\tt arXiv:2104.10309}}].

\bibitem{ElMoumni:2021woq}
H.~El~Moumni and J.~Khalloufi, {\it {Nonlinear-Maxwell-Yukawa de-Sitter black
  hole thermodynamics in a cavity: I\ensuremath{-}Canonical ensemble}},  {\em
  Nucl. Phys. B} {\bf 973} (2021) 115593.

\bibitem{Tannukij:2020njz}
L.~Tannukij, P.~Wongjun, E.~Hirunsirisawat, T.~Deesuwan, and C.~Promsiri, {\it
  {Thermodynamics and phase transition of spherically symmetric black hole in
  de Sitter space from R\'enyi statistics}},  {\em Eur. Phys. J. Plus} {\bf
  135} (2020), no.~6 500, [\href{http://xxx.arxiv.org/abs/2002.00377}{{\tt
  arXiv:2002.00377}}].

\bibitem{Du:2022jcb}
Y.-Z. Du, H.-F. Li, and R.~Zhao, {\it {Overview of thermodynamical properties
  for Reissner\textendash{}Nordstr\"om\textendash{}de Sitter spacetime in
  induced phase space}},  {\em Eur. Phys. J. C} {\bf 82} (2022) 850,
  [\href{http://xxx.arxiv.org/abs/2207.03126}{{\tt arXiv:2207.03126}}].

\bibitem{Zhang:2020odg}
Y.~Zhang, W.-q. Wang, Y.-b. Ma, and J.~Wang, {\it {Phase Transition and Entropy
  Force between Two Horizons in ($n+2$)-Dimensional de Sitter Space}},  {\em
  Adv. High Energy Phys.} {\bf 2020} (2020) 7263059,
  [\href{http://xxx.arxiv.org/abs/2004.06796}{{\tt arXiv:2004.06796}}].

\bibitem{Ma:2020aab}
Y.~Ma, Y.~Zhang, L.~Zhang, L.~Wu, Y.~Gao, S.~Cao, and Y.~Pan, {\it {Phase
  transition and entropic force of de Sitter black hole in massive gravity}},
  {\em Eur. Phys. J. C} {\bf 81} (2021), no.~1 42,
  [\href{http://xxx.arxiv.org/abs/2009.12726}{{\tt arXiv:2009.12726}}].

\bibitem{Ma:2022vwt}
Y.~Ma, Y.~Zhang, L.~Zhang, and Y.~Pan, {\it {Extended thermodynamics and
  entropic force of de Sitter space\textendash{}time with charged
  Gauss\textendash{}Bonnet Black Hole}},  {\em Chin. J. Phys.} {\bf 77} (2022)
  1854--1862.

\bibitem{Gregory:2017sor}
R.~Gregory, D.~Kastor, and J.~Traschen, {\it {Black Hole Thermodynamics with
  Dynamical Lambda}},  {\em JHEP} {\bf 10} (2017) 118,
  [\href{http://xxx.arxiv.org/abs/1707.06586}{{\tt arXiv:1707.06586}}].

\bibitem{Gregory:2018ghc}
R.~Gregory, D.~Kastor, and J.~Traschen, {\it {Evolving Black Holes in
  Inflation}},  {\em Class. Quant. Grav.} {\bf 35} (2018), no.~15 155008,
  [\href{http://xxx.arxiv.org/abs/1804.03462}{{\tt arXiv:1804.03462}}].

\bibitem{Beyen:2023rca}
A.~Beyen, E.~Hamamc\i{}, K.~Meerts, and D.~Van~den Bleeken, {\it {Dynamical de
  Sitter black holes in a quasi-stationary expansion}},  {\em Class. Quant.
  Grav.} {\bf 41} (2024), no.~9 095012,
  [\href{http://xxx.arxiv.org/abs/2310.08127}{{\tt arXiv:2310.08127}}].

\bibitem{Li:2020khm}
R.~Li and J.~Wang, {\it {Thermodynamics and kinetics of Hawking-Page phase
  transition}},  {\em Phys. Rev. D} {\bf 102} (2020), no.~2 024085.

\bibitem{Zwanzig}
R.~Zwanzig, {\em {Nonequilibrium Statistical Mechanics}}.
\newblock Oxford University Press, 2001.

\bibitem{Dai:2022mko}
H.~Dai, Z.~Zhao, and S.~Zhang, {\it {Thermodynamic phase transition of
  Euler-Heisenberg-AdS black hole on free energy landscape}},  {\em Nucl. Phys.
  B} {\bf 991} (2023) 116219, [\href{http://xxx.arxiv.org/abs/2202.14007}{{\tt
  arXiv:2202.14007}}].

\bibitem{Ali:2023wkq}
M.~S. Ali, H.~El~Moumni, J.~Khalloufi, and K.~Masmar, {\it {Born-Infeld-AdS
  black hole phase structure: Landau theory and free energy landscape
  approaches}},  \href{http://xxx.arxiv.org/abs/2303.11711}{{\tt
  arXiv:2303.11711}}.

\bibitem{Liu:2022sot}
F.~Liu, Y.-Z. Du, R.~Zhao, and H.-F. Li, {\it {Phase equilibrium and
  microstructure of topological AdS black holes in massive gravity *}},  {\em
  Chin. Phys. C} {\bf 46} (2022), no.~8 085102.

\bibitem{Luo:2022gss}
Z.~Luo, H.~Yu, and J.~Li, {\it {Effects of a global monopole on the
  thermodynamic phase transition of a charged AdS black hole*}},  {\em Chin.
  Phys. C} {\bf 46} (2022), no.~12 125101,
  [\href{http://xxx.arxiv.org/abs/2206.09729}{{\tt arXiv:2206.09729}}].

\bibitem{Ma:2024ysf}
C.~Ma, P.-P. Zhang, B.~Wu, and Z.-M. Xu, {\it {The Kramers escape rate of phase
  transitions for the 6-dimensional Gauss-Bonnet AdS black hole with triple
  phases}},  \href{http://xxx.arxiv.org/abs/2407.20512}{{\tt
  arXiv:2407.20512}}.

\bibitem{Li:2022yti}
R.~Li and J.~Wang, {\it {Non-Markovian dynamics of black hole phase
  transition}},  {\em Phys. Rev. D} {\bf 106} (2022), no.~10 104039,
  [\href{http://xxx.arxiv.org/abs/2205.00594}{{\tt arXiv:2205.00594}}].

\bibitem{Li:2024hje}
R.~Li, C.~Liu, and J.~Wang, {\it {Phase space path integral approach to the
  kinetics of black hole phase transition}},  {\em Phys. Rev. D} {\bf 110}
  (2024), no.~2 024079, [\href{http://xxx.arxiv.org/abs/2401.02260}{{\tt
  arXiv:2401.02260}}].

\bibitem{Wei:2021vdx}
S.-W. Wei and Y.-X. Liu, {\it {Topology of black hole thermodynamics}},  {\em
  Phys. Rev. D} {\bf 105} (2022), no.~10 104003,
  [\href{http://xxx.arxiv.org/abs/2112.01706}{{\tt arXiv:2112.01706}}].

\bibitem{Wei:2022dzw}
S.-W. Wei, Y.-X. Liu, and R.~B. Mann, {\it {Black Hole Solutions as Topological
  Thermodynamic Defects}},  {\em Phys. Rev. Lett.} {\bf 129} (2022), no.~19
  191101, [\href{http://xxx.arxiv.org/abs/2208.01932}{{\tt arXiv:2208.01932}}].

\bibitem{Cunha:2020azh}
P.~V.~P. Cunha and C.~A.~R. Herdeiro, {\it {Stationary black holes and light
  rings}},  {\em Phys. Rev. Lett.} {\bf 124} (2020), no.~18 181101,
  [\href{http://xxx.arxiv.org/abs/2003.06445}{{\tt arXiv:2003.06445}}].

\bibitem{Duan:1979ucg}
Y.-S. Duan and M.-L. Ge, {\it {SU(2) Gauge Theory and Electrodynamics with N
  Magnetic Monopoles}},  {\em Sci. Sin.} {\bf 9} (1979), no.~11.

\bibitem{Duan:1984ws}
Y.~Duan, {\it {THE STRUCTURE OF THE TOPOLOGICAL CURRENT}}, .

\bibitem{Wei:2020rbh}
S.-W. Wei, {\it {Topological Charge and Black Hole Photon Spheres}},  {\em
  Phys. Rev. D} {\bf 102} (2020), no.~6 064039,
  [\href{http://xxx.arxiv.org/abs/2006.02112}{{\tt arXiv:2006.02112}}].

\bibitem{Schouton:1951}
J.~Schouton, {\em {Tensor analysis for Physicists}}.
\newblock Claredon, Oxford, 1951.

\bibitem{Fu:2000pb}
L.-B. Fu, Y.-S. Duan, and H.~Zhang, {\it {Evolution of the Chern-Simons
  vortices}},  {\em Phys. Rev. D} {\bf 61} (2000) 045004,
  [\href{http://xxx.arxiv.org/abs/hep-th/0112033}{{\tt hep-th/0112033}}].

\bibitem{Wang:2024fhf}
H.~Wang and Y.-Z. Du, {\it {Topology of charged AdS black hole in restricted
  phase space*}},  {\em Chin. Phys. C} {\bf 48} (2024), no.~9 095109,
  [\href{http://xxx.arxiv.org/abs/2406.08793}{{\tt arXiv:2406.08793}}].

\bibitem{Chen:2024kmy}
H.~Chen, M.-Y. Zhang, H.~Hassanabadi, B.~C. L\"utf\"uo\u{g}lu, and Z.-W. Long,
  {\it {Topology of dyonic AdS black holes with quasitopological
  electromagnetism in Einstein-Gauss-Bonnet gravity}},
  \href{http://xxx.arxiv.org/abs/2403.14730}{{\tt arXiv:2403.14730}}.

\bibitem{Bhattacharya:2024bjp}
K.~Bhattacharya, K.~Bamba, and D.~Singleton, {\it {Topological interpretation
  of extremal and Davies-type phase transitions of black holes}},  {\em Phys.
  Lett. B} {\bf 854} (2024) 138722,
  [\href{http://xxx.arxiv.org/abs/2402.18791}{{\tt arXiv:2402.18791}}].

\bibitem{Zhu:2024jhw}
X.-D. Zhu, D.~Wu, and D.~Wen, {\it {Topological classes of thermodynamics of
  the rotating charged AdS black holes in gauged supergravities}},  {\em Phys.
  Lett. B} {\bf 856} (2024) 138919,
  [\href{http://xxx.arxiv.org/abs/2402.15531}{{\tt arXiv:2402.15531}}].

\bibitem{Wu:2024rmv}
D.~Wu, S.-Y. Gu, X.-D. Zhu, Q.-Q. Jiang, and S.-Z. Yang, {\it {Topological
  classes of thermodynamics of the static multi-charge AdS black holes in
  gauged supergravities: novel temperature-dependent thermodynamic topological
  phase transition}},  {\em JHEP} {\bf 06} (2024) 213,
  [\href{http://xxx.arxiv.org/abs/2402.00106}{{\tt arXiv:2402.00106}}].

\bibitem{Zhang:2023svu}
M.-Y. Zhang, H.~Chen, H.~Hassanabadi, Z.-W. Long, and H.~Yang, {\it
  {Thermodynamic topology of Kerr-Sen black holes via R\'enyi statistics}},
  {\em Phys. Lett. B} {\bf 856} (2024) 138885,
  [\href{http://xxx.arxiv.org/abs/2312.12814}{{\tt arXiv:2312.12814}}].

\bibitem{Sadeghi:2023dsg}
J.~Sadeghi, M.~A.~S. Afshar, S.~Noori~Gashti, and M.~R. Alipour, {\it
  {Thermodynamic topology of black holes from bulk-boundary, extended, and
  restricted phase space perspectives}},  {\em Annals Phys.} {\bf 460} (2024)
  169569, [\href{http://xxx.arxiv.org/abs/2312.04325}{{\tt arXiv:2312.04325}}].

\bibitem{Chen:2023pqk}
D.~Chen, Y.~He, J.~Tao, and W.~Yang, {\it {Topology of
  Ho\v{r}ava\textendash{}Lifshitz black holes in different ensembles}},  {\em
  Eur. Phys. J. C} {\bf 84} (2024), no.~1 96,
  [\href{http://xxx.arxiv.org/abs/2311.11606}{{\tt arXiv:2311.11606}}].

\bibitem{Tong:2023kob}
C.-W. Tong, B.-H. Wang, and J.-R. Sun, {\it {Topology of black hole
  thermodynamics via R\'enyi statistics}},  {\em Eur. Phys. J. C} {\bf 84}
  (2024), no.~8 826, [\href{http://xxx.arxiv.org/abs/2310.09602}{{\tt
  arXiv:2310.09602}}].

\bibitem{Shahzad:2023cis}
M.~U. Shahzad, A.~Mehmood, S.~Sharif, and A.~\"Ovg\"un, {\it {Criticality and
  topological classes of neutral Gauss\textendash{}Bonnet AdS black holes in
  5D}},  {\em Annals Phys.} {\bf 458} (2023), no.~3 169486.

\bibitem{Chen:2023elp}
Z.-Q. Chen and S.-W. Wei, {\it {Thermodynamics, Ruppeiner geometry, and
  topology of Born-Infeld black hole in asymptotic flat spacetime}},  {\em
  Nucl. Phys. B} {\bf 996} (2023) 116369.

\bibitem{Wu:2023meo}
D.~Wu, {\it {Topological classes of thermodynamics of the four-dimensional
  static accelerating black holes}},  {\em Phys. Rev. D} {\bf 108} (2023),
  no.~8 084041, [\href{http://xxx.arxiv.org/abs/2307.02030}{{\tt
  arXiv:2307.02030}}].

\bibitem{Wang:2023qxw}
Y.-S. Wang, Z.-M. Xu, and B.~Wu, {\it {Thermodynamic phase transition and
  winding number for the third-order Lovelock black hole*}},  {\em Chin. Phys.
  C} {\bf 48} (2024), no.~9 095101,
  [\href{http://xxx.arxiv.org/abs/2307.01569}{{\tt arXiv:2307.01569}}].

\bibitem{Sadeghi:2023aii}
J.~Sadeghi, S.~Noori~Gashti, M.~R. Alipour, and M.~A.~S. Afshar, {\it {Bardeen
  black hole thermodynamics from topological perspective}},  {\em Annals Phys.}
  {\bf 455} (2023) 169391, [\href{http://xxx.arxiv.org/abs/2306.05692}{{\tt
  arXiv:2306.05692}}].

\bibitem{Wu:2023fcw}
D.~Wu, {\it {Consistent thermodynamics and topological classes for the
  four-dimensional Lorentzian charged Taub-NUT spacetimes}},  {\em Eur. Phys.
  J. C} {\bf 83} (2023), no.~7 589,
  [\href{http://xxx.arxiv.org/abs/2306.02324}{{\tt arXiv:2306.02324}}].

\bibitem{Zhang:2023tlq}
M.-Y. Zhang, H.~Chen, H.~Hassanabadi, Z.-W. Long, and H.~Yang, {\it {Topology
  of nonlinearly charged black hole chemistry via massive gravity}},  {\em Eur.
  Phys. J. C} {\bf 83} (2023), no.~8 773,
  [\href{http://xxx.arxiv.org/abs/2305.15674}{{\tt arXiv:2305.15674}}].

\bibitem{Gogoi:2023qku}
N.~J. Gogoi and P.~Phukon, {\it {Topology of thermodynamics in R-charged black
  holes}},  {\em Phys. Rev. D} {\bf 107} (2023), no.~10 106009.

\bibitem{Alipour:2023uzo}
M.~R. Alipour, M.~A.~S. Afshar, S.~Noori~Gashti, and J.~Sadeghi, {\it
  {Topological classification and black hole thermodynamics}},  {\em Phys. Dark
  Univ.} {\bf 42} (2023) 101361,
  [\href{http://xxx.arxiv.org/abs/2305.05595}{{\tt arXiv:2305.05595}}].

\bibitem{Gogoi:2023xzy}
N.~J. Gogoi and P.~Phukon, {\it {Thermodynamic topology of 4D dyonic AdS black
  holes in different ensembles}},  {\em Phys. Rev. D} {\bf 108} (2023), no.~6
  066016, [\href{http://xxx.arxiv.org/abs/2304.05695}{{\tt arXiv:2304.05695}}].

\bibitem{Li:2023men}
R.~Li and J.~Wang, {\it {Generalized free energy landscapes of charged
  Gauss-Bonnet-AdS black holes in diverse dimensions}},  {\em Phys. Rev. D}
  {\bf 108} (2023), no.~4 044057,
  [\href{http://xxx.arxiv.org/abs/2304.03425}{{\tt arXiv:2304.03425}}].

\bibitem{Du:2023wwg}
Y.~Du and X.~Zhang, {\it {Topological classes of black holes in de-Sitter
  spacetime}},  {\em Eur. Phys. J. C} {\bf 83} (2023), no.~10 927,
  [\href{http://xxx.arxiv.org/abs/2303.13105}{{\tt arXiv:2303.13105}}].

\bibitem{Du:2023nkr}
Y.~Du, H.~Li, and X.~Zhang, {\it {Topological classes of BTZ black holes}},
  {\em Symmetry} {\bf 16} (2024) 1577,
  [\href{http://xxx.arxiv.org/abs/2302.11189}{{\tt arXiv:2302.11189}}].

\bibitem{Wu:2023xpq}
D.~Wu, {\it {Classifying topology of consistent thermodynamics of the
  four-dimensional neutral Lorentzian NUT-charged spacetimes}},  {\em Eur.
  Phys. J. C} {\bf 83} (2023), no.~5 365,
  [\href{http://xxx.arxiv.org/abs/2302.01100}{{\tt arXiv:2302.01100}}].

\bibitem{Wu:2023sue}
D.~Wu and S.-Q. Wu, {\it {Topological classes of thermodynamics of rotating AdS
  black holes}},  {\em Phys. Rev. D} {\bf 107} (2023), no.~8 084002,
  [\href{http://xxx.arxiv.org/abs/2301.03002}{{\tt arXiv:2301.03002}}].

\bibitem{Ahmed:2022kyv}
M.~B. Ahmed, D.~Kubiznak, and R.~B. Mann, {\it {Vortex-antivortex pair creation
  in black hole thermodynamics}},  {\em Phys. Rev. D} {\bf 107} (2023), no.~4
  046013, [\href{http://xxx.arxiv.org/abs/2207.02147}{{\tt arXiv:2207.02147}}].

\bibitem{Wu:2022whe}
D.~Wu, {\it {Topological classes of rotating black holes}},  {\em Phys. Rev. D}
  {\bf 107} (2023), no.~2 024024,
  [\href{http://xxx.arxiv.org/abs/2211.15151}{{\tt arXiv:2211.15151}}].

\bibitem{Fang:2022rsb}
C.~Fang, J.~Jiang, and M.~Zhang, {\it {Revisiting thermodynamic topologies of
  black holes}},  {\em JHEP} {\bf 01} (2023) 102,
  [\href{http://xxx.arxiv.org/abs/2211.15534}{{\tt arXiv:2211.15534}}].

\bibitem{Fan:2022bsq}
Z.-Y. Fan, {\it {Topological interpretation for phase transitions of black
  holes}},  {\em Phys. Rev. D} {\bf 107} (2023), no.~4 044026,
  [\href{http://xxx.arxiv.org/abs/2211.12957}{{\tt arXiv:2211.12957}}].

\bibitem{Bai:2022klw}
N.-C. Bai, L.~Li, and J.~Tao, {\it {Topology of black hole thermodynamics in
  Lovelock gravity}},  {\em Phys. Rev. D} {\bf 107} (2023), no.~6 064015,
  [\href{http://xxx.arxiv.org/abs/2208.10177}{{\tt arXiv:2208.10177}}].

\bibitem{Yerra:2022coh}
P.~K. Yerra, C.~Bhamidipati, and S.~Mukherji, {\it {Topology of critical points
  and Hawking-Page transition}},  {\em Phys. Rev. D} {\bf 106} (2022), no.~6
  064059, [\href{http://xxx.arxiv.org/abs/2208.06388}{{\tt arXiv:2208.06388}}].

\bibitem{Wei:2024gfz}
S.-W. Wei, Y.-X. Liu, and R.~B. Mann, {\it {Universal topological
  classifications of black hole thermodynamics}},  {\em Phys. Rev. D} {\bf 110}
  (2024), no.~8 L081501, [\href{http://xxx.arxiv.org/abs/2409.09333}{{\tt
  arXiv:2409.09333}}].

\bibitem{Wu:2024asq}
D.~Wu, W.~Liu, S.-Q. Wu, and R.~B. Mann, {\it {Novel Topological Classes in
  Black Hole Thermodynamics}},  \href{http://xxx.arxiv.org/abs/2411.10102}{{\tt
  arXiv:2411.10102}}.

\bibitem{Traschen:1985}
J.~Traschen, {\it {Constraints on stress energy perturbations in general
  relativity}},  {\em Phys.Rev.} {\bf D31} (1985) 283.

\bibitem{SudarskyWald:1992}
D.~Sudarsky and R.~M. Wald, {\it {Extrema of mass, stationarity, and staticity,
  and solutions to the Einstein Yang-Mills equations}},  {\em Phys.Rev.} {\bf
  D46} (1992) 1453--1474.

\bibitem{TraschenFox:2004}
J.~Traschen and D.~Fox, {\it {Tension perturbations of black brane
  space-times}},  {\em Class.Quant.Grav.} {\bf 21} (2004) 289--306,
  [\href{http://xxx.arxiv.org/abs/gr-qc/0103106}{{\tt gr-qc/0103106}}].

\bibitem{Kastor:2008xb}
D.~Kastor, {\it {Komar Integrals in Higher (and Lower) Derivative Gravity}},
  {\em Class.Quant.Grav.} {\bf 25} (2008) 175007,
  [\href{http://xxx.arxiv.org/abs/0804.1832}{{\tt arXiv:0804.1832}}].

\bibitem{Wu:2016auq}
S.-F. Wu, X.-H. Ge, and Y.-X. Liu, {\it {First law of black hole mechanics in
  variable background fields}},
  \href{http://xxx.arxiv.org/abs/1602.08661}{{\tt arXiv:1602.08661}}.

\bibitem{Bueno:2019ycr}
P.~Bueno, P.~A. Cano, and R.~A. Hennigar, {\it {(Generalized) quasi-topological
  gravities at all orders}},  {\em Class. Quant. Grav.} {\bf 37} (2020), no.~1
  015002, [\href{http://xxx.arxiv.org/abs/1909.07983}{{\tt arXiv:1909.07983}}].

\bibitem{Bueno:2016xff}
P.~Bueno and P.~A. Cano, {\it {Einsteinian cubic gravity}},  {\em Phys. Rev. D}
  {\bf 94} (2016), no.~10 104005,
  [\href{http://xxx.arxiv.org/abs/1607.06463}{{\tt arXiv:1607.06463}}].

\bibitem{Bueno:2016lrh}
P.~Bueno and P.~A. Cano, {\it {Four-dimensional black holes in Einsteinian
  cubic gravity}},  {\em Phys. Rev. D} {\bf 94} (2016), no.~12 124051,
  [\href{http://xxx.arxiv.org/abs/1610.08019}{{\tt arXiv:1610.08019}}].

\bibitem{Ahmed:2017jod}
J.~Ahmed, R.~A. Hennigar, R.~B. Mann, and M.~Mir, {\it {Quintessential Quartic
  Quasi-topological Quartet}},  {\em JHEP} {\bf 05} (2017) 134,
  [\href{http://xxx.arxiv.org/abs/1703.11007}{{\tt arXiv:1703.11007}}].

\bibitem{Myers:2010jv}
R.~C. Myers, M.~F. Paulos, and A.~Sinha, {\it {Holographic studies of
  quasi-topological gravity}},  {\em JHEP} {\bf 08} (2010) 035,
  [\href{http://xxx.arxiv.org/abs/1004.2055}{{\tt arXiv:1004.2055}}].

\bibitem{Mir:2019ecg}
M.~Mir, R.~A. Hennigar, J.~Ahmed, and R.~B. Mann, {\it {Black hole chemistry
  and holography in generalized quasi-topological gravity}},  {\em JHEP} {\bf
  08} (2019) 068, [\href{http://xxx.arxiv.org/abs/1902.02005}{{\tt
  arXiv:1902.02005}}].

\bibitem{Mir:2019rik}
M.~Mir and R.~B. Mann, {\it {On generalized quasi-topological cubic-quartic
  gravity: thermodynamics and holography}},  {\em JHEP} {\bf 07} (2019) 012,
  [\href{http://xxx.arxiv.org/abs/1902.10906}{{\tt arXiv:1902.10906}}].

\bibitem{Hennigar:2018hza}
R.~A. Hennigar, M.~B.~J. Poshteh, and R.~B. Mann, {\it {Shadows, Signals, and
  Stability in Einsteinian Cubic Gravity}},  {\em Phys. Rev. D} {\bf 97}
  (2018), no.~6 064041, [\href{http://xxx.arxiv.org/abs/1801.03223}{{\tt
  arXiv:1801.03223}}].

\bibitem{Poshteh:2018wqy}
M.~B.~J. Poshteh and R.~B. Mann, {\it {Gravitational Lensing by Black Holes in
  Einsteinian Cubic Gravity}},  {\em Phys. Rev. D} {\bf 99} (2019), no.~2
  024035, [\href{http://xxx.arxiv.org/abs/1810.10657}{{\tt arXiv:1810.10657}}].

\bibitem{Khodabakhshi:2020hny}
H.~Khodabakhshi, A.~Giaimo, and R.~B. Mann, {\it {Einstein Quartic Gravity:
  Shadows, Signals, and Stability}},  {\em Phys. Rev. D} {\bf 102} (2020),
  no.~4 044038, [\href{http://xxx.arxiv.org/abs/2006.02237}{{\tt
  arXiv:2006.02237}}].

\bibitem{Khodabakhshi:2020ddv}
H.~Khodabakhshi and R.~B. Mann, {\it {Gravitational Lensing by Black Holes in
  Einstein Quartic Gravity}},  {\em Phys. Rev. D} {\bf 103} (2021), no.~2
  024017, [\href{http://xxx.arxiv.org/abs/2007.05341}{{\tt arXiv:2007.05341}}].

\bibitem{Lovelock:1971yv}
D.~Lovelock, {\it {The Einstein tensor and its generalizations}},  {\em J.
  Math. Phys.} {\bf 12} (1971) 498--501.

\bibitem{Zanelli:2005sa}
J.~Zanelli, {\it {Lecture notes on Chern-Simons (super-)gravities. Second
  edition (February 2008)}},  in {\em {Proceedings, 7th Mexican Workshop on
  Particles and Fields (MWPF 1999)}}, 2005.
\newblock \href{http://xxx.arxiv.org/abs/hep-th/0502193}{{\tt hep-th/0502193}}.

\bibitem{Jacobson:1993xs}
T.~Jacobson and R.~C. Myers, {\it {Black hole entropy and higher curvature
  interactions}},  {\em Phys. Rev. Lett.} {\bf 70} (1993) 3684--3687,
  [\href{http://xxx.arxiv.org/abs/hep-th/9305016}{{\tt hep-th/9305016}}].

\bibitem{Liberati:2015xcp}
S.~Liberati and C.~Pacilio, {\it {Smarr Formula for Lovelock Black Holes: a
  Lagrangian approach}},  {\em Phys. Rev.} {\bf D93} (2016), no.~8 084044,
  [\href{http://xxx.arxiv.org/abs/1511.05446}{{\tt arXiv:1511.05446}}].

\end{thebibliography}\endgroup

\end{document}